\documentclass[12pt,twoside,fleqn]{book}
\usepackage[isolatin]{inputenc}
\usepackage{epsfig}
\usepackage{fancyhdr} 
\usepackage{amsfonts}
\usepackage{amsmath}
\usepackage{here}
\usepackage{float}
\usepackage{afterpage}
\usepackage{bm}
\usepackage[font=sf,labelfont=bf,parskip=3pt,width=.97\textwidth]{caption}
\usepackage[hang,sf,SF]{subfigure}
\usepackage{cellspace}
\usepackage{longtable}
\usepackage[ruled,vlined]{algorithm2e}

\makeindex

\makeatletter
\def\thickhrulefill{\leavevmode \leaders \hrule height 1ex \hfill \kern \z@}
\def\@makechapterhead#1{%
  \vspace*{6\p@}%
  {\parindent \z@ \raggedleft \reset@font
            \scshape \@chapapp{} \thechapter
        \par\nobreak
        \interlinepenalty\@M
    \Huge \bfseries #1\par\nobreak
    \hrulefill
    \par\nobreak
    \vskip 30\p@
  }}
\def\@makeschapterhead#1{%
  \vspace*{6\p@}%
  {\parindent \z@ \raggedleft \reset@font
            \scshape \vphantom{\@chapapp{} \thechapter}
        \par\nobreak
        \interlinepenalty\@M
    \Huge \bfseries #1\par\nobreak
    \hrulefill
    \par\nobreak
    \vskip 30\p@
  }}

\begin{document}

\newcommand{\rr}[1]{\mathbb{#1}}
\newcommand{\R}{\rr{R}}
\newcommand{\N}{\rr{N}}
\newcommand{\C}{\rr{C}}
\newcommand{\PP}{\rr{P}}
\newcommand{\bfxis}{\mbox{\scriptsize\boldmath$\xi$}}
\newcommand{\bfxi}{\mbox{\boldmath$\xi$}}

\def \chaptername {Chapter}
\def \figurename {Figure}
\def \bibname {References}
\def \appendixname {Appendix}
\newtheorem{lemma}{Lemma}[chapter]
\newtheorem{deff}{Definition}[chapter]
\newtheorem{theorem}{Theorem}[chapter]
\newtheorem{remark}{Remark}[chapter]
\newtheorem{example}{Example}[chapter]
\def \contentsname {Contents}


\pagestyle{empty}
\begin{titlepage}

\begin{center}
\phantom{t}
{\Huge \bf 
 Inverse Problems in Classical \\ \vspace{0.5cm} and Quantum Physics } \\
\vspace{5cm}
{\bf \large Dissertation zur Erlangung des Grades\\
  \vspace{-0.1cm}
  ,,Doktor der Naturwissenschaften''\\
  \vspace{-0.1cm}
  am Fachbereich Physik\\
der Johannes Gutenberg-Universit\"at in Mainz} \\
\vspace{4cm}
 {\bf \Large Andrea Amalia Almasy}\\
 {\small geb. in Sighetu Marma\c tiei, Rum\"anien}
\end{center}
\vfill
\begin{center}
{\large \bf Mainz, den 29. Juni 2007}
\end{center}
\end{titlepage}

\newpage
\vspace*{16cm}
\noindent Datum der m\"undlische Pr\"ufung: 29. Juni 2007\\
D77 (Diss. Universit\"at Mainz)

\clearpage{\thispagestyle{empty}\cleardoublepage}
\vspace*{3cm}
\hfill{\sf To my family}

\vspace*{13cm}
\hfill
\begin{tabular}{r}
'{\em Far better an approximate answer to the right question,}\\
{\em which is often vague,}\\
{\em than an exact answer to the wrong question,}\\
{\em which can always be made precise.}'\\
\\
\hfill John W. Tukey
\end{tabular}
\clearpage{\thispagestyle{empty}\cleardoublepage}

\section*{Acknowledgements}

\vspace{3cm}
First of all, I would like to express my deep gratitude to my supervisors Prof.\ Dr.\ Karl Schilcher and PD Dr.\ Hubert Spiesberger for their support and encouragement. Especially I would like to thank PD Dr.\ Hubert Spiesberger for excellent guidance and fruitfull discussions during this work. I also thank him for carefully reading my manuscript and giving valuable suggestions.

I am very grateful to the theoretical physics working group ThEP for the hospitality. For that I would like to thank Mrs.\ Monique Engler and PD Dr.\ Hubert Spiesberger for all their support and help in solving administrative problems. I would also like to thank Prof.\ Dr.\ Florian Scheck, Prof.\ Dr.\ Martin Reuter, Prof.\ Dr.\ Nikolaos Papadopoulos, Prof.\ Dr.\ J\"urgen G.\ K\"orner, and my dear colleges Dr.\ Astrid Bauer, Dr.\ Isabella Bierenbaum, Dr.\ Roxana \c Schiopu, Ulrich Seul, Jan-Eric Daum and other members of the group. Also, I will never forget my friends Dr.\ Markus Knodel, Eric Tuiran Otero and Dr.\ Alimjan Kadeer. We have spent together a pleasent time not only for science discussions but also in private {\em unphysical} life.

My love and appreciation go to my husband Sorin Tanase and to my parents Maria Almasy and \c Stefan Almasy. I wish to thank them for everything.

Finally, I would like to thank the Graduiertenkolleg ``Eichtheorien -- Experimentelle Tests und theoretische Grundlagen'' for the financial support during my Ph.D.

\vspace{2cm}

Mainz, June 2007 \hfill Andrea Almasy

\clearpage{\thispagestyle{empty}\cleardoublepage}

\section*{Notations and symbols}

\begin{center} 
\begin{longtable}{ll}
\begin{minipage}{4cm}$\in$\end{minipage} & \begin{minipage}[t]{10cm}Element of \end{minipage} \\[0.3cm]

\begin{minipage}{4cm}$\subset$\end{minipage} & \begin{minipage}[t]{10cm}Subset of\end{minipage} \\[0.3cm]

\begin{minipage}{4cm}$\equiv$\end{minipage} & \begin{minipage}[t]{10cm}Identiaclly equals\end{minipage} \\[0.3cm]

\begin{minipage}{4cm}$\sim$\end{minipage} & \begin{minipage}[t]{10cm}Essentially equal to or equivalent\end{minipage} \\[0.3cm]

\begin{minipage}{4cm}$\dagger$\end{minipage} & \begin{minipage}[t]{10cm}Hermitian conjugate\end{minipage} \\[0.3cm]

\begin{minipage}{4cm}$f^+$, $A^+$\end{minipage} & \begin{minipage}[t]{10cm}Generalised solution and generalised inverse respectively\end{minipage} \\[0.3cm]

\begin{minipage}{4cm}$f'$\end{minipage} & \begin{minipage}[t]{10cm}First derivative of $f$\end{minipage} \\[0.3cm]

\begin{minipage}{4cm}$\langle f\rangle$\end{minipage} & \begin{minipage}[t]{10cm}Mean value of $f$\end{minipage} \\[0.3cm]

\begin{minipage}{4cm}${\bm x}=(x^j)$\end{minipage} & \begin{minipage}[t]{10cm}Vector in Euclidean $n$-dimensional space $\R^n$. The components are labelled by latin letters ($j=1,2,3$) \end{minipage}\\[0.8cm]

\begin{minipage}{4cm}$\hat{\bm x}=\displaystyle\frac{\bm x}{|{\bm x}|}$\end{minipage} & \begin{minipage}[t]{10cm}Unit vector in the direction of ${\bm x}$ \end{minipage}\\[0.8cm]

\begin{minipage}{4cm}$x=(x^\mu)$\end{minipage} & \begin{minipage}[t]{10cm}4-Vector in $4$-dimensional Minkowski space. The components are labelled by greek letters ($\mu=0,1,2,3$) \end{minipage}\\[0.8cm]

\begin{minipage}{4cm}$d^nx,d{\bm x}$\end{minipage} & \begin{minipage}[t]{10cm}Volume element in Euclidean $n$-dimensional space $\R^n$ \end{minipage}\\[0.3cm]

\begin{minipage}{4cm}$d^4x=dx^0dx^1dx^2dx^3$\end{minipage} & \begin{minipage}[t]{10cm}Volume element in $4$-dimensional Minkowski space \end{minipage}\\[0.3cm]

\begin{minipage}{4cm}$I$ \end{minipage} & \begin{minipage}[t]{10cm}Unit operator or matrix \end{minipage}\\[0.3cm]

\begin{minipage}{4cm}$\overline\Omega$\end{minipage} & \begin{minipage}[t]{10cm}Closure of the domain $\Omega$ \end{minipage}\\[0.3cm]

\begin{minipage}{4cm}$\partial\Omega$\end{minipage} & \begin{minipage}[t]{10cm}Boundary of the domain $\Omega$ \end{minipage}\\[0.3cm]

\begin{minipage}{4cm}$O(E)$\end{minipage} & \begin{minipage}[t]{10cm}Terms of order $E$ \end{minipage}\\[0.3cm]

\begin{minipage}{4cm}${\cal N}(A)$\end{minipage} & \begin{minipage}[t]{10cm}Null-space of operator $A$ \end{minipage}\\[0.3cm]

\begin{minipage}{4cm}${\cal R}(A)$\end{minipage} & \begin{minipage}[t]{10cm}Range of operator $A$ \end{minipage}\\[0.3cm]

\begin{minipage}{4cm}${\cal D}(A)$\end{minipage} & \begin{minipage}[t]{10cm}Domain of operator $A$ \end{minipage}\\[0.3cm]

\begin{minipage}{4cm}$A^*$\end{minipage} & \begin{minipage}[t]{10cm}Adjoint of $A$ \end{minipage}\\[0.3cm]

\begin{minipage}{4cm}$A^T$\end{minipage} & \begin{minipage}[t]{10cm}Transposed of $A$ \end{minipage}\\[0.3cm]

\begin{minipage}{4cm}${\rm Tr}A$\end{minipage} & \begin{minipage}[t]{10cm}Trace of $A$ \end{minipage}\\[0.3cm]

\begin{minipage}{4cm}$|.|$\end{minipage} & \begin{minipage}[t]{10cm}Absolute value \end{minipage}\\[0.3cm]

\begin{minipage}{4cm}$||.||_{\cal X}$\end{minipage} & \begin{minipage}[t]{10cm}Norm defined on the space ${\cal X}$ \end{minipage}\\[0.3cm]

\begin{minipage}{4cm}${\rm Re}\, x$\end{minipage} & \begin{minipage}[t]{10cm}Real part of $x$ \end{minipage}\\[0.3cm]

\begin{minipage}{4cm}${\rm Im}\, x$\end{minipage} & \begin{minipage}[t]{10cm}Imaginary part of $x$ \end{minipage}\\[0.3cm]

\begin{minipage}{4cm}$\PP$\end{minipage} & \begin{minipage}[t]{10cm}Principal part \end{minipage}\\[0.3cm]

\begin{minipage}{4cm}$\sum$\end{minipage} & \begin{minipage}[t]{10cm}Summation \end{minipage}\\[0.3cm]

\begin{minipage}{4cm}$[{\cal O}]$\end{minipage} & \begin{minipage}[t]{10cm}Mass dimesion of operator ${\cal O}$, i.e., $[{\cal O}]=M^{d({\cal O})}$ \end{minipage}\\[0.3cm]

\begin{minipage}{4cm}${\cal F}[f](x)$\end{minipage} & \begin{minipage}[t]{10cm}Fourier transform of $f$ \end{minipage}\\[0.3cm]

\begin{minipage}{4cm}$\left(f,g\right)_{\cal X}$\end{minipage} & \begin{minipage}[t]{10cm}Scalar product defined on the space ${\cal X}$ \end{minipage}\\[0.3cm]

\begin{minipage}{4cm}$\Delta$\end{minipage} & \begin{minipage}[t]{10cm}Laplace operator \end{minipage}\\[0.3cm]

\begin{minipage}{4cm}$\bm\nabla$\end{minipage} & \begin{minipage}[t]{10cm}Nabla operator \end{minipage}\\[0.3cm]

\begin{minipage}{4cm}$\delta(x)$\end{minipage} & \begin{minipage}[t]{10cm}Dirac $\delta$-function ($-\infty<x<\infty$); $\delta(x)=0$ for $x\ne0$, $\int_{-\infty}^\infty\! dx\,\delta(x)=1$\end{minipage}\\[0.8cm]

\begin{minipage}{4cm}$\theta(x)$\end{minipage} & \begin{minipage}[t]{10cm}Heaviside-function ($-\infty<x<\infty$); $\theta(x)=0$ for $x<0$, $\theta(x)=1$ for $x>0$\end{minipage}\\[0.8cm]

\begin{minipage}{4cm}$q_\alpha^i,\bar q_\alpha^i$\end{minipage} & \begin{minipage}[t]{10cm} Quark fields. Flavours are labelled by latin letters, here $i$, and colours by greek letters, here $\alpha$ \end{minipage}\\[0.8cm]

\begin{minipage}{4cm}$G^{\mu\nu}_a$\end{minipage} & \begin{minipage}[t]{10cm} Gluon field strength tensor \end{minipage}\\[0.3cm]

\begin{minipage}{4cm}$\gamma^\mu,\gamma_5$\end{minipage} & \begin{minipage}[t]{10cm} Dirac matrices \end{minipage}\\[0.3cm]

\begin{minipage}{4cm}$\lambda^a$\end{minipage} & \begin{minipage}[t]{10cm} Gell-Mann colour matrices \end{minipage}\\[0.3cm]

\begin{minipage}{4cm}$f_{abc}$\end{minipage} & \begin{minipage}[t]{10cm} Structure constant of $SU(3)$ \end{minipage}\\[0.3cm]

\begin{minipage}{4cm}$:A\cdot B\cdot ...:$\end{minipage} & \begin{minipage}[t]{10cm}Normal ordering of the operators $A,B,...$ \end{minipage}\\[0.3cm]

\end{longtable}
\end{center}

\clearpage

\section*{Abbreviations}

\vspace{0.8cm}
\begin{center}
\begin{tabular}{ll}
\begin{minipage}{4cm}MC\end{minipage} & \begin{minipage}[t]{10cm}Monte Carlo \end{minipage}\\[0.3cm]

\begin{minipage}{4cm}LO\end{minipage} & \begin{minipage}[t]{10cm}Leading order \end{minipage}\\[0.3cm]

\begin{minipage}{4cm}NLO\end{minipage} & \begin{minipage}[t]{10cm}Next-to-leading order \end{minipage}\\[0.3cm]

\begin{minipage}{4cm}SVD\end{minipage} & \begin{minipage}[t]{10cm}Singular value decomposition \end{minipage}\\[0.3cm]

\begin{minipage}{4cm}QCD\end{minipage} & \begin{minipage}[t]{10cm}Quantum chromodynamics \end{minipage}\\[0.3cm]

\begin{minipage}{4cm}QED\end{minipage} & \begin{minipage}[t]{10cm}Quantum electrodynamics \end{minipage}\\[0.3cm]

\begin{minipage}{4cm}OPE\end{minipage} & \begin{minipage}[t]{10cm}Operator product expansion \end{minipage}\\[0.3cm]

\begin{minipage}{4cm}CKM\end{minipage} & \begin{minipage}[t]{10cm}Cabibbo-Kobayashi-Maskawa \end{minipage}\\[0.3cm]

\begin{minipage}{4cm}PCAC\end{minipage} & \begin{minipage}[t]{10cm}Partial conservation of axial-vector current \end{minipage}\\[0.3cm]

\begin{minipage}{4cm}EIT\end{minipage} & \begin{minipage}[t]{10cm}Electrical impedance tomography \end{minipage}\\[0.3cm]

\begin{minipage}{4cm}FEM\end{minipage} & \begin{minipage}[t]{10cm}Finite element method \end{minipage}\\[0.3cm]

\begin{minipage}{4cm}CL\end{minipage} & \begin{minipage}[t]{10cm}Confidence level \end{minipage}\\[0.3cm]

\begin{minipage}{4cm}LS\end{minipage} & \begin{minipage}[t]{10cm}Least-squares \end{minipage}\\[0.3cm]

\begin{minipage}{4cm}CR\end{minipage} & \begin{minipage}[t]{10cm}Confidence region\end{minipage}\\[0.3cm]

\begin{minipage}{4cm}p.d.f.\end{minipage} & \begin{minipage}[t]{10cm}probability distribution function\end{minipage}\\[0.3cm]
\end{tabular}
\end{center}

\clearpage{\thispagestyle{empty}\cleardoublepage}

\pagenumbering{arabic}
\pagestyle{plain}
\tableofcontents
\clearpage{\thispagestyle{empty}\cleardoublepage}
\listoffigures
\clearpage{\thispagestyle{empty}\cleardoublepage}
\listoftables
\clearpage{\thispagestyle{empty}\cleardoublepage}
\listofalgorithms
\clearpage{\thispagestyle{empty}\cleardoublepage}

\pagestyle{fancy}
\renewcommand{\chaptermark}[1]{\markboth{\thechapter.\ #1}{}}
\renewcommand{\sectionmark}[1]{\markright{\thesection.\ #1}}
\fancyhead{}
\fancyfoot{}
\fancyhead[LO]{\rm \rightmark}
\fancyhead[RE]{\rm \leftmark}
\fancyhead[LE,RO]{\rm \thepage}

\chapter*{Introduction}
\addcontentsline{toc}{chapter}{Introduction}

We call two problems {\em inverse} to each other if the formulation of each of them requires full or partial knowledge of the other. By this definition, it is obviously arbitrary which of the two problems we call the direct and which we call the inverse problem. But usually, one of the problems has been studied earlier and, perhaps, in more detail. This one is usually called the {\em direct} problem, whereas the other is the {\em inverse} problem. However, there is often another, more important difference between these two problems. Hadamard (see \cite{hada2}) introduced the concept of a {\em well-posed problem}, originating from the philosophy that the mathematical model of a physical problem has to have the properties of uniqueness, existence, and stability of the solution. If one of the properties fails to hold, he called the problem {\em ill-posed}. It turns out that many interesting and important inverse problems in science lead to ill-posed problems, while the corresponding direct problems are well-posed. Often, existence and uniqueness can be forced by enlarging or reducing the solution space (the space of ``models''). For restoring stability, however, one has to change the topology of the space, which is in many cases impossible because of the presence of measurement errors. At first glance, it seems to be impossible to compute the solution of a problem numerically if the solution of the problem does not depend continuously on the data, i.e., for the case of ill-posed problems. Under additional {\em a priori} information about the solution, such as smoothness and bounds on the derivatives, however, it is possible to restore stability and construct efficient numerical algorithms.

The thesis contains three main parts. The aim of the first part is to introduce the basic notations and difficulties encountered with ill-posed problems and then study the basic properties of regularisation methods for linear ill-posed problems. In the second and third part we aim to find stable solutions to two inverse problems arising in different fields of physics. 

The first is an inverse problem of quantum chromodynamics (QCD). QCD is widely considered to be a good candidate for a theory of the strong interactions. Asymptotic freedom allows us to perform a perturbative treatment of strong interactions at short distances. Long distance behaviour is not fully understood: it is commonly believed that, due to the nontrivial structure of the physical vacuum, the perturbation expansion does not completely define the theory and that one has to add non-perturbative effects as well. In order to make a comparison with experiment possible, even in the resonance energy range, Shifman, Vainshtein and Zakharov \cite{shifman1} have proposed to use the Operator Product Expansion (OPE) and to introduce the vacuum expectation values of the operators occurring in the OPE, the so called {\em condensates}, as phenomenological parameters. The knowledge of these parameters is useful for studying if one can indeed obtain a consistent  description of the low energy hadronic physics and get more insight into the properties of the QCD vacuum. It is therefore important not only to determine the values of the condensates from experimental data, but also the accuracy of the determination, i.e. to determine their allowed range.

In QCD, perturbation expansions, together with some non-perturbative effects, allows us to approximate the two point functions of hadronic currents, i.e., vacuum expectation values of the time ordered product of two hadronic currents, in the distant space-like region in terms of a few parameters (the values of condensates). On the other hand, the discontinuity of these amplitudes in the time-like region is related to more directly measurable quantities. Although analyticity strongly correlates the values of the amplitudes in these two regions, the errors affecting both types of {\em data} make the correlation much looser. Any procedure aimed to build up acceptable amplitudes must take into account these errors in a reasonable way.

There are several methods, generically called {\em QCD sum rules}, dealing with this problem of analytic extrapolation \cite{bell,bertlmann,shifman1}. Most of them include the {\em theoretical errors} in the space-like region only at a qualitative level, and/or need (explicit and implicit) assumptions on the derivatives of the amplitudes. The application of fully controlled analytic extrapolation techniques should remedy these effects. There are a few methods of this sort, in which the error channels in the space-like region are defined through $L^2$-norms \cite{almasy,ciulli} or $L^\infty$-norms \cite{auberson,auberson2,causse}.

The functional method we use \cite{almasy} allows us to extract within rather general assumptions the condensates from a comparison of the time-like experimental data with the asymptotic space-like results from theory. We will see that the price to be paid for the generality of assumptions is relatively large errors in the values of the extracted parameters. Although we do not claim that our method is superior to other approaches, we hope that our results lend additional confidence to the numerical results obtained with the help of methods based on QCD sum rules \cite{bell,bell1,bell2,bertlmann1,bertlmann,dominguez0,ioffe2,kremer,launer,narison8,narison3,narison9,narison7,narison6,narison2,reinders,shifman1,yndurain}.  

For the experimental data on the time-like region, one has more possibilities at hand. We have chosen to work with the final $\tau$ data provided by the ALEPH collaboration \cite{aleph05}, because they have the smallest experimental errors. One may also ask why $\tau$-decays and not $e^+e^-$-annihilation? The answer lies in the fact that the physics of hadronic $\tau$-decays has been the subject of much progress in the last decade, both at the experimental and theoretical level. Somewhat unexpectedly, hadronic $\tau$-decays provide one of the most powerful testing grounds for QCD. This situation results from a number of favourable conditions:
\begin{itemize}
 \item The $\tau$ lepton is heavy enough to decay into a variety of hadrons, with net strangeness 0 and $\pm1$;
 \item $\tau$ leptons are copiously produced in pairs at $e^+e^-$ colliders, leading to simple event topologies with little background;
 \item Experimental studies of $\tau$-decays could be performed with large data samples;
 \item As a consequence, $\tau$-decays are experimentally known to great detail and their rates measured with high precision;
 \item The theoretical description of hadronic $\tau$-decays is based on solid ground.
\end{itemize}

The second inverse problem we aimed to solve is in the field of Electrical Impedance Tomography (EIT). EIT is a technology developed to image the electrical conductivity distribution of a conductive medium. It is of interest because of its low cost and also because the electrical conductivity gives direct information about the internal composition of the conductive medium. The technique works by performing simultaneous measurements of direct or alternating electric currents and voltages on the boundary of an object. These are the data used by an image reconstruction algorithm to determine the electrical conductivity distribution within the object. This problem is also called the inverse conductivity problem and has applications as a method of industrial, geophysical and medical imaging.

It is well known that the inverse conductivity problem is a highly  ill-posed, non-linear inverse problem and that the images produced are very sensitive to errors which can occur in practice. There has been much interest in determining the class of conductivity distributions that can be recovered from the boundary data, as well as in the development of related reconstruction algorithms. The interest in this problem has been generated by both difficult theoretical challenges and by the important medical, geophysical and industrial application of it. Much theoretical work has been related to the approach of Calder\'on concerning the bijection between the the conductivity inside the region and the Neumann-to-Dirichlet operator, which relates the distribution of the injected currents to the boundary values of the induced electrical potential \cite{cald1,kohn2,nachman,sylv1}. The reconstruction procedures that have been proposed include a wide range of iterative methods based on formulating the inverse problem as a nonlinear optimisation problem. These techniques are quite demanding computationally, particularly when addressing the three-dimensional problem. This drawback has encouraged the search for reconstruction algorithms which reduce the computational demands either by using some {\em a priori} information e.g. \cite{buhl1,buhl2,cedi} or by developing non-iterative procedures. Some of these methods \cite{azzouz,buhl1,buhl2} use a factorisation approach while others are based on reformulating the inverse problem in terms of integral equations \cite{ciulli1,ciulli2,sebu1}.

One of the approaches presented here is based on reformulating the inverse problem in terms of integral equations. A feature of this method is that many of the calculations involve analytical expressions containing the eigenfunctions of the kernel of these equations, the computational part being restricted to the introduction of the data, the numerical evaluation of some of the analytic formul\ae\ and the solution of a final integral equation. The method consists in the determination of $Y(\bm x)=-\Delta\phi(\bm x)$, in the sense of a generalised solution of inverse problems, by a single measurement of the potential $\phi(\bm x)$, and its normal derivative on the boundary of the domain. The result $Y(\bm x)$ can either be used directly to obtain rough information on the conductivity or may be processed further to determine the conductivity by solving a first order partial differential equation. This can be done in a straightforward way by using the method of characteristics. We have applied this method for a two-dimensional domain, the unit disc, with no a priori information. However, since the problem is ill-posed, one needs to regularise it, i.e., search for approximate solutions satisfying additional constraints suggested by the physics of the problem. In our case we have used as a regularisation algorithm the truncated singular value decomposition. Unfortunately, the information gained on $Y(\bm x)$, and the conductivity respectively, is restricted to their angular dependence (no radial information is present). One can hope that using some a priori information could improve the reconstruction. 

Also an algorithm based on linearisation will be discussed. The aim of the algorithm is to perform reconstructions based on real data measured by the tomograph constructed in collaboration with Dr. K.H. Georgi and N. Schuster. For this, a belt of electrodes was placed around the chest of a human volunteer, voltages applied and currents measured. An important result of this algorithm is that the low conductivity (lungs) and high conductivity (heart) regions are well reconstructed. This is important for monitoring for lung problems such as accumulating fluid or a collapsed lung.  

The EIT research is still active. There are about 30 groups worldwide who are actively performing research and it is still seen as an exciting area of medical physics. However, EIT has not yet made the transition from an exciting medical physics discipline into wide spread routine clinical use. The technique still needs to break into widespread clinical acceptance and effort is continuing actively into the clinical trials and pilot studies which will achieve this. Technical advantages may allow us to obtain more accurate tissue characterisation and image quality and this will undoubtedly help to advance clinical acceptance.

\section*{Outline of the thesis}

As already stated, the first part of the thesis is an introduction to the basic concepts of inverse and ill-posed problems. The aim of the thesis is not to study inverse problems in general so that this part is just a general introduction collected from literature \cite{baumeister,bertero,bertero1,lectures422,louis}. In the first chapter one can find the definitions of inverse and ill-posed problems, together with a few examples. Chapter 2 presents some regularisation methods of ill-posed problems. First, the generalised solution of inverse problems is presented and then Tikhonov's regularisation method and the truncated singular value decomposition are considered.  

The second part of the thesis is concerned with an inverse problem in QCD: the determination of QCD condensates from $\tau$-decay data. Here, we will start in Chapter 3 with a theoretical description of $\tau$-decays: leptonic and hadronic decay width, hadronic polarisation function and its operator product expansion, and the dispersion relation satisfied by the polarisation function. This chapter represents the theory underlying the method we'll use to extract values for the condensates. The experimental description of $\tau$-decays is disscussed in Chapter 4. In Chapter 5 we present a functional approach which allows us to extract within rather general assumptions values for the condensates from a comparison of the time-like $\tau$-decay experimental data measured by the ALEPH Collaboration at LEP, with the asymptotic space-like QCD prediction. Results for condensates of dimesion $d=4,6,8$ and correlations between them for the $V-A$ and $V, A, V+A$ channels are presented in Chapters 6 and 7 respectively.    

The third part of the thesis is concerned with the inverse conductivity problem. In Chapter 8 we present the technique of EIT and its practical applications in medicine, industry and geophysics as well as a history of the problem. In Chapter 9 we formulate the forward problem and aim to solve it by means of finite elements. Two reconstruction algorithms are discussed in Chapter 10. The first is based on reformulating the problem in terms of integral equations and aim to perform the reconstruction from a single set of measurements, while the latter is a linearisation type of algorithm which uses more sets of measurements. The second algorithm was also used on real data in Chapter 11, where reconstructions of a phantom immersed in a test tank filled with a conducting liquid are presented. Also measurements on the chest of a human object were taken and reconstructions performed with the aim of monitoring the lungs.

There are also 4 appendices which contain supplementary mathematical informations. In Appendix A we present the basic concepts of the theory of linear integral equations. In Appendix B we give the definition of Green's functions and we illustrate how they can be used to reduce the differential equations to integral ones. Appendix C presents the singular value decomposition method for integral operators. And last, in Appendix D, some basic concepts of statistics are presented: least-squares parameter estimation, confidence regions and the $\chi^2$ test of goodness-of-fit.

\clearpage{\thispagestyle{empty}\cleardoublepage}

\thispagestyle{empty}

\begin{center}
\vspace*{9cm}
{\Huge\bf Inverse problems}

\addcontentsline{toc}{part}{Inverse problems}

\end{center}

\clearpage{\thispagestyle{empty}\cleardoublepage}

\chapter {Inverse and ill-posed problems}

Inverse problems of mathematical physics may be broadly described as problems of determining the internal structure or past state of a system from indirect measurements. Such problems would include for example the determination of diffusivities, conductivities, densities, sources, geometry of scatterers and absorbers and prior temperature distributions, to name just a few typical applications. Only recently a systematic treatment of such problems begun to emerge. The past few decades have witnessed a remarkable growth in inverse problems \cite{sabatier}.

Inverse problems most often do not fulfil Hadamard's principle of well-posedness: they might not have a solution in the strict sense, solutions might not be unique and/or might not depend continuously on data. Hence their mathematical analysis is subtle. The belief of Hadamard that problems motivated by physical reality should be well-posed is essentially generated by physics of the nineteenth century. The requirements of existence, uniqueness and continuity of the solution are deeply inherent in the idea of a unique, complete and stable determination of the physical events. As a consequence of this point of view, ill-posed problems were considered, for many years, as mathematical anomalies and were not seriously investigated. The discovery of the ill-posedness of inverse problems has completely modified this conception.

\section{Inverse problems}

\begin{figure}[h]
\centering
\includegraphics[height=.30\textwidth,angle=0]{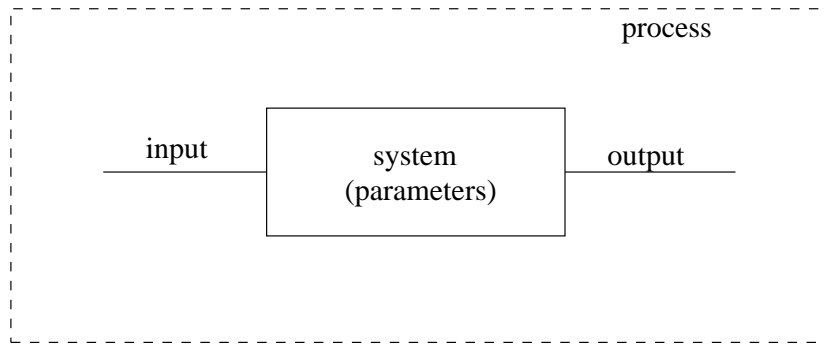}
\caption{Mathematical model of a physical process.}
\label{mathmodel}
\end{figure}

Suppose that we have a mathematical model of a physical process. We assume that this model gives a description of the system behind the process and its operating conditions and explains the principal quantities of the model (see Fig.\ref{mathmodel}): input, system parameters, output.

In most cases the description of the system is given in terms of a set of equations (ordinary and/or partial differential equations, integral equations,...), containing certain parameters.

The analysis of a given physical process via the mathematical model may be separated into three distinct types of problems.
\begin{itemize}
 \item[(A)] The {\em direct problem}: Given the input and the system parameters, find out the output of the model.
 \item[(B)] The {\em reconstruction problem}: Given the system parameters and the output, find out which input has led to this output.
 \item[(C)] The {\em identification problem}: Given the input and the output, determine the system parameters which are in agreement with the relation between input and output.
\end{itemize}
We call a problem of type (A) a {\em direct} (or {\em forward}) problem since it is oriented along a cause-effect sequence. In this sense problems of type (B) and (C) are called {\em inverse} problems because they are problems of finding out unknown causes of known consequences. It is immediately clear that the solution of one of the problems above involves a treatment also of the other problems.

We give a mathematical description of the input, the output and the system in functional analytic terms.
\begin{itemize}
 \item[] ${\cal X}$   \ \ \ \ space of input quantities;
 \item[] ${\cal Y}$   \ \ \ \ space of output quantities;
 \item[] ${\cal P}$   \ \ \ \ space of system parameters;
 \item[] $A(p)$   operator from ${\cal X}$ into ${\cal Y}$ associated to $p\in{\cal P}$.
\end{itemize}
In these terms we may reformulate the problems above in the following way:
\begin{itemize}
 \item[(A)] Given $f\in{\cal X}$ and $p\in{\cal P}$, find $g=A(p)f$.
 \item[(B)] Given $g\in{\cal Y}$ and $p\in{\cal P}$, solve the equation
\begin{equation}
 Af=g\ \ \ (f\in{\cal X})
\label{inversproblem}
\end{equation}
where $A=A(p)$.
 \item[(C)] Given $g\in{\cal Y}$ and $f\in{\cal X}$, find $p\in{\cal P}$ such that 
\begin{equation}
 A(p)f=g.
\end{equation}

\end{itemize}
At first glance, the direct problem seems to be solved much more easier than the inverse problems. However, for the computation of $g=A(p)f$ it may be necessary to solve a differential or integral equation, a task which may be of the same complexity as the solution of the equations in the inverse problem.

In certain simple examples inverse problems can be converted formally into a direct problem. For example, if $A$ has a known inverse then the reconstruction problem is solved by $f=A^{-1}g$. However, the explicit determination of the inverse does not help if the output $y$ is not in the domain of definition of $A^{-1}$. This situation is typical in applications due to the fact that the output may be only imprecisely known and/or distorted by noise.

As we stated above, a direct problem is a problem oriented along a cause-effect sequence; it is also very often a problem directed towards a loss of information: its solution defines a transition from a physical quantity with a certain information content to another quantity with smaller information content. In general it implies that the 
solution is much smoother than the data: the image provided by a bandlimited system is smoother than the corresponding object, the scattered wave due to an obstacle is smooth even if the obstacle is rough, and so on. A more rigorous description of the loss of information typical for direct problems can be found in Section \ref{illposedprob}.

The conceptual difficulty common to most inverse problems is that by solving these problems, we would like to accomplish a transformation which should correspond to a gain of information. This provides the explanation of a typical mathematical property of inverse problems which is known as ill-posedness.
 
\section{Some examples of inverse problems}

Inverse problems fall mainly into three different but intimately related categories:
\begin{itemize}
 \item Inverse scaterring problems,
\item Inverse boundary value problems,
\item Inverse spectral problems.
\end{itemize}
In the early 90's, the active research carried out in the field of inverse problems has brought a lot of new insight into the deeper nature of these problems and especially to the interrelation between them. In the following, we describe and discuss briefly the main features of each of these classes, give examples and further references.

\subsubsection*{Inverse scattering problems}

Inverse scattering problems form undoubtedly one of the most studied set of inverse problems. The setting is the following: Far away from the target having unknown physical properties, a wave field is sent in. It is assumed that the interaction mechanism of the wave field with the target is qualitatively known. The scattered field is measured, and from this data one attempts to reconstruct the properties of the scatterer. 

A classical example of inverse scattering problems arises in quantum mechanics. Assume that we have a scattering potential $q$ in $\R^3$. The quantum mechanical scattering with fixed energy $E=k^2$, $k>0$ ($\hbar=c=1$), is described by the Schr\"odinger equation
\begin{equation}
 (-\Delta-k^2+q({\bm x}))\psi({\bm x})=0.
\label{schrodinger}
\end{equation}
The potential $q$ should decrease fast enough as ${\bm x}$ tends to infinity. The typical assumption about the field $\psi$ is that it is a superposition of the incoming plane wave and the scattered radiation field satisfying Sommerfeld's radiation condition at infinity, i.e.,
\begin{equation}
 \psi({\bm x})=e^{ik{\bm\theta}\cdot{\bm x}}+\psi_{\rm sc}({\bm x}),
\end{equation}
where
\begin{equation}
 \lim_{|{\bm x}|\rightarrow\infty}|{\bm x}|(\hat{\bm x}\cdot\bm\nabla-ik)\psi_{\rm sc}({\bm x})=0.
\label{radiationcond}
\end{equation}
An equivalent way of formulating the radiation condition (\ref{radiationcond}) is to assume that
\begin{equation}
 \psi_{\rm sc}({\bm x})=\frac{e^{ik|{\bm x}|}}{|{\bm x}|}A(\hat{\bm x},{\bm\theta},k)+O\left(\frac{1}{|{\bm x}|^2}\right).
\end{equation}
The function $A(\hat{\bm x},\bm\theta,k)$ is called the scattering amplitude, and it is related to the scattering potential and the scattered field through
\begin{equation}
 A(\hat{\bm x},\bm\theta,k)=-\frac{1}{4\pi}\int_{\R^3}\ e^{-ik\hat{\bm x}\cdot\bm y}q(\bm y)\psi(\bm y)d^3y.
\end{equation}
Depending on the type of measurements, one can now pose different inverse scattering problems, of which we list the following:
\begin{itemize}
 \item Reconstruct the potential from the knowledge of the scattering amplitude at any energy
\begin{equation}
 \{A(\hat{\bm x},\bm\theta,k)\ |\ \hat{\bm x},\bm\theta\in S^2,\ k\in\R_+\};
\end{equation}
 \item Reconstruct the potential from the knowledge of the scattering amplitude at fixed energy
\begin{equation}
 \{A(\hat{\bm x},\bm\theta,k)\ |\ \hat{\bm x},\bm\theta\in S^2\},\ k>0\ \mbox{fixed};
\end{equation}
 \item Reconstruct the potential from the knowledge of the backscattering amplitude at any energy
\begin{equation}
 \{A(-\bm\theta,\bm\theta,k)\ |\ \bm\theta\in S^2,\ k\in\R_+\}.
\end{equation}

\end{itemize}

The first one of the above inverse problems is the most classical one. It is formally over-determined in the sense that the data set is indexed over a five dimensional space $S^2\times S^2\times\R_+$, while the unknown function $q$ is over a three-dimensional space. Based on this over-determinacy, there is a rather simple way of seeing the uniqueness of the solution to this problem. Indeed, one can show that the scattering solution $\psi$ to (\ref{schrodinger}) behaves as
\begin{equation}
 \psi(\bm x)=e^{ik\bm\theta\cdot\bm x}+{\cal O}\left(\frac{1}{k}\right), \ \mbox{ as } k\rightarrow\infty.
\end{equation}
Therefore, if we choose $\bm\xi\in\R^3$ and let $k$ tend to infinity while keeping the vector $k(\bm\theta-\bm x)=\bm\xi$ fixed, we find that
\begin{equation}
 \lim_{{k\rightarrow\infty}\atop{k(\bm\theta-\bm x)=\bm\xi}}A(\hat{\bm x},\bm\theta,k)={\cal F}[q](\bm\xi),
\end{equation}
i.e., the scattering amplitude tends towards the Fourier transform of the potential. Therefore, the data of the problem determine the potential $q$ uniquely.

For the second problem posted, the over-determinacy of the data is one dimension less and consequently the problem to show the uniqueness of the solution is more difficult. As for the third one (called the inverse backscattering problem) we may mention that there are still open problems related to the uniqueness of the solution and reconstruction of the potential.

Another classical, very important and closely related type of inverse scattering problems deals with obstacle scattering. A typical inverse obstacle scattering can be formulated as follows: Assume that in $\R^3$, there is an obstacle $\Omega$, whose shape one tries to recover from far field measurements. Assume that the medium outside the obstacle is governed by the equations of linear acoustics, i.e., the pressure field $u$ satisfies the Helmholtz equation
\begin{equation}
 \Delta u(\bm x)+k^2u(\bm x)=0, \ \ \bm x\in\R^3/\bar\Omega.
\end{equation}
The pressure field is assumed to satisfy a boundary condition at $\partial\Omega$, typically the Dirichlet (``soft sound''), Neumann (``hard sound'') or a mixed (``impedance'') condition. For penetrable obstacles, the appropriate boundary condition is a transmission condition. Again, one probes the target by sending in an initial field $u_0$, and the interacting total field is
\begin{equation}
 u(\bm x)=u_0(\bm x)+u_{\rm sc}(\bm x),
\end{equation}
where the scattered field satisfies the outgoing radiation condition (\ref{radiationcond}), or
\begin{equation}
 u_{\rm sc}(\bm x)=\frac{e^{ik|\bm x|}}{|\bm x|}u_\infty(\hat{\bm x})+{\cal O}\left(\frac{1}{|\bm x|^2}\right),
\end{equation}
the function $u_\infty$ being the far pattern of the field which corresponds to the scattering amplitude here. The inverse scattering problem now consists in reconstructing the shape of the object from the far field patterns generated by a set of incoming fields.

Besides the acoustic inverse obstacle scattering, one can look at the similar problem when the unknown target is illuminated with electromagnetic radiation. Since the electromagnetic fields satisfy the Helmholtz equation in vacuum, the transition from acoustics to electromagnetism does not seem so large. The boundary conditions, however, have vectorial nature and the field components will be coupled.

\subsubsection*{Inverse boundary value problems}

Let us move now to the second large area of inverse problems, the inverse boundary value problems. The change in the setting when moving from inverse scattering problems to inverse boundary value problems is by no means sharp. Again, one has an object with unknown physical parameters and the objective is to find out these parameters in a non-invasive way.

Let us start with a concrete example of impedance tomography: We ask whether it is possible to make an image of the internal electromagnetic structure of a body (e.g.\ human body) by injecting electric currents into the body and measuring the voltages needed to maintain the current. In practice, one attaches a number of electrodes on the surface of the body and measures the voltages needed to maintain the current configuration. By the linearity of the governing equations, dependence of the voltages on the currents is linear, so effectively the boundary data consists of a linear boundary map (or a matrix for the discretized version of the problem).

The governing equation for the potential $u$ in the body $\Omega$ is simply the equation of continuity for the current $\bm j=\sigma\bm\nabla u$:
\begin{equation}
 \bm\nabla\cdot\sigma\bm\nabla u(\bm x)=0,\ \bm x\in\Omega,
\label{continuityeq}
\end{equation}
where $\sigma(\bm x)>0$ is the conductivity distribution, assumed to be a scalar function of $\bm x$. The current density through the boundary of the body is
\begin{equation}
 j(\bm x)=\left.\sigma(\bm x)\frac{\partial u}{\partial n}(\bm x)\right|_{\partial\Omega}.
\label{currentdensity}
\end{equation}
Assuming that the current density $j$ is specified, one can solve the Neumann problem (\ref{continuityeq} - \ref{currentdensity}). In this way, one gets a complete collection of pairs
\begin{equation}
 \left\{(j,u|_{\partial\Omega})|\ u\ \mbox{ satisfies (\ref{continuityeq} - \ref{currentdensity})}\right\},
\end{equation}
or, equivalently, one knows the Neumann-to-Dirichlet boundary map
\begin{equation}
 \Lambda\ :\ \sigma\left.\frac{\partial u}{\partial n}\right|_{\partial\Omega}\rightarrow u|_{\partial\Omega}. 
\end{equation}
The inverse problem is to reconstruct $\sigma$ in $\Omega$ from the knowledge of $\Lambda$.

The close connection with the inverse scattering problems becomes more obvious if we modify Eq.(\ref{continuityeq}) slightly. Let us introduce the function $\psi$ defined as
\begin{equation}
 \psi(\bm x)=\sigma(\bm x)^{1/2}u(\bm x).
\end{equation}
One can show that Eq.(\ref{continuityeq}) can be rewritten for $\psi$ as
\begin{equation}
 (\Delta-q)\psi=0, \mbox{  where  } q=\frac{\Delta\sigma^{1/2}}{\sigma^{1/2}}.
\end{equation}
Thus, we are back to the Schr\"odinger equation with zero energy.

As the discussion shows, there is a strong interrelation between the inverse scattering and inverse boundary value problems. In many cases they can be shown to be equivalent: The knowledge of the boundary map determines the far field data uniquely and vice versa. 

Inverse boundary value problems get considerably more complicated if one allows for anisotropies in the medium. In fact, there are known limitations on the uniqueness for anisotropic inverse problems while the corresponding isotropic problems allow a unique solution. As an example, consider the anisotropic counterpart of Eq.(\ref{continuityeq}),
\begin{equation}
 \sum_{i,j=1}^n\frac{\partial}{\partial x_i}\sigma^{i,j}\frac{\partial}{\partial x_j}u=0, \ \ \mbox{in}\ \Omega\in\R^n,
\end{equation}
the potential $u$ satisfying the boundary condition
\begin{equation}
 \sigma^{i,j}\left.\frac{\partial u}{\partial x_j}\right|_{\partial\Omega}=g^i.
\end{equation}
These equations can be written in coordinate free form using differential forms as
\begin{equation}
 d\sigma du=0\ \ \mbox{in}\ \Omega,
\end{equation}
\begin{equation}
 \left.\left(\sigma du\right)\right|_{\partial\Omega}=g.
\end{equation}
From this formulation, it can be shown that one can transform $\sigma$ by a diffeomorphism that leaves the boundary $\partial\Omega$ untouched without affecting the boundary data. The question whether this is the only limitation on uniqueness is still open.

\subsubsection*{Inverse spectral problems}

In the inverse spectral problems the input is the spectrum of an operator and one wishes to determine an unknown parameter of the operator. A classical example, also one of the simplest inverse problems in pure mathematics, is the one formulated by Mark Kac: {\em ``Can one hear the shape of the drum?''} \cite{kac}. Mathematically, the question is formulated as follows: Let $\Omega$ be a simply connected, plane domain (the drumhead) bounded by a smooth curve $\gamma$, and consider the wave equation for $u(x,t)$ (the displacement of the drumhead) on $\Omega$ with a Dirichlet boundary condition on $\gamma$ (the drumhead is clamped at the boundary):
\begin{equation}
 \begin{array}{r}
\displaystyle \Delta u(x,t)=\frac{1}{c^2}\frac{\partial^2u}{\partial t^2}(x,t) \ \ \mbox{in }\Omega,\\
\\
u(x,t)=0\ \ \mbox{on }\gamma.
\end{array}
\end{equation}
Looking for solutions of the form $u(x,t)={\rm Re}\, e^{i\omega t}v(x)$ (normal modes) leads to an eigenvalue problem for the Dirichlet Laplacian on $\Omega$:
\begin{equation}
\begin{array}{r}
\Delta v(x)+\lambda v(x)=0 \ \ \mbox{in }\Omega,\\
v(x)=0\ \ \mbox{on }\gamma,
\end{array}
\end{equation}
where $\lambda=\omega^2/c^2$. Kac's question means the following: is it possible to distinguish ``drums'' $\Omega_1$ and $\Omega_2$ with distinct bounding curves $\gamma_1$ and $\gamma_2$, simply by ``hearing'' all of the eigenvalues of the Dirichlet Laplacian? 

Kac has showed that the asymptotic behaviour of $\lambda_k$ (the resonance frequencies) at large $k$ yields the volume and the total scalar curvature of $\Omega$ or the length of $\gamma$ \cite{kac}. This kind of inverse problem has not been given real data applications, but, in some way, it is the pillars of the so-called {\em geometric scattering theory}.

\section{Ill-posed problems} 
\label{illposedprob}

In the previous section we mentioned that a typical property of inverse problems is ill-posedness, a property which is opposite to that of well-posedness.

The basic concept of a well-posed problem was introduced by the French mathematician Jacques Hadamard in a paper published in 1902 on boundary-value problems for partial differential equations and their physical interpretation 
\cite{hada1}. In this first formulation, a problem is called well-posed when its solution is unique and exists for arbitrary data. In subsequent work Hadamard emphasises the requirement of continuous dependence of the solution 
on the data \cite{hada2}, claiming that a solution which varies considerably for a small variation of the data is not really a solution in the physical sense. Indeed, since the physical data are never known exactly, this should 
imply that the solution is not known at all.

From an analysis of several cases Hadamard concludes that only problems motivated by physical reality are well-posed. An example is provided by the initial value problem for the D'Alembert equation which is fundamental in the description of wave propagation
\begin{equation}
\frac{\partial^2u}{\partial x^2}(x,t)-\frac{1}{c^2}\frac{\partial^2u}{\partial t^2}(x,t)=0,
\end{equation}
where $c$ is the wave velocity. If we consider, for instance, the following Cauchy initial data at $t=0$
\begin{equation}
u(x,0)=f(x),\ \ \frac{\partial u}{\partial t}(x,0)=0,
\end{equation}
then there exists a unique solution given by
\begin{equation}
 u(x,t)=\frac{1}{2}\left[f(x-ct)+f(x+ct)\right].
\end{equation}
This is a solution for any continuous function $f(x)$. Moreover it is obvious that a small variation of $f(x)$ produces a small variation of $u(x,t)$.

The previous problem is well-posed and, of course, basic in the description of physical phenomena. It is an example of a direct problem. An impressive example of an ill-posed problem and, in particular, of a non-continuous 
dependence on the data, was also provided by Hadamard \cite{hada2}. This problem which, at that time, was deprived of a physical motivation, is the Laplace equation in two variables
\begin{equation}
\frac{\partial^2u}{\partial x^2}(x,y)+\frac{\partial^2u}{\partial y^2}(x,y)=0.
\end{equation}
If we consider the following Cauchy initial conditions at $y=0$
\begin{equation}
u(x,0)=\frac{1}{n}\cos(nx),\ \ \frac{\partial u}{\partial y}(x,0)=0,
\end{equation}
then the unique solution is given by
\begin{equation}
u(x,y)=\frac{1}{n}\cos(nx)\cosh(ny).
\label{solution}
\end{equation}
The factor $\cos(nx)$ produces an oscillation of the surface representing the solution of the problem. This oscillation is imperceptible near $y=0$ but becomes enormous at any given finite distance from the $x$-axis when $n$ is sufficiently large. More precisely, when $n\rightarrow\infty$, the data of the problem tend to zero but, for any finite value of $y$, the solution tends to infinity.

This is now a classical example illustrating the effects produced by a non-continuous dependence of the solution on the data. If the oscillating function describes the experimental errors affecting the data of the problem then the error propagation from the data to the solution is described by Eq.(\ref{solution}) and its effect is so dramatic that the solution corresponding to real data is deprived of physical meaning. Moreover it is also possible to show that the solution does not exist for arbitrary data but only for data with specific analyticity properties.   

A problem satisfying the requirements of existence, uniqueness and continuity is now called {\it well-posed} in the sense of Hadamard, even if the complete formulation in terms of the three requirements was first given by R. Courant \cite{cour2}. The problems which are not well-posed are called {\it ill-posed} or also {\it incorrectly posed} or {\it improperly posed}. Therefore an ill-posed problem is a problem whose solution is not unique or does not exist for arbitrary data or does not depend continuously on the data.

The previous observations and considerations can justify now the following general statement: A direct problem, i.e., a problem oriented along a cause-effect sequence, is well-posed while the corresponding inverse problem, which implies a reversal of the cause-effect sequence, is in general ill-posed. This statement, however, is meaningful only if we provide a suitable mathematical setting for the description of direct and inverse problems.

The first point is to define the class of objects to be imaged, which will be described by suitable functions with certain properties. In this class we also need a {\it distance}, in order to establish when two objects are close and when they are not. In such a way our class of objects takes the structure of a {\it metric space} of functions. We denote this space by $\cal X$ and we call it the {\it object space}. 

The second point is to solve the direct problem, i.e. to compute, for each object, the corresponding image which can be called the computed image or the {\it noise-free image}. Since the direct problem is well-posed, to each object
 we associate one, and only one, image. As we already mentioned, this image may be rather smooth as a consequence of the fact that its information content is smaller than the information content of the corresponding object. This 
property of smoothness, however, may not be true for the measured images, also called {\it noisy images}, because they correspond to some noise-free image corrupted by the noise affecting the measurement process. 

Therefore the third point is to define the class of the images in such a way that it contains both the noise-free and the noisy images. It is convenient to introduce a distance also in this class. We denote the corresponding function
 space by $\cal Y$ and we call it the {\it image space}.

\begin{figure}[ht]
\begin{center}
\epsfig{file=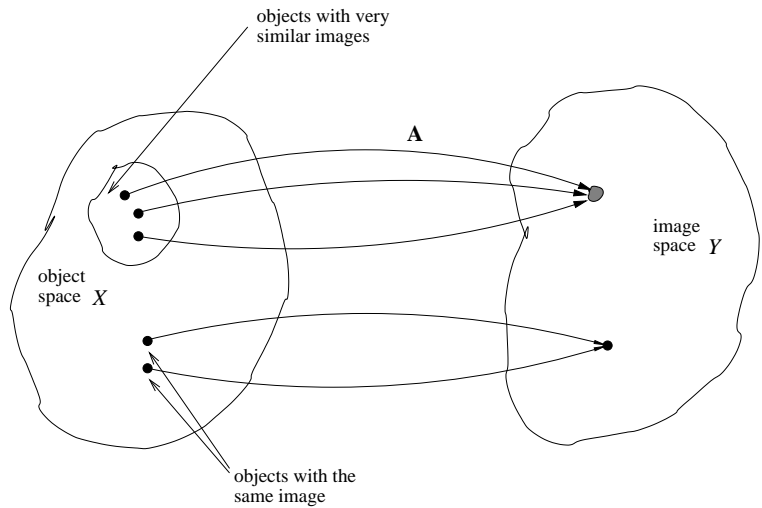,width=100mm}
\end{center}
\caption{Schematic representation of the relationship between objects and images.}
\label{operator}
\end{figure} 

In conclusion, the solution of the direct problem defines a mapping (operator), denoted by $A$, which transforms any object of the space $\cal X$ into a noise-free image of the space $\cal Y$. This operator is continuous, i.e.\ the images of two close objects are also close, because the direct problem is well-posed. The set of the noise-free images is usually called, in mathematics, the {\it range} of the operator $A$, and, as follows from our previous remark, this range does not coincide with the image space $\cal Y$ because this space contains also the noisy images.

By means of this mathematical scheme it is possible to describe the loss of information which, as we said, is typical in the solution of the direct problem. It has two consequences. First, it may be possible that two, or even more, objects have exactly the same image. In the case of a linear operator this is related to the existence of objects whose image is exactly zero. These objects will be called {\it invisible objects}. Then, given any object of the space $\cal X$, if we add to it an invisible object, we obtain a new object which has exactly the same image. Secondly, and this fact is much more general than the previous one, it may be possible that two very distant objects have images which are very close. In other words there exist very broad sets of distinct objects such that the corresponding sets of images are very small. All these properties are illustrated in Fig.\ref{operator}.

If we consider now the inverse problem, i.e.\ the problem of determining the objects corresponding to a given image, we find that this problem is ill-posed as a consequence of the loss of information intrinsic to the solution of the 
direct one. Indeed, if we have an image corresponding to two distinct objects, the solution of the inverse problem is not unique. If we have a noisy image, which is not in the range of the operator $A$, then the solution of the 
inverse problem does not exist. If we have two neighbouring images such that the corresponding objects are very distant, then the solution of the inverse problem does not depend continuously on the data.

\section{A few examples of ill-posed problems}

\subsubsection*{Fredholm integral equations of the first kind}

A Fredholm integral equation of the first kind is an equation of the form
\begin{equation}
 y(s)=\int_a^bK(s,t)x(t)dt,
\label{fredint1}
\end{equation}
where $y$ is a given function (usually called the data), $K(\cdot,\cdot)$ is the kernel of the equation and the solution $x$ is the unknown function which is sought. There are several observations concerning this equation. The first one is that the function $y$ inherits some of the smoothness of the kernel $K$ and therefore a solution may not exist if $y$ is too roughly behaved. For example, if the kernel $K$ is continuous and $x$ is integrable, then the function $y$ defined by Eq.(\ref{fredint1}) is also continuous and hence if the given function $y$ is not continuous while the kernel is, then Eq.(\ref{fredint1}) can not have an integrable solution. Consequently, the question of existence of solutions is not trivial and requires more detailed knowledge of the properties of $K$.

Another point to be considered is the uniqueness of solutions. For example, if $K(s,t)=s\sin t$, then the function $x(t)=1/2$ is a solution of 
\begin{equation}
 s=\int_0^\pi K(s,t)x(t)dt
\end{equation}
but so is each of the functions $x_n(t)=1/2+\sin(nt)$, for $n=1,2,3,...$.

A more serious concern arises from the Riemann-Lebesgue lemma which states that if $K(\cdot,\cdot)$ is any square integrable kernel, then
\begin{equation}
 \int_0^\pi K(s,t)\sin(nt)dt\rightarrow 0\ \mbox{ as } \ n\rightarrow\infty.
\end{equation}
From this it follows that if $x$ is a solution of Eq.(\ref{fredint1}) and $A$ is arbitrary, then
\begin{equation}
 \int_0^\pi K(s,t)\left(x(t)+A\sin(nt)\right)dt\rightarrow y(s) \ \mbox{ as }\ n\rightarrow\infty.
\end{equation}
Therefore for large values of $n$ the slightly perturbed data
\begin{equation}
 \tilde y(s)=y(s)+A\int_0^\pi K(s,t)\sin(nt)dt
\end{equation}
corresponds to a solution $x(t)+A\sin(nt)$ which differs markedly from $x(t)$. Hence, for Fredholm equations of the first kind, solutions generally depend discontinuously upon the data.\footnote{\sf For more details related to the solutions of the Fredholm integral equations of the first kind see Appendix A.}

\subsubsection*{Cauchy problem for the Laplace equation}

The simplest example of ill-posed problems for the Laplace equation is a mixed boundary value problem in two dimensions. The problem is to determine in a rectangle $\{0\le x\le\pi,\ 0\le y\le y_0\}$ a function of two variables $u(x,y)$ satisfying the following conditions
\begin{equation}
 \begin{array}{l}
  \Delta u(x,y)=0,\\
u(0,y)=u(\pi,y)=0,\\
u(x,0)=f_0(x),\\
u_y(x,0)=f_1(x).
 \end{array}
\label{mixed2dprob}
\end{equation}
The solution of the Laplace equation satisfying the homogeneous conditions on the edges of the strip can be represented in the form
\begin{equation}
 u(x,y)=\sum_{k=1}^\infty\left(a_ke^{ky}+b_ke^{-ky}\right)\sin kx.
\end{equation}
From the initial conditions one finds
\begin{equation}
a_k=\frac{1}{\pi}\left(\int_0^\pi f_0(x)\sin kxdx+\frac{1}{k}\int_0^\pi f_1(x)\sin kxdx\right),
\end{equation}
\begin{equation}
b_k=\frac{1}{\pi}\left(\int_0^\pi f_0(x)\sin kxdx-\frac{1}{k}\int_0^\pi f_1(x)\sin kxdx\right).
\end{equation}
Thus, the solution of the problem (\ref{mixed2dprob}) is unique but its existence is not guaranteed and depends on the integrability properties of $f_0$ and $f_1$ and on the properties which $u(x,y)$ should have.

Let us now consider the following solutions of (\ref{mixed2dprob})
\begin{equation}
 u_n(x,y)=a_ne^{ny}\sin nx.
\end{equation}
It is clear that the Cauchy data which lead to these solutions are:
\begin{equation}
 f_{0n}=a_n\sin nx,
\end{equation}
\begin{equation}
 f_{1n}=na_n\sin nx.
\end{equation}
Obviously, an appropriate choice of $n$ and $a_n$ may render these Cauchy data arbitrarily small in the norm\footnote{\sf Independent of the specific choice of the norm.}, while the function $u_n$ will be arbitrarily large for any fixed $y$.

\subsubsection*{Analytic continuation}

There are several settings of the analytic continuation problem. We present here one classical example of analytic continuation for functions of a complex variable.

Let $f(z)$ be an analytic\footnote{\sf An analytic function is an infinitely differentiable function such that the Taylor series at any point $x_0$ in its domain, $$T(x)=\sum_{n=0}^\infty\frac{f^{(n)}(x_0)}{n!}(x-x_0)^n,$$ is convergent for $x$ close enough to $x_0$ and its value equals $f(x)$.} function of a complex variable, which is regular\footnote{\sf A function is termed regular if and only if it is analytic and single-valued throughout a region $\Omega$.} within a bounded region $\Omega$ on the complex plane and continuous in the closure $\bar\Omega$,
\begin{equation}
 |f(z)|\le c,\ z\in\Omega.
\end{equation}
Let $\partial\Omega$ be the boundary of $\Omega$, $\Gamma_1$, $\Gamma_2$ be parts of $\partial\Omega$: $\Gamma_1\cup\Gamma_2=\partial\Omega$, $\Gamma_1\cap\Gamma_2=\emptyset$. Suppose also that $f(z)$ is specified on $\Gamma_1$, and the problem is to determine $f(z)$ within the interior part of $\Omega$. If $\partial\Omega=\Gamma_1$, the solution is provided by the Cauchy integral,
\begin{equation}
 f(z)=\frac{1}{2\pi i}\int_{\partial\Omega}\frac{f(\xi)}{\xi-z}d\xi.
\end{equation}
If $\Gamma_1\neq\partial\Omega$, then to find the analytic continuation of $f(z)$ is equivalent to the Cauchy problem for the Laplace equation.

Let the value of a harmonic function $u$ and its normal derivative ${\partial u}/{\partial n}$ be given on $\Gamma_1$. Denote by $f(z)$ the function
\begin{equation}
 f(z)=u(z)+iv(z),\ z=x+iy,
\end{equation}
where $v$ is the function conjugate to $u$. It is known that on $\Gamma_1$
\begin{equation}
 v(z)=\int_{z_0}^z\frac{\partial}{\partial n}u(s)ds+C_1,
\end{equation}
where $z_0$ is one of the ends of $\Gamma_1$ and $C_1$ a constant. Hence, if $u(z)$, ${\partial u(z)}/{\partial n}$ are known on $\Gamma_1$, then the analytic function $f(z)$ may be deemed given on $\Gamma_1$.

From the Cauchy-Riemann conditions it follows that
\begin{equation}
 \left.\frac{\partial u}{\partial n}\right|_{\Gamma_1}=\left.\frac{\partial v}{\partial s}\right|_{\Gamma_1},
\end{equation}
where ${\partial v}/{\partial s}$ is the derivative of $v$ along $\Gamma_1$. Taking the derivative of $f(z)$ and $\bar f(z)$ along $\Gamma_1$ yields\footnote{\sf Here, $\bar f$ denotes the complex conjugate of $f$.}
\begin{equation}
 \left.\frac{\partial u}{\partial n}\right|_{\Gamma_1}=\left.\frac{1}{2}\frac{\partial}{\partial s}(f-\bar f)\right|_{\Gamma_1}
\end{equation}
and we arrive at the Cauchy initial conditions for $u(x,y)$.

Thus, if $\Gamma_1\neq\partial\Omega$, the problem of analytic continuation is equivalent to the Cauchy problem for the Laplace equation and so is ill-posed.

\section{How to cure ill-posedness}

The property of a non-continuous dependence of the solution on the data strictly applies only to ill-posed problems formulated in infinite dimensional spaces like the ones discussed in the previous section. In practice one has discrete data and one has to solve discrete problems. These, however, are obtained by discretizing problems with very bad mathematical properties. What happens in these cases?

If we consider a linear inverse problem, its discrete version is a linear algebraic system, apparently a rather simple mathematical problem. Many methods exist for solving numerically this problem. However the solution often does not work. A description  of the first attempts of data inversions is given by S. Twomey in the preface of his book \cite{twom}: 'The crux of the difficulty was that numerical inversions were producing results which were physically 
unacceptable but were mathematically acceptable (in the sense that {\it had} they existed they should have given measured values identical or almost identical with what was measured)'. These results were 'rejected as impossible or ridiculous by the recipient of the computer's answer. And yet the computer was often blamed, even though it had done all that had been asked of it'. ...'Were it possible for computers to have ulcers or neuroses there is little 
doubt that most of those with which early numerical inversion attempts were made would have required both afflictions' \cite{twom}.

The explanation can be found having in mind the examples discussed in the previous section, where small oscillating data produce large oscillating solutions. In any inverse problem, data are always affected by noise which can be viewed as a small randomly oscillating function. Therefore the solution method amplifies the noise producing a large and wildly oscillating function which completely hides the physical solution corresponding to the noise-free data. This property holds true also for the discrete version of the ill-posed problem. Then one says that the corresponding linear algebraic system is {\it ill-conditioned}: even if the solution exists and is unique, it may be, and is in general, completely corrupted by a small error on the data. 

In conclusion, we have the following situation: we can compute one, and only one, solution of our algebraic system but this solution may be unacceptable for the reasons indicated above; the physically acceptable solution we are looking for is not a solution of the problem but only an approximate solution in the sense that it does reproduce the data not exactly but only within the experimental errors. However, if we look for approximate solutions, we find that they constitute a set which is extremely broad and contains completely different functions, a consequence of the loss of information in the direct problem. Then the question arises: how can we choose the good ones?

We can state now the 'golden rule' for solving inverse problems which are ill-posed: search for approximate solutions satisfying additional constraints coming from the physics of the problem.

The set of the approximate solutions corresponding to the same data function is just the set of objects with images close to the measured one. The set of objects is too broad, as a consequence of the loss of information due to the 
imaging process. Therefore we need some additional information to compensate this loss. This information, which is also called {\it a priori} or {\it prior} information, is additional in the sense that it cannot be derived from the image or from the proprieties of the mapping $A$ which describes the imaging process but expresses some expected physical properties of the object. Its role is to reduce the set of the objects compatible with the given image or also to discriminate between interesting objects and spurious objects, generated by uncontrolled propagation of the noise affecting the image.

The idea of using prescribed bounds to produce approximate and stable solutions was introduced by C. Pucci in the case of the Cauchy problem for the Laplace equation \cite{pucc}, i.e., the first example of an ill-posed problem discussed by Hadamard. A general version of similar ideas was formulated independently by V.K. Ivanov \cite{ivan}. His method and the method of D.L. Phillips for Fredholm integral equations of the first kind \cite{phil} were the first examples of {\it regularisation methods} for the solution of ill-posed problems. The theory of these methods was formulated by A.N. Tikhonov one year later \cite{tikh}.

The principle of the regularisation methods is to use the additional information explicitly, at the start, to construct families of approximate solutions, i.e. of objects compatible with the given image. These methods are now one of the most powerful tools for the solution of inverse problems, another one being provided by the so-called Bayesian methods, where the additional information used is of statistical nature. 

We will continue the discussion on regularisation methods in the next chapter. First, we will present the generalised solution of inverse problems, describe the regularisation method introduced by Tikhonov and also the truncated singular value decomposition. In the last section, we will give a general definition of a regularisation algorithm and give some properties a regulariser should have so that it gives approximate and stable solutions to inverse problems. 

\clearpage{\thispagestyle{empty}\cleardoublepage} 

\chapter {Regularisation of ill-posed problems}
\label{regularisationmethods}

\section{The generalised solution}
\label{generalisedsolution}

Given the noisy image $g\in{\cal Y}$ and the linear operator $A$ describing the imaging system, we are interested in solving the linear equation
\begin{equation}
Af=g
\label{imagingprob}
\end{equation}
for $f\in{\cal X}$.

We assume that the operator $A$ has a singular value decomposition so that we can write (see Appendix \ref{appSVD})
\begin{equation}
Af=\sum_{j=1}^\infty\sigma_j(f,v_j)_{\cal X}u_j,
\label{opsvd}
\end{equation}
where $\sigma_j$ are the singular values of $A$ and $v_j$, $u_j$ are singular functions on ${\cal X}$ and ${\cal Y}$ respectively.

The problem (\ref{imagingprob}) is, in general, ill-posed in the sense that the solution is not unique, does not exist, or else, does not depend continuously on the data.

Uniqueness does not hold when the null-space of the operator $A$, ${\cal N}(A)$, i.e. the set of the invisible objects $f$ such that $Af=0$, is not trivial.

The procedure most frequently used for restoring uniqueness is the following one. Any element $f$ of the object space ${\cal X}$ can be represented by
\begin{equation}
f=\sum_{j=1}^\infty (f,v_j)_{\cal X}v_j+v,
\label{object}
\end{equation}
where $v$ is the projection of $f$ onto ${\cal N}(A)$, while the first term is the component of $f$ orthogonal to ${\cal N}(A)$. The term $v$ can be called the invisible component of the object $f$ because it does not contribute to the image of $f$, $Af$. Since the invisible component cannot be determined from Eq.(\ref{imagingprob}), it may be natural to look for a solution of this equation whose invisible component is zero. Such a solution is unique because from Eqs.(\ref{opsvd}) and (\ref{object}), with $v=0$, we easily deduce that $Af=0$ implies $f=0$. If this solution exists, it is denoted by $f^+$ and called {\em minimal norm solution}. Indeed, any solution of (\ref{imagingprob}) is 
given by
\begin{equation}
f=f^++v,
\end{equation}
where $v$ is an arbitrary element of ${\cal N}(A)$. Since $v$ is orthogonal to $f^+$, we have 
\begin{equation}
||f||^2_{\cal X}=||f^+||^2_{\cal X}+||v||^2_{\cal X}
\end{equation}
and therefore the solution with $v=0$, i.e., $f^+$, is the solution of minimal norm.

As concerns the existence of a solution of (\ref{imagingprob}) and, in particular, of $f^+$, we first have to distinguish between the following two cases.

\begin{itemize}

\item The null space of the adjoint $A^*$ of $A$, ${\cal N}(A^*)$, contains only the zero element. Then the singular functions (vectors) $u_j$ constitute an orthonormal basis in ${\cal Y}$ and any noisy image $g$ can be represented by
\begin{equation}
g=\sum_{j=1}^\infty (g,u_j)_{\cal Y}u_j.
\end{equation}
By comparing this representation with the SVD of $A$, Eq.(\ref{opsvd}), we see that a solution of (\ref{imagingprob})
may exist.  

\item The null space ${\cal N}(A^*)$ contains non-zero elements. In such a case the singular functions (vectors) $u_j$ do not constitute an orthonormal basis in ${\cal Y}$ and the noisy image $g$ can be represented as follows 
\begin{equation}
g=\sum_{j=1}^\infty (g,u_j)_{\cal Y}u_j+u
\label{noisyimage}
\end{equation}
where $u$ is the component of $g$ in ${\cal N}(A^*)$, i.e. the component of $g$ orthogonal to the range of $A$. Notice that, if the mathematical model of the imaging system is physically correct, the presence of this term is an effect due to the noise. If $u\neq0$, by comparing the representation (\ref{opsvd}) of $Af$ with the representation (\ref{noisyimage}) of $g$, we see that there does not exist any object $f$ such that $Af$ coincides with $g$. Then we can look for objects $f$ such that $Af$ is as close as possible to $f$, i.e. for objects which minimise the discrepancy functional
\begin{equation}
||Af-g||_{\cal Y}=\mbox{minimum}.
\label{discrepancyfunc}
\end{equation}
Any solution of this variational problem is called a {\em least-squares solution}.
\end{itemize}

The concept of least-squares solutions is more general than the concept of solution because a solution of (\ref{imagingprob}) is also a least-squares solution. More precisely, the set of the least-squares solutions coincides with the set of the solutions if and only if the minimum of the discrepancy functional (\ref{discrepancyfunc}) is zero. This remark shows that, without loss of generality, we can investigate the problem of existence in the case of the least-squares solutions.

Solving problem (\ref{discrepancyfunc}) is equivalent to solving its Euler equation, which is given by
\begin{equation}
A^*Af=A^*g.
\label{euler}
\end{equation}
From the SVD of the operator $A$, Eq.(\ref{opsvd}), and of the operator $A^*$
\begin{equation}
A^*g=\sum_{j=1}^\infty\sigma_j(g,u_j)_{\cal Y}v_j,
\label{a*svd}
\end{equation}
we obtain
\begin{equation}
A^*Af=\sum_{j=1}^\infty\sigma^2_j(f,v_j)_{\cal X}v_j.
\label{a*asvd}
\end{equation}
If we insert these representations into Eq.(\ref{euler}) and compare the coefficients of $v_k$, we find that the components of any solution $f$ of (\ref{euler}) are given by
\begin{equation}
\sigma^2_j(f,v_j)_{\cal X}=\sigma_j(g,u_j)_{\cal Y},
\end{equation}
and therefore
\begin{equation}
(f,v_j)_{\cal X}=\frac{1}{\sigma_j}(g,u_j)_{\cal Y}.
\label{objectcomponents}
\end{equation}

In such a way the existence of least-squares solutions has been reduced to the existence of elements of the object space ${\cal X}$ whose components with respect to the singular functions $v_j$ are given by Eq.(\ref{objectcomponents}). 

According to (\ref{objectcomponents}) we can introduce the following formal solution
\begin{equation}
f^+=\sum_{j=1}^\infty\frac{1}{\sigma_j}(g,u_j)_{\cal Y}v_j.
\label{seriessolution}
\end{equation}
We say that this solution is formal because it is given by a series expansion so that the solution exists if and only if the series is convergent.

If we consider the convergence in the sense of the norm on ${\cal X}$, then this convergence is assured if and only if the sum of the squares of the coefficients of the eigen-functions $v_j$ is convergent. We obtain the following condition
\begin{equation}
\sum_{j=1}^\infty\frac{1}{\sigma^2_j}\left|(g,u_j)_{\cal Y}\right|^2< \infty
\label{picard}
\end{equation}
which is also called the {\em Picard criterion} for the existence of solutions or least-squares solutions of the linear inverse problem we are considering \cite{groet}.

It is important to point out that, if the singular values $\sigma_j$ accumulate to zero then condition (\ref{picard}) may not be satisfied by any arbitrary noisy image $g$. If the condition (\ref{picard}) is not satisfied, then {\em no solution or least-squares solution of the inverse problem exists}.

The functions $g$ satisfying the Picard criterion are images in the range of $A$, ${\cal R}(A)$. For any one of these functions, the series (\ref{seriessolution}) defining $f^+$ is convergent. Then we can conclude that {\em for any image $g$ satisfying the Picard criterion there exists a unique generalised solution $f^+$, whose singular function expansion is given by Eq.(\ref{seriessolution})}.

The generalised solution defines a {\em generalised inverse operator} $A^+$. This operator is not defined everywhere on ${\cal Y}$ but only on the set of functions $g$ satisfying the Picard criterion. This set is the domain of the operator $A^+$, ${\cal D}(A^+)$. Moreover the operator $A^+$ is not continuous or, in other words, the generalised solution $f^+$ does not depend continuously on the image $g$. In order to prove this statement, let us assume that $g$ is an image satisfying the Picard criterion and let us consider a sequence of images given by
\begin{equation}
g_j=g+\sqrt{\sigma_j}u_j.
\end{equation}
If we assume that the singular values tend to zero, which is in general the case, it is obvious that
\begin{equation}
||g_j-g||_{\cal Y}=\sqrt{\sigma_j}\rightarrow0.
\end{equation}
On the other hand, if we denote by $f^+$ the generalised solution associated with $g$, and by $f_j^+$ the generalised solution associated with $g_j$, we have
\begin{equation}
f_j^+=f^++\frac{1}{\sqrt{\sigma_j}}v_j
\end{equation}
so that
\begin{equation}
\left|\left|f_j^+-f^+\right|\right|=\frac{1}{\sqrt{\sigma_j}}\rightarrow\infty.
\end{equation}
In such a way we have found a sequence of images, converging to $g$, such that the sequence of the corresponding generalised solutions does not converge to $f^+$.

\section{Tikhonov's regularisation method}
\label{tikhonovreg}

The generalised solution of an ill-posed or ill-conditioned problem is not physically meaningful because it may be completely corrupted by the noise propagation from data to solution. For this reason we must look for approximate solutions satisfying additional constraints suggested by the physics of the problem and regularisation is a way for obtaining such solutions. 

The starting point is to define a family of {\em regularised solutions} $f_\lambda$, depending on a {\em regularisation parameter} $\lambda>0$, as the family of the functions minimising the functionals
\begin{equation}
\Phi_\lambda(f;g)=\left|\left|Af-g\right|\right|^2_{\cal Y}+\lambda||f||^2_{\cal X}
\label{functional}
\end{equation}
where $g$ is the given image. The meaning of the regularisation parameter $\lambda$ will become clear in Section \ref{regularizationalgorithms}, where a general theory of regularisation algorithms will be presented. Let us first discuss how can one find a unique solution to the inverse problem (\ref{imagingprob}) which simultaneously minimise the functional (\ref{functional}).

The Euler equation associated with the minimisation of this functional is given by
\begin{equation}
(A^*A+\lambda I)f=A^*g.
\label{euler2}
\end{equation}
An object is a minimum point $f_\lambda$ of the functional (\ref{functional}) if and only if it is a solution of (\ref{euler2}).

In order to solve this equation, let us represent an arbitrary element $f$ of ${\cal X}$ in terms of the singular functions $v_j$ of the operator $A$ and of the elements $v$ (orthogonal to all $v_j$) of the null space of $A$, as in Eq.(\ref{object}). If we insert this representation in (\ref{euler2}) and we take into account Eqs.\ (\ref{a*svd}) and (\ref{a*asvd}), we obtain
\begin{equation}
\sum_{j=1}^\infty(\sigma^2_j+\lambda)(f,v_j)_{\cal X}v_j+\lambda v=\sum_{j=1}^\infty\sigma_j(g,u_j)_{\cal Y}v_j.
\end{equation}
It follows that there exists a unique solution $f_\lambda$ of (\ref{euler2}), which can be obtained from (\ref{object}) with $v=0$ and with coefficients $(f,v_j)_{\cal X}$ given by
\begin{equation}
(\sigma^2_j+\lambda)(f,v_j)_{\cal X}=\sigma_j(g,u_j)_{\cal Y}.
\end{equation}
In conclusion we find
\begin{equation}
f_\lambda=\sum_{j=1}^\infty\frac{\sigma_j}{\sigma^2_j+\lambda}(g,u_j)_{\cal Y}v_j.
\label{regsol}
\end{equation}  
The series at the r.h.s. of this equation does always converge (thanks to the factors $\sigma_j$, the coefficients tend to zero more rapidly than the components $(g,u_j)_{\cal Y}$ of $g$) and therefore the regularised solution $f_\lambda$ exists for any noisy image $g$.

\section{Truncated SVD}
\label{truncatedsvd}
The representation (\ref{regsol}) of the regularised solution can be recast in the following form
\begin{equation}
f_\lambda=\sum_{j=1}^\infty\frac{W_{\lambda,j}}{\sigma_j}(g,u_j)_{\cal Y}v_j,
\label{regsol1}
\end{equation}
where
\begin{equation}
W_{\lambda,j}=\frac{\sigma^2_j}{\sigma^2_j+\lambda}.
\label{filter}
\end{equation}
This expression shows that the regularised solution $f_\lambda$ can be obtained by a filtering of the singular value decomposition of the generalised solution: the components of $f^+$ corresponding to singular values much larger than $\lambda$ are taken without any significant modification, whereas the components corresponding to singular values much smaller than $\lambda$ are essentially removed.

One possibility is to replace the smooth filter given in Eq.(\ref{filter}) by a sharp one, i.e., to take in the singular function expansion of the generalised solution only the terms corresponding to singular values greater than a certain threshold value. Since the singular values are ordered to form a decreasing sequence, those greater than the threshold value are those corresponding to values of the index less than a certain maximum integer.

Let us denote by $J$ the number of singular values satisfying the condition
\begin{equation}
\sigma^2_j\geq\lambda,\ \ j\le J,
\end{equation}
then the approximate solution provided by the truncated SVD is as follows
\begin{equation}
F_J=\sum_{j=1}^J\frac{1}{\sigma_j}(g,u_j)_{\cal Y}v_j.
\end{equation}
This equation can be obtained from (\ref{regsol1}) by taking $W_{\lambda,j}=1$ when $\sigma^2_j\geq\lambda$ and $W_{\lambda,j}=0$ when $\sigma^2_j<\lambda$.

\section{Regularisation algorithms}
\label{regularizationalgorithms}

The methods investigated in the previous sections can be embedded in a more general approach called by Tikhonov the {\em regularisation method} \cite{tikh,tikhonov}. It consists in the introduction of families of continuous approximations to the generalised inverse of the operator $A$.

A one parameter family of operators $\{R_\lambda\}_{\lambda>0}$ is called a {\em regularisation algorithm} or a {\em regulariser} for the solution or general solution of Eq.(\ref{inversproblem}), if the following conditions are satisfied \cite{tikhonov}:
\begin{itemize}
 \item[i)] for any $\lambda>0$, $R_\lambda:{\cal Y}\rightarrow{\cal X}$ is continuous;
 \item[ii)] for any $g$ such that $g\in {\cal R}(A)$
\begin{equation}
 \lim_{\lambda\searrow0}||R_\lambda g-f^+||_{\cal X}=0
\label{regdef1}
\end{equation}
 where $f^+$ is the generalised solution (or solution when $A^{-1}$ exists) of Eq.(\ref{inversproblem}).
\end{itemize}

When $R_\lambda$ is linear we have a linear regularisation algorithm; the variable $\lambda$ is called the regularisation parameter.

Eq.(\ref{regdef1}) can also be written as follows
\begin{equation}
 \lim_{\lambda\searrow0}||R_\lambda Af^+-f^+||_{\cal X}=0.
\label{regdef2}
\end{equation}
Therefore the operator $T_\lambda:{\cal X}\rightarrow{\cal X}$, defined by
\begin{equation}
 T_\lambda=R_\lambda A
\label{talphaop}
\end{equation}
is an approximation of the orthogonal projection onto ${\cal N}(A)^\perp$ or of the identity operator when ${\cal N}(A)=\{0\}$.

Conditions i) and  ii) imply that, for any $\lambda$ and for any exact data $g\in {\cal R}(A)$, $R_\lambda g$ is a continuous approximation of the solution or generalised solution of Eq.(\ref{inversproblem}). However, the important case is that of noisy data $g_\varepsilon$, with $g_\varepsilon\not\in {\cal R}(A)$, which are close to exact data $g\in {\cal R}(A)$, $||g_\varepsilon-g||_{\cal Y}\le\varepsilon$, if $\varepsilon$ is small enough. In this case, no solution of the equation $Af=g_\varepsilon$ may exist. 

If we consider the functions $f_\lambda=R_\lambda g_\varepsilon$, it is easy to see that there must exist an optimum value of $\lambda$ such that $f_\lambda$ is as close as possible to $f^+$, the generalised solution associated with the exact data $g$. Assuming that $R_\lambda$ is linear and that the noisy data $g_\varepsilon$ are written in the form $g_\varepsilon=Af+w_\varepsilon$, we get
\begin{equation}
 R_\lambda g_\varepsilon-f^+=(R_\lambda Af^+-f^+)+R_\lambda w_\varepsilon,
\end{equation}
and therefore
\begin{equation}
 ||R_\lambda g_\varepsilon-f^+||_{\cal X}\le \omega(\lambda;f^+)+\varepsilon N(\lambda),
\label{regnorm}
\end{equation}
where 
\begin{equation}
 N(\lambda)=||R_\lambda||, \ \ \omega(\lambda; f^+)=||R_\lambda Af^+-f^+||_{\cal X}\ \ \mbox{and}\ \ \varepsilon=||w_\varepsilon||_{\cal Y}.
\label{normofR}
\end{equation}
The first term of the r.h.s.\ in Eq.(\ref{regnorm}) represents the approximation error introduced by the choice of a non-zero value of the regularisation parameter; it tends to zero when $\lambda\rightarrow 0$. The second term represents the error on the approximate solution induced by the error on the data; it tends to infinity when $\lambda\rightarrow 0$. Therefore it is necessary to find a compromise between approximation and error magnification. Assume, for simplicity that $N(\lambda)$ and $\omega(\lambda;f^+)$ are monotonous functions of $\lambda$ (this condition is satisfied by all the regularisation algorithms used in practice) and, more precisely, that $N(\lambda)$ is a decreasing function, with $N(0+)=+\infty$, while $\omega(\lambda;f^+)$ is an increasing function, with $\omega(0+;f^+)=0$. Under these assumptions, there exists a unique value of $\lambda$, $\lambda(\varepsilon)$, which minimises the r.h.s.\ of Eq.(\ref{regnorm}) and which represents the optimum compromise between approximation and error magnification. Furthermore $\lambda=\lambda(\varepsilon)\rightarrow0$, when $\varepsilon\rightarrow 0$, and $R_\lambda g_\varepsilon\rightarrow f^+$.

The previous argument implies that a regularisation algorithm can give approximate and stable solutions which converge to the exact solution when the error on the data tends to zero. 

\subsubsection*{The Tikhonov regulariser}

The most intensively investigated example of a regularisation algorithm is the so-called Tikhonov regulariser, given by
\begin{equation}
 R_\lambda=(A^*A+\lambda I)^{-1}A^*.
\label{tikhonovregulariser}
\end{equation}
Its remarkable properties derive from the fact that it can be obtained by minimising the functional (\ref{functional}) (see Section \ref{tikhonovreg}). In particular it is easy to show that $R_\lambda g\in {\cal N}(A)^\perp$, for any $g\in{\cal Y}$ and this implies that $\{R_\lambda\}_{\lambda>0}$ is a regularisation algorithm for $f^+$. Using the spectral representation of $A^*A$, indeed, it is easy to show that
\begin{equation}
 \omega(\lambda;f^+)=||R_\lambda Af^+-f^+||_{\cal X}=\lambda||(A^*A+\lambda I)^{-1}f^+||_{\cal X}
\end{equation}
tends to zero when $\lambda\rightarrow 0$. Furthermore, for any $f^+$, $\omega(\lambda;f^+)$ is an increasing function of $\lambda$.

It is also easy to find an estimate for $N(\lambda)$, Eq.(\ref{normofR}),
\begin{equation}
 ||R_\lambda g||^2_{\cal X}=\left(AR_\lambda g\ ,\ (A^*A+\lambda I)^{-1}g\right)_{\cal Y}\le \frac{1}{\lambda}||g||^2_{\cal Y}
\end{equation}
since $||AR_\lambda||\le 1$, and therefore
\begin{equation}
 N(\lambda)\le \frac{1}{\sqrt{\lambda}}.
\end{equation}
From this bound and Eq.(\ref{regnorm}) one can derive results concerning possible choices of the function $\lambda(\varepsilon)$.

\subsubsection*{Spectral windows}

The regulariser (\ref{tikhonovregulariser}) can be written in the following formal way
\begin{equation}
 R_\lambda=U_\lambda(A^*A)A^*,
\label{spectralregulariser}
\end{equation}
with the help of the function
\begin{equation}
 U_\lambda(\mu)=(\mu+\lambda)^{-1}.
\label{spectralex1}
\end{equation}
This remark suggests a way for defining a wide class of regularisation algorithms which can be applied whenever the spectral representation of $A^*A$ is known \cite{bakushinskii,groetsch}.

Consider a family of real-valued, piecewise continuous functions $\{U_\lambda\}_{\lambda>0}$, defined on the interval $[0,||A||^2]$ and assume that they satisfy the following conditions:
\begin{itemize}
 \item[i)] for each $\lambda>0$, there exists a constant $c_\lambda$ such that
\begin{equation}
 |U_\lambda(\mu)|\le c_\lambda; \ \ \forall\ \mu\in[0,||A||^2];
\end{equation}
 \item[ii)] for each $\lambda>0$
\begin{equation}
 0\le\mu U_\lambda(\mu)\le1; \ \ \forall\ \mu\in[0,||A||^2];
\end{equation}
\item[iii)] \begin{equation}
             \lim_{\lambda\searrow0}\mu U_\lambda(\mu)=1; \ \ \forall\ \mu\in(0,||A||^2].
            \end{equation}
\end{itemize}
Then the family of operators $R_\lambda$, defined by Eq.(\ref{spectralregulariser}) is a regularisation algorithm for $f^+$. This result derives from the following remarks. Thanks to condition i), for each $\lambda>0$, the operator $R_\lambda$ is bounded. Furthermore the following relation holds:
\begin{equation}
 U_\lambda(A^*A)A^*=A^*U_\lambda(AA^*)
\end{equation}
since it is true for polynomials and therefore it is also true for continuous functions. This relation implies that, for any $\lambda>0$ and any $g\in{\cal Y}$
\begin{equation}
 f_\lambda=R_\lambda g\in {\cal N}(A)^\perp.
\end{equation}
Finally, from the spectral representation of $A^*A$, from conditions ii) and iii) and the dominated convergence theorem\footnote{\sf The dominated convergence theorem states: Let $X$ be a measurable space and let $g$, $f_1,f_2,...$ be measurable functions such that $\int_Xg<\infty$ and $|f_n|\le g$ for each $n$. If $f_n\rightarrow f$ almost everywhere, then $f$ is integrable and $$\lim_{n\rightarrow\infty}\int_Xf_n=\int_Xf.$$}, property (\ref{regdef2}) can be derived.

The function $U_\lambda(\mu)$ defined in Eq.(\ref{spectralex1}) satisfies the previous conditions. Another important example is given by
\begin{equation}
U_\lambda(\mu)=\left\{
\begin{array}{lcl}
 0&,&0\le\mu\le\lambda \\
\mu^{-1}&,&\mu>\lambda
\end{array}.
\right.
\label{spectralex2}
\end{equation}

For a regularisation algorithm of the type (\ref{spectralregulariser}), the operator $T_\lambda$, Eq.(\ref{talphaop}), is
\begin{equation}
 T_\lambda=W_\lambda(A^*A),
\end{equation}
where $W_\lambda(\mu)=\mu U_\lambda(\mu)$. The function $W_\lambda(\mu)$ can be called a spectral window. The justification of this name derives from the example (\ref{spectralex2}) since in such a case the function $W_\lambda(\mu)$ is zero in a neighbourhood of 0 and is unity elsewhere. 

In the case of a compact operator, the regulariser (\ref{spectralregulariser}) can be expressed in terms of the singular value decomposition
\begin{equation}
 R_\lambda g=\sum_{k=0}^\infty \mu_k^{-1/2}W_\lambda(\mu_k)(g,u_k)v_k.
\end{equation}
In particular, in the case (\ref{spectralex2}) we have
\begin{equation}
 R_\lambda g=\sum_{\sqrt{\mu_k}} \mu_k^{-1/2}(g,u_k)v_k.
\end{equation}
and this is the well-known method of the truncated singular value decomposition discussed in Section \ref{truncatedsvd}.

\section{Choice of regularisation parameter}

The choice of the regularisation parameter is a crucial and difficult problem in the theory of regularisation. This point has been widely discussed in the mathematical literature. No precise recipe has been discovered which could be used for any problem. The existence of a recipe depends a lot on the application and the actual information one has at hand. In what follows we will restrict ourselves to the case of Tikhonov's regularisation algorithm discussed in Section \ref{tikhonovreg} and summarise the main methods which are used in practice.

As we know from the discussion of Section \ref{regularizationalgorithms}, for any image $g$ there exists an optimum value, $\lambda_{\rm opt}$, of the regularisation parameter. For this value of $\lambda$, the corresponding regularised solution $f_\lambda$ has minimal distance from the true object $f$. The problem is that the determination of this optimal value requires the knowledge of $f$.

\subsubsection*{Regularised solution with prescribed energy}

If we do not know $f$ but know its norm $E=||f||$, also called 'energy', one may try the value $\lambda=\lambda(E)$ such that the corresponding regularised solution has the same energy as the true object, i.e.
\begin{equation} 
 ||f_\lambda||=E.
\end{equation}
$\lambda(E)$ is a decreasing function of $E$. Therefore, if we overestimate the energy of the object, we obtain a value of the regularisation parameter smaller than that corresponding to the exact energy of the object. In such a case the regularised solution will show a higher noise contamination.

\subsubsection*{Regularised solution with prescribed discrepancy}

If we know a precise estimate $\varepsilon$ of the energy of the noise, then the estimate is the value $\lambda=\lambda(\varepsilon)$ such that the discrepancy of the corresponding regularised solution is just equal to $\varepsilon$, i.e.
\begin{equation}
 ||Af_\lambda-g||=\epsilon.
\end{equation}
This method is also known as the discrepancy principle \cite{morozov}. $\lambda(\varepsilon)$ is an increasing function of $\varepsilon$. Therefore, if we overestimate the energy of the noise, we get a value of the regularisation parameter which is larger than that corresponding to the exact energy of the noise. In such a case the regularised solution will show a smaller noise contamination.

\subsubsection*{The Miller method}

An approach to ill-posed problems proposed by Miller \cite{miller} can also be considered as a method for estimating the value of the regularisation parameter. In this approach it is assumed that one knows both a bound on the energy and a bound on the discrepancy of the unknown object $f$. Then the set of all objects $f$ satisfying the two conditions
\begin{equation}
 ||Af-g||^2\le\varepsilon^2, \ \ ||f||^2\le E^2,
\label{millercond}
\end{equation}
is called the set of admissible approximate solutions. This set corresponds to the intersection of the ball of the objects with squared energy smaller than $E^2$ and of the ellipsoid of the objects with discrepancy smaller than $\varepsilon^2$. If this intersection is not empty, then the pair $\{\varepsilon,E\}$ is said to be permissible.

Now, if the pair $\{\varepsilon,E\}$ is permissible, it has been shown by Miller \cite{miller} that the regularised solution corresponding to the following value of the regularisation parameter
\begin{equation}
 \lambda=\left(\frac{\varepsilon}{E}\right)^2
\end{equation}
satisfies the condition (\ref{millercond}) except for a factor of $\sqrt{2}$ and therefore it is essentially an admissible approximate solution.

\subsubsection*{Generalised cross-validation}

The methods considered previously require the knowledge of $\varepsilon$ or of $E$ or of both. In many cases one does not have a sufficiently accurate estimate of these quantities and therefore it is important to have methods which do not require this kind of information. One such method is that of cross-validation \cite{wahba} which can only be used in problems with discrete data. It is based on the idea of letting the data themselves choose the value of the regularisation parameter. In other words one requires that a good value of the regularisation parameter should predict missing data values. 

The mathematical formulation of this method is more complicated and will not be presented here. For more details one can look in \cite{bertero}.

\subsubsection*{L-curve method}

This graphically motivated method, introduced by Hansen \cite{hansen}, is another method which does not require information about the energy of the noise or of the true object. The starting point is the plot of $E(f_\lambda)$ versus $\varepsilon(f_\lambda;g)$, introduced in connection with Miller's method. This curve has, in many cases, a rather characteristic L-shaped behaviour in a log-log plot. 

A qualitative explanation of this behaviour is the following. We recall that $E(f_\lambda)$ is large for small $\lambda$ and small for large $\lambda$ while $\varepsilon(f_\lambda;g)$ has the opposite behaviour. Therefore $E(f_\lambda)$ is large when $\varepsilon(f_\lambda;g)$ is small, and conversely. This is the trade-off between noise-propagation error and approximation error. Now the vertical part of the L-curve corresponds to values of the regularisation parameter such that $f_\lambda$ is dominated by the noise propagation error. As a consequence $E(f_\lambda)$ is very sensitive to variations of $\lambda$ while $\varepsilon(f_\lambda;g)$ is not. Analogously the horizontal part of the L-curve corresponds to values of the regularisation parameter such that $f_\lambda$ is dominated by the approximation error. As a consequence $\varepsilon(f_\lambda;g)$ is very sensitive to variations of $\lambda$ while $E(f_\lambda)$ is not.

Now the L-curve method consists of taking as an estimate of the regularisation parameter the value of $\lambda$ corresponding to the corner of the L-curve. In fact this point should correspond to the compromise between approximation error and noise-propagation error. From the computational point of view a convenient definition of the L-curve corner is the point with maximum curvature. 

The L-curve method, even if it can be useful in some cases, does not work in all cases and has some theoretical and practical inconveniences. It has been shown \cite{engl,vogel} that, in certain cases, it does not provide a regularised solution converging to the exact one when the noise tends to zero. Moreover, examples can be found where the L-curve does not even have an L-shape so that the method cannot be used. 

\subsubsection*{The interactive method}

The speed and versatility of modern digital computers allows us to restore images interactively: the user controls the restorations obtained by means of several values of the regularisation parameter and, by tuning $\lambda$, he selects the best restoration on the base of his intuition or of the attainment of some specific purpose.

\clearpage{\thispagestyle{empty}\cleardoublepage}

\thispagestyle{empty}

\begin{center}
\vspace*{9cm}
{\Huge\bf QCD condensates from $\tau$-decay data}

\addcontentsline{toc}{part}{QCD condensates from $\tau$-decay data}

\end{center}

\clearpage{\thispagestyle{empty}\cleardoublepage}

\chapter{The theory of $\tau$-decays}

The $\tau$ is the only lepton heavy enough ($m_\tau=1.777\mbox{ GeV}$) to decay not only into other leptons, but 
into final states involving hadrons as well. These decays offer an ideal laboratory for the study of strong interactions, including the transition from the perturbative to the non-perturbative regime of QCD in the simplest possible reaction. This might explain the tremendous efforts ongoing in both theoretical and experimental studies of $\tau$ physics (for a review see \cite{stahl}).

\begin{figure}[H]
\centering
\includegraphics[height=.30\textwidth,angle=0]{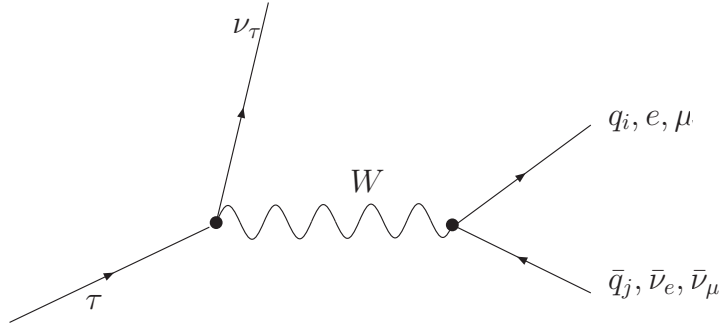}
\caption{Leptonic and hadronic decays of $\tau$ at zeroth order in $\alpha_s$.}
\label{taudecay}
\end{figure}

The {\em leptonic} decays are:
\begin{equation}
\begin{array}{l}
\tau\rightarrow\nu_\tau+W\\
\hspace{1.9cm} \searrow \\
\hspace{2.4cm} l^-+\bar\nu_l
\end{array}
\hspace{1cm} l=\mu\mbox{ or }e.
\label{lepdecay}
\end{equation}
We will write formulas for $\tau= \tau^-$; the formulas for $\tau^+$ are obtained with obvious changes. Besides the decays (\ref{lepdecay}), we have {\em hadronic} decays. At the parton level these are given by the processes
\begin{equation}
\begin{array}{l}
\tau\rightarrow\nu_\tau+W\\
\hspace{1.9cm}\searrow\\
\hspace{2.4cm} q_i+\bar q_j
\end{array}
\label{haddecay}
\end{equation}
where the flavour indices $i,j$ stand for the light quarks $(u,d,s)$. The Feynman diagram for Eqs.(\ref{lepdecay}) and (\ref{haddecay}) is shown in Fig.\ref{taudecay}.

The permitted hadronic decays may be split into decays that involve only non-strange particles
\begin{equation}
\begin{array}{l}
\tau\rightarrow\nu_\tau+W\\
\hspace{1.9cm}\searrow\\
\hspace{2.4cm} d+\bar u,
\end{array}
\end{equation}
or decays involving strange particles
\begin{equation}
\begin{array}{l}
\tau\rightarrow\nu_\tau+W\\
\hspace{1.9cm}\searrow\\
\hspace{2.4cm} s+\bar u.
\end{array}
\end{equation}
Although the theory will be presented in general, i.e., for both types of decay, we are especially interested in the non-strange ones.

\section{Hadronic $\tau$-decays}

The decay of the $\tau$ lepton into hadrons (see Fig.\ref{taudecay}) was calculated by Paul Tsai \cite{tsai1}, even before the discovery of the $\tau$. Ignoring the propagator of the $W^{\pm}$ boson, the matrix element is given by the product of the leptonic and the hadronic current:
\begin{equation}
i{\cal M}=\frac{G_F}{\sqrt{2}}\langle\tau\left|J_{\rm lep}^\mu J_{{\rm had},\mu}\right| {\rm had},\nu_\tau\rangle,
\label{matrixel}
\end{equation}
where $G_F$ is the Fermi constant. The leptonic current is the standard left-handed one, $J_{\rm lep}^\mu=\bar\tau\gamma^\mu(1-\gamma_5)\nu_\tau$, while the hadronic current can be any of the vector $J_{\rm had}^\mu=\sum_{i,j}V_{ij}\bar q_i\gamma^\mu q_j$, axial-vector current  $J_{\rm had}^\mu=\sum_{i,j}V_{ij}\bar q_i\gamma^\mu\gamma_5 q_j$ or combinations of them. By convention, we have included in the definition of the hadronic current part of its coupling to the $W$ boson, the CKM mixing matrix element $V_{ij}$. Other factors were explicitly taken into account when writing out the matrix element (\ref{matrixel}).

\begin{figure}[H]
\centering
\includegraphics[height=.20\textwidth,angle=0]{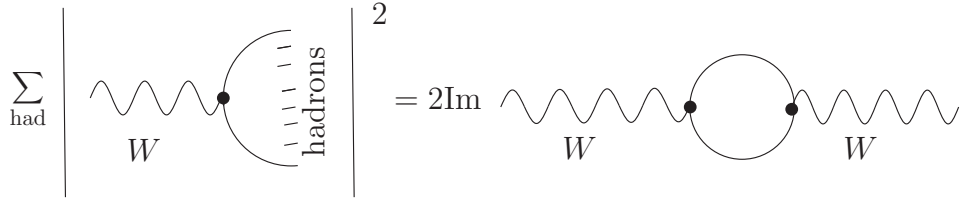}
\caption[The optical theorem applied to $\tau$-decays.]{The optical theorem relates the production of hadrons from $\tau$ decays (left diagram) to the imaginary part of the forward scattering amplitude of the diagram on the right.}
\label{tau-optical}
\end{figure}

With this matrix element, one can calculate the decay width $\Gamma(\tau\rightarrow\nu_\tau+{\rm had})$ as \cite{braaten,narison}

\begin{equation}
\begin{array}{l}
\displaystyle d\Gamma(\tau\rightarrow\nu_\tau+{\rm had})= \frac{1}{2m_\tau}(2\pi)^4\delta^{(4)}(p_{\rm had}-q)\frac{G_F^2}{2}L^{\mu\nu}\\
\\
\hspace{1.8cm}\displaystyle\times\sum_{\rm had}i\!\int d^4x\,e^{iqx}
\langle 0|J_{{\rm had},\mu}(x)|{\rm had}\rangle\langle {\rm had}|J_{{\rm had},\nu}^\dagger(0)|0\rangle d\phi_{\rm had}d\phi_{\nu_\tau},
\end{array}
\label{decaywidth}
\end{equation}
where $d\phi$ denotes the invariant phase space elements of the hadrons and the neutrino, and $L^{\mu\nu}$ is the leptonic tensor. The total 4-momentum of the hadronic system is written as $q\ (q^2\equiv s)$. Now the optical theorem (Fig.\ref{tau-optical}) can be used to write the matrix element for the production of hadrons in terms of the imaginary part of the forward scattering amplitude:
\begin{equation}
d\Gamma(\tau\rightarrow\nu_\tau+{\rm had})=\frac{1}{2m_\tau}\frac{G_F^2}{\sqrt{2}}L^{\mu\nu}2\,\mbox{Im}\ i\!\int d^4x\,e^{iqx}\langle 0|TJ_{{\rm had},\mu}(x)J_{{\rm had},\nu}^\dagger(0)|0\rangle d\phi_{\nu_\tau}.
\label{newwidth}
\end{equation}
This is quite an important step. While (\ref{decaywidth}) requires the calculation of matrix elements of exclusive final states (and their summation), Eq.(\ref{newwidth}) contains no explicit reference to hadronic final states. The first one cannot be handled by perturbative QCD, but the second one can.

We may split the hadron tensor into a transverse and a longitudinal part, writing
\begin{equation}
\begin{array}{lcl}
\Pi_{\mu\nu}(q)&=&\displaystyle i\int d^4x\,e^{iqx}\langle 0|T\,J_{{\rm had},\mu}(x)J_{{\rm had},\nu}^\dagger(0)|0\rangle\\
\\
&=&(q_\mu q_\nu-g_{\mu\nu}q^2)\Pi^{(1)}(q^2)+q_\mu q_\nu\Pi^{(0)}(q^2).
\end{array}
\label{Lorentzdecomp}
\end{equation}

The two functions $\Pi^{(J)}$ introduced here are called the two-point correlators of the quark currents (see Section \ref{ope}). They describe the creation of hadronic states with total angular momentum $J$ from the vacuum. The integration over the phase space of the neutrino can now be performed. The result is
 
\begin{equation}
\begin{array}{l}
\Gamma(\tau\rightarrow\nu_\tau+{\rm had})=\\
\\
\hspace{0.8cm}\displaystyle\frac{G_F^2m_\tau^5}{16\pi^2}\int_0^{m_\tau^2}\ \frac{ds}{m_\tau^2}\left(1-\frac{s}{m_\tau^2}\right)^2\left\{\left(1+2\frac{s}{m_\tau^2}\right)\mbox{Im}\Pi^{(1)}(s)+\mbox{Im}\Pi^{(0)}(s)\right\}.
\label{hardwidth}
\end{array}
\end{equation}

\section{Leptonic $\tau$-decays}

The leptonic decay width can be calculated in a straightforward way from the Feynman diagram of Fig.\ref{taudecay} as well. It is sufficient to treat the process as an effective four-fermion interaction and add the effect from the $W^\pm$ propagator as a correction later. The result is
\begin{equation}
\Gamma(\tau\rightarrow\nu_\tau\bar\nu_ll)=\frac{G_F^2m_\tau^5}{192\pi^3}(1+\Delta_l).
\label{lepwidth}
\end{equation}

\begin{table}[h]
\begin{center}
\begin{tabular}{@{\extracolsep{\fill}}||Sl@{\vrule height 14pt depth4pt width0pt\hskip\arraycolsep}|Sc||}\hline
\multicolumn{2}{||l||}{Standard Model corrections}\\\hline
\parbox[c]{9cm}{phase space correction due to the finite mass $m_l$ of the charged daughter lepton \cite{barish}} & $\displaystyle-8\left(\frac{m_l}{m_\tau}\right)^2$  \\[0.5cm]
\parbox[c]{9cm}{QED radiative corrections \cite{berman,kinoshita,kallen,marciano,roos,sirlin}} & $\displaystyle\frac{\alpha(m_\tau^2)}{2\pi}\left(\frac{25}{4}-\pi^2\right)$\\[0.5cm]
correction due to the $W^\pm$ propagator \cite{lee} & $\displaystyle\frac{3}{5}\left(\frac{m_\tau}{m_{W^\pm}}\right)^2$\\\hline
\multicolumn{2}{||l||}{Corrections due to new physics}\\\hline
neutrino mass \cite{barish} & $\displaystyle-8\left(\frac{m_{\nu_\tau}}{m_\tau}\right)^2$\\[0.5cm]
scalar current \cite{stahl1} & $\displaystyle 4\eta\frac{m_l}{m_\tau}$\\[0.5cm]
mixing with a 4th generation $\nu$ \cite{swain} & $-\sin^2\theta$\\[0.5cm]
magnetic dipole moment \cite{rizzo} & $\displaystyle\frac{\kappa}{2}+\frac{\kappa^2}{10}$\\[0.5cm]
electric dipole moment \cite{rizzo} & $\displaystyle\frac{\tilde\kappa^2}{10}$\\\hline
\end{tabular}
\caption[Corrections to the leptonic width of the $\tau$ lepton.]{Corrections (contributions to $\Delta_l$) to the leptonic width of the $\tau$ lepton.}
\label{corrections}
\end{center}
\end{table}
The quantity $\Delta_l$ collects a number of corrections shown in Table \ref{corrections}. The table also shows corrections which are not present in the Standard Model but will modify the decay width if new physics is present. Indeed, a massive neutrino would change the phase space of the decays and consequently the decay width. The direct limits on the mass of $\nu_e$ and $\nu_\mu$ are so strict that these cannot have a measurable impact, but the $\nu_\tau$ could. Also, mixing with a yet undiscovered, heavy, fourth-generation neutrino would reduce the decay width. The authors of \cite{dova,swain} derived from universality limits for:
\begin{equation}
 m_{\nu_\tau}<38{\rm MeV},\  \sin^2\theta<0.008
\end{equation}
at the $95\%$ CL.

A charged Higgs boson would introduce a scalar coupling and therefore a nonvanishing Michel parameter $\eta$, which in turn modifies the decay width. A limit on these bosons has been found \cite{stahl1,stahl2}:
\begin{equation}
 m_{H^\pm}>2.3\tan\beta,
\end{equation}
at the $95\%$ CL. An estimate of the Michel parameter is \cite{stahl1}:
\begin{equation}
 \eta=0.01\pm 0.05.
\end{equation}

An anomalous magnetic moment or an electric dipole moment of the charged weak current, usually parametrised with the help of parameters $\kappa$ and $\tilde\kappa$ \cite{stahl}, would modify the decay width. The correction is given in Table \ref{corrections}. Limits have been derived in \cite{dova}. They are ($95\%$ CL)
\begin{equation}
 -0.015<\kappa<0.017,\  |\tilde\kappa|<0.31.
\end{equation}

Nevertheless, the corrections are small ($<4\%$) and can be safely neglected. 

\section{The hadronic branching ratio $R_\tau$} 

$R_\tau$ is defined as the ratio of the total hadronic decay width $\Gamma(\tau\rightarrow\nu_\tau+{\rm had})$ to the leptonic one:
\begin{equation}
R_\tau\equiv\frac{\Gamma(\tau\rightarrow\nu_\tau+{\rm had})}{\Gamma(\tau\rightarrow\nu_\tau
\bar\nu_ee)}.
\end{equation}
With the earlier computed decay widths of Eqs.\ (\ref{hardwidth}) and (\ref{lepwidth}), the hadronic branching ratio becomes:
\begin{equation}
R_\tau=12\pi\int_0^{m_\tau^2}\frac{ds}{m_\tau^2}
\left(1-\frac{2}{m_\tau^2}\right)^2\left[\left(1+2\frac{s}{m_\tau^2}\right)
\mbox{Im}\Pi^{(1)}(s)+\mbox{Im}\Pi^{(0)}(s)\right].
\end{equation}

\section{Operator Product Expansion (OPE)}
\label{ope}

As a consequence of asymptotic freedom the theoretical results obtained from QCD can 
be compared with the experimental situation for so-called hard, i.e.\ high energy, processes: 
at short distances the effective coupling constant $\bar\alpha_s$ becomes small and 
the interaction can be treated perturbatively. On the other hand, any complete 
theory of the strong interaction must include large-distance dynamics as well. In 
particular quarks interact strongly when forming hadronic bound states. 

A great deal of effort has been made towards the construction of new tools for reliable computations in the non-perturbative region of QCD. Most of the efforts to obtain quantitative results can be divided into two categories \cite{parisi}: numerical computations and analytic calculations. Numerical computations are usually time consuming and require powerful computers (sometimes specially designed). They are mainly based on lattice gauge theories \cite{kogut,wilson1} and are producing promising results \cite{hamber}.

There are several approaches using analytical calculations. For several years much effort has been devoted to the search for classical solutions of non-abelian field theories \cite{chakrbarti} with the hope that a semiclassical approach may shed some light on the underlying quantum world and that classical configurations of 
fields that make the action stationary play an important role in the problem of 
confinement.

An interesting new approach based on the operator product expansion was opened in 1979 \cite{shifman1,shifman2,shifman3}. This approach is less fundamental in the sense that it does not try to solve the problem of confinement but assumes that confinement exists. In practice the effects of confinement can be described through the use of a few parameters, the so called condensates, and this allows one to investigate many hadronic properties.

As stated above, one of the ingredients of this approach is the operator product expansion \cite{wilson2}. 
Wilson proposed a short distance operator product expansion of the following form
\begin{equation}
A(x)B(0)
\underset{\mbox{\small $x^\mu\rightarrow0$}\  } {\mbox{\Large$\sim$}} \sum_nC_n(x){\cal O}_n(0),
\end{equation}
where $A$ and $B$ are local operators. The $C_n(x)$ are $\C$-number functions which 
can have singularities on the light cone of the form $[x^2-i\varepsilon x^0]^{-p}$, 
$p$ being any real number. They can also involve logarithms of $x^2$. In general, 
the complete expansion involves an infinite number of non-singular operators 
${\cal O}_n$ but to any finite order in $x$ only a finite number of these operators 
contributes. The expansion is valid in the weak sense: one must sandwich the product 
$A(x)B(0)$ between fixed initial and final states. Similar expansions exist for time-ordered products or commutators. The OPE can be verified explicitly  for the free 
scalar and spinor field theories and for renormalised interacting fields to any finite order in perturbation theory. In every case they are valid for any elementary or 
composite local fields.

The nature of the singularities of the functions $C_n(x)$ is determined, in general, 
by exact and broken symmetries of the theory. The most crucial of these symmetries 
is broken scale invariance. The free scalar and spinor field theories with zero mass 
are exactly scale invariant. Mass terms and renormalizable interactions treated in 
perturbation theory break the symmetry but some remainder of scale invariance still governs 
the behaviour of the singular functions \cite{wilson2}. Exact scale invariance means that the field 
theory is invariant under a one-parameter group of transformations $U(\lambda)$. 
A local operator ${\cal O}_n(x)$ transforms as
\begin{equation}
U^\dagger(\lambda){\cal O}_n(x)U(\lambda)=\lambda^{d({\cal O}_n)}{\cal O}_n(\lambda x).
\end{equation}
In free field theories the constant $d({\cal O}_n)$ is the canonical dimension of the 
field, i.e. $[{\cal O}_n(x)]=M^{d({\cal O}_n)}$, which can be determined for example from the 
commutation relations.

In an exactly scale-invariant theory the behaviour of the functions $C_n(x)$ is 
determined, except for some constants, by scale invariance. Performing a scale 
transformation in   
\begin{equation}
A(x)B(y)
\underset{\mbox{\small $x^\mu\rightarrow y^\mu$}\ }{\mbox{\Large$\sim$}}\sum_nC_n(x-y){\cal O}_n(y),
\end{equation}
we obtain
\begin{equation}
\lambda^{d(A)+d(B)}A(\lambda x)B(\lambda y)
\underset{\mbox{\small $x^\mu\rightarrow y^\mu$}\ }{\mbox{\Large$\sim$}}
\sum_nC_n(\lambda x-\lambda y)\lambda^{d({\cal O}_n)}{\cal O}_n(y).
\end{equation}
If the fields ${\cal O}_n(x)$ are linearly independent, which can always be arranged, 
one must have
\begin{equation}
C_n(\lambda x-\lambda y)=\lambda^{d({\cal O}_n)-d(A)-d(B)}C_n(x-y).
\end{equation}
This equation tells us that $C_n(x-y)$ must be a homogeneous function of order 
$d({\cal O}_n)-d(A)-d(B)$ in $(x-y)$. This property as well as the behaviour under Lorentz 
transformations determines $C_n(x-y)$ completely, up to a finite number of constants \cite{schreier}.
When we turn on the interactions \cite{wilson2}, it is still true that the $\C$-number functions 
$C_n(x)$ have a scaling behaviour, as $x^\mu\rightarrow0$, which can be summarised 
with the rules
\begin{equation}
C_n(x)
\underset{\mbox{\small $x^\mu\rightarrow y^\mu$}\ }{\mbox{\Large$\sim$}}
x^{-\lambda_n},\ \ \lambda_n\equiv d(A)+d(B)-d({\cal O}_n), 
\end{equation}
except that the operator dimensions are no longer given by naive counting of the 
mass dimensions of the operators: they become {\em anomalous} in general \cite{coleman}. There are, 
nevertheless, some operators which retain their naive dimensions: these include 
the identity $I$, and the operators generating symmetries of the theory, such as 
currents $J^\mu(x)$ or the field operator $G^{\mu\nu}_a$. 

The operators ${\cal O}_n$ are conveniently classified according to their spin and dimension $d({\cal O}_n)$. In particular, if $A$ and $B$ are scalars, and gauge invariant, only scalar, gauge-invariant operators have to be considered. 

Interesting for us is the case of QCD where the two operators $A$, $B$ are constructed from quark and gluon fields. In this case, the lowest dimension scalar, gauge-invariant operators are: 
\begin{itemize}
\item dimension 0: $I$ (the unit operator)

\item dimension 4: $:m_i \bar q^i_\alpha q^i_\alpha:$ (no summation in flavours)

\hspace{2.3cm} $:G^{\mu\nu}_a(x)G_{\mu\nu}^a(x):$

\item dimension 6: $:\bar q_\alpha(x)\Gamma q_\alpha(x)\bar q_\beta(x)\Gamma q_\beta(x):$

\hspace{2.3cm} $:\bar q_\alpha(x)\Gamma (\lambda^a)_{\alpha\beta} q_\beta(x)\bar q_\gamma(x)\Gamma 
(\lambda^a)_{\gamma\sigma} q_\sigma(x):$

\hspace{2.3cm} $:m_i\bar q^i(x)\lambda^a \sigma_{\mu\nu}q^i(x)G^{\mu\nu}_a(x):$

\hspace{2.3cm} $:f_{abc}G^{\mu\nu,a}(x)G^b_{\nu\rho}(x)G^{c\ \rho}_\mu(x):$
\end{itemize}
where the first two operators of dimension 6 have been given only for one flavour. $\Gamma$ denotes any combination of Dirac matrices and therefore for a given flavour 
there are 16 such quantities 
$(I,\gamma^\mu,\sigma^{\mu\nu},\gamma^\mu\gamma_5,\gamma_5)$ that render a scalar 
operator. Any other gauge invariant scalar operator of the same dimension can be 
reduced to these with the help of the equations of motion. Notice that the use 
of the equations of motion is legitimate because expectation 
values in physical states will be taken and the equations of motion are satisfied if restricted to the space of  physical states.

Within standard perturbation theory only the unit operator would have a non-zero vacuum expectation value, but non-perturbative effects induce non-vanishing vacuum expectation values for operators of higher dimensions as well. These are the so-called {\em condensates}. Therefore non-perturbative effects of QCD introduce power corrections of the type 
$1/[q^2]^N$, $N\ge1$, to the perturbative calculation. 

\section{Hadronic vacuum polarisation tensor}
\label{hadronicpolarization}

The hadronic vacuum polarisation tensor is defined as (cf.\ Eq.(\ref{Lorentzdecomp})):
\begin{equation}
\Pi_{ij}^{\mu\nu}=i\int d^4x\ e^{iqx}\langle0| TJ^\mu_{ij}(x)J^\nu_{ij}(0)^\dagger|0\rangle,
\end{equation}
where $J^\mu_{ij}=\bar q_i\gamma^\mu(\gamma_5)q_j$ are the vector (axial-vector) hadronic currents describing electroweak interactions, $i,j$ are flavour indices and $T$ stands for time ordering. Performing a Lorentz decomposition, we can split $\Pi^{\mu\nu}_{ij}$ into a transverse and a longitudinal part (cf. Eq.(\ref{Lorentzdecomp})):
\begin{equation}
\Pi^{\mu\nu}_{ij}(q)=(q_\mu q_\nu-g_{\mu\nu}q^2)\Pi_{ij}^{(1)}(q^2)+q_\mu q_\nu\Pi_{ij}^{(0)}(q^2),
\end{equation}
where the indices $J=0,1$ in $\Pi^{(J)}_{ij}$ refer to the total spin carried by the correlator. The conservation of the vector current implies $\Pi_{ij,V}^{(0)}=0$. The imaginary parts of $\Pi_{ij}$ give the spectral functions:
\begin{equation}
\begin{array}{l}
\displaystyle \mbox{Im}\Pi_{ij, V}^{(1)}(s)=\frac{1}{2\pi}v_1^{ij}(s),\\ \\
\displaystyle \mbox{Im}\Pi_{ij, A}^{(1)}(s)=\frac{1}{2\pi}a_1^{ij}(s),\\ \\
\displaystyle \mbox{Im}\Pi_{ij, A}^{(0)}(s)=\frac{1}{2\pi}a_0^{ij}(s),
\label{specfuncdef}
\end{array}
\end{equation}
which provide the basis for comparing short distance theory with hadronic data.

Defined as a time ordered product of local operators, the hadronic polarisation tensor can be rewritten using the OPE (see Section \ref{ope}):
\begin{equation}
\Pi_{ij}^{(J),V/A}(s)=\sum_{d\ge0}\frac{{\cal O}_d^{(J),V/A}}{(-s)^{d/2}},
\label{opeexp}
\end{equation}
where ${\cal O}_d\equiv {\cal C}_d\langle{\cal O}_d\rangle$ is the short hand notation for the QCD non-perturbative condensate $\langle{\cal O}_d\rangle$ of dimension $d$ and its associated perturbative Wilson coefficient ${\cal C}_d$; $s\equiv q^2$ is the momentum transfer. $\langle{\cal O}_d\rangle$ are vacuum expectation values of gauge invariant scalar operators constructed out of quark and gluon fields and called condensates. A more detailed discussion of them is found in Chapter \ref{condensates}.

The dimension $d=0$ contribution to (\ref{opeexp}) is entirely given by perturbation theory (recall that $\langle{\cal O}_0\rangle=I$, Section \ref{ope}). For that reason it is useful to separate the two contributions in (\ref{opeexp})
\begin{equation}  
\Pi_{ij}^{(J),V/A}(s)=\Pi_{ij,{\rm PT}}^{(J),V/A}(s)+\Pi_{ij,{\rm OPE}}^{(J),V/A}(s).
\label{corrseparation}
\end{equation}

A convenient way to calculate the perturbative part is to use the Adler function which is perturbatively constructed as an expansion in the coupling constant:
\begin{equation}
 D(s)\equiv -s\frac{d}{ds}\Pi(s)=\frac{1}{4\pi^2}\sum_{n\ge 0}K_na^n,\ \ a\equiv\frac{\alpha_s}{\pi}\ \left(\Pi(s)\equiv\Pi_{ij,{\rm PT}}^{(J),V/A}(s)\right).
\end{equation}
 The renormalization group equation for $a(s)$, $s=-Q^2$, reads:
\begin{equation}
\begin{array}{l}
 \displaystyle\frac{da}{d\ln Q^2}=-\beta(a)=-\sum_{n\ge0}\beta_na^{n+2}\\ \\
 \displaystyle\hspace{2cm}\Rightarrow\ \ln\frac{Q^2}{\mu^2}=-\int_{a(\mu^2)}^{a(Q^2)}\frac{da}{\beta(a)}.
\end{array}
\end{equation}
In the $\overline{\rm MS}$ scheme for 3 flavours $\beta_0=9/4$, $\beta_1=4$, $\beta_2=10.06$, $\beta_3=47.23$ \cite{larin,ritbergen,tarasov}. This allows one to get the perturbative contribution to the polarisation operator explicitly at any order of perturbation theory:
\begin{equation}
 \Pi(s)-\Pi(\mu^2)=\frac{1}{4\pi^2}\int_{a(\mu^2)}^{a(Q^2)}D(a)\frac{da}{\beta(a)}.
\end{equation}
 
The second term in Eq.(\ref{corrseparation}) is given by the OPE expansion, but now starting with dimension $d=2$.
\begin{figure}[H]
\centering
\includegraphics[height=5cm,angle=0]{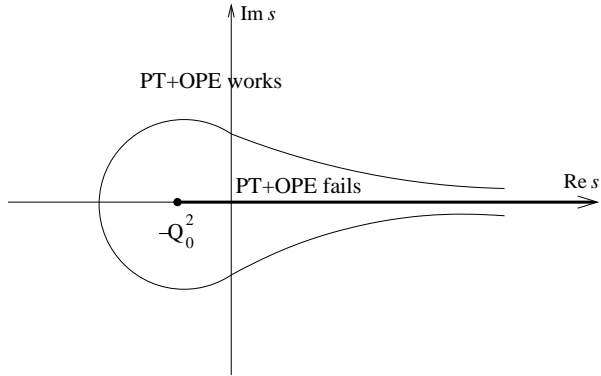}
\caption{Region of validity of the operator product expansion.}
\label{opevalidity}
\end{figure}

The Adler function constructed in this way has an unphysical cut from $s=-Q_0^2$ to $s=0$. It is an obvious indication of the fact that QCD is inapplicable at low $|s|$. Also, the expression (\ref{opeexp}) is not valid for all values of $s$, the series not converging for small $|s|$, where the effective degrees of freedom are hadrons rather than quarks. The exact polarisation tensor is known to be an analytical function of $s$ with a cut along the positive real semi-axis, while the OPE series with a finite number of operators has a very singular behaviour at $s=0$ and a possible cut, on the positive real semi-axis, starting at some $s_0>0$. Based upon these arguments one may draw schematically in Fig.\ref{opevalidity} the region of validity of the series (\ref{opeexp}).
\clearpage

\section{Dispersion relations}
\label{dispersionrelations}

A dispersion relation connects the real part of an analytic function to its absorbative (imaginary) part. The name is due to physicists and it is not found in the mathematical literature. Similar problems have been studied by the mathematicians under names such as Hilbert transforms, the Riemann-Hilbert problem, the Wiener-Hopf method and Carleman's method. The physicist's name arises from the original applications of the technique by Kramers and Kronig to the study of the index of refraction describing the dispersion of light waves. More recently, further refinements of the method have been employed extensively in elementary particle theory as an alternative to perturbative QCD.

We will first briefly state the mathematical basis without any indication to physics. Later on, an application to the two point functions of the previous section will be considered.

Let us consider a complex-valued function, $F(s)$ (which will later be identified with $\Pi(s)$), of complex argument, $s$, and assume that:
\begin{itemize}
 \item $F(s)$ is real for real $s<s_0$;
 \item $F(s)$ has a branch cut for real $s>s_0$;
 \item $F(s)$ is analytic for complex $s$ (except along the cut).
\end{itemize}

\begin{figure}[H]
\centering
\includegraphics[height=5cm,angle=0]{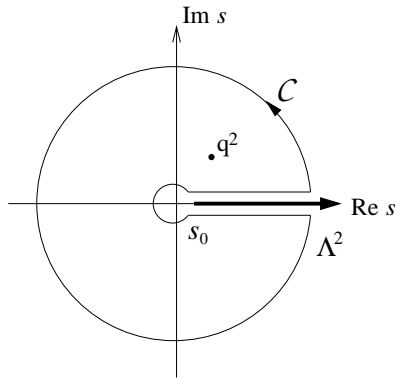}
\caption[Contour of integration for an analytic function in the complex $s$-plane, having a cut on the positive real semi-axis.] {Contour ${\cal C}$ of Eq.(\ref{disprel1}) in the complex $s$-plane.} 
\label{cauchycont}
\end{figure}

We fix the sign of the absorbative (imaginary) part of $F$ along the branch cut by
\begin{equation}
 F(s+i\varepsilon)={\rm Re}F(s)+i{\rm Im}F(s),
\end{equation}
where $\varepsilon>0$ is infinitesimal. Since $F$ is analytic at each point $q^2$ within the contour ${\cal C}$ (Fig.\ref{cauchycont}), we may apply Cauchy's theorem to find
\begin{equation}
 \begin{array}{lcl}
\displaystyle  F(q^2)&=&\displaystyle\frac{1}{2\pi i}\oint_{{\cal C}}ds\,\frac{F(s)}{s-q^2}\\
\\
&=&\displaystyle\frac{1}{2\pi i}\left(\int_{s_0}^{\Lambda^2}\!ds\,\frac{F(s+i\varepsilon)-F(s-i\varepsilon)}{s-q^2}+\oint_{|s|=\Lambda^2}\!ds\,\frac{F(s)}{s-q^2}\right)\\
\\
&=&\displaystyle\frac{1}{\pi}\int_{s_0}^{\Lambda^2}\!ds\,\frac{{\rm Im}F(s)}{s-q^2-i\varepsilon}+\frac{1}{2\pi i}\oint_{|s|=\Lambda^2}\!ds\,\frac{F(s)}{s-q^2},
 \end{array}
\label{disprel1}
\end{equation}
where, in the last step, we have used Schwartz' reflection principle:
\begin{equation}
 F(s+i\varepsilon)-F(s-i\varepsilon)=2i{\rm Im}F(s).
\end{equation}
Suppose that we only know ${\rm Im}F$ along the branch cut and wish to evaluate $F$ at some point $q^2$. Then, Eq.(\ref{disprel1}) is not useful since $F$ also appears on the r.h.s.\ under the integral along the circle. However, if
\begin{equation}
 \lim_{\Lambda^2\rightarrow\infty}\oint_{|s|=\Lambda^2}\!ds\,\frac{F(s)}{s-q^2}=0,
\label{cond0}
\end{equation}
then we obtain the unsubtracted dispersion relation
\begin{equation}
 F(q^2)=\frac{1}{\pi}\int_{s_0}^{\Lambda^2}\!ds\,\frac{{\rm Im}F(s)}{s-q^2-i\varepsilon}.
\end{equation}
This means that $F$ can be reconstructed at any point $q^2$ from the knowledge of its absorbative part along the cut. In particular, the dispersive (real) part of $F$ may be evaluated from 
\begin{equation}
 {\rm Re}F(q^2)=\frac{1}{\pi}\PP\int_{s_0}^{\Lambda^2}\!ds\,\frac{{\rm Im}F(s)}{s-q^2-i\varepsilon}.
\end{equation}

In general, Eq.(\ref{cond0}) is not satisfied, but instead one has
\begin{equation}
 \lim_{\Lambda^2\rightarrow\infty}\oint_{|s|=\Lambda^2}\!ds\,\frac{F(s)}{s-q^2}=a+bq^2+...\ ,
\end{equation}
and thus
\begin{equation}
 {\rm Re}F(q^2)=\frac{1}{\pi}\PP\int_{s_0}^{\Lambda^2}\!ds\,\frac{{\rm Im}F(s)}{s-q^2-i\varepsilon}+a+bq^2+...\ .
\label{disprel2}
\end{equation}
The coefficients of the polynomial in (\ref{disprel2}), $a$, $b$, $...$, depend on the properties of $F$. 

For the two point functions of the previous section, $\Pi(s)$, one will then have the dispersion relation  
\begin{equation}
\mbox{Re}\Pi(q^2)=\frac{1}{\pi}\PP\int_{s_0}^\infty\!ds\,\frac{1}{s-q^2}\,\mbox{Im}\Pi(s)+a+bq^2+...\ .
\label{realpi}
\end{equation}
Here, the physical meaning of the coefficients $a$, $b$, $...$, depends of course on the choice of the local 
operator $J(x)$ in the two-point function. In some cases the coefficients in question 
are fixed by low-energy theorems; e.g.\ if $\Pi(0)$ is known, we can trade the constant 
$a$ in (\ref{realpi}) for $\Pi(0)$:
\begin{equation}
\mbox{Re}\Pi(q^2)=\mbox{Re}\Pi(0)+\frac{1}{\pi}\PP\int_{s_0}^\infty\!ds\,\frac{q^2}{s}\,\frac{\mbox{Im}\Pi(s)}{s-q^2}+bq^2+...\ .
\end{equation}
In general, it is always possible to get rid of the polynomial terms by taking an appropriate number of derivatives with respect to $q^2$.

\clearpage{\thispagestyle{empty}\cleardoublepage}

\chapter{Hadronic spectral functions}

The $\tau$ lepton was discovered in 1974 from a handful of $e-\mu$ events. In the 30 years following the discovery, $\tau$ physics has matured into a field of high precision, testing many aspects of our current understanding of particle physics:

\begin{itemize}
 \item The coupling of the $\tau$ lepton to the charged weak current has been tested at the level of a few parts per thousand. Its couplings to the weak neutral current have been measured with similar precision.

\item Substantial contributions to precision tests of the electroweak theory at LEP and SLC have been derived from $\tau$ production data.

\item The coupling to the photon has been investigated. No anomalies have been found so far.

\item More than 100 different, mainly hadronic, branching ratios have been measured, allowing tests of QCD and many model predictions.

\item The spectral functions have been determined. They allow one to study perturbative QCD at low energy scales and provide one of the most precise measurements of the strong coupling constant $\alpha_s$.

\item The structure of the charged current in $\tau$ decays has been examined for deviations from $V-A$. These investigations are sensitive to many kinds of new physics.

\item CP violation in the production or decay of the $\tau$ lepton is a very interesting question which has been analysed in a search for a deeper understanding of the origin of CP violation.

\item In the light of neutrino oscillations, lepton-flavour-violating $\tau$ decays are expected to occur at some level.

\item The searches for a finite mass of the $\tau$ neutrino are approaching 10 MeV, thereby closing an interesting astrophysical window. 
\end{itemize}

We are especially interested in the hadronic spectral functions. They are the basic ingredients to the theoretical description of $\tau$ decays, since they represent the probability to produce a given hadronic system from the vacuum, as a function of its invariant mass squared $s$. The spectral functions are dominated by known resonances, but tend to approach the quark-level asymptotic regime at $m_\tau^2$. They have been determined for individual hadronic modes, however they are summed and separated into their inclusive vector and axial-vector components for the non-strange part. The Cabbibo-suppressed strange part cannot be separated at present, due to the lack of necessary experimental information.

The non-strange $\tau$ vector spectral functions can be compared to the corresponding quantities obtained in $e^+e^-$ annihilation by the virtue of isospin symmetry. The precision reached in the experimental data makes it necessary to correct for isospin-symmetry breaking. Beyond overall agreement, the detailed comparison unveils discrepancies with results from $e^+e^-$ data annihilation. The vector spectral functions are necessary ingredients to compute vacuum polarisation integrals, required for the evaluation of the running coupling constant $\alpha_s$ or the anomalous magnetic moment of the muon $a_\mu$. The disagreement leads to different results when using $\tau$ or $e^+e^-$ spectral functions. While the $e^+e^-$-based theoretical value of $a_\mu$ disagrees with the measured one by 3.4$\sigma$ \cite{millerJ}, possibly indicating contributions from physics beyond the SM, the $\tau$-based calculation is consistent with experiment. Although the use of $e^+e^-$ data is {\em a priori} more direct, the present situation must evolve, requiring more experimental and theoretical cross-checks. 

The observation that hadronic spectral functions derived from $\tau$-decay data can provide precision information on perturbative QCD is surprising, considering the moderate energy scale involved in these decays. Hence it is useful to recall briefly the reasons behind this success:

\begin{itemize}

\item The total hadronic decay rate normalised to the leptonic width, $R_\tau$, is obtained by integrating the total spectral function weighted by a known kinematic factor originating from the $V-A$ leptonic tensor. This integral can be transformed to a complex contour integral over a circle at $|s|=m_\tau^2$, hence involving only large complex $s$ values where perturbative QCD can presumably be applied;

\item One factor in the weight function, $(1-s/m_\tau^2)^2$, cuts off the integrand in the vicinity of the real axis where poles of the current correlator are located;

\item The invariant-mass spectra in the hadronic $\tau$ decay, dominated by resonances, are thus related to the quark-level contributions expected at large mass. This global quark-hadron duality is expected to work best in more inclusive situations. In the case of hadronic $\tau$ decays, this condition seems to be well met, with few 
resonance contributions in each of the equally important vector and axial-vector components;

\item The perturbative expansion of the spectral function is known to the third order in $\alpha_s$ \cite{chetyrkin2,celmaster,dine,gorishny,surguladze}; estimates of the fourth order coefficient are also available \cite{baikov,kataev};

\item The perturbative and non-perturbative contributions can be treated systematically using the OPE, giving corrections with inverse powers $d$ of $m_\tau$;

\item The {\em a priori} dominant non-perturbative term from the gluon condensate $(d=4)$ is suppressed by the second factor of the weight function, $(1+2s/m_\tau^2)^{-1}$. This accidental fact renders hadronic $\tau$ decays particularly favourable since non-perturbative effects are expected to be small;

\item The next $d=6$ term is subject to a partial cancellation between the vector $V$ and axial-vector $A$ contributions, due to the $(1-\gamma_5)$ factor of the weak current. 
\end{itemize}
 
\section{Overview of experiments}

Since its discovery, the $\tau$ lepton has been studied with ever-increasing precision at every new $e^+e^-$ collider that has gone into operation. The appearance of the events changed with increasing energy of the machines and improvements in detectors. The samples become more numerous and more and more decay modes become available.

The history of $\tau$ physics started with a handful of events at SPEAR (MARK I), confirmed by the events from DORIS (PLUTO). The energy of DORIS was increased in steps to the $\Upsilon(4s)$ resonance to produce $B$ mesons, but it also was the first machine to provide a large number of $\tau$ pairs, recorded by ARGUS. Meanwhile the next generation of accelerators, PEP and PETRA, were built with centre-of-mass energies in the continuum around 30 to 40 ${\rm GeV}$ and with a number of experiments. Also CESR started running at the $\Upsilon(4s)$ with the CLEO experiment, which today has the largest $\tau$ sample. With TRISTAN the Japanese joined, with a machine again in the continuum at 50 to 60 ${\rm GeV}$. Despite good luminosity the $\tau$ production rate is low, as the cross section falls like $1/s$. In 1989 SLC, and shortly afterwards, LEP began running at the $Z^0$ boson. BEPC was constructed in Beijing to go back to the $\tau$ production threshold and to precisely remeasure, amongst other things, the $\tau$ mass. Table \ref{accelerators} lists the accelerators that have produced $\tau$ pairs and Table \ref{experiments} summarises the experiments that have analysed them.

The set of comprehensive measurements of exclusive hadronic branching ratios from ALEPH \cite{aleph,aleph05} and the non-strange spectral functions from ALEPH \cite{aleph,aleph05}, CLEO \cite{cleo} and OPAL \cite{opal} have yielded important contributions to the measurements of $\alpha_s(m_\tau^2)$ and study of perturbative QCD at low energies. Recent measurements of a set of semi-exclusive branching ratios made by DELPHI are also available \cite{delphi}.

Although no spectral function measurements have yet been presented by the B-factories, BELLE and BABAR, a number of exclusive branching ratio measurements of 3-prong and 5-prong final states from BABAR \cite{sobie} are reported. It is evident with this data in hand that work on understanding decay mechanisms and form-factors for these higher multiplicity states is now required.

We have chosen to use the final data from the ALEPH collaboration \cite{aleph05} because, as compared to those available from OPAL \cite{opal} and CLEO \cite{cleo}, they have the smallest experimental errors. Their quality has increased a lot as compared to the earlier ones \cite{aleph}: first, the higher statistics has allowed the ALEPH collaboration to double the number of bins, and, secondly, the experimental errors decreased due to both higher statistics and a better understanding of systematic errors. 

\begin{table}[h]
 \begin{center}
  \begin{tabular}{SlSlScSl}\hline\hline
\parbox[c][1cm][c]{2.5cm}{Accelerator} & \parbox[c][1cm][c]{3.5cm}{Laboratory} & \parbox[c][1cm][c]{3.0cm}{Energy in ${\rm GeV}$} & \parbox[c][1cm][c]{3.5cm}{Years of operation}\\\hline
SPEAR & SLAC, USA & 3 - 8 & 73 - 88\\
DORIS & DESY, Germany& 8 - 11 & 77 - 92\\
PETRA & DESY, Germany & 10 - 47 & 78 - 86 \\
PEP & SLAC, USA & 29 & 80 - 90\\
CESR & Cornell, USA & 9 - 12 & 79 - 02 \\
TRISTAN & KEK, Japan & 50 - 62 & 86 - 95\\
SLC & SLAC, USA & 91 & 89 - present\\
LEP & CERN, Europe & 88 - 200 & 89 - 00\\
BEPC & Beijing, China & 3 - 4 & 91 - present \\
PEP-II & SLAC, USA & 11 & 99 - present\\
KEK-B & KEK, Japan & 11 & 99 - present \\\hline\hline
  \end{tabular}
 \end{center}
\caption{Accelerators for $\tau$ physics.}
\label{accelerators}
\end{table}

\begin{table}[hp]
 \begin{center}
  \begin{tabular}{SlSlSlScScScSc}\hline\hline
\parbox[c][1cm][c]{1.2cm}{Experiment} & \parbox[c][1cm][c]{1.2cm}{Accelerator} & \parbox[c][][c]{1.7cm}{Years\\ of\\ operation} & \parbox[c][][c]{1.7cm}{Typical \\$E_{\rm cm}$ \\in ${\rm GeV}$} & \parbox[c][][c]{1.3cm}{$N_{\tau^+\tau^-}$ produced} & \parbox[c][][c]{1.3cm}{Int. ${\cal L}$ \\in\\ ${\rm pb}^{-1}$} & \parbox[c][][c]{1.3cm}{Typical \\$\varepsilon$ in $\%$}\\\hline
ALEPH & LEP & 89 - 95 & 91 & 200 & 170 & 90 \\
AMY & TRISTAN & 86 - 94 & 50 - 62 & 4 & 150 & 40 \\
ARGUS & DORIS & 82 - 92 & 10.58 & 400 & 500 & 10 \\
BABAR & PEP-II & 99 - present& 10.58 & 50000 & $10^5$ &  \\
BELLE & KEK-B & 99 - present& 10.58 & 50000 & $10^5$ &  \\
BES & BEPC & 91 - present& 3.4 - 3.6 & 1.5 & 5 & 5  \\
CB & DORIS & 82 - 86 & 10.58 & 250 & 300 & 5  \\
CELLO & PETRA & 80-86 & 14 - 47 & 10 & 140 & 35  \\
CLEO & CESR & 79 - 02& 10.58 & 4300 & 4700 & 10  \\
DASP & DORIS & 78 - 78 & 3 -5 & 20 & 7 & 1  \\
DELCO & PEP & 81 - 84 & 29 & 15 & 150 & 20  \\
DELPHI & LEP & 89 - 95 & 91 & 200 & 170 & 90 \\
HRS & PEP & 78 - 86 & 29 & 30 & 300 & 20 \\
JADE & PETRA & 78 - 86 & 12 - 47 & 6 & 100 & 50\\
L3 & LEP & 89 - 95 & 91 & 200 & 170 & 90 \\
MAC & PEP & 80 - 86 & 29 & 20 & 200 & 30 \\
MARK I & SPEAR & 73 -77 & 3 - 8 & & & \\
MARK II & PEP & 79 - 84 & 29 & 20 & 200 & 20 \\
MARK III & SPEAR & 82 - 88 & 3.77 & 25 & 10 & 5 \\
MARK J & PETRA & 78 - 82 & 12 - 47 & 15 & 200 & 15 \\
OPAL & LEP & 89 - 95 & 91 & 200 & 170 & 90 \\
PLUTO & DORIS & 77 - 78 & 3 - 9 & & & \\
 & PETRA & 78 - 79 & 35 & 3 & 40 & 2 \\
SLD & SLC & 89 - 99 & 91 & 20 & 10 & 90 \\
TASSO & PETRA & 79 -86 & 14 - 47 & 10 & 200 & 8 \\
TOPAZ & TRISTAN & 90 - 95 & 52 - 62 & 7 & 280 & 20 \\
TPC/2$\gamma$ & PEP & 82 - 90 & 29 & 14 & 140 & 5 \\
VENUS & TRISTAN & 86 - 95 & 50 - 62 & 7 & 270 & 35 \\\hline\hline
 \end{tabular}
 \end{center}
\caption[Experiments in $\tau$ physics.]{Experiments in $\tau$ physics. The number of $\tau$ pairs produced, $N_{\tau^+\tau^-}$, is given in units of 1000. 'Int. ${\cal L}$' is the integrated luminosity recorded by the experiment. A typical efficiency for the identification of a $\tau$ pair is quoted as $\varepsilon$. Only the running periods relevant to the $\tau$ results are listed.}
\label{experiments}
\end{table}

\section{Definitions}
\label{spectralfuncdef}

The measurement of the $\tau$ vector and axial-vector current spectral functions requires the selection and identification of $\tau$ decay modes with fixed isotopic $G$-parity $G=+1$ and $G=-1$, and hence hadronic channels with an even and odd number of neutral or charged pions, respectively. Since hadronic final states of different $G$-parity differ also in their $J^P$ quantum numbers, there is no interference between these two states. Hence the total hadronic width separates into $\Gamma_{\rm had}=\Gamma_V+\Gamma_A$. 

The spectral functions are obtained by dividing the normalised invariant mass-squared distribution $dR_{\tau,V/A}/ds$ for a given hadronic mass $\sqrt s$ by the appropriate kinematic factor. They are then normalised to the branching fraction of the massless leptonic, i.e. electron, 
channel ${\cal B}_e=(17.810\pm0.039)\%$ \cite{aleph05}.  
\begin{equation}
\begin{array}{r}
\displaystyle v_1(s)=\frac{m_\tau^2}{6|V_{ud}|^2S_{\rm EW}}
\frac{dR_{\tau,V}}{{\cal B}_eds}\left[\left(1-\frac{2}{m_\tau^2}\right)^2
\left(1+2\frac{s}{m_\tau^2}\right)\right]^{-1},\\
\\
\displaystyle a_1(s)=\frac{m_\tau^2}{6|V_{ud}|^2S_{\rm EW}}
\frac{dR_{\tau,A}}{{\cal B}_eds}\left[\left(1-\frac{2}{m_\tau^2}\right)^2
\left(1+2\frac{s}{m_\tau^2}\right)\right]^{-1},\\
\\
\displaystyle a_0(s)=\frac{m_\tau^2}{6|V_{ud}|^2S_{\rm EW}}
\frac{dR_{\tau,A}}{{\cal B}_eds}\left(1-\frac{2}{m_\tau^2}\right)^{-2}.
\end{array}
\label{specdef}
\end{equation}
$S_{\rm EW}=1.0198\pm0.0006$ accounts for short distance electroweak radiative corrections \cite{marciano} and the CKM mixing matrix element has the value $|V_{ud}|=0.9746\pm0.0006$ \cite{davier}. Due to the conservation of the vector current, there is no $J=0$ contribution to the vector spectral function, while the only contribution to $a_0$ is assumed to come from the pion pole. $a_0$ is connected via partial conservation of the axial-vector current (PCAC) to the pion decay constant $f_\pi$ through $a_{0,\pi}(s)=2\pi^2f_\pi^2\delta(s-m_\pi^2)$.

\section{The mass spectra}

In this section we will sumarise the experimental analysis techniques used by the ALEPH Collaboration as described in Ref. \cite{davier3}. 

The measurement of the $\tau$ spectral functions defined in (\ref{specdef}) requires the determination of the invariant mass-squared distributions, obtained from the experimental distributions after unfolding the effects of measurement distortions. The unfolding procedure used by the ALEPH collaboration follows a method based on the regularised inversion of the simulated detector response matrix using the SVD technique \cite{hocker}. The regularisation function minimises the average curvature of the distribution. The optimal choice of the regularisation strength is found by means of a MC simulation where the true distribution is known and chosen to be close to the expected physical distribution.

To measure exclusive spectral functions, individual unfolding procedures with specific detector response matrices and regularisation parameters are applied for each $\tau$ decay channel considered. An iterative procedure is followed to correct the MC spectral functions used to subtract the cross-feed among the modes.

Each spectral function is determined in $140$ mass-squared bins of equal width ($0.025\mbox{ GeV}^2$).

All systematic uncertainties concerning the decay classification are contained in the covariance matrix of the branching ratios. Therefore only the systematic effects affecting the shape of the mass-squared distributions, and not its normalisation, need to be examined.

All effects affecting the decay classification and the calculation of the hadronic invariant mass are considered in turn. Comparisons of data and MC distributions are made and the corresponding biases are corrected, while the uncertainty in the correction is taken as input for the calculation of the systematic uncertainty. In the case of the spectral functions, the whole analysis including the unfolding procedure is repeated, for each systematic effect. This generates new mass distributions which are compared bin-by-bin to the nominal ones, hence providing the full $140\times 140$ covariance matrix of the spectral function for the studied effect.

The systematic studies include the effects from the photon and $\pi^0$ energy calibration and resolution, the photon detection efficiency, the shape of the identification probability distribution, the estimate of the number of fake photons, the proximity in the calorimeter of other photon showers and of energy deposition by charged particles, and the separation between radiative and $\pi^0$ decay photons for residual single photons.

In addition, systematic errors introduced by the unfolding procedure are tested by comparing known, true distributions to their corresponding unfolded ones.

Finally, systematic errors due to the limited MC statistics and to uncertainties in the branching ratios are added.

In order to illustrate the importance of these systematic uncertainties, one may perform an integration over the spectral functions with some given kernel, characteristic of a given physical problem. The integration error is then obtained by Gaussian error propagation taking into account the correlations. Using moderately $s$-dependent integration kernels, the integration error is dominated by normalisation uncertainties, i.e., the errors on the contributing $\tau$ branching ratios. However, the error on an integration with a strongly $s$-dependent weighting kernel enhancing the low energy parts of the spectral functions is dominated by systematics (mainly due to the fake photon rejection and the photon efficiency correction at threshold), while the central energy region (0.6 - 1.4 $\mbox{GeV}^2$) is statistically limited. When enhancing the higher part of the spectrum, the integration error is equally dominated by uncertainties due to the unfolding process, and by limited data and MC statistics.

\section{Inclusive non-strange spectral functions}

The ALEPH collaboration \cite{aleph05} have provided the spectral functions for the hadronic modes, separated into the vector and axial-vector contributions. 

\subsection*{Vector and axial-vector spectral functions}

\begin{figure}[ht]
\centering
\subfigure[]{\label{v1}\includegraphics{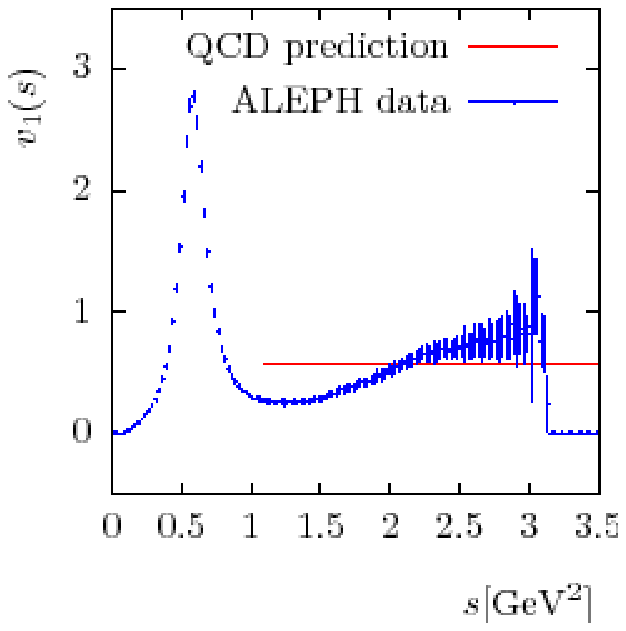}}
\subfigure[]{\label{a1}\includegraphics{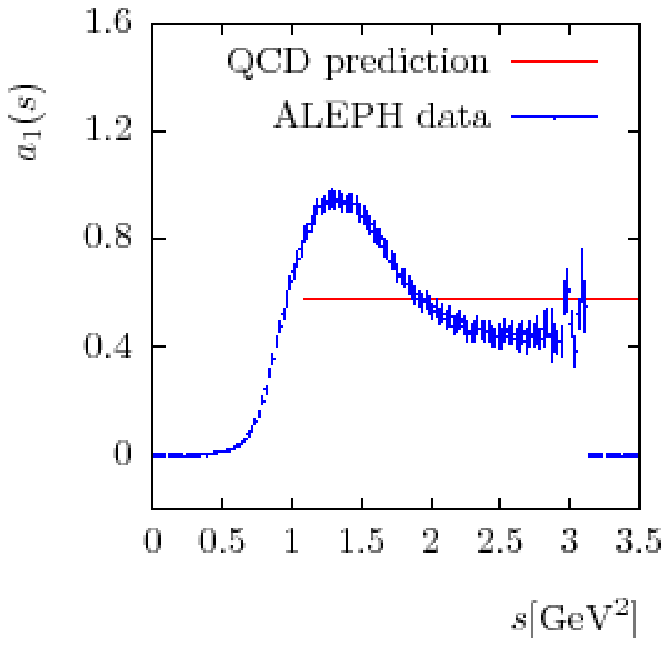}}
\caption[Inclusive vector and axial-vector spectral functions as measured by the ALEPH collaboration.]{Inclusive vector and axial-vector spectral functions as measured by the ALEPH collaboration \cite{aleph05}: (a) Vector spectral function $v_1(s)$; (b) Axial-vector spectral function $a_1(s)$.}
\label{v1&a1}
\end{figure}

The inclusive $\tau$ vector and axial-vector spectral functions are shown in Fig.\ref{v1&a1}. The solid line depicts the massless perturbative QCD prediction. Although the statistical power of the data is weak near the kinematic limit, the trend of the spectral functions clearly indicates that the asymptotic region \cite{gorishny,surguladze} is not reached. The QCD prediction lies roughly 25$\%$ (20$\%$) lower (higher) than the data at $m_\tau^2$ for the vector (axial-vector) channel.

To obtain the vector spectral function the two- and four-pion final states were measured exclusively, while the six-pion state was only partly measured. The total six-pion branching fraction has been determined using isospin symmetry from the two- and four-pion ones.

In complete analogy to the vector spectral function the inclusive axial-vector spectral function is obtained by summing up the exclusive axial-vector spectral functions (odd number of pions in the final state), with the addition of small unmeasured modes taken from the MC simulation. 

\subsection*{$V\pm A$ spectral functions}
 
For the total $V+A$ hadronic spectral function one does not have to distinguish the properties of the nonstrange hadronic $\tau$ decay channels. Hence the mixture of all contributing non-strange final states is measured inclusively.

\begin{figure}[ht]
\centering
\subfigure[]{\label{v+a}\includegraphics{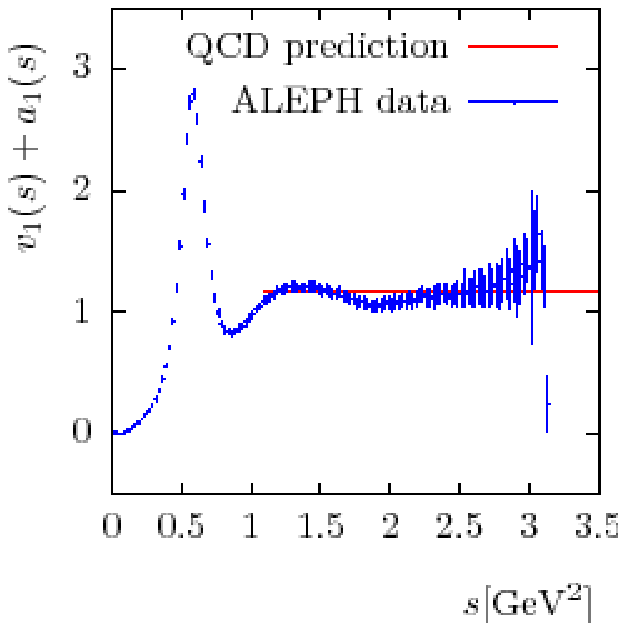}}
\subfigure[]{\label{v-a}\includegraphics{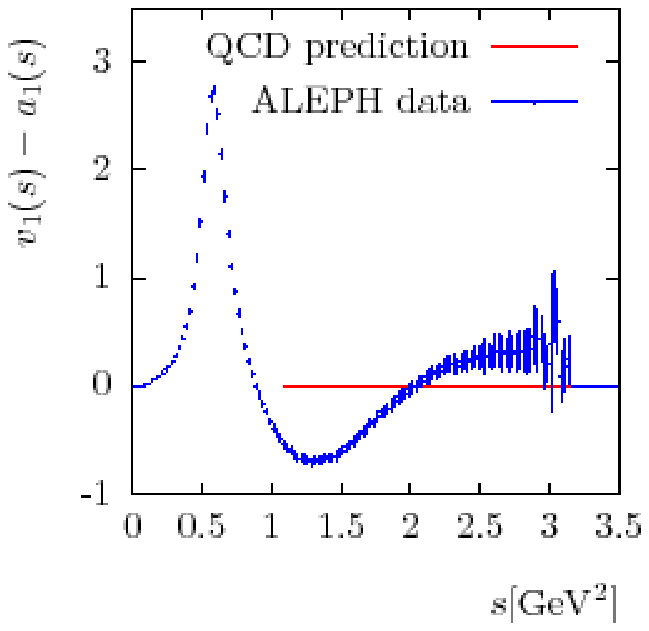}}
\caption[Inclusive vector plus axial-vector and vector minus axial-vector spectral functions as measured by the ALEPH collaboration.]{Inclusive vector plus axial-vector and vector minus axial-vector spectral functions as measured by the ALEPH collaboration \cite{aleph05}: (a) Vector plus axial-vector spectral function $v_1(s)+a_1(s)$; (b) Vector minus axial-vector spectral function $v_1(s)-a_1(s)$.}
\label{vpma}
\end{figure}

The one, two and three-pion final states dominate and their exclusive measurements are added with proper accounting for (anti-) correlated errors. The remaining contributing topologies are treated inclusively, 
i.e., without separation of the vector and axial-vector decay modes. The effect of the feed-through between $\tau$ final states on the invariant mass spectrum is described by MC simulation and thus corrected in the data 
unfolding. In this procedure the simulated mass distributions are iteratively corrected using the exclusive vector/axial-vector unfolded mass spectra. The $V+A$ spectral function is depicted in Fig.\ref{v+a}. 
The improvement in precision when comparing to the exclusive sum of the two parts in Fig.\ref{v1&a1} is significant at higher mass-squared values.

One nicely identifies an oscillating behaviour of the spectral function due to resonance contributions. It does also  roughly reach (within errors) the asymptotic limit predicted by perturbative QCD at $m_\tau^2$.

In the case of the $V-A$ spectral function, uncertainties on the V/A separation are reinforced due to their anti-correlation. In addition, anti-correlations given between $\tau$ final states with adjacent numbers of pions increase the errors. The $V-A$ spectral function is shown in Fig.\ref{v-a}. The oscillating behaviour of the $v_1$ and $a_1$ spectral functions is emphasised but the asymptotic regime seems not to be reached at $s=m_\tau^2$. However, the strong oscillation generated by the hadron resonances mostly averages out to zero, as predicted by perturbative QCD. 

\clearpage{\thispagestyle{empty}\cleardoublepage}
\chapter{Extraction of QCD condensates}
\label{condensates}

Nowadays, it is reliably established that the microscopic theory of the strong interaction is QCD, the gauge theory of interacting quarks and gluons. It is also established, that unlike, e.g., QED, the vacuum in QCD has a nontrivial structure: due to nonperturbative effects, i.e.\ not admitting the expansion in the intercation constant (even if it is small) there persist non-zero fluctuations of gluonic and quark fields in the QCD vacuum. The nontrivial structure of QCD manifests itself in the presence of vacuum condensates. It is very important to determine these condensates, to study if one can indeed obtain a consistent description of the low energy hadronic physics, to get more insight into the properties of the QCD vacuum, and to confront them to theoretical estimates (from instanton calculus or lattice computation) in order to test the validity of the condensates from experimental data, but also the accuracy of the determination, i.e.\ to determine their allowed range.

Condensates, in particular quark and gluonic ones, were investigated starting form the 70-ties. Here, first, one should mention the QCD sum rule approach by Shifman, Vainshtein and Zakharov \cite{shifman1,shifman2}, which emphasised the leading role of condensates in the calculation of masses of the low-lying hadronic states. Starting with this pioneering work, a lot of papers were published on the extraction of condensates from experiment. For that purpose one has to relate error affected data in the time-like region to asymptotic QCD in the space-like region. This task of analytic continuation constitutes, mathematically, an ill-posed problem. In fact, extracting condensates from data is highly sensitive to data errors. Not surprisingly, results form different collaborations have not been always consistent.

In what follows we will first discuss general properties of the condensates and briefly review previous extractions. Then, a functional approach for the extraction of condensates form $\tau$-decay data will be presented.

\section{Condensates: general properties and previous extractions}

As already stated in Section \ref{hadronicpolarization}, QCD condensates are vacuum expectation values of gauge invariant scalar operators constructed out of quark and gluon fields. We will review here the familiar ones, describe their properties and give numerical values extracted previously from low-energy and lattice data.
\clearpage
\begin{itemize}

\item {\bf Quark condensate: $\displaystyle\langle\bar qq\rangle$}
 
The quark condensate has the lowest dimension ($d=3$) and it is one of the condensates violating the chiral symmetry. We may rewrite it in the form
\begin{equation}
 \langle\bar qq\rangle=\langle\bar q_Lq_R+\bar q_Rq_L\rangle,
\label{quarkcond}
\end{equation}
where $q_L$, $q_R$ are the fields of left- and right-chiral quarks. As follows form (\ref{quarkcond}), a non-zero value of the quark condensate means the transition of left-handed quark fields into right-handed ones and would lead to chiral symmetry violation in QCD. (If chiral symmetry is not violated spontaneously, then non-zero quark masses will induce $\langle\bar qq\rangle\neq0$.) 

The quark condensate is related by the Gell-Mann-Oakes-Renner relation \cite{gell-mann} to measurable constants and the quark masses:
\begin{equation}
 \langle\bar qq\rangle=-\frac{1}{2}\frac{f_\pi^2m_\pi^2}{m_u+m_d}.
\end{equation}
Here $m_\pi$, $f_\pi$ are the mass and the decay constant of the $\pi^+$-meson ($m_\pi=140{\rm MeV}$, $f_\pi=131{\rm MeV}$), $m_u$ and $m_d$ are the masses of $u$ and $d$-quarks. For $m_u=4.2{\rm MeV}$, $m_d=7.5{\rm MeV}$, the quark condensate takes the value:
\begin{equation}
 \langle\bar qq\rangle=-1.4\times 10^{-2}{\rm GeV}^3.
\label{valquarkcond}
\end{equation}
 The value (\ref{valquarkcond}) has a characteristic hadronic scale. This shows that chiral symmetry, which is satisfied with a good accuracy in the light quark Lagrangian ($m_u,m_d/M\sim0.01$, $M$ is the hadronic mass scale, $M\sim0.5-1{\rm GeV}$), is spontaneously violated in hadronic state spectrum. Due to this fact, the quark condensate may be considered as an order parameter for spontaneous chiral symmetry breaking.

\item {\bf Gluon condensate: $\displaystyle\langle \frac{\alpha_s}{\pi}G_{\mu\nu}^aG_{\mu\nu}^a\rangle$} 

The gluon condensate has dimesion $d=4$ and was initially introduced by Shifman, Vainshtein and Zakharov in Ref.\cite{shifman1} and estimated phenomenologically by different groups in the literature \cite{bell,bell1,bell2,bertlmann1,bertlmann,dominguez0,ioffe2,kremer,launer,narison8,narison3,narison9,narison7,narison6,narison2,reinders,shifman1,yndurain}. Its value is found to be definitely non-zero and positive. However, results from different groups differ considerably. Some of the results, including the value from Ref.\cite{shifman1}, do not satisfy the lower bound derived by Bell and Bertlmann \cite{bell} from moment sum rules. Recent estimates in \cite{narison2}, satisfying this bound from light and heavy quark channels, are of order $(7.1\pm0.9)\times10^{-2}{\rm GeV}^4$. We may add the remark that the gluonic condensate conserves chirality.
\clearpage

\item {\bf Mixed quark-gluon condensate: $\displaystyle g\langle\bar qG_{\mu\nu}^a\frac{\lambda_a}{2}\sigma^{\mu\nu}q\rangle$}

The next in dimension ($d=5$) is the mixed quark-gluon condensate which has the form:
\begin{equation}
 g\langle\bar qG_{\mu\nu}^a\frac{\lambda_a}{2}\sigma^{\mu\nu}q\rangle\equiv m_0^2\langle\bar qq\rangle.
\end{equation}
The value of the parameter $m_0^2$ has been estimated from baryon sum rules \cite{belyaev,chung,chung1,ioffe3,ioffe4}, the $B-B^*$ mass splitting \cite{narison4}, a string model \cite{giacomo1} and lattice calculations, with a fair agreement. Note that the mixed quark-gluon condensate is also violating the chiral symmetry.

\item {\bf Four-quark condensate: $\displaystyle\langle\bar q\Gamma_1 q\bar q\Gamma_2 q\rangle$}

 The general form of the dimension $d=6$ four-quark condensate is as follows:
\begin{equation} 
 \langle\bar q\Gamma_1 q\bar q\Gamma_2 q\rangle,
\label{fourquark}
\end{equation}
where $\Gamma_1,\Gamma_2$ are combinations of Dirac matrices. Following \cite{shifman1,shifman2}, Eq.(\ref{fourquark}) is usually factorized, and in the sum over intermediate states in all channels only the vacuum state is taken into account. The accuracy of such an approximation is $\sim 1/N_c^2$, with $N_c$ the number of colours, i.e., $\sim 10\%$. After factorisation, Eq.(\ref{fourquark}) reduces to
\begin{equation}
 \langle\bar qq\rangle^2.
\end{equation}
 
The four-quark condensate was estimated from different channels \cite{ciulli,launer,narison2}. The corresponding results indicate a violation of the factorisation assumption by at least a factor of 2.  

\item {\bf Triple gluon condensate: $\displaystyle\langle g^3f_{abc}G^aG^bG^c\rangle$} 

The triple gluon condensate ($d=6$) was originally estimated by Shifman, Vainshtein and Zakharov \cite{shifman1} using an instanton liquid model. Its value was confirmed later by lattice calculations \cite{giacomo,gurci}.

\item {\bf Higher-order condensates} 

The dimension $d=8$ and higher dimension condensates were estimated in the $V$ \cite{narison3}, $V+A$ and $P+S$ \cite{narison5} channels using experimental and lattice data \cite{degrand}.

\end{itemize}

More recently, condensates of the $V-A$ channel have been estimated from $\tau$-decay data \cite{aleph,almasy,bordes,cirigliano1,ciulli,dominguez,ioffe,narison1,opal,rojo,zyablyuk} and from QCD at large $N_c$ \cite{friot}. Large violations of the vacuum saturation estimates of these condensates have been observed also in these analyses.

\section{The method: a functional approach}
\label{approach}

In what follows we will study a functional method\footnote{\sf The functional method underlying the analysis has first been described in \cite{auberson,auberson2,causse}.}, introduced in Ref.\ \cite{ciulli}, which allows, in principle, to extract the condensates within rather general assumptions.

For our purpose, let us consider a set of functions $F(s)$ which are admissible as a representation of the true amplitude if they are real analytic functions in the complex $s$-plane cut along the time-like interval $[s_0,\infty)$. The asymptotic behaviour of $F(s)$ is restricted by fixing the number of subtractions in the dispersion relation between $F(s)$ and its imaginary part $f(s)=\mbox{Im} F(s+i0)$ along the cut 
\begin{equation}
F(s)=\frac{1}{\pi}\int_{s_0}^\infty K(s,z)f(z)dz+\mbox{subtractions}\ .
\label{disprel}
\end{equation}
In general, $K(s,z)$ is the usual Cauchy kernel but we prefer this more general notation since later on we will use also more complex kernels, e.g.\ derivatives of the Cauchy kernel.

In order to determine $F(s)$ and $f(s)$ we use the following two available sources of information:
\begin{itemize}
\item experimental data	measured in the time-like interval $\Gamma_{\rm exp}=[s_0,s_{\rm max}]$, $s_0>0$: $f_{\rm exp}(s)$;
\item theoretical model given by perturbative QCD, i.e.,
\begin{itemize}
\item the prediction for $F(s)$ in the space-like interval $\Gamma_L=[s_2,s_1]$: $F_{QCD}(s)$
\item and $\left.f_{QCD}(s)=\mbox{Im}F_{QCD}(s+i0)\right|_{s\in(s_{\rm max},\infty)}$ 
since QCD is expected to be reliable for large energies. 
\end{itemize}
\end{itemize}

As a next step in extracting values for the condensates, we split the integral on the r.h.s.\ of the dispersion 
relation (\ref{disprel}) into two parts: one that can be described by the experiment and the other one by the theoretical model, i.e., QCD:
\begin{equation}
\underbrace{F_{QCD}(s)-\frac{1}{\pi}\int_{s_{\rm max}}^\infty
K(s,z)f_{QCD}(z)dz}=\underbrace{\frac{1}{\pi}\int_{s_0}^{s_{\rm max}}K(s,z)f(z)dz}.
\label{dispsource}
\end{equation}
\vspace{0.2 cm}
\hspace{2.0 cm}  {QCD prediction: $\tilde F_{QCD}(s)$ }\hspace {2.0 cm} {experiment}

The goal of the method is to check if there exists a function $F(s)$ which is in accord with both the data on $\Gamma_{\rm exp}$ and the model on $\Gamma_L$. For doing this, one can use an $L^2$-norm\footnote{\sf The $L^2$-norm is defined as: $\displaystyle||f||^2\equiv (f,f)=\int |f(x)|^2\,dx$.} approach and define two functionals $\chi_L^2[f]$ and $\chi_R^2[f]$. $\chi_R^2[f]$ compares the true amplitude $f(s)$ with the data. Using its covariance matrix as a weight function, we can define:
\begin{equation}
\chi^2_R[f]=\frac{1}{|\Gamma_{\rm exp}|}\int_{s_0}^{s_{\rm max}}dx\int_{s_0}^{s_{\rm max}}dx'
V^{-1}(x,x')(f(x)-f_{\rm exp}(x))(f(x')-f_{\rm exp}(x')).
\end{equation}

As a measure for the agreement of the true function $f(s)$ with the theory, we define $\chi_L^2[f]$ by comparing the left and right hand sides of (\ref{dispsource})
\begin{equation}
\chi^2_L[f]=\displaystyle\frac{1}{|\Gamma_L|}\int_{\Gamma_L}w_L(x)
\left(\tilde F_{QCD}(x)-\displaystyle\frac{1}{\pi}\int_{s_0}^{s_{\rm max}}
K(x,x')f(x')dx'\right)^2dx,
\label{chiL2}
\end{equation}
where $w_L$ is a weight function for the space-like interval, i.e.\ an {\em a-priori} estimate of the accuracy of the QCD predictions and identified with $1/\sigma_L^2(s)$. $\sigma_L(s)$ is a continuous, strictly positive function of $s\in\Gamma_L$ which should encode errors due to the truncation of the perturbative series and the OPE. It is expected to decrease as $|s|\rightarrow\infty$ and diverge for $s\rightarrow0$. 

In order to find the true function $f(s)$ one can combine the information contained in these two functionals by means of Lagrange multipliers and find the unrestricted minimum of
\begin{equation}
{\cal F}[f]=\chi^2_L[f]+\mu\chi^2_R[f],
\end{equation}
subject to the condition
\begin{equation}
\chi_R^2[f]\le\chi_{\rm exp}^2=\frac{1}{N}\sum_{i,j}\sqrt{V(s_i,s_i)V(s_j,s_j)}V^{-1}(s_i,s_j),
\end{equation}
which will be the criteria to determine the Lagrange multiplier $\mu$. This procedure leads to an integral equation for the imaginary part of the true amplitude, $f(x;\mu)$ \cite{ciulli}:
\begin{equation}
\begin{array}{r}
f(x;\mu)=\displaystyle f_{\rm exp}(x)+\frac{\lambda|\Gamma_{\rm exp}|}{\pi|\Gamma_L|}
\int_{s_0}^{s_{\rm max}}dy\,V(y,x)\int_{\Gamma_L}dx'w_L(x')K(x',y)\tilde F_{QCD}(x')\\
\\
+\displaystyle\lambda\int_{s_0}^{s_{\rm max}}dx'K_2(x,x')f(x';\mu),
\end{array}
\label{inteq}
\end{equation}
where $\lambda=1/\mu$ and
\begin{equation}
K_2(x,x')=-\frac{|\Gamma_{\rm exp}|}{\pi^2|\Gamma_L|}\int_{s_0}^{s_{\rm max}}dyV(y,x)
\int_{\Gamma_L}dz\,w_L(z)K(z,y)K(z,x').
\end{equation}

The size of $\sigma_L$ determines the minimal value of $\chi_L^2$ according to Eq.(\ref{chiL2}) that can be reached by this algorithm. Obviously, widening the error corridor (increasing $\sigma_L$) will lead to values for $\chi_{L,{\rm min}}^2$ as small as desired. In such a case, the information obtained from the fit is not conclusive, since any model function $f(s)$ can be made consistent with the data if one allows for a wide enough error corridor. On the other hand, narrowing the error corridor will increase $\chi_{L,{\rm min}}^2$, signalling a bad fit, i.e., bad consistency of theory with data if the model function is not perfectly describing the data. However, a nontrivial result of our approach is the fact, that there exists a choice for the error corridor that leads to values $\chi_{L,{\rm min}}^2={\cal O}(1)$. We shall assume that the underlying probability distribution is Gaussian, and choose $\chi_{L,{\rm min}}^2$ and accordingly $\sigma_L$ such that the fit result corresponds to a $1\sigma$CL. In practice this is done by adjusting the value of the parameters in $\sigma_L(s)$. The underlying theory for $\chi^2$ minimisation can be found in Appendix \ref{statistics}.

The algorithm to determine acceptable values for the condensates is described below. The entire code, programmed in FORTRAN 77, can be found in \cite{almasy1}.
\vspace{1cm}

\begin{algorithm}[H]
\caption[Algorithm to determine acceptable values for the condensates.]{\sf Algorithm to determine acceptable values for the condensates.}
\dontprintsemicolon
\KwData{$f_{\rm exp}(x)$, $f_{QCD}(x)$, $F_{QCD}(x)$, $K(x,y)$.}
\KwResult{values for the condensates ${\bm{{\cal O}}}_0$ and confidence regions around them.}
\Begin{
initialise $M$ \tcc*{nr.\ of free parameters $\equiv$ dimension of ${\bm{{\cal O}}}_0$}\;
initialise $\mu$ \tcc*{Lagrange multiplier}\;
\Repeat{$\chi_R^2[f]=\chi_{\rm exp}^2$}{
	reinitialise $\mu$\;
	solve Eq.(\ref{inteq}) $\Rightarrow$ $f(x;\mu)$\;}
$f_{\rm opt}(x)\leftarrow f(x;\mu)$\;
$\mu_{\rm opt}\leftarrow \mu$\;
$\chi_L^2({\bm{{\cal O}}})\leftarrow\chi_L^2[f_{\rm opt}]({\bm{{\cal O}}})$\;
$\chi_{L,{\rm min}}^2$ $\leftarrow$ minimise $\chi_L^2({\bm{{\cal O}}})$ with respect to ${\bm{{\cal O}}}$\;
solve $\chi_L^2({\bm{{\cal O}}})=\chi_{L,{\rm min}}^2$ for ${\bm{{\cal O}}}$\;
${\bm{{\cal O}}}_0\leftarrow{\bm{{\cal O}}}$\;
solve $\chi_L^2({\bm{{\cal O}}})=\chi_{L,{\rm min}}^2+\Delta\chi^2$ for ${\bm{{\cal O}}}$\;
${\bm{{\cal O}}}-{\bm{{\cal O}}}_0$ defines a confidence region around ${\bm{{\cal O}}}_0$}
\end{algorithm}

\chapter{$V-A$ analysis}
\label{V-Aanalysis}

The $V-A$ correlator, $\Pi_{V-A}^{(0+1)}=\Pi_V^{(1)}-\Pi_A^{(1)}-\Pi_A^{(0)}$, is of particular interest, since it contains no truly perturbative contribution in the massless quark limit and it is entirely described by the OPE expansion.

To apply the method described in Section \ref{approach} to the $V-A$ channel, we identify $F(s)$ with $\Pi_{V-A}^{(0+1)}(s)$. We will use the appropriate combination of spectral functions, $v_1(s)-a_1(s)-a_0(s)$ as the experimental information. The spin-0 axial spectral function, $a_0(s)$, is basically saturated by the $\tau\rightarrow\pi\nu_\tau$ channel (see Section \ref{spectralfuncdef}) and can thus be taken into account separately.
Inserting this into the dispersion relation (\ref{disprel}) one finds: 
\begin{equation}
F(s)=\frac{1}{\pi}\int_{s_0}^\infty dzK(s,z)f(z)+\frac{f_\pi^2}{s},
\end{equation}
with
\begin{equation}
 K(s,z)=\frac{1}{z-s}.
\end{equation}
Note that for the $V-A$ correlator no subtractions are needed. The term $f_\pi^2/s$ comes from the pion pole$; f_\pi=0.1307\mbox{ GeV}$ \cite{partdatagroup} is the pion decay constant.

So, one has the information:
\begin{equation}
 f_{\rm exp}(s)=\frac{1}{2\pi}\left[v_1(s)-a_1(s)\right],
\end{equation}
\begin{equation}
 F_{QCD}(s)=\Pi_{V-A}^{(0+1)}(s)=\sum_{d\ge 6}\frac{{\cal O}_d^{V-A}}{(-s)^{d/2}}\left(1+c_D^{NLO}\frac{\alpha_s(\mu^2)}{\pi}+O(\alpha_s^2)\right)
\end{equation}
which can be used for the extraction of condensates. The QCD prediction, $\tilde F_{QCD}(s)$, will then be:
\begin{equation}
 \tilde F_{QCD}(s)=F_{QCD}(s)-\frac{f_\pi^2}{s}-\frac{1}{\pi}\int_{s_{\rm max}}^\infty K(s,z)f_{QCD}(z)dz.
\end{equation}

The complete expression for ${\cal O}_6^{V-A}$ is known to involve two operators \cite{cirigliano0,cirigliano}.  Assuming vacuum dominance or the factorisation approximation which holds, e.g., in the large-$N_C$ limit, the matrix elements can be written:
\begin{equation}
\begin{array}{l}
{\cal O}_6^{V-A}=\displaystyle-\frac{64\pi}{9}\alpha_s\langle\bar qq\rangle^2,\ c_6^{NLO}=\frac{1}{4}\left[\tilde c_6+\ln\left(\frac{\mu^2}{-s}\right)\right],\\
\\
{\cal O}_8^{V-A}=-4\pi\alpha_s i(1-N_C^{-2})\langle\bar qq\rangle\langle\bar q\gamma^{\alpha\beta}G_{\alpha\beta}q\rangle,\\
\\
{\cal O}_{10}^{V-A}=-\frac{8}{9}\pi\alpha_s\langle\bar qq\rangle^2\left(\frac{50}{9}m_0^4+32\pi\alpha_s\langle G_{\alpha\beta}G^{\alpha\beta}\rangle\right).
\end{array}
\label{v-a-cond}
\end{equation}
However, our analysis does not rely on the factorisation hypothesis since we are going to determine ${\cal O}_d^{V-A}$, but not the condensates separately. In the result for ${\cal O}_{10}^{V-A}$ \cite{zyablyuk}, $m_0^2$ is defined through the 5-dimensional quark-gluon mixed condensate. Starting from the second order, coefficients in perturbation theory depend on the regularisation scheme implying that the value of the condensates are scheme-dependent quantities. The NLO corrections $c_6^{NLO}$ for ${\cal O}_6^{V-A}$ were computed in \cite{lanin} and the constant $\tilde c_6$ was found equal to $247/12$. This calculation was based on the Breitenlohner-Maison definition of $\gamma^5$ in dimensional regularisation. A different treatment of $\gamma^5$ as used in Ref. \cite{adam} leads to $\tilde c_6=89/12$.

Since the $V-A$ correlator is entirely given by OPE terms and has no truly perturbative contribution, we will take as an error estimate the next higher dimension contribution in the OPE, depending on the condensates one wants to determine. 

\section{1-parameter fit: determination of ${\cal O}_6^{V-A}$}

We start with a discussion of results obtained by 1-parameter fits of the dimension $d=6$ condensate. We have chosen the dimension $d=8$ contribution to define the error channel on the space-like region, i.e. $\sigma_L(s)=\left|{\cal O}_8^{V-A}\right|_{\rm max}/s^4$ with $\left|{\cal O}_8^{V-A}\right|_{\rm max}$ in the order of $10^{-3}{\rm GeV}^8$.

\subsection{${\cal O}_6^{V-A}$ at leading-order}
\label{LOO6}
 
We have performed first the fit for the leading-order (LO) dimension $d=6$ condensate, i.e., we have treated ${\cal O}_6^{V-A}$ in (\ref{v-a-cond}) as a constant and took $c_6^{NLO}=0$. The corresponding QCD amplitude, and its imaginary part, are then:
\begin{equation}
F_{QCD}(s)=\frac{{\cal O}_6^{V-A}}{(-s)^3}, \ \ \ f_{QCD}(s)=0.
\end{equation}

\begin{figure}[H]
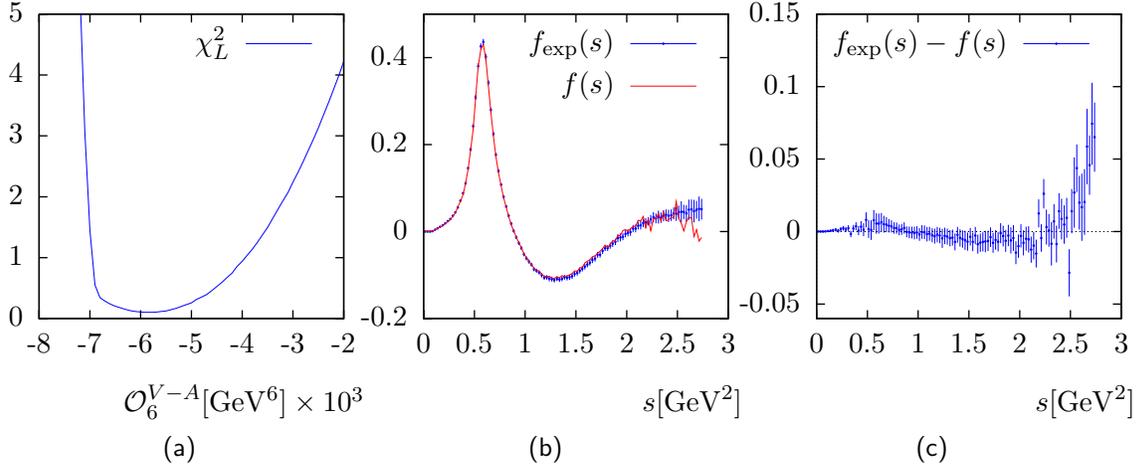

\begin{center}
\subfigure[]{\label{LOchiL}\includegraphics{figures/V-A/oneparam/lo/chiL.epsi}}\subfigure[]{\label{LOfreg}\includegraphics{figures/V-A/oneparam/lo/freg.epsi}}\subfigure[]{\label{LOdiff}\includegraphics{figures/V-A/oneparam/lo/diff.epsi}}
\caption[$V-A$ analysis,1-parameter fit at LO: a typical result.]{A typical result of the algorithm: (a) $\chi_L^2$ [Eq.(\ref{chiL2})] as a function of ${\cal O}_6^{V-A}$; (b) 
the regularised function $f(s)$ [Eq.(\ref{inteq})] compared with data \cite{aleph05}; (c) difference between data and the regularised function. We have chosen for the error parameter ${\cal O}_8^{V-A}=1.5\times 10^{-3}\mbox{GeV}^8$.}
\label{freg+chiL}
\end{center}
\end{figure}

\begin{figure}[H]
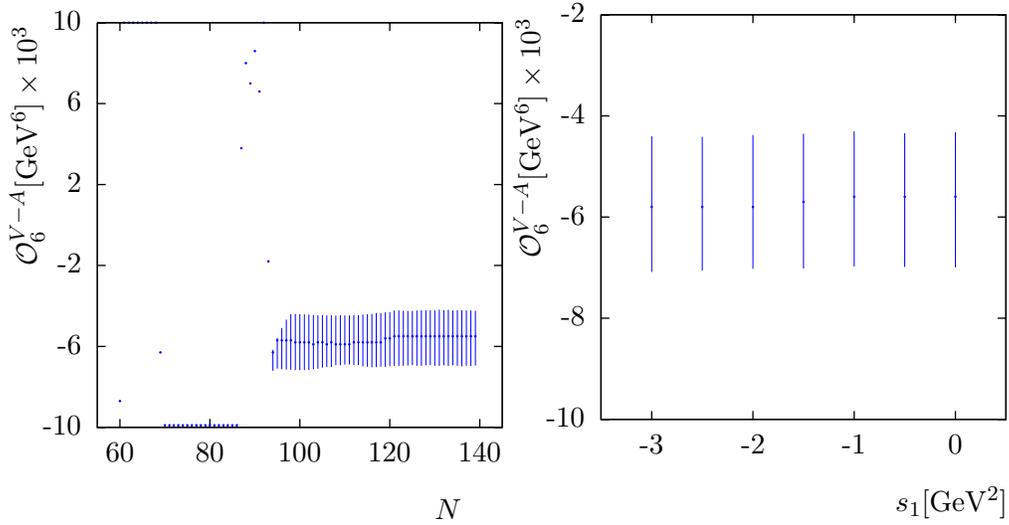

\centering
\includegraphics{figures/V-A/oneparam/lo/depN.epsi}
\includegraphics{figures/V-A/oneparam/lo/deps1.epsi}
\caption[$V-A$ analysis, consistency checks at LO: dependence on the number of data points used and the upper limit of the space-like interval.] {\label{pointdep}${\cal O}_6^{V-A}$ as a function of the number $N$ of data points used in the analysis (e.g.\ $N=120$ means that the data points used in the analysis are the first 120) (left) and of $s_1$, the upper limit of the space-like interval $\Gamma_L$ (right).}
\end{figure}

A typical situation resulting from this algorithm is shown in Fig.\ref{freg+chiL}. 
$\chi_L^2$ has indeed a minimum and at least in the vicinity of this minimum it has the expected quadratic dependence on ${\cal O}_6^{V-A}$ (Fig.\ref{LOchiL}). The regularised function (Fig.\ref{LOfreg}) follows nicely the data 
points, except at large $s$, as it is well illustrated by Fig.\ref{LOdiff} where the difference between experimental data and the regularised function is plotted. One can show that the information on the condensate is contained in the 
lower part of the spectrum by adding or removing data-points contained in the 
saturated upper part: this would not change the result (Fig.\ref{pointdep}, left panel). Thus, the 
discrepancies between data and the regularised function at large $s$, mainly due to 
larger errors in the data, play no role for the determination of ${\cal O}_6^{V-A}$.

\clearpage

We have restricted the integration range in $\chi_L^2$ within the limits 
$s_2\le s\le s_1\le 0$. We have observed a well-defined plateau for the value of 
${\cal O}_6^{V-A}$ as a function of $s_1$ between $-1.5$ and $0$ $\mbox{GeV}^2$ (Fig.\ref{pointdep}, right panel)  
and quote the results for $s_1=-1.0\ \mbox{GeV}^2$.

\begin{figure}[H]
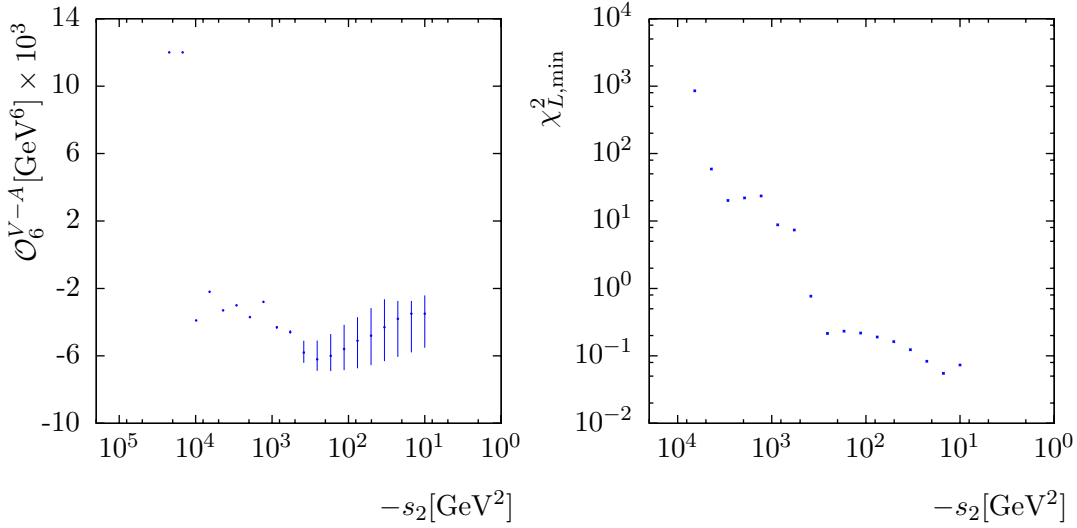

\centering
\includegraphics{figures/V-A/oneparam/lo/deps2.epsi}
\includegraphics{figures/V-A/oneparam/lo/deps2-chiL.epsi}
\caption[$V-A$ analysis, consistency checks at LO: dependence on the lower limit of the space-like interval.]{Dependence on the lower end of the space-like interval, $s_2$: ${\cal O}_6^{V-A}$ as a function of $s_2$ (left); $\chi_{L,{\rm min}}^2$ as a function of $s_2$, on a log-log scale (right).}
\label{deps2}
\end{figure}

We have also studied the behaviour of the algorithm with respect to changes of the lower limit, $s_2$. In principle, one could choose $s_2$ to be minus infinity but, as one can see in Fig.\ref{deps2}, we are forced to choose a finite value. The figure on the right shows us that the value of $|s_2|$ for which we still have agreement between data and experiment is in the order of $10^2$. For larger values, the minimum of $\chi_L^2$ (Eq.(\ref{chiL2})) becomes larger then 1 and thus signalling a bad fit (for more details see Appendix \ref{statistics}). The reason for this inconsistency, for values of $|s_2|$ larger than $O(10^2)$, could lie in the assumption of the decoupling of heavy quarks. The heavy quarks could play an important role at sufficiently high energies. There are also other reasons, like numerical instabilities and rounding errors. However, there exists a plateau for the value of ${\cal O}_6^{V-A}$ as a function of $s_2$ between $-10{\rm GeV}^2$ and $-300{\rm GeV}^2$. In the rest of the analysis we have quoted results for $s_2=-150\mbox{GeV}^2$.

\begin{figure}[H]
\centering
\includegraphics{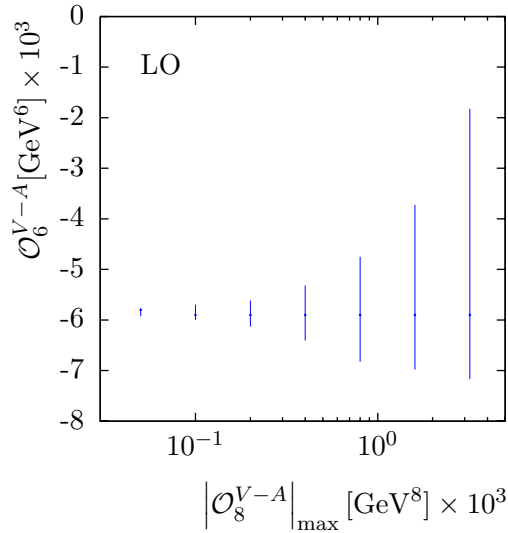}
\caption[$V-A$ analysis, 1-parameter fit at LO: dependence on the error parameter.]{Dependence on the error parameter for the LO analysis: ${\cal O}_6^{V-A}$ as a function of ${\cal O}_8^{V-A}$.}
\label{depO8}
\end{figure}

The dimension $d=8$ condensate ${\cal O}_8^{V-A}$ was used only to define the error channel in the space-like region and thus we expect that the resulting central value of 
${\cal O}_6^{V-A}$ does not depend strongly on $|{\cal O}_8^{V-A}|_{\rm max}$, which is indeed 
reflected in the plot of Fig.\ref{depO8}. However, increasing 
$|{\cal O}_8^{V-A}|_{\rm max}$, i.e., opening the error channel, leads to larger uncertainties 
for ${\cal O}_6^{V-A}$. A value of $\chi_{L,{\rm min}}^2$ corresponding to a $1\sigma$CL can be fixed by choosing an error corridor described by the contribution from an ${\cal O}_8^{V-A}$ condensate of the expected size of $\simeq 1.3\times 10^{-3}{\rm GeV^8}$. The results of the 1-parameter fit at LO can be summarised by quoting the value for ${\cal O}_6^{V-A}$ \cite{almasy}:
 \begin{equation}
  {\cal O}_6^{V-A}=-5.9_{-1.0}^{+1.7}\times 10^{-3}{\rm GeV}^6\ \mbox{for}\ \tilde c_6=0.
\label{resO6LO} 
\end{equation}

The value of ${\cal O}_6^{V-A}$ can be translated into values for the 4-quark condensate  $\alpha_s\langle\bar qq\rangle^2$: 
\begin{equation}
 \alpha_s\langle\bar qq\rangle^2=2.64_{-0.76}^{+0.44}\times 10^{-4}{\rm GeV}^6.
\label{4quarkLO}
\end{equation}
as given by Eq.(\ref{v-a-cond}). This value can be compared with the lowest order vacuum saturation expression
\begin{equation}
 \alpha_s\langle\bar qq\rangle^2\simeq2.78\times 10^{-4}{\rm GeV}^6,
\label{vacsat}
\end{equation}
where we have used for the quark condensate the value of Eq.(\ref{valquarkcond}).

\subsection{${\cal O}_6^{V-A}$ at next-to-leading-order}

In what follows we are interested in studying the influence of NLO corrections to ${\cal O}_6^{V-A}$ and check if the results based on the NLO formula are in agreement with the ones found in the last section. For this purpose one needs to consider the NLO corrections in (\ref{v-a-cond}) and treat ${\cal O}_6^{V-A}$ as a constant like before. With the NLO corrections included, the corresponding QCD amplitude and its imaginary part are:
\begin{equation}
\begin{array}{l}
\displaystyle F_{QCD}(s)=\frac{{\cal O}_6^{V-A}}{(-s)^3}\left\{1+\tilde c_6\frac{\alpha_s(-s)}{4\pi}\right\},\\
\\
\displaystyle f_{QCD}(s)=\frac{{\cal O}_6^{V-A}}{(-s)^3}\frac{\alpha_s(s)}{4}.\\
\end{array}
\end{equation}
The renormalization scale $\mu^2$ was conveniently chosen to be $|s|$. 

\begin{figure}[H]
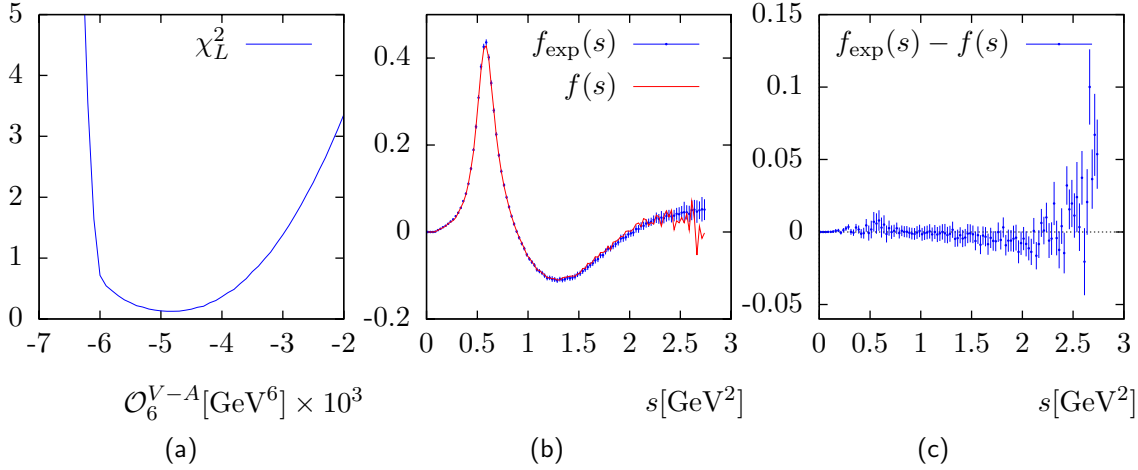

\centering
\subfigure[]{\label{NLOchiLs}\includegraphics{figures/V-A/oneparam/nlo/chiL.epsi}}\subfigure[]{\label{NLOfregs}\includegraphics{figures/V-A/oneparam/nlo/freg.epsi}}\subfigure[]{\label{NLOdiffs}\includegraphics{figures/V-A/oneparam/nlo/diff.epsi}}
\caption[$V-A$ analysis,1-parameter fit at NLO: a typical result.]{A typical result of the algorithm for the NLO analysis: (a) $\chi_L^2$ [Eq.(\ref{chiL2})] as a function of ${\cal O}_6^{V-A}$; (b) 
the regularised function $f(s)$ [Eq.(\ref{inteq})] compared with data \cite{aleph05}; (c) difference between data and the regularised function. We have chosen for the error parameter ${\cal O}_8^{V-A}=1.5\times 10^{-3}\mbox{GeV}^8$ and $\tilde c_6=89/12$.}
\label{freg+chiLs}
\end{figure}

\begin{figure}[H]
\centering
\includegraphics{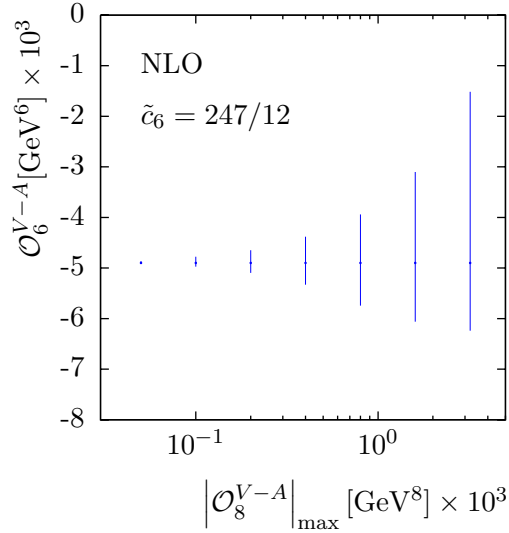}
\caption[$V-A$ analysis, 1-parameter fit at NLO: dependence on the error parameter.]{Dependence on the error parameter for the NLO analysis with $\tilde c_6=247/12$: ${\cal O}_6^{V-A}$ as a function of ${\cal O}_8^{V-A}$.}
\label{depO8s}
\end{figure}

A typical result is seen in Fig.\ref{freg+chiLs}. Here, one should remark that the discrepancies between experimental data and the regularised function are still present. One can thus conclude that including NLO corrections does not improve the quality of our fit. The central value of ${\cal O}_6^{V-A}$ is now shifted with respect to the one from the LO fit, Eq.(\ref{resO6LO}), which was to be expected (Fig.\ref{depO8s}). All the consistency checks (see Figs.\ \ref{pointdep} and \ref{deps2}) performed for the LO fit remain valid.

\clearpage

We can now summarise the results of the NLO fit and quote values for ${\cal O}_6^{V-A}$ for the two available values of the NLO coefficients \cite{almasy}:
\begin{equation}
 \begin{array}{l}
\displaystyle  {\cal O}_6^{V-A}=-4.9_{-1.1}^{+1.5}\times 10^{-3}{\rm GeV}^6\ \mbox{for}\ \tilde c_6=89/12,\\
\\
\displaystyle {\cal O}_6^{V-A}=-3.6_{-1.2}^{+1.1}\times 10^{-3}{\rm GeV}^6\ \mbox{for}\ \tilde c_6=247/12.
 \end{array}
\label{resO6NLO}
\end{equation}

The results are based on the 4-loop expression for $\alpha_s$ with $\Lambda_{\overline{MS}}(N_f=3)=0.326{\rm GeV}$. They are not sensitive to changing $\Lambda_{\overline{MS}}$ within the present experimental error of $\pm0.030{\rm GeV}$. Moreover, the fit results based on the two different values of the NLO coefficients $\tilde c_6$ agree within errors and their difference with respect the LO result, Eq.(\ref{resO6LO}), is consistent with a shift calculated from the correction term choosing a typical value of ${\cal O}(1.5){\rm GeV}$ for the renormalization scale in $\alpha_s$. This is to be expected since our method would work for any $s$-dependent ansatz for ${\cal O}_6^{V-A}$ as well.

We have found agreement of theory and data at the $1\sigma$CL for the dimension $d=8$ contribution describing the error with the value $\simeq 1.3\times 10^{-3}{\rm GeV^8}$.

The values of ${\cal O}_6^{V-A}$ from Eq.(\ref{resO6NLO}) translate into values for the 4-quark condensate $\alpha_s\langle\bar qq\rangle^2$ according to Eq.(\ref{v-a-cond}): 
\begin{equation}
\begin{array}{l}
\displaystyle \alpha_s\langle\bar qq\rangle^2=2.19_{-0.67}^{+0.49}\times 10^{-4}{\rm GeV}^6\ \mbox{for}\ \tilde c_6=89/12,\\
\\
\displaystyle \alpha_s\langle\bar qq\rangle^2=1.61_{-0.49}^{+0.54}\times 10^{-4}{\rm GeV}^6\ \mbox{for}\ \tilde c_6=247/12.
\end{array}
\end{equation}
The values are consistent with the one from the LO analysis, Eq.(\ref{4quarkLO}), as well as with that of the vacuum saturation expression (\ref{vacsat}).

\section{2-parameter fit: ${\cal O}_6^{V-A}$ -- ${\cal O}_8^{V-A}$ correlation}

An interesting and new result is obtained from a 2-parameter fit of the dimension $d=6$ and $d=8$ condensates. Here we do not include the NLO contribution to ${\cal O}_6^{V-A}$ since the corresponding NLO coefficient for ${\cal O}_8^{V-A}$ is not known. To define the error corridor on the space-like region we have used the dimension $d=10$ contribution to the operator product expansion and set:
\begin{equation}
\sigma_L(s)=\left|{\cal O}_{10}^{V-A}\right|_{\rm max}/(-s)^5
\end{equation}
with $\left|{\cal O}_{10}^{V-A}\right|_{\rm max}$ in the order of $10^{-3}{\rm GeV}^{10}$.

\begin{figure}[!p]
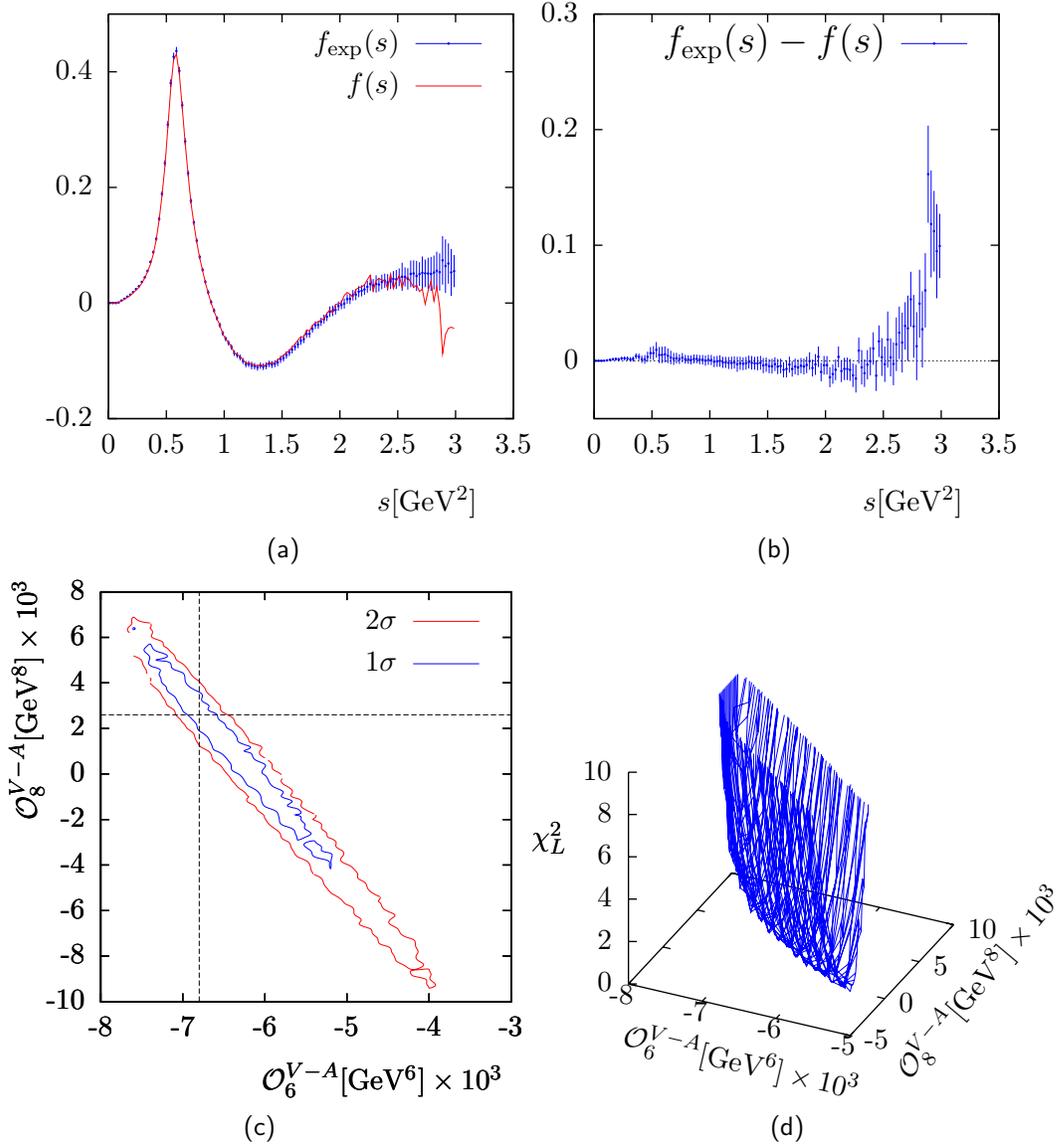

\centering
\subfigure[]{\label{2freg}\includegraphics{figures/V-A/twoparam/lo/freg4.epsi}}
\subfigure[]{\label{2diff}\includegraphics{figures/V-A/twoparam/lo/diff4.epsi}}
\subfigure[]{\label{2chi2D}\includegraphics{figures/V-A/twoparam/lo/chiL4-1.epsi}}
\subfigure[]{\label{2chi3D}\includegraphics{figures/V-A/twoparam/lo/chiL4-3D.epsi}}
\caption[$V-A$ analysis, 2-parameter fit at LO: a typical result.]{A typical result of the 2-parameter fit (LO): (a) the regularised function $f(s)$ [Eq.(\ref{inteq})] compared with data \cite{aleph05}; (b) difference between data and the regularised function; (c) 1- and 2$\sigma$ confidence regions in the $\left({\cal O}_6^{V-A},{\cal O}_8^{V-A}\right)$-plane; The central values are marked by dashed lines. (d) $\chi_L^2$ [Eq.(\ref{chiL2})] as a function of ${\cal O}_6^{V-A}$ and ${\cal O}_8^{V-A}$. The error parameter was set to $\left|{\cal O}_{10}^{V-A}\right|_{\rm max}=4\times 10^{-3} \mbox{GeV}^{10}$.}
\label{2paramtypical}
\end{figure}

We will not repeat all the consistency checks performed in the 1-parameter case. We may argue that they will be qualitatively not changed. Keeping the parameters at the quoted values (see Section \ref{LOO6}), one finds a typical result as the one of Fig.\ref{2paramtypical}. A direct comparison of experimental data with the regularised function $f(s)$ (see Fig.\ref{2freg}) shows a nice agreement over the full range of $s$ with the exception of the highest $s$-bins. Nevertheless, we have learned from the 1-parameter fits that this is due mainly to large experimental errors and that it does not play a sensible role in determining the condensates. 
\clearpage

\begin{figure}[!p]
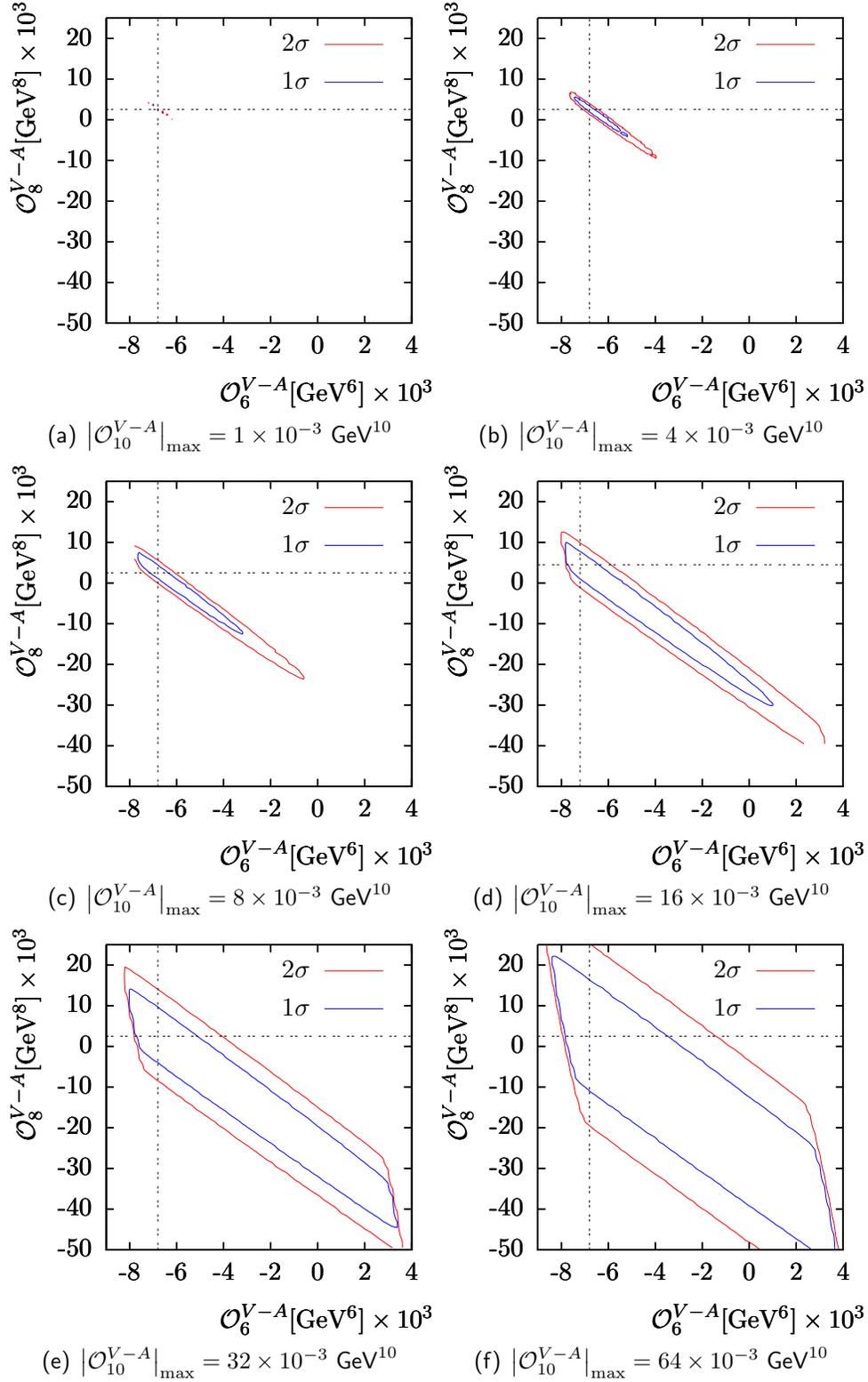

\centering
\subfigure[$\left|{\cal O}_{10}^{V-A}\right|_{\rm max}=1\times 10^{-3} \mbox{ GeV}^{10}$]{\includegraphics{figures/V-A/twoparam/lo/chiL1.epsi}}
\subfigure[$\left|{\cal O}_{10}^{V-A}\right|_{\rm max}=4\times 10^{-3} \mbox{ GeV}^{10}$]{\includegraphics{figures/V-A/twoparam/lo/chiL4.epsi}}
\subfigure[$\left|{\cal O}_{10}^{V-A}\right|_{\rm max}=8\times 10^{-3} \mbox{ GeV}^{10}$]{\includegraphics{figures/V-A/twoparam/lo/chiL8.epsi}}
\subfigure[$\left|{\cal O}_{10}^{V-A}\right|_{\rm max}=16\times 10^{-3} \mbox{ GeV}^{10}$]{\includegraphics{figures/V-A/twoparam/lo/chiL16.epsi}}
\subfigure[$\left|{\cal O}_{10}^{V-A}\right|_{\rm max}=32\times 10^{-3} \mbox{ GeV}^{10}$]{\includegraphics{figures/V-A/twoparam/lo/chiL32.epsi}}
\subfigure[$\left|{\cal O}_{10}^{V-A}\right|_{\rm max}=64\times 10^{-3} \mbox{ GeV}^{10}$]{\includegraphics{figures/V-A/twoparam/lo/chiL64.epsi}}
\caption[$V-A$ analysis, 2-parameter fit at LO: dependence on the error parameter.]{1- and 2$\sigma$ confidence regions in the $\left({\cal O}_6^{V-A},{\cal O}_8^{V-A}\right)$-plane for different values of the error parameter $\left|{\cal O}_{10}^{V-A}\right|_{\rm max}$. The central values are marked by dashed lines.}
\label{depO102D}
\end{figure}
\clearpage

The plot in Fig.\ref{2chi3D} shows that $\chi_L^2$ has the expected quadratic dependence on the parameters ${\cal O}_6^{V-A}$ and ${\cal O}_8^{V-A}$, at least in the vicinity of the minimum. However, a more convenient way to visualise $\chi_L^2$ are contour plots, i.e., lines of constant values of $\chi_L^2$ projected on the $\left({\cal O}_6^{V-A},{\cal O}_8^{V-A}\right)$-plane. Fig.\ref{2chi2D} shows such contour lines corresponding to 1- and 2$\sigma$ confidence regions (CR) for the two parameters (see Appendix \ref{confreg} for the way the confidence regions are defined and constructed).

As expected, the confidence region becomes larger as we open the error corridor, i.e. increase $\left|{\cal O}_{10}^{V-A}\right|_{\rm max}$, but the central values of ${\cal O}_6^{V-A}$ and ${\cal O}_8^{V-A}$ corresponding to the absolute minimum of $\chi_L^2$ remain the same within errors. This is well illustrated in Fig.\ref{depO102D}. One can observe that the islands which appear for $\left|{\cal O}_{10}^{V-A}\right|_{\rm max}=1\times 10^{-3} \mbox{ GeV}^{10}$ become larger and transform into approximative ellipses as we increase $\left|{\cal O}_{10}^{V-A}\right|_{\rm max}$ and for sufficiently large $\left|{\cal O}_{10}^{V-A}\right|_{\rm max}$ they become parallelogram like. This specific change in the shape has no physical meaning, but ruther shows that at large values of ${\cal O}_6^{V-A}$ and ${\cal O}_8^{V-A}$ the underlying probability distribution function is not Gaussian, $\chi_L^2$ having a quadratic dependence on the two parameters only in a small neighbourhood of its minimum.

\begin{figure}[H]
\centering
\includegraphics{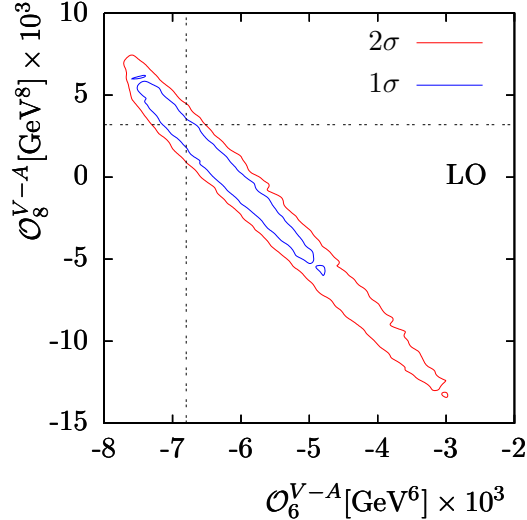}
\caption[$V-A$ analysis, result of the 2-parameter fit at LO.]{1- and 2$\sigma$ confidence regions in the $\left({\cal O}_6^{V-A},{\cal O}_8^{V-A}\right)$-plane from the LO 2-parameter fit at the 1$\sigma$CL agreement of theory and data. The central values are marked by dashed lines.} 
\label{2chiL2DLO}
\end{figure}

The nontrivial result is that we find agreement of theory and data at the $1\sigma$CL if we choose $\left|{\cal O}_{10}^{V-A}\right|_{\rm max}\simeq 5.7\times 10^{-3}{\rm GeV}^{10}$. The result presented in Fig.\ref{2chiL2DLO} shows a strong correlation of ${\cal O}_6^{V-A}$ and ${\cal O}_8^{V-A}$. Both the central values as well as the errors from the 2-parameter fit are consistent with those from the LO 1-parameter analysis: The $1\sigma$ range allowed for ${\cal O}_6^{V-A}$ for fixed ${\cal O}_8^{V-A}=0$ (which is the assumption underlying the 1-parameter fit) agrees with (\ref{resO6LO}) and (\ref{resO6NLO}). However, leaving the value of ${\cal O}_8^{V-A}$ unconstrained, as is the case here, one finds a much larger range for ${\cal O}_6^{V-A}$. The minimum value of $\chi_L^2$ is located at the values:
\begin{equation}
\begin{array}{l}
\displaystyle{\cal O}_6^{V-A}=-6.8_{-0.8}^{+2.0}\times 10^{-3}\mbox{ GeV}^6,\\
\\
\displaystyle{\cal O}_8^{V-A}=3.2_{-9.2}^{+2.8}\times 10^{-3}\mbox{ GeV}^8.
\end{array}
\label{resO6O8LO}
\end{equation}

The errors on ${\cal O}_6^{V-A}$ for fixed ${\cal O}_8^{V-A}$ are small, but the allowed range for ${\cal O}_8^{V-A}$ is not very restrictive (note the different scales for the two coordinate axes in Fig.\ref{2chiL2DLO}). However, the strong correlation allows one to determine a linear combination of ${\cal O}_6^{V-A}$ and ${\cal O}_8^{V-A}$ with a rather small error \cite{almasy}:
\begin{equation} 
{\cal O}_8^{V-A}+2.22\mbox{ GeV}^2\ {\cal O}_6^{V-A}=-18.30_{-0.25}^{+0.38}\times 10^{-3} \mbox{ GeV}^8
\end{equation}
This is an important result. If one would know one of the parameters from an independent determination, one could specify the value of the other one with well defined small errors in the order of $3\%$.

\begin{figure}[H]
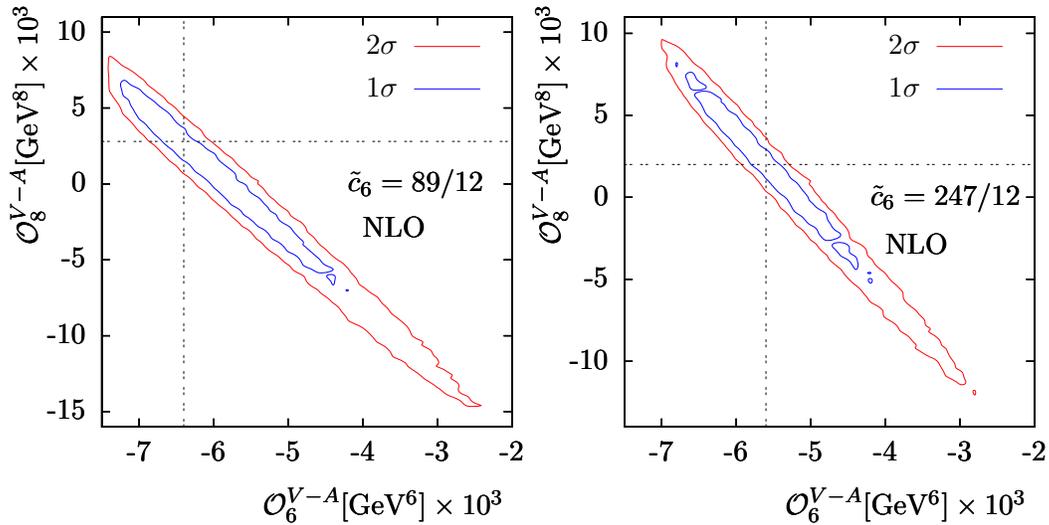

\centering
\includegraphics{figures/V-A/twoparam/nlo/chiL5.6s89.epsi}
\includegraphics{figures/V-A/twoparam/nlo/chiL4.8s247.epsi}
\caption[$V-A$ analysis, result of the 2-parameter fit at NLO.]{1- and 2$\sigma$ confidence regions in the $\left({\cal O}_6^{V-A},{\cal O}_8^{V-A}\right)$-plane from the NLO 2-parameter fit at the 1$\sigma$CL agreement of theory and data. The central values are marked by dashed lines.} 
\label{2chiL2DNLO}
\end{figure}

We have also checked that we obtain consistent results when including the NLO correction to ${\cal O}_6^{V-A}$: The 1- and 2$\sigma$ contours are shifted, essentially without changing their form (Fig.\ref{2chiL2DNLO}), to larger values of ${\cal O}_6^{V-A}$ exactly as can be inferred from the 1-parameter fits: for $\tilde c_6=0(89/12,247/12)$ we find the minimum of $\chi_L^2$ at ${\cal O}_6^{V-A}\times 10^3/ {\rm GeV^6}=-6.8(-6.6,-5.8)$.

The linear combinations of ${\cal O}_6^{V-A}$ and ${\cal O}_8^{V-A}$ for the two available NLO coefficients are then:
\begin{equation} 
\begin{array}{lcl}
{\cal O}_8^{V-A}+2.69\mbox{ GeV}^2\ {\cal O}_6^{V-A}=-14.40_{-0.29}^{+0.16}\times 10^{-3} \mbox{ GeV}^8 & {\rm for} & \tilde c_6=89/12,\\
\\
{\cal O}_8^{V-A}+4.07\mbox{ GeV}^2\ {\cal O}_6^{V-A}=-20.80_{-0.20}^{+0.20}\times 10^{-3} \mbox{ GeV}^8 & {\rm for} & \tilde c_6=247/12.
\end{array}
\end{equation}
They are consistent with the LO result and have even smaller errors.

\section{Review and comparison of results}

There exists a number of previous extractions of QCD condensates in the literature, mainly based on sum rule approaches and at LO. They are listed in Table \ref{LOdet}. There also exist previous extractions at NLO, based on the same functional approach as in this work, listed in Table \ref{NLOdet}.

The results for ${\cal O}_6^{V-A}$ cover values between $(-2.27\pm0.51)\times 10^{-3}\,{\rm GeV}^6$ \cite{cirigliano1} and $(-8.7\pm2.3)\times 10^{-3}\,{\rm GeV}^6$ \cite{narison1}. These values correspond to a scale of about $200\,{\rm MeV}$ which is comparable to $\Lambda_{\rm QCD}$. Although errors given by the respective authors are typically in the order of $25\%$, their central values are only in rough agreement. The observed variations of these results represent the ambiguities inherent in the QCD sum rule approach. Our result nicely falls into the same range, also with an error estimate of the same size.

For ${\cal O}_8^{V-A}$, previous results range from $(-10.8\pm6.6)\times 10^{-3}\,{\rm GeV}^8$ \cite{bordes} to $(15.6\pm4.0)\times 10^{-3}\, {\rm GeV}^8$ \cite{narison1}. A recent conservative estimate \cite{rojo} is ${\cal O}_8^{V-A}=\left(-12_{-11}^{+7}\right)\times 10^{-3}\,{\rm GeV}^8$. Again, our result agrees within the estimated precision. The same is true for ${\cal O}_{10}^{V-A}$ for which previous results range from $(-17.1\pm4.4)\times 10^{-3}\, {\rm GeV}^{10}$ \cite{narison1} to $(78\pm24)\times 10^{-3}\, {\rm GeV}^{10}$ \cite{rojo}. The spread of values found in the literature for the $d=12$ condensate , ${\cal O}_{12}^{V-A}$, is even larger: from $(-240\pm100)\times 10^{-3}\, {\rm GeV}^{12}$ \cite{bordes} to $(14.7\pm3.7)\times 10^{-3}\, {\rm GeV}^{10}$ \cite{narison1} as one can read off from Table \ref{LOdet}. These values are also consistent with the values we had to choose for the error corridor.

Moreover, it is also interesting to note the agreement of the correlation between ${\cal O}_6^{V-A}$ and ${\cal O}_8^{V-A}$ with corresponding results from \cite{narison1,zyablyuk}. In Ref.\ \cite{narison1}, this correlation is extracted from weighted finite energy sum rules but no errors are given while in Ref.\ \cite{zyablyuk} Borel sum rules were used. In the latter, 1-, 2- and 3$\sigma$ confidence regions for the correlations ${\cal O}_6^{V-A}$--${\cal O}_8^{V-A}$ and ${\cal O}_6^{V-A}$--${\cal O}_{10}^{V-A}$ are presented. The 1$\sigma$ allowed ranges for ${\cal O}_6^{V-A}$ and ${\cal O}_8^{V-A}$ are shifted as compared to ours, but the slope agrees well within errors. 

Obviously, our numerical results are, from a practical point of view, not superior to approaches based on finite energy sum rules. However, the fact that we find agreement within errors is not trivial. Since this approach is based on much more general assumptions, the results obtained in our analysis give additional confidence to the numerical values obtained with the help of QCD sum rules.   
\clearpage

\begin{table}[H]
\begin{center}
\begin{tabular}{||l@{\vrule height 11pt depth4pt width0pt\hskip\arraycolsep}|c|c|c|c||c}\hline
 & ${\cal O}_6^{V-A}$ & ${\cal O}_8^{V-A}$ & ${\cal O}_{10}^{V-A}$ & ${\cal O}_{12}^{V-A}$ \\ \hline
\cite{rojo} & $-4\pm2.0$ & $-12_{-11}^{+7}$ & $78\pm24$ & $-2.6\pm0.8$\\ \hline 
\cite{bordes}\footnotemark[1] & $-4.52\pm1.1$ & $-10.8\pm6.6$ & $72\pm28$ & $-240\pm100$\\ \hline 
\cite{cirigliano1}\footnotemark[1] & $-2.27\pm0.51$ & $-2.85\pm2.18$ & $24.1\pm6.1$ & $-80\pm16$\\ \hline 
\cite{narison1} & $-8.7\pm2.3$ & $15.6\pm4.0$ & $-17.1\pm4.4$ & $14.7\pm3.7$\\ \hline 
\cite{friot} & $-7.9\pm1.6$ & $11.7\pm2.6$ & $-13.1\pm3.0$ & $13.2\pm3.3$\\ \hline 
\cite{zyablyuk} & $-7.2\pm1.2$ & $7.8\pm2.5$ & $-4.4\pm2.8$ & \\ \hline 
\cite{dominguez}\footnotemark[1] & $-8\pm2$ & $-2\pm12$ &  & \\ \hline 
\cite{ioffe} & $-6.8\pm2.1$ & $7\pm4$ &  & \\ \hline 
\cite{aleph} & $-7.7\pm0.8$ & $11.0\pm1.0$ &  & \\ \hline 
\cite{opal} & $-6\pm0.6$ & $7.5\pm1.3$ &  & \\ \hline \hline 
\multicolumn{5}{||l||}{This work}\\\hline
Eq.(\ref{resO6LO}) & $-5.9_{-1.0}^{+1.7}$ &  &  & \\ \hline 
Eq.(\ref{resO6O8LO}) & $-6.8_{-0.8}^{+2.0}$ & $3.2_{-9.2}^{+2.8}$ &  & \\ \hline 
\end{tabular}
\caption[$V-A$ analysis: estimated values of the dimension $d\le12$ condensates at LO.]{Estimated values of the $d\le12$ ${\cal O}_d^{V-A}$ condensates in units of $10^{-3}{\rm GeV}^d$ at leading order.}
\label{LOdet}
\end{center}
\end{table}

\begin{table}[H]
\begin{center}
\begin{tabular}{||l@{\vrule height 11pt depth4pt width0pt\hskip\arraycolsep}|c|c||c|c||c}\hline
 & \multicolumn{2}{|c||}{${\cal O}_6^{V-A}$} &\multicolumn{2}{|c||}{ ${\cal O}_8^{V-A}$}\\ \hline
 & $\tilde c_6=89/12$ & $\tilde c_6=247/12$ & $\tilde c_6=89/12$ & $\tilde c_6=247/12$ \\ \hline
\cite{ciulli}\footnotemark[1] & $-4.0\pm2.8$ & $-3.0\pm1.8$ & & \\ \hline
\multicolumn{5}{||l||}{This work}\\\hline
Eq.(\ref{resO6LO}) & $-4.9_{-1.1}^{+1.5}$ & $-3.6_{-1.2}^{+1.1}$ &  & \\ \hline 
Eq.(\ref{resO6O8LO}) & $-6.4_{-0.9}^{+2.0}$ & $-5.6_{-1.2}^{+1.2}$ & $2.8_{-9.6}^{+4.1}$ & $2.0_{-6.4}^{+6.2}$ \\ \hline 
\end{tabular}
\caption[$V-A$ analysis: estimated values of the dimension $d\le8$ condensates at NLO.]{Estimated values of the $d\le8$ ${\cal O}_d^{V-A}$ condensates in units of $10^{-3}{\rm GeV}^d$ at next-to-leading order.}
\label{NLOdet}
\end{center}
\end{table}

\footnotetext[1]{\sf Note the different normalisation of spectral functions. The values shown are adjusted so that they can be compared to those from this work.}

\clearpage{\thispagestyle{empty}\cleardoublepage}
\chapter{$V$, $A$ and $V+A$ analysis}
\label{V+Aanalysis}

The final precise data on the non-strange hadronic spectral functions from ALEPH \cite{aleph05} allows us to perform also an analysis in the $V+A$, $V$ and $A$ channels separately.

Since the $V$, $A$ and $V+A$ correlators are dominated by their perturbative contributions, there are subtractions needed in the dispersion relation (\ref{disprel}), the starting point of our analysis. However, as stated in Section \ref{dispersionrelations}, one can eliminate these subtractions by taking an appropriate number of derivatives of the correlator in question. For $V$, $A$ and $V+A$ there is only one subtraction and thus one needs to take only the first derivative and write
\begin{equation}
 F(s)=-\frac{1}{\pi}\int_{s_0}^\infty\ dx K(s,x)f(x),
\end{equation}
where now
\begin{equation}
 F(s)=-s\frac{d}{ds}\Pi_{V,A,V+A}(s),
\end{equation}
\begin{equation}
 K(s,x)=\frac{s}{(x-s)^2},
\end{equation}
and
\begin{equation}
 f(x)={\rm Im}\Pi_{V,A,V+A}(x)
\end{equation}
is the same as before.

According to (\ref{corrseparation}), one has:
\begin{equation}
 F(s)=D_{V,A,V+A}(s)+\sum_{d\ge4}\frac{d}{2}\frac{{\cal O}^{V,A,V+A}_d}{(-s)^{d/2}}\left(1+O(\alpha_s)\right).
\end{equation}

The Adler function, $D(s)$, is known in the massless-quark limit up to terms of order $\alpha_s^4$ and reads:
\begin{equation}
 D_{V,A}(s)=\frac{1}{4\pi^2}\sum_{n\ge0}^4\ K_n\left(\frac{\alpha_s(s)}{\pi}\right)^n.
\end{equation}
The coefficients $K_n$ are the same for both $A$ and $V$ channels. For 3 flavours, in $\overline{\rm MS}$ regularisation, $K_0=K_1=1$, $K_2=1.64$ \cite{chetyrkin2,celmaster,dine}, $K_3=6.37$ \cite{gorishny,surguladze} and for $K_4$ there are two estimates $K_4=25\pm25$ \cite{kataev} and $K_4=27\pm16$ \cite{baikov}.

The experimental information we use are the appropriate spectral functions, i.e., $v_1(s)$ for the $V$ channel, $a_1(s)+a_0(s)$ for the $A$ channel and $v_1(s)+a_1(s)+a_0(s)$ for the $V+A$ one. As seen in Section \ref{spectralfuncdef}, $a_0(s)$ is basically saturated by the pion pole and thus we will take it into account separately, as we did before.

The QCD prediction on the time-like interval $(s_{\rm max},\infty)$ is given by the perturbative expansion:
\begin{equation}
 f_{QCD}^{V,A}=\frac{1}{4\pi}\left(1+\frac{\alpha_s(x)}{\pi}+O(\alpha_s^2)\right).
\end{equation}
To summarise, the QCD information that we have at hand, i.e., the l.h.s.\ of Eq.(\ref{dispsource}), is:
\begin{equation}
 F_{QCD}^V(x)=D^V(x)+\sum_{d\ge4}\frac{d}{2}\frac{{\cal O}_d^V}{(-x)^{d/2}}\left(1+c_{d,V}^{NLO}\frac{\alpha_s(\mu^2)}{\pi}\right),
\end{equation}
\begin{equation}
 F_{QCD}^A(x)=D^A(x)+\sum_{d\ge4}\frac{d}{2}\frac{{\cal O}_d^A}{(-x)^{d/2}}\left(1+c_{d,A}^{NLO}\frac{\alpha_s(\mu^2)}{\pi}\right)+\frac{f_\pi^2}{x}.
\end{equation}
Obviously, for $F_{QCD}^{V+A}(x)$ one has to sum the two equations.

Assuming vacuum dominance, as in the case of the $V-A$ correlator, the $V+A$ matrix elements can be written as:
\begin{equation}
 \begin{array}{ll}
\displaystyle  {\cal O}_4^{V+A}=\frac{\alpha_s}{6\pi}\langle G_{\mu\nu}^aG_{\mu\nu}^a\rangle, & \displaystyle c_{4,V+A}^{NLO}=\frac{7}{6},\\
\\
\displaystyle  {\cal O}_6^{V+A}=\frac{128}{81}\pi\alpha_s\langle\bar qq\rangle^2, & \displaystyle c_{6,V+A}^{NLO}=\frac{29}{24}+\frac{17}{18}\ln\frac{-s}{\mu^2}.
\end{array}
\label{nlocorrections}
\end{equation}
The coefficients for the $\alpha_s$-corrections were calculated in \cite{chetyrkin} and \cite{adam}, respectively. Considering these matrix elements together with those from the $V-A$ channel, Eq.(\ref{v-a-cond}), one can easily find similar representations for ${\cal O}_d^{V,A}$. However, as before, our analysis does not rely on such representations since we aim to determine ${\cal O}_d^{V,A,V+A}$ and not the condensates $\langle G_{\mu\nu}^aG_{\mu\nu}^a\rangle$, $\langle\bar qq\rangle^2$ separately.

For the error corridor one can use the last known term of the perturbation series possibly combined with the first omitted term in the series over condensates.

\section{$A$ analysis}

Let us start with the analysis of the axial-vector channel and perform 1- and 2-parameter fits in order to determine acceptable values and ranges for the dimension $d=4$ and $d=6$ condensates, as well as their correlation. 

\subsection{1-parameter fit: determination of ${\cal O}_4^A$}
\label{A1LO}

By a 1-parameter fit we aim to determine the dimension $d=4$ condensate ${\cal O}_4^A$ at leading order. Even though NLO corrections are calculated (see Eq.(\ref{nlocorrections})), we will not take them into account. We have learned from similar fits performed in the $V-A$ channel that they are not changing qualitatively the result.

For defining an estimate for the error corridor in the space-like region one can choose to use the last known term of the perturbation series, or the first omitted term in the OPE, or a combination of the two of them:
\begin{equation}
 \sigma_L(x)=\left\{
\begin{array}{ll}
 \displaystyle\frac{1}{4\pi^2}K_3\left(\frac{\alpha_s(-x)}{\pi}\right)^3,\\
\\
\displaystyle3\frac{{\cal O}_6^{A}}{(-x)^3}, \\
\\
 \displaystyle\sqrt{\left[\frac{1}{4\pi^2}K_3\left(\frac{\alpha_s(-x)}{\pi}\right)^3\right]^2+\left[3\frac{{\cal O}_6^{A}}{(-x)^3}\right]^2}. \\
\end{array}
\right.
\label{A3error}
\end{equation}

\begin{figure}[H]
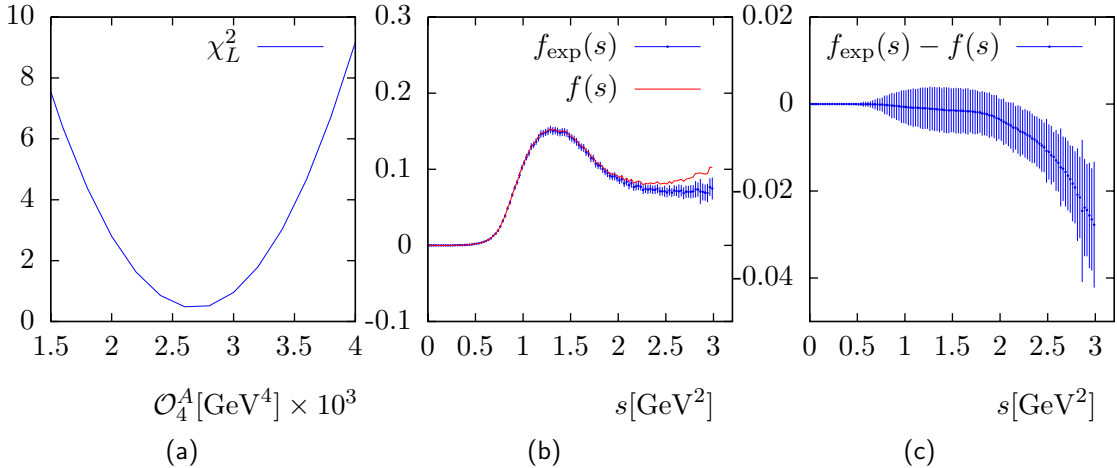

\begin{center}
\subfigure[]{\label{A-LOchiL}\includegraphics{figures/A/oneparam/lo/chiL.epsi}}\subfigure[]{\label{A-LOfreg}\includegraphics{figures/A/oneparam/lo/freg.epsi}}\subfigure[]{\label{A-LOdiff}\includegraphics{figures/A/oneparam/lo/diff.epsi}}
\caption[$A$ analysis, 1-parameter fit at LO: a typical result.]{A typical result of the algorithm: (a) $\chi_L^2$ as a function of ${\cal O}_4^A$; (b) 
the regularised function $f(s)$ [Eq.(\ref{inteq})] compared with data \cite{aleph05}; (c) difference between data and the regularised function. For the error corridor the first omitted term in the OPE was used with $\left|{\cal O}_6^A\right|_{\rm max}=1\times 10^{-3}\mbox{GeV}^6$.}
\label{A-freg+chiL}
\end{center}
\end{figure}

A typical result of the algorithm is shown in Fig.\ref{A-freg+chiL}. One can see that the regularised function (Fig. \ref{A-LOfreg}) follow nicely the data points, except at large $s$ and, as expected, $\chi_L^2$ has the quadratic dependence on ${\cal O}_4^A$.

\begin{figure}[H]
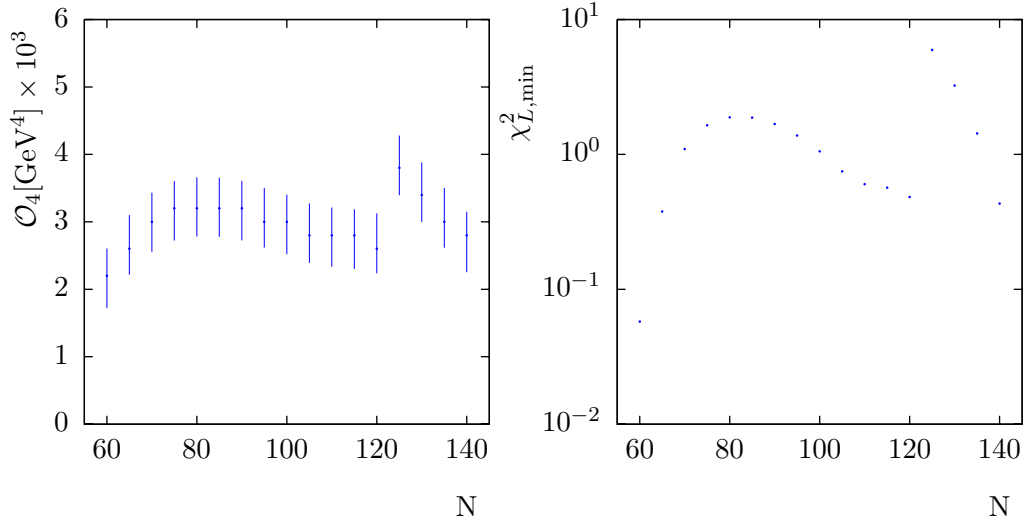

\begin{center}
\subfigure{\includegraphics{figures/A/oneparam/lo/depN.epsi}}
\subfigure{\includegraphics{figures/A/oneparam/lo/depNL.epsi}}
\caption[$A$ analysis, consistency checks at LO: dependence on the number of data points used.]{${\cal O}_4^A$ (left) and $\chi_{L,{\rm min}}^2$ (right) as a function of the first $N$ data points used in the analysis.}
\label{AdepN}
\end{center}
\end{figure}

\begin{figure}[H]
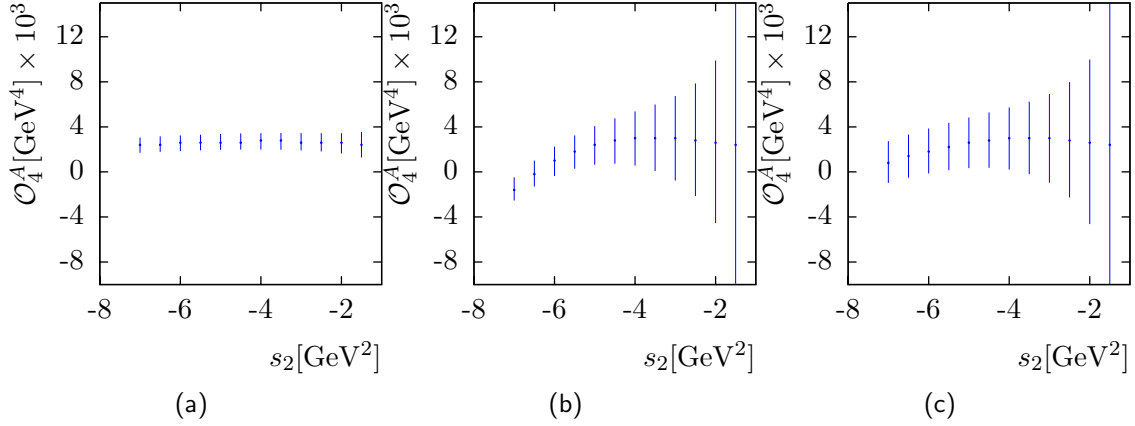

\begin{center}
\subfigure[]{\includegraphics{figures/A/oneparam/lo/deps2K3.epsi}}\subfigure[]{\includegraphics{figures/A/oneparam/lo/deps2O6.epsi}}\subfigure[]{\includegraphics{figures/A/oneparam/lo/deps2O6K3.epsi}}
\caption[$A$ analysis, consistency checks at LO: dependence of ${\cal O}_4^A$ on the lower limit of the space-like interval.]{${\cal O}_4^A$ as a function of the lower end of the space-like interval $\Gamma_L$, $s_2$, for the three possible error corridors: (a) error corridor defined by the last known term in the perturbation expansion, (b) error corridor defined by the first omitted contribution in the OPE, (c) error corridor defined by the combination of the two previous cases. For the cases (b) and (c), $\left|{\cal O}_6^A\right|_{\rm max}=3\times 10^{-3}{\rm GeV}^6$.}
\label{Adeps2}
\end{center}
\end{figure}

We have performed a series of consistency checks summarised in Figs.\ \ref{AdepN}, \ref{Adeps2} and \ref{Adeps2L}, i.e., we have studied the behaviour of the algorithm with respect to various parameters: the number of experimental data points $N$ used in the analysis, the end-points of the time-like interval $\Gamma_L$, $s_1$ and $s_2$, as well as the dependence on the error parameter ${\cal O}_6^A$.

Like for the $V-A$ channel, one can show that the information on the condensate is contained in the lower part of the spectrum by adding or removing data-points contained in the saturated upper part of the spectrum: this would not change the result (Fig.\ref{AdepN} , left panel). However, one can observe that there exists also a region which deviates from the saturated one, for $N\le125$. In this region, the high oscillation in data points as well as small experimental errors play an important role. A good argument to remove these data points from the spectrum, i.e., not to use them in the analysis, is to check wether the consistency between theory and data is good enough when including them. This is not the case (see Fig.\ref{AdepN} , right panel) and thus for the rest of the analysis we will use only the first $N=120$ data points. 

Studying the behaviour of the algorithm with respect to changes of $s_2$, we found that the best simultaneous description of experimental data and theory is when we choose to define the error corridor with the help of the last known term in the perturbation series, i.e.,
\begin{equation}
 \sigma_L(s)=\frac{1}{4\pi^2}K_3\left(\frac{\alpha_s(-x)}{\pi}\right)^3.
\label{errorA}
\end{equation}
This is well illustrated in Figs.\ \ref{Adeps2} and \ref{Adeps2L}, where the dependence of ${\cal O}_4^A$ and respectively $\chi_{L,{\rm min}}^2$ on $s_2$ is plotted for all three possible error corridors. Based on these plots we further use the error corridor defined by Eq.(\ref{errorA}) and quote results for $s_2=-3.5{\rm GeV}^2$.

We have also observed a well-defined plateau for the value of ${\cal O}_4^A$ as a function of $s_1$ between $-1.5$ and $0{\rm GeV}^2$ and quote the results for $s_1=-0.4{\rm GeV}^2$.

\begin{figure}[H]
\begin{center}
\subfigure[]{\includegraphics{figures/A/oneparam/lo/deps2K3L.epsi}}\subfigure[]{\includegraphics{figures/A/oneparam/lo/deps2O6L.epsi}}\subfigure[]{\includegraphics{figures/A/oneparam/lo/deps2O6K3L.epsi}}
\caption[$A$ analysis, consistency checks at LO: dependence of $\chi_{L,{\rm min}}^2$ on the lower limit of the space-like interval.]{$\chi_{L,{\rm min}}^2$ as a function of the lower end of the space-like interval $\Gamma_L$, $s_2$, for the three possible error corridors: (a) error corridor defined by the last known term in the perturbation expansion, (b) error corridor defined by the first omitted contribution in the OPE, (c) error corridor defined by the combination of the two previous cases. For the cases (b) and (c), $\left|{\cal O}_6^A\right|_{\rm max}=3\times 10^{-3}{\rm GeV}^6$.}
\label{Adeps2L}
\end{center}
\end{figure}

Now, the results of the 1-parameter fit at LO can be summarised by quoting the value for ${\cal O}^A_4$:
\begin{equation}
 {\cal O}_4^A=2.8_{-0.8}^{+0.6}\times 10^{-3}{\rm GeV}^4\ \ \mbox{at}\ \ 66.57\%\mbox{CL}.
\label{1Aresults}
\end{equation}
 According to Eq.(\ref{nlocorrections}) and keeping in mind the chiral symmetry underlying our analysis, i.e., assuming ${\cal O}_4^{V-A}=0$, one can translate the value of ${\cal O}^A_4$ into values for the gluon condensate
\begin{equation}
 \langle\frac{\alpha_s}{\pi}G_{\mu\nu}^aG_{\mu\nu}^a\rangle=3.36_{-0.96}^{+0.72}\times 10^{-2}{\rm GeV}^4.
\label{1Agluonresults}
\end{equation}
This result is in good agreement with the first extraction by Shifman, Vainshtein and Zakharov in \cite{shifman1}.

\subsection{2-parameter fit: ${\cal O}_4^A$ -- ${\cal O}_6^A$ correlation}

A two parameter fit is also possible. We aim to fit the condensates of dimension $d=4$ and $d=6$ at leading order. Based on the results from the 1-parameter fit we will take the last known term of the perturbative series as a sensible estimate for the error corridor on the space-like region, Eq.(\ref{errorA}). All the consistency checks performed for the 1-parameter fit remain valid and thus we keep the corresponding parameters, i.e., $s_1$, $s_2$ and $N$, at the quoted values (see Section \ref{A1LO}).

\begin{figure}[!p]
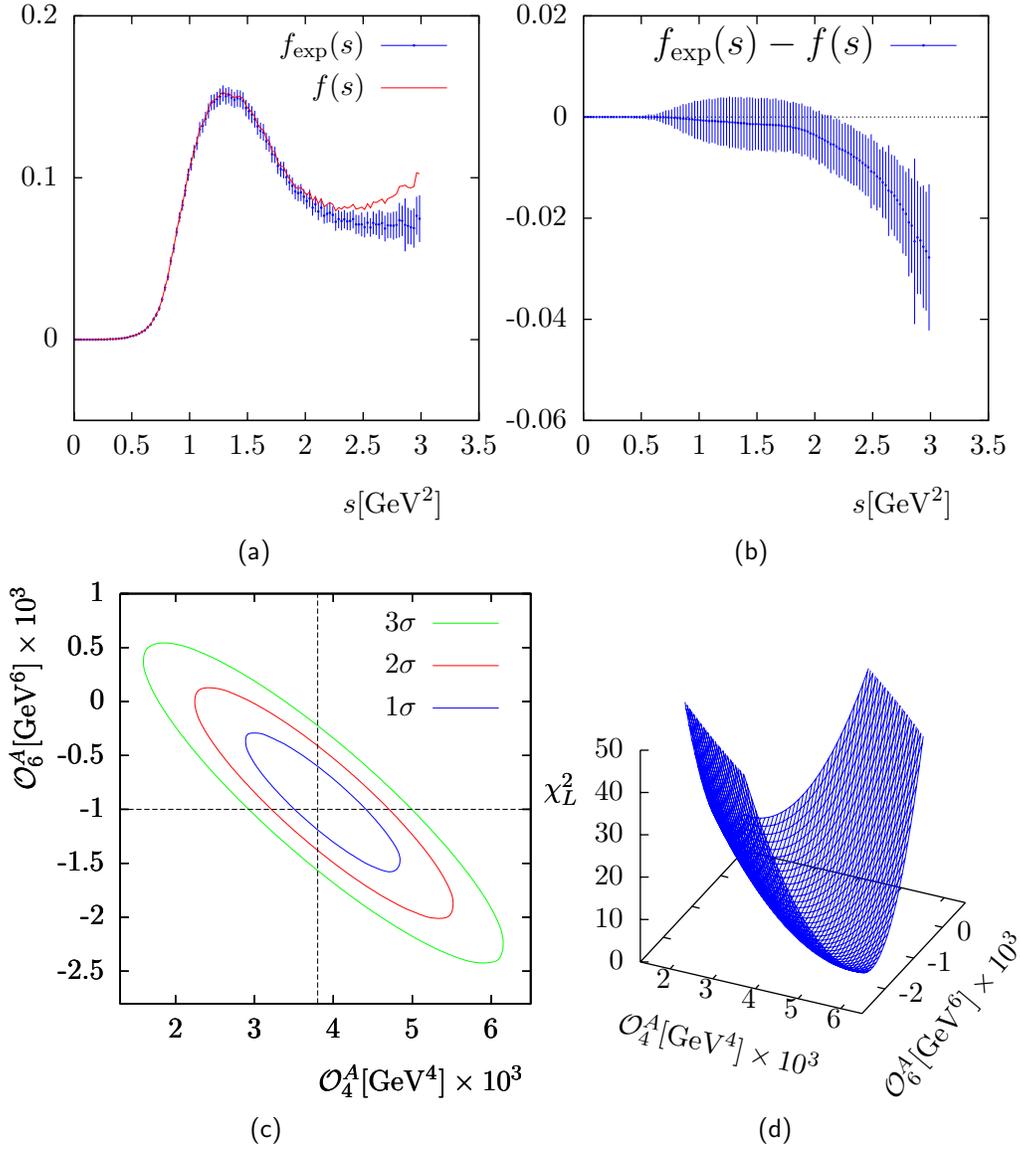

\centering
\subfigure[]{\label{A2freg}\includegraphics{figures/A/twoparam/freg.epsi}}
\subfigure[]{\label{A2diff}\includegraphics{figures/A/twoparam/diff.epsi}}
\subfigure[]{\label{A2chi2D}\includegraphics{figures/A/twoparam/chiL.epsi}}
\subfigure[]{\label{A2chi3D}\includegraphics{figures/A/twoparam/chiL-3D.epsi}}
\caption[$A$ analysis, result of the 2-parameter fit at LO.]{$A$ analysis, 2-parameter fit (LO) at $62.58\%$CL for the agreement between theory and data: (a) the regularised function $f(s)$ [Eq.(\ref{inteq})] compared with data \cite{aleph05}; (b) difference between data and the regularised function; (c) 1-, 2- and 3$\sigma$ confidence regions in the $\left({\cal O}_4^{A},{\cal O}_6^{A}\right)$-plane; The central values are marked by dashed lines. (d) $\chi_L^2$ [Eq.(\ref{chiL2})] as a function of ${\cal O}_4^{A}$ and ${\cal O}_6^{A}$.}
\label{A2paramresult}
\end{figure}

The result of the 2-parameter analysis is presented in Fig.\ref{A2paramresult}. This result corresponds to a $62.58\%$CL agreement of experimental data with theory. As in the 1-parameter case, a direct comparison of experimental data with the regularised function $f(s)$ (see Figs.\ \ref{A2freg} and \ref{A2diff}) shows a nice agreement over the full range of $s$ with the exception of the highest $s$-bins. We can then conclude that including the contribution from the next higher-order condensate in the OPE does not improve the quality of the fit.

Also $\chi_L^2$ (Fig.\ref{A2chi3D}) has the expected quadratic dependence on the two free parameters, ${\cal O}_4^A$ and ${\cal O}_6^A$. Fig.\ref{A2chi2D} shows lines of constant values of $\chi_L^2$ projected on the $\left({\cal O}_4^A,{\cal O}_6^A\right)$-plane corresponding to 1-, 2- and 3$\sigma$CR around the two parameters (for a definition of these confidence regions one can consult Appendix \ref{confreg}). As seen, there exists a strong negative correlation of ${\cal O}_4^A$ and ${\cal O}_6^A$, the central values, i.e., the values corresponding to $\chi_{L,{\rm min}}^2$, being located at:
\begin{equation}
 \begin{array}{l}
  {\cal O}_4^A=3.8_{-0.9}^{+1.1}\times 10^{-3}{\rm GeV}^4,\\
\\
  {\cal O}_6^A=-1.0_{-0.7}^{+0.6}\times 10^{-3}{\rm GeV}^6.
 \end{array}
\label{2paramfitresult}
\end{equation}
These values are at the same $62.58\%$CL for the agreement between theory and data. The errors presented are the $1\sigma$ errors.

The strong correlation allows one to determine a linear combination of ${\cal O}_4^A$ and ${\cal O}_6^A$ with rather small errors:
\begin{equation}
 {\cal O}_6^A+0.65{\rm GeV^2}\cdot{\cal O}_4^A=1.60_{-0.25}^{+0.26}\times 10^{-3}{\rm GeV}^6,
\end{equation}
and thus if one of the parameters is known from an independent determination, one could specify the value of the other one with well defined errors.

These results are in agreement with those from the 1-parameter fit. However, since here the value of ${\cal O}_6^A$ was left unconstrained, we have found a larger range for ${\cal O}_4^A$. 

\clearpage

\section{$V$ and $V+A$ analysis}

We will now move to the analysis of the  $V$ and $V+A$ channels. We will again perform 1- and 2-parameter fits and quote their results.

\subsection{1-parameter fits}

Let us first start with the 1-parameter fits and aim to determine the dimension $d=4$ condensates ${\cal O}_4^V$ and ${\cal O}_4^{V+A}$ at leading order. Like for the $A$ channel, since the correlators are dominated by the perturbative regime, there are three possible choices for the error corridor on the space-like region, i.e.,
\begin{equation}
 \sigma_L(x)=\left\{
\begin{array}{ll}
 \displaystyle\frac{1}{4\pi^2}K_3\left(\frac{\alpha_s(-x)}{\pi}\right)^3,\\
\\
\displaystyle3\frac{{\cal O}_6^{V}}{(-x)^3}, \\
\\
 \displaystyle\sqrt{\left[\frac{1}{4\pi^2}K_3\left(\frac{\alpha_s(-x)}{\pi}\right)^3\right]^2+\left[3\frac{{\cal O}_6^{V}}{(-x)^3}\right]^2}, \\
\end{array}
\right.
\label{V3error}
\end{equation}
for the $V$ channel. For the $V+A$ channel one needs to add the separate contributions coming from the $V$, Eq.(\ref{V3error}), and $A$, Eq.(\ref{A3error}), channels. One should keep in mind that ${\cal O}_d^{V+A}={\cal O}_d^V+{\cal O}_d^V$ for any dimension $d$. 

\begin{figure}[H]
\begin{center}
\subfigure[]{\label{V-LOchiL}\includegraphics{figures/V/oneparam/lo/chiL.epsi}}\subfigure[]{\label{V-LOfreg}\includegraphics{figures/V/oneparam/lo/freg.epsi}}\subfigure[]{\label{V-LOdiff}\includegraphics{figures/V/oneparam/lo/diff.epsi}}
\caption[$V$ analysis, 1-parameter fit at LO: a typical result.]{A typical result for the $V$ analysis: (a) $\chi_L^2$ as a function of ${\cal O}_4^V$; (b) 
the regularised function $f(s)$ [Eq.(\ref{inteq})] compared with data \cite{aleph05}; (c) difference between data and the regularised function. We have chosen for the error parameter ${\cal O}_6^V=1\times 10^{-3}\mbox{GeV}^6$.}
\label{V-freg+chiL}
\end{center}
\end{figure}

\begin{figure}[H]
\begin{center}
\subfigure[]{\label{VpA-LOchiL}\includegraphics{figures/V+A/oneparam/lo/chiL.epsi}}\subfigure[]{\label{VpA-LOfreg}\includegraphics{figures/V+A/oneparam/lo/freg.epsi}}\subfigure[]{\label{VpA-LOdiff}\includegraphics{figures/V+A/oneparam/lo/diff.epsi}}
\caption[$V+A$ analysis, 1-parameter fit at LO: a typical result.]{A typical result for the $V+A$ analysis: (a) $\chi_L^2$ as a function of ${\cal O}_4^{V+A}$; (b) 
the regularised function $f(s)$ [Eq.(\ref{inteq})] compared with data \cite{aleph05}; (c) difference between data and the regularised function. We have chosen for the error parameter ${\cal O}_6^{V+A}=1\times 10^{-3}\mbox{GeV}^6$.}
\label{V+Afreg+chiL}
\end{center}
\end{figure}

Typical results of the algorithm, for both channels, are shown in Figs.\ \ref{V-freg+chiL} and \ref{V+Afreg+chiL}. Again, as we have seen in the analysis of the $V-A$ and $A$ channels, the regularised function follows nicely the data points, except at large $s$, and $\chi_L^2$ has the expected quadratic dependence.

We have performed all the consistency checks discussed in the analysis of the $A$ channels and found no justification to change the number of data points $N$ used in the analysis or $s_1$, the upper limit of the space-like interval $\Gamma_L$. Also, the dependence on the lower limit $s_2$ shows again that the best simultaneous description of theory and data corresponds to an error corridor defined by the last known term in the perturbative series. Unfortunately this agreement is very poor now for the $V$ and $V+A$ channels, the corresponding $\chi_{L,{\rm min}}^2$ being in the order of $O(10)$ while a value in the order of $O(1)$ is expected.

The results of these 1-parameter fits are:
\begin{equation}
 \begin{array}{ll}
 {\cal O}_4^V=2.6_{-0.9}^{+0.9}\times 10^{-3}{\rm GeV}^4\ \ \mbox{at}\ \ 0.28\%\mbox{CL},\\
\\
 {\cal O}_4^{V+A}=5.4_{-1.9}^{+1.7}\times 10^{-3}{\rm GeV}^4\ \ \mbox{at}\ \ 9.17\%\mbox{CL}.
 \end{array}
\label{1VV+Aresults}
\end{equation}

As before, these values can be translated into values for the gluon condensate:
\begin{equation}
 \begin{array}{ll}
 \langle\frac{\alpha_s}{\pi}G_{\mu\nu}^aG_{\mu\nu}^a\rangle=3.12_{-1.08}^{+1.08}\times 10^{-2}{\rm GeV}^4\ \ \mbox{from the}\ V\ \mbox{channel},\\
\\
\langle\frac{\alpha_s}{\pi}G_{\mu\nu}^aG_{\mu\nu}^a\rangle=3.24_{-1.14}^{+1.02}\times 10^{-2}{\rm GeV}^4\ \ \mbox{from the}\ V+A\ \mbox{channel}.
 \end{array}
\label{1VV+Agluonrsults}
\end{equation}

Both results, Eqs.\ \ref{1VV+Aresults} and \ref{1VV+Agluonrsults}, agree with the corresponding results from the $A$ analysis, Eqs.\ \ref{1Aresults} and \ref{1Agluonresults}, respectively. Even though in the present analysis there is a poor agreement between theory and data, the extracted values for the condensates seem to be conclusive. As expected, the values of ${\cal O}_4^A$ and ${\cal O}_4^V$ are almost equal and their sum agrees nicely with the one extracted from the $V+A$ channel. 

\subsection{2-parameter fits}

\begin{figure}[!p]
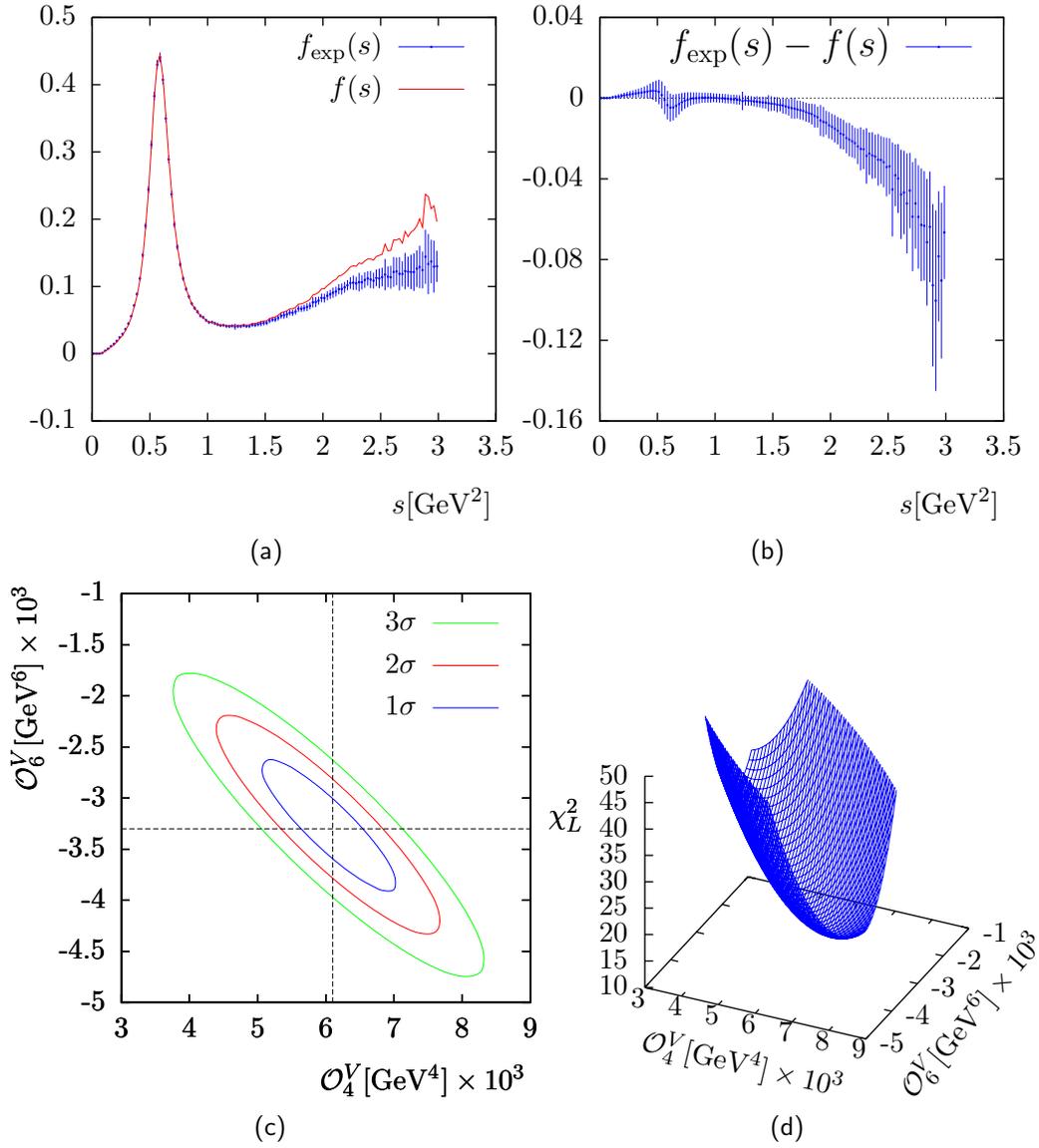

\centering
\subfigure[]{\label{V2freg}\includegraphics{figures/V/twoparam/freg.epsi}}
\subfigure[]{\label{V2diff}\includegraphics{figures/V/twoparam/diff.epsi}}
\subfigure[]{\label{V2chi2D}\includegraphics{figures/V/twoparam/chiL.epsi}}
\subfigure[]{\label{V2chi3D}\includegraphics{figures/V/twoparam/chiL-3D.epsi}}
\caption[$V$ analysis, result of the 2-parameter fit at LO.]{$V$ analysis, 2-parameter fit (LO) at $<0.01\%$CL ($\chi_{L,{\rm min}}^2=20.42$) for the agreement between theory and data: (a) the regularised function $f(s)$ [Eq.(\ref{inteq})] compared with data \cite{aleph05}; (b) difference between data and the regularised function; (c) 1-, 2- and 3$\sigma$ confidence regions in the $\left({\cal O}_4^{V},{\cal O}_6^{V}\right)$-plane; The central values are marked by dashed lines. (d) $\chi_L^2$ [Eq.(\ref{chiL2})] as a function of ${\cal O}_4^{V}$ and ${\cal O}_6^{V}$.}
\label{V2paramresult}
\end{figure}

\begin{figure}[!p]
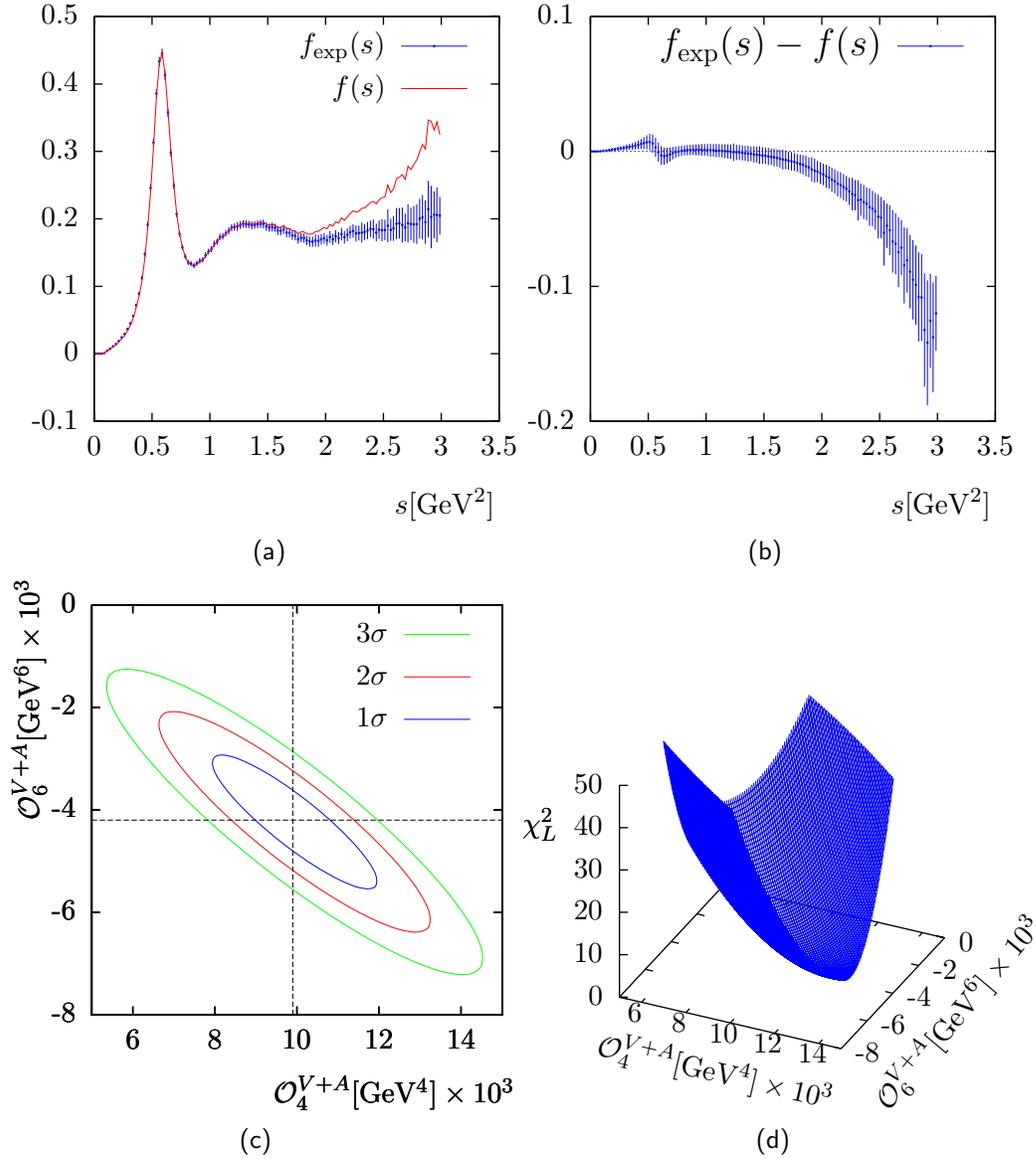

\centering
\subfigure[]{\label{V+A2freg}\includegraphics{figures/V+A/twoparam/freg.epsi}}
\subfigure[]{\label{V+A2diff}\includegraphics{figures/V+A/twoparam/diff.epsi}}
\subfigure[]{\label{V+A2chi2D}\includegraphics{figures/V+A/twoparam/chiL.epsi}}
\subfigure[]{\label{V+A2chi3D}\includegraphics{figures/V+A/twoparam/chiL-3D.epsi}}
\caption[$V+A$ analysis, result of the 2-parameter fit at LO.]{$V+A$ analysis, 2-parameter fit (LO) at $0.08\%$CL ($\chi_{L,{\rm min}}^2=7.11$) for the agreement between theory and data: (a) the regularised function $f(s)$ [Eq.(\ref{inteq})] compared with data \cite{aleph05}; (b) difference between data and the regularised function; (c) 1-, 2- and 3$\sigma$ confidence regions in the $\left({\cal O}_4^{V+A},{\cal O}_6^{V+A}\right)$-plane; The central values are marked by dashed lines. (d) $\chi_L^2$ [Eq.(\ref{chiL2})] as a function of ${\cal O}_4^{V+A}$ and ${\cal O}_6^{V+A}$.}
\label{V+A2paramresult}
\end{figure}

We have also performed 2-parameter fits to determine the correlation between the dimension $d=4$ and $d=6$ condensates. For this purpose, we have kept all the other parameters at the quoted values (see consistency checks for the 1-parameter fits) and found the results presented in Figs.\ \ref{V2paramresult} and \ref{V+A2paramresult}. Even though these results are as expected, i.e., the regularised function $f(s)$ follows nicely the data points and $\chi_L^2$ has a quadratic behaviour, the problem of having $\chi_{L,{\rm min}}^2$ in the order $O(10)$ still persits and thus the corresponding CL, taken from Fig.\ref{dofpvaluechi} in Appendix \ref{statistics}, is very small.

One can see that also in these two channels one finds a strong correlation between the two parameters with the central values:
\begin{equation}
\begin{array}{l}
  {\cal O}_4^V=6.1_{-1.1}^{+0.9}\times 10^{-3}{\rm GeV}^4,\\
\\
  {\cal O}_6^V=-3.3_{-0.6}^{+0.7}\times 10^{-3}{\rm GeV}^6,
\end{array}\ \mbox{at}\ <0.01\%\mbox{CL}\ (\chi_{L,{\rm min}}^2=20.42) ,
\end{equation}
\begin{equation}
\begin{array}{l}
  {\cal O}_4^{V+A}=9.9_{-2.0}^{+2.1}\times 10^{-3}{\rm GeV}^4,\\
\\
  {\cal O}_6^{V+A}=-4.2_{-1.3}^{+1.3}\times 10^{-3}{\rm GeV}^6,
\end{array}\ \mbox{at}\ 0.08\%\mbox{CL}\ (\chi_{L,{\rm min}}^2=7.11).\\
\end{equation}

Due to the negative correlation, Figs.\ \ref{V2chi2D} and \ref{V+A2chi2D} contain a lot of information about a specific combination of the two parameters, which allows the determination of one of the parameters, with well defined errors, if the other one is known from an independent analysis. These combinations are:
\begin{equation}
\begin{array}{l}
 {\cal O}_6^V+0.65{\rm GeV^2}\cdot{\cal O}_4^V=0.66_{-0.25}^{+0.25}\times 10^{-3}{\rm GeV}^6,\\
\\
{\cal O}_6^{V+A}+0.65{\rm GeV^2}\cdot{\cal O}_4^{V+A}=2.20_{-0.51}^{+0.50}\times 10^{-3}{\rm GeV}^6.
\end{array}
\end{equation}

Despite the poor agreement of theory with data, present in the analysis of the $V$ and $V+A$ channels, it is important to remark that the values and allowed ranges for the condensates are in good agreement with those found from the $A$ and $V-A$ channels. Note that the slope of the correlations in the $V$, $A$ and $V+A$ channels is identical, and thus allowing one to see that the sum of the $V$ and $A$ channels agrees with that found in the $V+A$ case. Also the central values agree within errors. For the $V-A$ case, the values for both condensates, of dimension $d=4$ and $d=6$, agree with the values found by actually taking the difference between the results from the $V$ and $A$ channels separately.  

\clearpage

\section{Review and comparison of results}

Besides a consistency check of the results presented in this work, one should also compare them to those from other groups. There exist in the literature some previous extractions of QCD condensates in the $V$ and $A$ channels. They are based on sum rules approaches and the analysis is performed at leading order. However, the normalisation of spectral functions is different from ours and there are also other factors absorbed in the definition of the condensates. We have translated the results so that they can be compared to ours. They can be found in Table 7.1. 

\begin{table}[H]
\begin{center}
\begin{tabular}{||l@{\vrule height 11pt depth4pt width0pt\hskip\arraycolsep}|c|c||c|c||c}\hline
 & \multicolumn{2}{|c||}{$V$ channel} &\multicolumn{2}{|c||}{$A$ channel }\\ \hline
 & ${\cal O}_4^V$ & ${\cal O}_6^V$ & ${\cal O}_4^A$ & ${\cal O}_6^A$ \\ \hline
\cite{bertlmann1} & (1.8 -- 4.8) & -- (12.9 -- 8.4) & & \\ \hline
\cite{dominguez0} & (0.6 -- 2.8) & -- (8.1 -- 4.1) & (1.2 -- 2.5) & (4.1 -- 7.1)\\ \hline 
\cite{dominguez1} & (1.3 -- 4.8) & -- (19.0 -- 5.9) & (1.3 -- 4.8) & -- (16.5 -- 1.3) \\ \hline 
\end{tabular}
\caption[$V$, $A$ and $V+A$ analysis: estimated ranges of dimension $d\le6$ condensates found in the literature.]{Estimated ranges for the dimension $d=4$ and $d=6$ condensates of the $V$ and $A$ channels in units of $10^{-3}{\rm GeV}^d$ at leading order. Existing results from the literature are presented. Note that the normalisation, for all of them, was adjusted so that they can be compared to those from this work.}
\end{center}
\label{VpAlit}
\end{table}

\begin{table}[H]
\begin{center}
\begin{tabular}{||l@{\vrule height 11pt depth4pt width0pt\hskip\arraycolsep}|c|c||c|c||c|c||c}\hline
 & \multicolumn{2}{|c||}{$V$ channel} &\multicolumn{2}{|c||}{$A$ channel } & \multicolumn{2}{|c||}{$V+A$ channel}\\ \hline
 & ${\cal O}_4^V$ & ${\cal O}_6^V$ & ${\cal O}_4^A$ & ${\cal O}_6^A$ & ${\cal O}_4^{V+A}$ & ${\cal O}_6^{V+A}$ \\ \hline
1-parameter fit & $2.6_{-0.9}^{+0.9}$ & & $2.8_{-0.8}^{+0.6}$ & & $5.4_{-1.9}^{+1.7}$ & \\ \hline
2-parameter fit & $6.1_{-1.1}^{+0.9}$ & $-3.3_{-0.6}^{+0.7}$ & $3.8_{-0.9}^{+1.1}$ & $-1.0_{-0.7}^{+0.6}$ & $9.9_{-2.0}^{+2.1}$ & $-4.2_{-1.3}^{+1.3}$\\ \hline 
\end{tabular}
\caption[$V$, $A$ and $V+A$ analysis: estimated values of dimension $d\le6$ condensates found in this work.]{Results of this work for the dimension $d=4$ and $d=6$ condensates in units of $10^{-3}{\rm GeV}^d$ at leading order.}
\end{center}
\label{VpAtable}
\end{table}

One can remark that the values found in this work (Table 7.2) are consistent with those from the literature. The sign of ${\cal O}_6^A$, though, disagrees with the vacuum saturation approximation, provided ${\cal O}_6^V$ is negative, as we have found here. The negative sign of ${\cal O}_6^A$ disagrees also with the results from \cite{dominguez0}. 

As a conclusion, we can state that the values and ranges found for the QCD condensates are all consistent among themselves and with previous extractions found in the literature even though the agreement between theory and data is very poor in the case of the $V$ and $V+A$ channels. However, one can hope that when eliminating some (or all) of the restrictions imposed when performing the analysis the agreement between theory and data will improve. 

When analysing all four channels, we have assumed chiral symmetry, decoupling of heavy quarks and no duality violation. If the chiral symmetry is broken, there are also lower order terms entering the OPE and also mass terms would be present both in the OPE and the perturbative expansion. As a consequence, there will be also a perturbative contribution to the $V-A$ correlator. These terms could change the results, however remains to be seen if the effect is negligible or not. 

Also the inclusion of heavy quarks could change the results. Heavy quarks are expected to play an important role at high energies, i.e., they could play a sensible role for the lower end of the space-like interval, $s_2$, as well as for $\Gamma_{\rm exp}:(s_{\rm max},\infty)$ where several resonances are known. For a detailed discussion of these resonances one can look for example in \cite{eidelman} and the references therein.

Duality refers to the fact that, if the approximation $\Pi(s)\rightarrow\Pi_{OPE}(s)$ were exact and this substitution carried no error, one can say that the experimental spectral function $\mbox{Im}\Pi(t)$ is dual to $\Pi_{OPE}(s)$. The term {\em duality violation}, consequently, refers to any contribution missed by this substitution. 

As stated above, in our analysis we have assumed no duality violation. However it is interesting to check if, and how much, the results would change if one would consider duality violation contributions. Unfortunately, the theory of duality violation in QCD is almost non-existing. Ref.\cite{shifman4} was the first to point out the importance of duality violations. That the OPE may also miss other important contributions has recently been suggested in \cite{zakharov}. A model of duality violation can be found in \cite{cata}.

\clearpage{\thispagestyle{empty}\cleardoublepage}

\thispagestyle{empty}

\begin{center}
\vspace*{9cm}
{\Huge\bf The inverse conductivity problem}

\addcontentsline{toc}{part}{The inverse conductivity problem}

\end{center}

\clearpage{\thispagestyle{empty}\cleardoublepage}

\chapter{Electrical impedance tomography}
\label{EITintorduction}

Magnetic resonance imaging (MRI), functional MRI (fMRI), X-ray tomography or computed tomography (CT), magnetic encephalography (MEG) and ultrasound imaging are examples of medical imaging methods that produce a picture of a patient's inner organs without intruding into the body. Images coming from different systems complement each other since each of these methods measures the distribution of some specific physical parameter, such as mass density (CT) or
 sound speed (ultrasound imaging). 

\begin{figure}[H]
\begin{center}
\epsfig{file=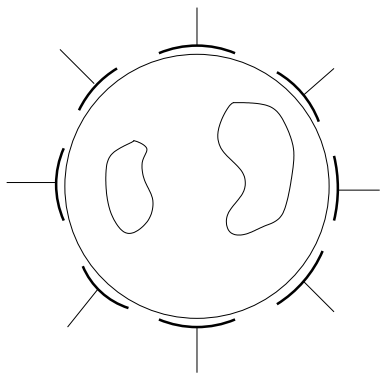,width=5cm}
\end{center}
\caption{Electrodes attached on the boundary of an object for current injection and voltage measurement.}
\end{figure}

Some of the methods are more harmful to the patient than others: X-ray measurements expose her to dangerous radiation whereas ultra sound imaging has little or no health risks. Another difference between the systems is that some of the machines are very expensive. For these reasons it is valuable to have as many different imaging methods available as possible. 

Electrical impedance tomography (EIT) is a medical imaging system that produces a picture of the electric conductivity distribution inside the patient. For the EIT measurement an array of electrodes is attached for instance around
 the chest of the patient. Measurements are done by feeding electric currents through the electrodes and measuring the corresponding voltages at the electrodes. This measurement is repeated with several different choices of current patterns. The resulting image is an approximate map of the electric conductivity. 

Electrical properties such as the electrical conductivity $\sigma$ and the electric susceptivity $\epsilon$, determine the behaviour of materials under the influence of external electric fields. For example, conductive materials have a high electrical conductivity and both direct and alternating currents flow easily through them. Dielectric materials have a large electric susceptivity and they allow passage of only alternating electric currents. 

\begin{figure}[ht]
\begin{center}
\epsfig{file=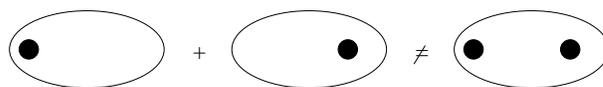,width=80mm}
\end{center}
\caption{EIT is a nonlinear problem.}
\end{figure}

The mathematical task of interpreting EIT measurements is an inverse boundary-value problem called the {\em inverse electric conductivity problem}. It is a challenging question in the theory of partial differential equations since it 
is nonlinear and ill-posed. Designing a practical algorithm for EIT is hard because non-linearity makes the mathematical solution theoretically difficult and ill-posedness means that a small error in the practical measurements may cause a large error in the reconstructed image. 

Let us consider a bounded, simply connected region $\Omega\subset{\R}^n$, for $n\geq 2$ and, at frequency $\omega$, let $\gamma$ be the complex admittivity function
\begin{equation}
\gamma({\bm x},\omega)=\sigma({\bm x})+i\omega\epsilon({\bm x}),
\end{equation}
where $\sigma(\bm x)$ is called the conductivity and $\epsilon(\bm x)$ the susceptivity.

The electrical impedance is the inverse of $\gamma({\bm x})$ and it measures the ratio between the electric field and the electric current at location ${\bm x}\in\Omega$. Electrical impedance tomography is the inverse problem of determining the impedance in the interior of $\Omega$, given simultaneous measurements of direct or alternating electric currents and voltages at the boundary $\partial\Omega$.

Different materials have different electrical properties, as shown in Tables \ref{tissue1} and \ref{tissue2}, so a map of $\sigma({\bm x})$ and $\epsilon({\bm x})$, for ${\bm x}\in\Omega$, can be used to infer the internal structure in $\Omega$. Due to this fact, EIT is an imaging tool with important applications in vastly different fields such as medicine, geophysics, environmental sciences and non-destructive testing of materials. Examples of medical applications of EIT are the detection of pulmonary emboli \cite{chen1,harr,hold}, monitoring of apnoea \cite{akba}, monitoring of heart function and blood-flow \cite{coll,isaa1} and breast cancer detection \cite{chen1}. In geophysics and environmental sciences, EIT can be useful for locating underground mineral deposits \cite{park}, detection of leaks in underground storage tanks \cite{ram1} and for monitoring flows of injected fluids into the earth, for the purpose of oil extraction or environmental cleaning \cite{ram2}. Finally, in non-destructive testing, EIT can be used for the detection of corrosion \cite{sant1} and of small defects, such as cracks or voids, in metals \cite{aless1,aless2,cedi,eggl,frie,sant2}. 

\begin{table}[H]
\centering
\begin{tabular}{lc} \hline\hline
Rock or fluid & $1/\sigma$ ($\Omega$ cm)  \\ \hline
Marine sand, shale & 1-10  \\
Terrestrial sands, claystone & 15-50 \\
Volcanic rocks, basalt & 10-200 \\
Granite & 500-2000 \\
Limestone dolomite, anhydrite & 50-5000 \\ 
Chloride water from oil fields & 0.16 \\
Sulfate water from oil fields & 1.2 \\ \hline\hline
\end{tabular}
\caption[Resistivity of rocks and fluids.]{Resistivity of rocks and fluids \cite{kell}.}
\label{tissue1}
\end{table}
\begin{table}[H]
\centering
\begin{tabular}[b]{lcc} \hline\hline
Tissue & $1/\sigma(\Omega\mbox{cm})$ & $\epsilon(\mu\mbox{Fm}^{-1})$ \\ \hline
Lung & 950 & 0.22 \\
Muscle & 760 & 0.49 \\
Liver & 685 & 0.49 \\
Heart & 600 & 0.88 \\
Fat & $>$1000 & 0.18 \\ \hline\hline
\end{tabular}
\caption[Electrical properties of biological tissue measured at frequency $10{\rm kHz}$.]{Electrical properties of biological tissue measured at frequency $10{\rm kHz}$ \cite{barb,schw}.}
\label{tissue2}  
\end{table}

EIT has been studied extensively in the last two decades and substantial progress has been made in both the theoretical and applied aspects of the problem. At the same time, EIT remains an area of active research and it continues to pose a variety of challenging questions for theoreticians, numerical analysts and experimentalists alike. 

\section{The mathematical model}
\label{eitprob}

Electric and magnetic fields with harmonic time dependence
\begin{equation}
\begin{array}{l}
{\cal {\bm E}}({\bm x},t)=\mbox{Re}\{{\bm E}({\bm x},\omega)e^{i\omega t}\}, \\
\\
{\cal {\bm H}}({\bm x},t)=\mbox {Re}\{{\bm H}({\bm x},\omega)e^{i\omega t}\}, 
\end{array}
\end{equation}
satisfy Maxwell's equations in the following form:
\begin{equation}
\begin{array}{l}
\bm\nabla\times {\bm H}({\bm x},\omega)=\gamma({\bm x},\omega){\bm E}({\bm x},\omega),  \\
\\
\bm\nabla\times {\bm E}({\bm x},\omega)=-i\omega\mu({\bm x}){\bm H}({\bm x},\omega), 
\end{array}
\label{maxwell}
\end{equation}
where $\mu({\bm x})$ is the magnetic permeability. EIT operates at low frequencies $\omega$, in a regime with admittivities $\gamma$ and length scales $L$ satisfying $\omega\mu\mid\gamma\mid L^2 \ll 1$, such that, after a simple scaling analysis \cite{chen1}, equations (\ref{maxwell}) are approximated by
\begin{equation}
\begin{array}{l}
\bm\nabla\times {\bm H}({\bm x},\omega)=\gamma({\bm x},\omega){\bm E}({\bm x},\omega),  \\
\\
\bm\nabla\times {\bm E}({\bm x},\omega)=0.
\end{array}
\label{maxwell1}
\end{equation} \par
We define the scalar electric potential $\phi$ and the vector-valued, electric current density with harmonic time dependence ${\cal {\bm I}}({\bm x},t)=\mbox{Re}\{{\bm j}({\bm x},\omega)e^{i\omega t}\}$, as
\begin{eqnarray}
{\bm E}({\bm x},\omega)=-\bm\nabla\phi({\bm x},\omega), & \bm\nabla\times{\bm H}({\bm x},\omega)={\bm j}({\bm x},\omega),
\end{eqnarray}
such that the first equation in (\ref{maxwell1}) becomes Ohm's law
\begin{equation}
{\bm j}({\bm x})=-\gamma({\bm x},\omega)\bm\nabla\phi({\bm x},\omega).
\label{ohm}
\end{equation}

Note that the density of dissipated energy, averaged over a period of oscillations, 
\begin{eqnarray}
\lefteqn{{\frac{\omega}{2\pi}}\int_t^{t+{\frac{2\pi}{\omega}}}{\cal {\bm I}}
({\bm x},\tau)\cdot{\cal {\bm E}}({\bm x},\tau)d\tau=}\nonumber\\ 
& &={\frac{1}{2}}[\mbox{Re}\{{\bm j}({\bm x},\omega)\}\cdot\mbox{Re}\{{\bm E}
({\bm  x},\omega)\}+\mbox{Im}\{{\bm j}({\bm x},\omega)\}\cdot\mbox{Im}\{
{\bm E}({\bm x}  ,\omega)\} \nonumber\\
 & &=\frac{1}{2}\sigma({\bm x})|\bm\nabla\phi({\bm x},\omega)|^2,
\end{eqnarray}
must be strictly positive, so we require that
\begin{equation}
\sigma({\bm x})=\mbox{Re}\{\gamma({\bm x},\omega)\}\geq m>0,
\label{sigma}
\end{equation}
with $m$ a positive constant.

If there are no electrical sources within $\Omega$, by definition, ${\bm j}$ is divergence free so Ohm's law (\ref{ohm}) gives the partial differential equation 
\begin{equation}
\bm\nabla\cdot[\gamma({\bm x},\omega)\bm\nabla\phi({\bm x},\omega)]=0 \ \ \ 
\mbox{in}\ \Omega,
\label{forword}
\end{equation}
which one should consider together with either Dirichlet boundary conditions
\begin{equation}
\begin{array}{ll}
\phi({\bm x},\omega)=V({\bm x},\omega), & \mbox{for} \ {\bm x}\in\partial
\Omega,
\end{array}
\label{forword:bound}
\end{equation}
or Neumann boundary conditions
\begin{equation}
\gamma({\bm x},\omega)\bm\nabla\phi({\bm x},\omega)\cdot{\bm  n}({\bm x})\equiv
\gamma({\bm x},\omega)\frac{\partial\phi({\bm x},\omega)}{\partial
  n}=I({\bm x},\omega)\ \  \mbox{at}\ \partial\Omega,
\label{neumannbound}
\end{equation}
such that
\begin{equation}
\int_{\partial\Omega}I({\bm x},\omega)ds({\bm x})=0,
\label{current}
\end{equation}
where ${\bm n}({\bm x})$ is the outer-pointing normal at $\bm x\in\partial\Omega$.

The boundary value problems (\ref{forword}--\ref{forword:bound}) or (\ref{forword}--\ref{neumannbound}) for a known function $\gamma({\bm x},\omega)$ in $\Omega$ and data $I({\bm x},\omega)$ or $V({\bm x},\omega)$, given for all ${\bm x}\in \partial\Omega$, are referred to as {\em continuum, forward} mathematical problems for EIT\footnote{\sf Analytic 
and semi-analytic solutions to a number of forward problems related to the image reconstruction problem of EIT in two and three dimensions can be found in \cite{pidc1}.}.

\section{Modelling the electrodes}
\label{electrodesmodeling}

In practice, we do not know the boundary current $I({\bm x},\omega)$ for all ${\bm x}\in\partial\Omega$. What we actually know are currents flowing along wires which are attached to $N$ discrete electrodes, which in turn are attached to the boundary $\partial\Omega$ \cite{some1}. Then, the question is how to model the electrodes? 

The {\em gap model} approximates the current density by a constant at the surface of each electrode and by zero in the gaps between the electrodes. This model is appealing because of its simplicity but it is not accurate \cite{isaa3}. 

A better choice is the {\em complete model} proposed in \cite{some1}. Suppose that $I_l(\omega)$ is the electric current sent through the wire attached to the $l$-th electrode. At the surface $S_l$ of this electrode, the normal current density satisfies
\begin{equation}
\int_{S_l}\gamma({\bm x},\omega)\frac{\partial\phi({\bm x},\omega)}{\partial n}ds({\bm x})=I_l(\omega).
\label{complete1}
\end{equation}
In the gaps between the electrodes, we have
\begin{equation}
\gamma({\bm x},\omega)\frac{\partial\phi({\bm x},\omega)}{\partial n}=0.
\end{equation}
At the contact of $S_l$ with $\partial\Omega$, there is in general (often non-negligible) an electro-chemical effect which gives rise to a thin, highly resistive layer. This is taken into account by the surface impedance $z_l(\bm x,\omega)$ and
\begin{equation}
\phi({\bm x},\omega)+z_l(\bm x,\omega)\gamma({\bm x},\omega)\frac{\partial\phi({\bm x},\omega)}{\partial  n}=V_l(\omega)\ \  \mbox{for}\ {\bm x}\in S_l,\ l=1,...N,
\end{equation}
where $V_l(\omega)$ is the measured voltage at the $l$-th electrode. Finally, due to conservation of charge one has
\begin{equation}
\sum_{l=1}^N I_l(\omega)=0,
\label{complete2}
\end{equation}
and by choice of ground,
\begin{equation}
 \sum_{l=1}^N V_l(\omega)=0.
\label{complete3}
\end{equation}

It is proved in \cite{some1} that equations (\ref{forword}), (\ref{complete1}-\ref{complete3}) have a unique solution and that they predict experimental data relatively well. However, the inverse problem based on this complete model remains essentially unstudied from the theoretical and practical points of view. 

In our analysis and the applications we discuss, we will use the approximation that the electrodes are point-like and take the current density to be zero in the gaps between them. Such a model is not as accurate as the complete one, but good enough for our purpose to first find qualitative results. Later on, one can try to implement also more complex models, e.g.\ gap or complete ones.
\clearpage

\section{Formulation of the inverse problem}
\label{inverseformulation}

In EIT, the admittivity function $\gamma({\bm x},\omega)$ is unknown and is to be determined from simultaneous measurements of boundary voltages $V({\bm x},\omega)$ and current densities $I({\bm x},\omega)$, respectively. 
There are two maps, the Dirichlet-to-Neumann and Neumann-to-Dirichlet maps, which relate $V({\bm x},\omega)$ to $I({\bm x},\omega)$. These maps depend non-linearly on the unknown $\gamma({\bm x},\omega)$.

The Dirichlet-to-Neumann map $\Lambda_\gamma:H^{1/2}(\partial\Omega)\rightarrow H^{-{1/2}}(\partial\Omega)$ is defined as\footnote{\sf $H^{1/2}(\partial\Omega)$ are special Sobolev spaces. $H^{-{1/2}}(\partial\Omega)$ is the dual of $H^{1/2}(\partial\Omega)$. For more details one can consult for example \cite{gilbarg}.} 
\begin{equation}
\Lambda_\gamma V({\bm x},\omega)=\gamma({\bm x},\omega)\frac{\partial\phi({\bm x},\omega)}{\partial n} \ \ \mbox{for} \ {\bm x}\in\partial\Omega,
\label{DtNmap}
\end{equation}
where  $V({\bm x},\omega)$ is arbitrary in $H^{1/2}$ and $\phi({\bm x},\omega)$ solves the forward problem (\ref{forword}-\ref{forword:bound}). In the static case $\omega=0$, we have $\gamma({\bm x},\omega)=\sigma({\bm x})$ and $\Lambda_\gamma=\Lambda_\sigma$.

The mathematical formulation of EIT, as first posed by Calder\'on \cite{cald1}, is as follows: {\em Find the admittivity function $\gamma({\bm x},\omega)$ in $L^\infty(\bar\Omega)$\footnote{\sf $L^\infty(\bar\Omega)$ denotes the Banach space of bounded functions on $\bar\Omega$ with the norm $$\displaystyle||u||_{L^\infty(\bar\Omega)}={\rm ess}\,\sup_{\bar\Omega}|u|.$$}, with a strictly positive real part $\sigma({\bm x})$, given the Dirichlet-to-Neumann map $\Lambda_\gamma$.}

In practice, it is not advisable to work with the Dirichlet-to-Neumann map. Instead, one uses the Neumann-to-Dirichlet map $(\Lambda_\gamma)^{-1}$ which is smooth and therefore better behaved for noisy measurements. Nevertheless, in principle, both maps contain the same information and, usually, the Dirichlet-to-Neumann map is used for convenience. 

The Neumann-to-Dirichlet map $(\Lambda_\gamma)^{-1}:{\sf I}\rightarrow H^{1/2}(\partial\Omega)$ is defined on the restricted space of currents
\begin{equation}
{\sf I}=\left\{I({\bm x},\omega)\in H^{-{1/2}}(\partial\Omega)\ \ \
\mbox{such that} \ \int_{\partial\Omega}I({\bm x},\omega)ds({\bm x})=0\right\}
\end{equation}
and, for any $I({\bm x},\omega)\in{\sf I}$,
\begin{equation}
(\Lambda_\gamma)^{-1}I({\bm x},\omega)=\phi({\bm x},\omega)\ \ \mbox{ at}\ \ \partial\Omega,
\label{NtDmap}
\end{equation}
 where
$\phi({\bm x},\omega)$ is the solution of the Neumann boundary value problem (\ref{forword},\ref{neumannbound}). For $\omega=0$, we have $(\Lambda_\gamma)^{-1}=(\Lambda_\sigma)^{-1}$. We note that $(\Lambda_\gamma)^{-1}$ is the generalised inverse, as defined in Section \ref{generalisedsolution}, of $\Lambda_\gamma$. 
\clearpage

\section{Imaging with incomplete, noisy data}

\begin{figure}[ht]
\begin{center}
\resizebox{9cm}{!}{
\rotatebox{-0}{
\epsfig{file=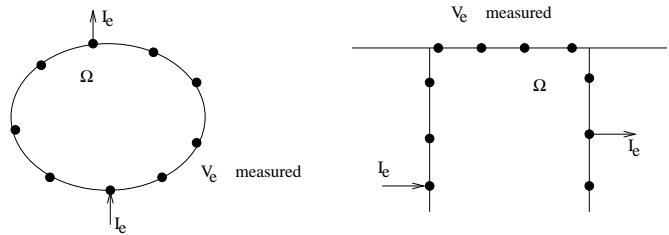}
}}
\end{center}
\caption{Illustration of experimental setups for gathering partial data about the Neumann-to-Dirichlet map.}
\label{s5}
\end{figure}

In practice, we do not have full knowledge of the maps $(\Lambda_\gamma)^{-1}$ or $\Lambda_\gamma$. Instead, we have a set of $N$ experiments, where we define an excitation pattern $I_e({\bm x},\omega)\in{\sf I}$ and we measure the resulting voltage $V_e({\bm x},\omega)$, at discrete locations ${\bm x}_p\in\partial\Omega$ of the electrodes, along the boundary. 

In Fig.\ref{s5} we illustrate two experimental setups: An excitation current $I_e$ is injected (extracted) at some electrodes and the resulting voltage is measured at all or some of the electrodes. The first setup is typical of medical applications, where one has access to all points of the boundary (i.e.\ the surface of the body or parts of it). The second setup is typical of geophysics applications, where measurements can be made at the earth's surface or in some boreholes. In any case, we say that we have partial knowledge of the Neumann-to-Dirichlet map and, furthermore, the data $V_e$ are usually contaminated with noise. The practical EIT problem is the following: {\em Find the admittivity function $\gamma({\bm x},\omega)$, given a partial, noisy knowledge of $(\Lambda_\gamma)^{-1}$.}

\section {A brief history of the problem}

Research on the inverse conductivity problem has two aspects: a purely mathematical one providing answers to the questions of uniqueness and reconstruction and a practical approach having as its goal algorithms that perform well on physical data. 

The theoretically oriented line of research was started by Calder\'on\footnote{\sf He studied first electrical engineering before changing to mathematics.} \cite{cald1}. He stated the inverse conductivity problem for the first time (in dimensions $n\ge 2$) and gave first results for conductivities close to a constant. Mathematically his work continued the tradition of Gel'fand, Levitan and B\"org \cite{borg,gelfand}. 

Once Calder\'on asked his question, new results started to appear. Kohn and Vogelius showed that for piecewise analytic conductivities the Dirichlet-to-Neumann map uniquely determines the conductivity in dimensions 2 or more ($n\ge2$) \cite {kohn}. Sylvester and Uhlmann showed in dimensions $n\ge 3$ that if $\partial\Omega\subset{\cal C}^\infty$ then $\Lambda_\sigma$ uniquely determines $\sigma\in {\cal C}^\infty(\bar\Omega)$ \cite{sylvester}\footnote{\sf ${\cal C}^k$ is the space of functions having $k$ continuous derivatives with a norm defined as: $$||f||_{{\cal C}^k(\Omega)}=\sum_{n=0}^k\sup_{x\in\Omega}|f^{(n)}(x)|.$$ Thus, ${\cal C}^\infty$ are infinitely derivable functions with continuous derivatives.}. 

The first reconstruction method was published by A. Nachman \cite{nachman}; this method is valid in dimensions strictly greater than two ($n\ge3$). 

Meanwhile the case $n=2$ remained unsolved. Why is the two-dimensional case more difficult than the higher-dimensional ones? A naive argument for this is that in $n$ dimensions the Schwartz kernel of $\Lambda_{\sigma}$ depends on $2(n-1)$ variables while $\sigma$ is a function of $n$ variables; this means that there is no over-determinacy in two dimensions and all the data have to be used for reconstruction. 

Since the inverse conductivity problem is very complicated mathematically, we can expect the practical EIT imaging task to be difficult. In addition to the non-linearity and ill-posedness of the theoretical problem there are further 
problems: the Dirichlet-to-Neumann map is not given; instead its inverse operator has to be approximated from noisy current-to-voltage measurements with finite-precision. Furthermore, only a finite number of current patterns 
can be applied through a finite number of electrodes. This situation places restrictions on the resolution of the images and requires the physical model of the electrode system to be refined. This leads to the notion of {\em distinguishability} describing the resolution it is possible to obtain in principle with a given system. 

With these restrictions, many algorithms have been developed and tested on real data. They fall roughly in three categories:
\begin{itemize}
\item{} Solve the linearised inverse problem. This approach works for conductivities close to constant; examples include back-projection and one-step Newton schemes like NOSER.
\item{} Solve the nonlinear problem by finding iteratively a conductivity distribution that matches the physical measurements best in the least squares sense. These methods require regularisation due to the ill-posedness of the 
inverse problem.
\item{} Solve the problem directly. The methods of this category are called direct methods as opposed to iterative ones; they aim as well to solve the full non-linear problem.
\end{itemize} 
Some of these ideas combined with elaborate electrical engineering make it possible to build reasonably working EIT devices. This has been done in several laboratories around the world.

\clearpage{\thispagestyle{empty}\cleardoublepage}

\chapter{The forward problem}
\label{directproblem}

The attention given to the direct problem of EIT is due to the fact that many image reconstruction algorithms are iterative and make many comparisons between the boundary data predicted by an estimate of the internal conductivity and the data that are measured. Also, data produced by solving the direct problem for known conductivities is very usefull in checking and studying reconstruction algorithms. In the following we discuss method for solving the direct problem often used in practice. 

For an isotropic conductivity distribution $\sigma({\bm x})$ the potential $\phi({\bm x})$ satisfies the equation (see Section \ref{eitprob})
\begin{equation}
\bm\nabla\cdot\left[\sigma({\bm x})\bm\nabla\phi({\bm x})\right]=0,
\label{basiceq2}
\end{equation}
where $\sigma({\bm x})$ and $\phi({\bm x})$ are defined in a volume $\Omega$ bounded by a closed surface $\partial\Omega$. The forward problem described here can be formulated as: {\em If the conductivity $\sigma$ is known in $\Omega$, together with either the potential $\phi$ or the current $\sigma\partial\phi/\partial n$ on the boundary $\partial\Omega$, find the potential $\phi$ everywhere inside $\Omega$}.

Eq.(\ref{basiceq2}) can be written in the form
\begin{equation}
\Delta\phi({\bm x})=-Y({\bm x}),
\label{eqy}
\end{equation}
where
\begin{equation}
Y({\bm x})=\frac{\bm\nabla\sigma({\bm x})}{\sigma({\bm x})}\cdot\bm\nabla\phi({\bm x})= \bm\nabla\ln\sigma({\bm x})\cdot\bm\nabla\phi({\bm x}).
\label{characteristics}
\end{equation}
A simplification occurs if $\sigma({\bm x})$ differs only slightly from a constant conductivity distribution
\begin{equation}
\sigma({\bm x})=\sigma_0+\delta\sigma({\bm x}).
\end{equation}
Then a linear approximation can be used in (\ref{eqy})
\begin{equation}
\Delta\phi({\bm x})=\frac{\bm\nabla\delta\sigma({\bm x})}{\sigma_0}\cdot\bm\nabla\phi_0({\bm x})
\end{equation}
where $\phi_0({\bm x})$ is the solution of (\ref{basiceq2}) for a constant conductivity $\sigma_0$, i.e. of Laplace's equation.

An elegant way to solve the differential equation (\ref{eqy}) is by changing it into an integral equation by means of the appropriate Green's function (how this is actually done see Appendix \ref{app:green}). Using the Neumann Green's function one gets:
\begin{equation}
\phi({\bm x})=\psi({\bm x})+\int_\Omega Y({\bm x'})G_N({\bm x},{\bm x'})d^nx',
\label{inteq1}
\end{equation} 
where $\psi({\bm x})$ is a harmonic function constructed from the measured Neumann boundary data:
\begin{equation}
\psi({\bm x})=\langle\phi\rangle_{\partial\Omega}+\oint_{\partial\Omega}\frac{\partial}{\partial n'}\phi({\bm x'})G_N({\bm x},{\bm x'})d^{n-1}x'.
\end{equation}

\section{Methods based on integral equations}

As it was shown in \cite{ciulli1,ciulli2}, one may use the change of variable $\tau=\sqrt{\sigma}$ to transform Eq.(\ref{basiceq2}) into an integral equation for the function $\Psi=\tau\phi$. Indeed, using this change of variable, Eq.(\ref{basiceq2}) becomes
\begin{equation}
 \Delta\Psi(\bm x)=-V(\bm x)\Psi(\bm x),\ \ \mbox{ where}\ V(x)\equiv-\frac{\Delta\tau(\bm x)}{\tau(\bm x)},
\end{equation}
 and, with the help of Green's theorem (see Appendix \ref{app:green}), one finds the desired integral equation:
\begin{equation}
 \Psi(\bm x)=\langle\Psi\rangle_{\partial\Omega}+\oint_{\partial\Omega}\frac{\partial}{\partial n'}\Psi(\bm x')G_N(\bm x,\bm x')d^{n-1}x'+\int_\Omega V(\bm x')\Psi(\bm x')G_N(\bm x,\bm x')d^nx'.
\end{equation}
Once this equation is solved, the potential $\phi(\bm x)$ can be found anywhere in the domain $\Omega$ by a simple division of $\Psi$ by $\tau$.

Another way to solve the forward problem consists of applying the operator $\bm\nabla_{\bm x}(\ln\sigma({\bm x}))\cdot\bm\nabla_{\bm x}$ to (\ref{inteq1}) to find
\begin{equation}
Y({\bm x})=\bm\nabla_{\bm x}(\ln\sigma({\bm x}))\cdot\bm\nabla_{\bm x}\psi({\bm x})+\int_\Omega K({\bm x},{\bm x}')Y({\bm x}')d^n{\bm x}'
\label{direct1}
\end{equation}
where $K({\bm x},{\bm x}')=\bm\nabla_{\bm x}(\ln\sigma({\bm x}))\cdot\bm\nabla_{\bm x}G_N({\bm x},{\bm x}')$.

This is an integral equation for $Y({\bm x})$, the Laplacian of $\phi({\bm x})$. Once $Y({\bm x})$ is known, one can compute $\phi({\bm x})$ by means of a quadrature using again formula (\ref{inteq1}).

Since $\bm\nabla_{\bm x}G_N({\bm x},{\bm x}')$ is divergent, $K({\bm x},{\bm x}')$ is not a compact operator, that is, the  integral equation is not of Fredholm type (see Appendix \ref{integrlequation}). However, by considering its first iteration, that is:
\begin{eqnarray}
Y({\bm x})=\bm\nabla_{\bm x}(\ln\sigma({\bm x}))\cdot\bm\nabla_{\bm x}\psi({\bm x})+\int_\Omega K({\bm x},{\bm x}')\bm\nabla_{{\bm x}'}(\ln\sigma({\bm x}'))\cdot\bm\nabla_{{\bm x}'}\psi({\bm x}')d^n{\bm x}'\nonumber\\
+\int_\Omega K_2({\bm x},{\bm x}')Y({\bm x}')d^n{\bm x}'\hspace{3.1cm}
\label{direct2}
\end{eqnarray}
with
\begin{equation}
K_2({\bm x},{\bm x}')=\int_\Omega\bm\nabla_{\bm x}(\ln\sigma({\bm x}))\cdot\bm\nabla_{\bm x}G_N({\bm x},{\bm y})\bm\nabla_{\bm y}(\ln\sigma({\bm y}))\cdot\bm\nabla_{\bm y}G_N({\bm y},{\bm x}')d^n{\bm y},
\end{equation}
one finds a second kind integral equation with a Hilbert-Schmidt kernel (see Appendix \ref{integrlequation}). A proof of this statement can be found in Ref.\ \cite{sebu1}, where a similar approach for solving the forward problem has been presented. The integral equation (\ref{direct2}) can now be solved by means of usual numerical procedures for Fredholm equations. 

\section{Finite element method} 
\label{direct:FEM}

Another way of solving the direct problem is to use the Finite Element Method (FEM). But one should first transform the partial differential equation into a variational problem. This is simply done by taking the weighted integral on both sides of (\ref{basiceq2})
\begin{equation}
0=\int_\Omega w({\bm x})\bm\nabla\cdot\left[\sigma({\bm x})\bm\nabla\phi({\bm x})\right]d^nx
\end{equation}
with $w({\bm x})$ any square integrable, once differentiable, continuous function, i.e.\ $w\in H^1_\diamond(\Omega)$. Then, by partial integration, one finds an integral condition for the potential inside $\Omega$
\begin{equation}
0=-\int_\Omega\sigma({\bm x})\bm\nabla w({\bm x})\cdot\bm\nabla\phi({\bm x})d^nx+\oint_{\partial\Omega}\sigma({\bm x})w({\bm x})\frac{\partial}{\partial n}\phi({\bm x})d^{n-1}x.
\end{equation}
It is to be remarked that here the Neumann boundary conditions appear naturally.

In this way, the problem of finding the potential $\phi$ for a known conductivity is changed: {\em Find the potential $\phi$ which, for any function $w\in H^1_\diamond(\Omega)$, satisfies the condition:}
\begin{equation}
\int_\Omega\sigma({\bm x})\bm\nabla w({\bm x})\cdot\bm\nabla\phi({\bm x})d^nx=\oint_{\partial\Omega}w({\bm x})\sigma({\bm x})\frac{\partial}{\partial n}\phi({\bm x})d^{n-1}x.
\label{FEMeq}
\end{equation}

Such a formulation is also called a {\em weak} formulation. It is {\em weak} in the sense that the local condition for the potential given by the differential equation is now changed into an integral one which is definitely less demanding.

\begin{figure}[ht]
\begin{center}
\epsfig{file=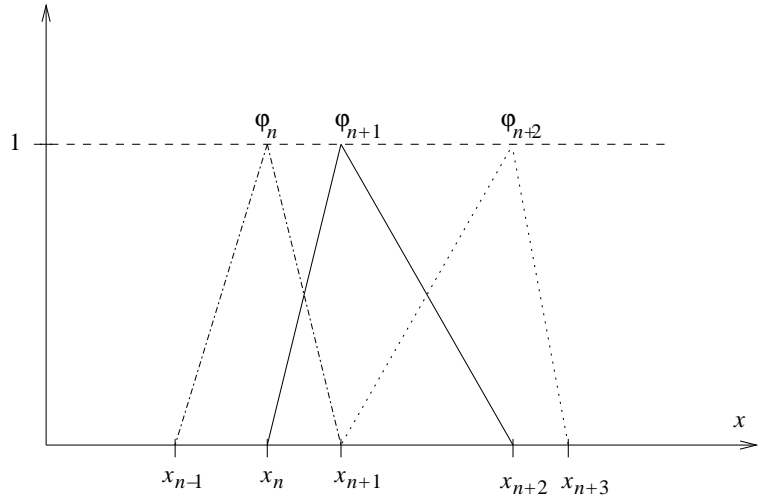,width=100mm}
\end{center}
\caption{Tent functions for the one-dimensional case.}
\label{tent}
\end{figure}

In order to find a numerical solution of Eq.(\ref{FEMeq}), let us rewrite it in the form of a variational problem:
\begin{equation}
a(\phi,w)=l(w),\ \ \forall w
\end{equation}
with the bilinear form
\begin{equation}
a(u,v)=\int_\Omega \sigma\bm\nabla u\bm\nabla v\, d^nx,
\end{equation}
and the linear functional
\begin{equation}
l(w)=\oint_{\partial\Omega}fw\,d^{n-1}x,\ \ f=\sigma\frac{\partial u}{\partial n}.
\end{equation}
Choosing $\{\varphi_1,\varphi_2,...\}$ as a base in the space of the functions $w$ and postulating the expansion
\begin{equation}
\phi=\sum_k z_k\varphi_k
\label{potexpansion}
\end{equation}
for the potential, we find the following system of linear equations to be solved:
\begin{equation}
\sum_k z_ka(\varphi_k,\varphi_i)=l(\varphi_i).
\label{femsystem}
\end{equation}
For the computation of (otherwise time consuming) integrals we use linear finite element spaces \cite{pepper}. It is practical to do so since the bases of these spaces are formed out of linear tent-functions (Fig.\ref{tent}) and thus, the integrals differ from zero only for nearest neighbour nodal points or when they are taken over the same index.

\begin{eqnarray}
\int_\Omega\varphi_k\varphi_i\ d^nx=0 & \forall\ |k-i|>1, \\
\int_\Omega\bm\nabla\varphi_k\bm\nabla\varphi_i\ d^nx=0 & \forall\ |k-i|>1. 
\end{eqnarray}

The matrix inversion is safe since the matrix is positive-definite. Indeed,
\begin{eqnarray}
z^TAz&=&\sum_{i,k} z_iA_{ik}z_k\nonumber\\
&=&a\left(\sum_i z_i\varphi_i,\sum_k z_k\varphi_k\right)\nonumber\\
&=&a(u,u)\geq\rho||u||^2\nonumber.
\end{eqnarray}

Once the system (\ref{femsystem}) is solved, the direct problem is also solved and thus we have the potential $\phi$ in the whole domain and we can further use it to check our reconstruction algorithms (see next Chapter).

For numerically solving the forward problem we have used MATLAB \cite{kwon}, since the FEM is already implemented, for two-dimensional domains, via a toolbox (PDE-Toolbox). The main steps of this algorithm are described below, while the entire code can be found in \cite{almasy1}.

\begin{algorithm}[H]
\dontprintsemicolon
\caption[Algorithm for solving the 2-dimensional forward problem by means of FEM.]{\sf Algorithm for solving the 2-dimensional forward problem by means of FEM in MATLAB.}
\KwData{Geometry of the problem, boundary conditions, conductivity distribution, desired number of electrodes.}
\KwResult{Data files containing the potential and current on the electrodes.}
\Begin{
solve the system of equations (\ref{femsystem}) for $z_k$\;
by means of Eq.(\ref{potexpansion}) calculate the potential everywhere inside $\Omega$\;
calculate the data on the boundary $\partial\Omega$\;
write out data files\;}
\end{algorithm}
 
\clearpage{\thispagestyle{empty}\cleardoublepage}

\chapter{Reconstruction algorithms}
\label{reconstructions}

It is well known that the inverse conductivity problem is a highly  ill-posed, non-linear inverse problem and that the images produced are very sensitive to errors which can occur in practice. There has been much interest in determining the class of conductivity distributions that can be recovered from the boundary data, as well as in the development of related reconstruction algorithms. The interest in this problem has been generated by both difficult theoretical challenges and by the important medical, geophysical and industrial application of it. Much theoretical work has been related to the approach of Calder\'on concerning the bijection between the conductivity inside the region and the Neumann-to-Dirichlet operator, which relates the distribution of the injected currents to the boundary values of the induced electrical potential, \cite{cald1,kohn2,nachman,sylv1}. The reconstruction procedures that have been proposed include a wide range of iterative methods based on formulating the inverse problem as a nonlinear optimisation problem. These techniques are quite demanding computationally particularly when addressing the three dimensional problem. This  difficulty has encouraged the search for reconstruction algorithms which reduce the computational demands either by using some {\em a priori} information e.g. \cite{buhl1,buhl2,cedi} or by developing non-iterative procedures. Some of these methods \cite{buhl1,buhl2} use a factorisation approach while others are based on reformulating the inverse problem in terms of integral equations \cite{ciulli1,ciulli2,sebu1}.

\section{Reconstruction from a single measurement}
\label{singelrec}

We will present here a reconstruction algorithm based on reformulating the problem in terms of integral equations. A feature of this method is that many of the calculations involve analytical expressions containing the eigenfunctions of the kernel of these integral equations, the computational part being restricted to the introduction of data, the numerical evaluation of some of the analytic formul\ae\ and the solution of a final integral equation.

However, this method requires the knowledge on the boundary $\partial\Omega$ of the domain $\Omega$ not only of the electrical potential $\phi$ and its normal derivative $\partial\phi/\partial n$, but also of the electrical conductivity $\sigma$. In geophysics the conductivity on the surface can be measured by taking small soil samples, while in medical applications this might be achieved using closely spaced electrodes.

The method consists in the approximate determination (in the sense of a generalised solution of inverse problems presented in Section \ref{generalisedsolution}) of $Y({\bm x})$ (Eq.(\ref{eqy})) by a single simultaneous measurement of the potential $\phi$ and its normal derivative $\partial\phi/\partial n$ on the boundary $\partial\Omega$.

\subsection{The method} 

One can use Eq.(\ref{inteq1}) to determine the potential on the boundary
\begin{equation}
\left.\phi({\bm x})\right|_{{\bm x}\in\partial\Omega}=\left.\psi({\bm x})\right|_{{\bm x}\in\partial\Omega}+\int_\Omega Y({\bm x'})\left.G_N({\bm x},{\bm x'})\right|_{{\bm x}\in\partial\Omega}d^nx' 
\label{inteq2}
\end{equation}
with
\begin{equation}
\left.\psi({\bm x})\right|_{{\bm x}\in\partial\Omega}=\langle\phi\rangle_{\partial\Omega}+\oint_{\partial\Omega}\frac{\partial}{\partial n'} \phi({\bm x}')\left.G_N({\bm x},{\bm x'})\right|_{{\bm x}\in\partial\Omega}d^{n-1}x'. 
\end{equation}
Eq.(\ref{inteq2}) represents a generalised Neumann-to-Dirichlet mapping. If the potential $\phi({\bm x})$ is measured on the boundary together with the current distribution $\sigma(\bm x)\frac{\partial\phi({\bm x})}{\partial n}$, one can determine $Y({\bm x})$ as a (generalised) solution of the integral equation (\ref{inteq2}). One can rewrite this equation in the form
\begin{equation}
\left.F({\bm x})\right|_{{\bm x}\in\partial\Omega}=\int_\Omega Y({\bm x'})\left.G_N({\bm x},{\bm x'})\right|_{{\bm x}\in\partial\Omega} d^nx',
\label{inteq3}
\end{equation}
where
\begin{equation}
\left.F({\bm x})\right|_{{\bm x}\in\partial\Omega}=-\langle\phi\rangle_{\partial\Omega}+\left.\phi({\bm x})\right|_{{\bm x}\in\partial\Omega}-\oint_{\partial\Omega}\frac{\partial}{\partial n'}\phi({\bm x'})\left.G_N({\bm x},{\bm x'})\right|_{{\bm x}\in\partial\Omega}d^{n-1}x'
\label{dataeq}
\end{equation}
contains the measured input. Typically a given current pattern would be applied to the surface $\partial\Omega$ and the resulting surface potential measured. The integral in (\ref{dataeq}) can be evaluated analytically if the applied current pattern is one of the eigenfunctions of the problem. Equivalently this integral can be viewed as the potential $\phi_0({\bm x})$ on the surface $\partial\Omega$ due to a constant conductivity density
\begin{equation}
\left.\phi_0({\bm x})\right|_{{\bm x}\in\partial\Omega}=\oint_{\partial\Omega}\frac{\partial}{\partial n'}\phi({\bm x'})\left.G_N({\bm x},{\bm x'})\right|_{{\bm x}\in\partial\Omega}d^{n-1}x'
\label{phi0}
\end{equation}

The data $\left.F({\bm x})\right|_{{\bm x}\in\partial\Omega}$ are defined on a $(n-1)$-dimensional manifold $\partial\Omega$, while the unknown functions $Y({\bm x})$ live in the $n$-dimensional world $\Omega$. As is usual in inverse problems, we seek only a generalised solution of (\ref{inteq3}), i.e., a function $Y({\bm x})$ which under the mapping of (\ref{inteq3}) produces an $\left.F({\bm x})\right|_{{\bm x}\in\partial\Omega}$ that is closest, in a least-squares sense, to the data input and has minimum norm. We rewrite (\ref{inteq3}) in the form
\begin{equation}
F={\bm K}Y,
\label{system}
\end{equation}
with ${\bm K}$ an integral operator. Its domain is the whole region $\Omega$ and its range the boundary $\partial\Omega$. If the operator $\bm K$ possesses a singular value decomposition (see Appendix \ref{appSVD})
\begin{equation}
{\bm K}Y=\sum_{j=1}^\infty\sigma_j(Y,v_j)u_j,
\label{svd}
\end{equation}
where $\{v_j\}$ is a complete set of basis functions on $\Omega$, $\{u_j\}$ a complete set of basis functions on $\partial\Omega$ and $\sigma_j$ are the singular values of $\bm K$, the least-squares solution of (\ref{system}) will be
\begin{equation}
Y^+=\sum_{j=1}^\infty\frac{1}{\sigma_j}(F,u_j)v_j.
\end{equation}

Due to ill-posedness, one needs to regularise the solution. A convenient tool to carry out this operation is the so called truncated singular value decomposition (Section \ref{truncatedsvd}). The regularised solution will then be
\begin{equation}
Y_{\rm reg}=\sum_{j=1}^J \frac{1}{\sigma_j}(F,u_j)v_j.
\end{equation}

Once $Y_{\rm reg}({\bm x})$ is known, the potential $\phi_{\rm reg}({\bm x})$ can be obtained by means of the integral
\begin{equation}
\phi_{\rm reg}({\bm x})=\psi({\bm x})+\int_\Omega Y_{\rm reg}({\bm x'})G_N({\bm x},{\bm x'})d^nx'.
\end{equation}

To determine the regularised conductivity $\sigma_{\rm reg}({\bm x})$, one can use the method of characteristics to solve the first order partial differential equation (see Eq.(\ref{characteristics})):
\begin{equation}
\bm\nabla(\ln\sigma_{\rm reg}({\bm x}))\cdot\bm\nabla\phi_{\rm reg}({\bm x})-Y_{\rm reg}({\bm x})=0, \ \ {\bm x}\in\Omega
\label{charsigma}
\end{equation}
subject to the known boundary values $\sigma({\bm x})|_{{\bm x}\in\partial\Omega}$.

\subsection{An example: the unit disc}
\label{ex:unitdisc}

For the unit disc, Eq.(\ref{inteq3}) takes the form
\begin{equation}
F(\theta)=\int_\Omega d^2x'K(\theta;{\bm x'})Y({\bm x'}),
\label{circleinteq}
\end{equation}
where
\begin{equation}
F(\theta)=-\langle\phi\rangle_{\partial\Omega}+\left.\phi(r,\theta)\right|_{r=R}-\int_0^{2\pi}\frac{\partial}{\partial r'}\left.\phi(r',\theta')\right|_{r'=1}G_N(1,\theta;1,\theta')d\theta',
\end{equation}
and
\begin{equation}
K(\theta,{\bm x}')=\sum_{l,m=1}^\infty\sum_{j=1}^2 \sigma_{lm}u_l^j(\theta)v_{lm}^j(r',\theta'),
\label{ksvd}
\end{equation}
with
\begin{equation}
\sigma_{lm}=\frac{1}{\lambda_{lm}}c_{lm}J_l(j^*_{lm}).
\end{equation}
The determination of the eigenvalue expansion for the kernel of Eq.(\ref{circleinteq}), can be found in Appendix \ref{app:neumann}.

Eq.(\ref{ksvd}) represents the singular value decomposition of the integral operator (Appendix \ref{appSVD}). The singular functions $\{u_l^j(\theta)\}$ form an orthonormal basis on the unit circle (the range of $K$) and the singular functions $\{v_{lm}^j(r,\theta)\}$ one on the unit disc (the domain of $K$) (see Appendix \ref{app:green}). The singular values $\sigma_{lm}$ vanish for large $m,l$ which shows the ill-posedness of the problem (a proof of this statement can be found in Appendix \ref{firstkindinteq}). The least-squares solution of (\ref{circleinteq}) reads
\begin{equation}
Y^+(r,\theta)=\sum_{l,m=1}^\infty\sum_{j=1}^2\frac{1}{\sigma_{lm}}\left(\oint_{\partial\Omega}d\theta'F(\theta')u_l^j(\theta')\right)v_{lm}^j(r,\theta).
\label{discsolution}
\end{equation}

An easy way to compute the integrals over the function $F(\theta')$ is to take advantage of the eigenfunction expansion of the Green's function, which on the boundary reads (see Appendix \ref{app:green}):
\begin{equation}
G_N(1,\theta;1,\theta')=\sum_{l=1}^\infty\sum_{j=1}^2\frac{1}{l}u_l^j(\theta)u_l^j(\theta'),
\end{equation}
and write for the data function
\begin{equation}
F(\theta)=-\langle\phi\rangle_{\partial\Omega}+\phi(r,\theta)|_{r=1}-\sum_{l=1}^\infty\sum_{j=1}^2\frac{1}{l}u_l^j(\theta)\int_0^{2\pi}\left.\frac{\partial\phi(r',\theta')}{\partial r'}\right|_{r'=1}u_l^j(\theta')d\theta'.
\end{equation}
The integration in (\ref{discsolution}) is now straightforward:
\begin{equation}
\int_0^{2\pi} d\theta' F(\theta')u_l^j(\theta')=\phi_l^j-\frac{1}{l}{\it I}_l^j,
\end{equation}
where we have introduced the abbreviations:
\begin{equation}
\phi_l^j=\int_0^{2\pi} d\theta \phi(r,\theta)|_{r=1}u_l^j(\theta),
\end{equation}
\begin{equation}
{\it I}_l^j=\int_0^{2\pi} d\theta \left.\frac{\partial\phi(r,\theta)}{\partial r}\right|_{r=1}u_l^j(\theta).
\end{equation}
Thus, the least-squares solution of (\ref{circleinteq}) will be:
\begin{equation}
Y^+(r,\theta)=\sum_{l,m=1}^\infty\sum_{j=1}^2\frac{1}{\sigma_{lm}}\left(\phi_l^j-\frac{1}{l}{\it I}_l^j\right)v_{lm}^j(r,\theta).
\end{equation}
Performing the regularisation, the final result takes the form:
\begin{equation}
Y_{\rm reg}(r,\theta)=\sum_{l=1}^L\sum_{m=1}^M\sum_{j=1}^2\frac{1}{\sigma_{lm}}\left(\phi_l^j-\frac{1}{l}{\it I}_l^j\right)v_{lm}^j(r,\theta).
\label{regularisedsolution}
\end{equation} 
An algorithm where we have implemented all these steps was written in MATLAB and can be found in \cite{almasy1}.

\subsection{Numerical results}

To illustrate the performance of such an algorithm, we present here a numerical example. We try to reconstruct a conductivity distribution consisting of a high/low conductivity region within a uniform background. An analytical form of such a conductivity can be chosen as:
\begin{equation}
\sigma(x,y)=1+a\exp[b((x-{\bar x})^2+(y-{\bar y})^2)^2],
\label{modelsigma}
\end{equation}
having a maximum, for $a>0$, or a minimum, for $a<0$, at $(x,y)=({\bar x},{\bar y})$. In our numerical simulations we have chosen two sets of parameters $(a,b,\bar x,\bar y)$: $\sigma_1\mbox{ with } (1,-1500,0,0.4)$ and $\sigma_2\mbox{ with }\ (-0.5,-1000,0.5,0)$. The two exact conductivities are displayed in the figure below.

\begin{figure}[H]
\centering
\subfigure[$\sigma_1^{\rm model}({\bm x})$]{\includegraphics[height=5cm,angle=0]{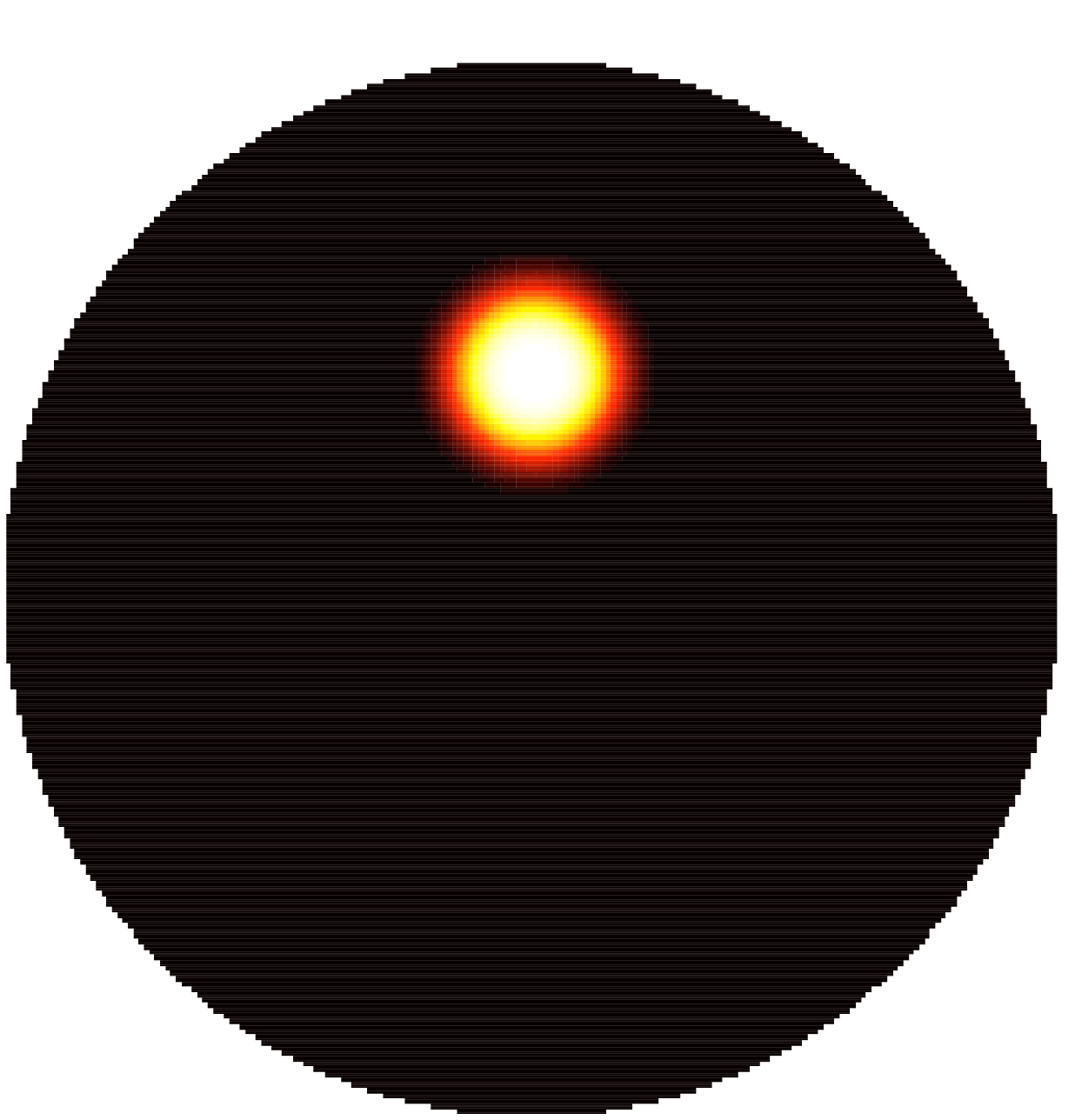}}
\subfigure[$\sigma_2^{\rm model}({\bm x})$]{\includegraphics[height=5cm,angle=0]{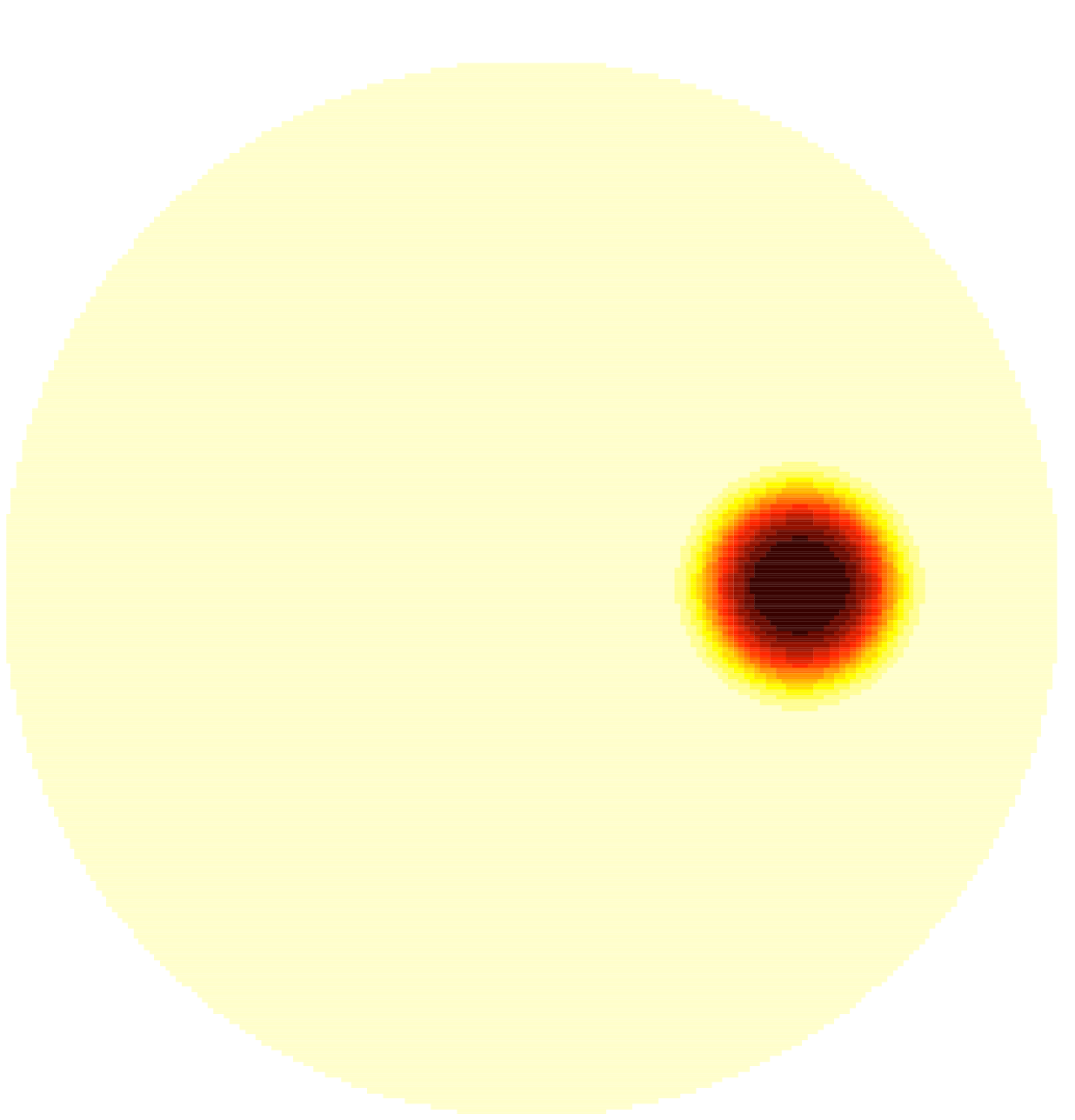}}
\caption[Reconstruction from a single set of measurements: Two examples of model conductivities.]{Density-plot of the two model conductivities considered (see text).}
\label{exactsig}
\end{figure}

First, one needs to solve the forward problem as in Chapter \ref{directproblem}. Let us consider the currents
\begin{equation}
\left.\sigma(r,\theta)\frac{\partial\phi}{\partial n}(r,\theta)\right|_{\partial\Omega}=\left\{
\begin{array}{l}
 \sin(m\theta)\\
\cos(m\theta)
\end{array}\right.
,\ \ m\in\N_+
\label{inputcurrents}
\end{equation}
as input and simulate the measured values of the potential on the boundary. With the simulated data we can easily compute the coefficients $\phi_l^j$ and ${\it I}_l^j$ needed for the regularised solution $Y_{\rm reg}(r,\theta)$. To fix the regularisation parameters, i.e., the two summation limits $L,M$ in Eq.(\ref{regularisedsolution}), the interactive method was used and we quote results for the values $L=10$ and $M=2$. 

\begin{figure}[H]
\centering
\subfigure[$Y_1^{\rm reg}({\bm x})$]{\includegraphics[height=5cm,angle=0]{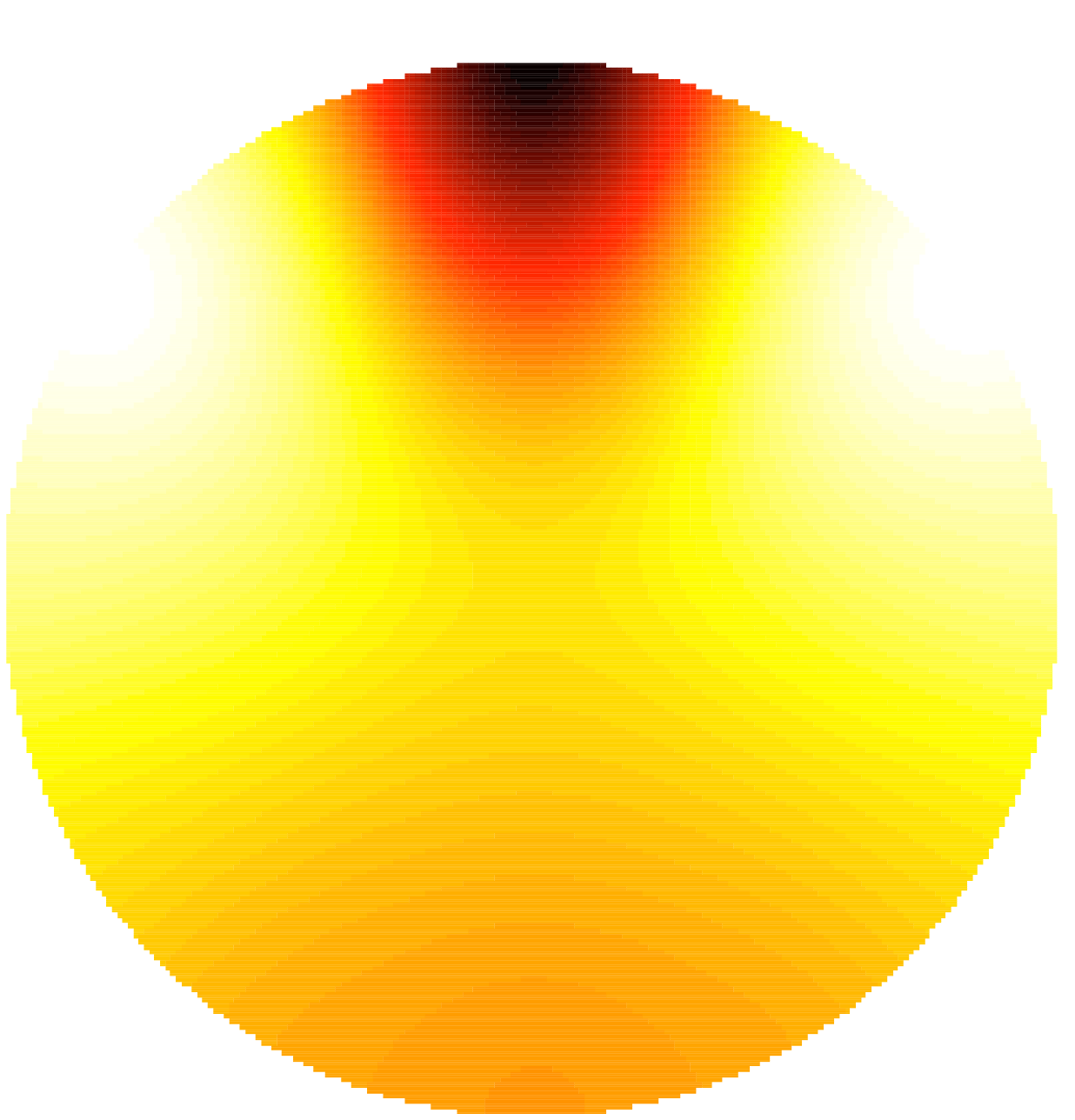}}
\subfigure[$Y_2^{\rm reg}({\bm x})$]{\includegraphics[height=5cm,angle=0]{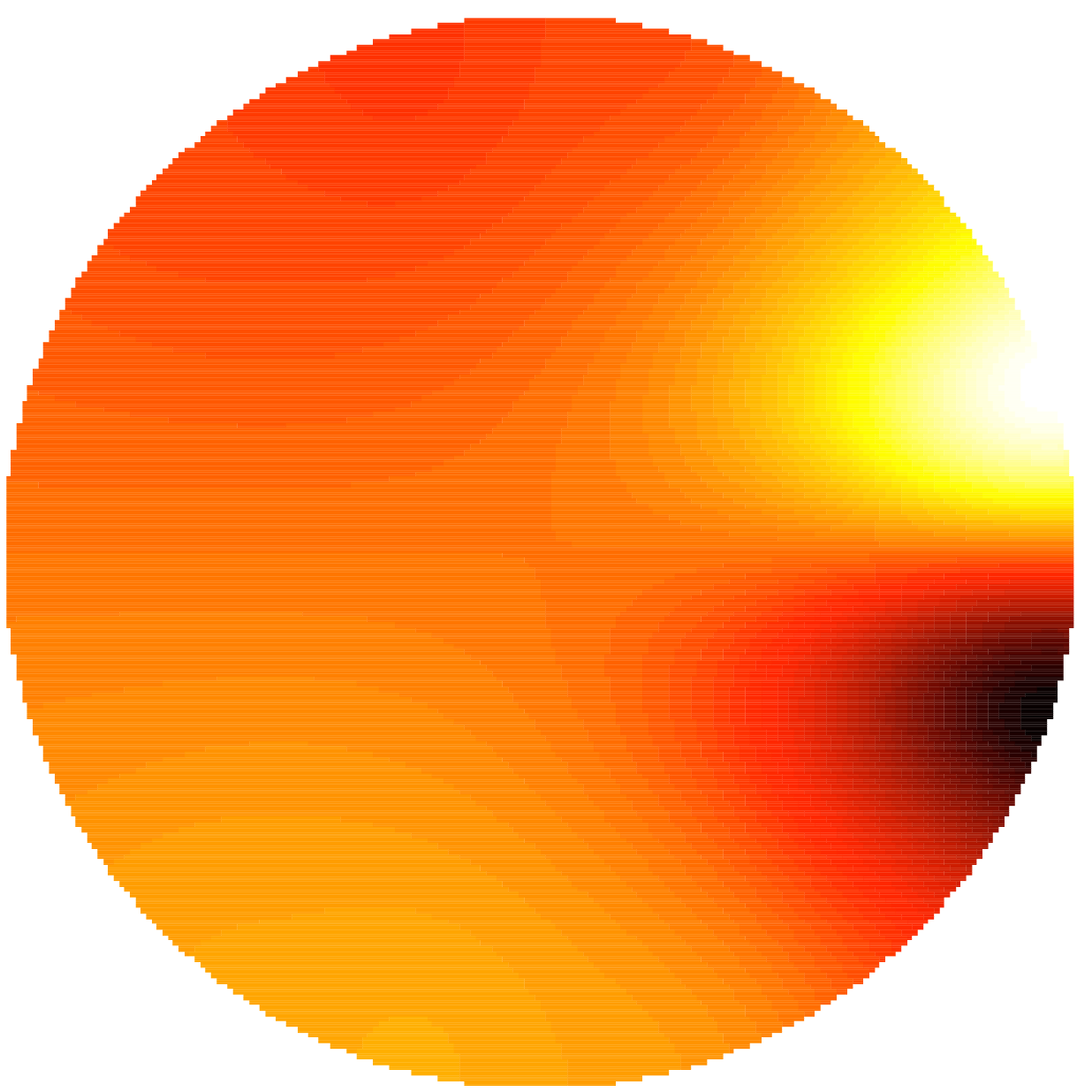}}
\caption[Reconstruction from a single set of measurements: Reconstructed $Y_{\rm reg}({\bm x})$ for the two considered examples.]{Reconstructed $Y_{\rm reg}({\bm x})$ for input current $\sin(\theta)$ [Eq.(\ref{inputcurrents})]. Reconstructions for the two model conductivities from Fig.\ref{exactsig} are shown.}
\label{recYplot}
\end{figure}

$Y_{\rm reg}({\bm x})$ can already be used directly to obtain rough information on the conductivity. One can observe in Fig.\ref{recYplot} how the effect of the high/low conductivity region can be observed on the boundary. The angular information is fine but unfortunately there is no radial information present.

It is still interesting to check whether one can get a better picture when including the next step in the reconstruction algorithm, i.e., computing the regularised potential:
\begin{equation}
\phi_{\rm reg}(r,\theta)=\psi(r,\theta)+\sum_{k=1}^{N}w_kG_N(r,\theta;r_k,\theta_k)Y_{\rm reg}(r_k,\theta_k),
\end{equation}
where $\{w_k;\ r_k,\ \theta_k\}$ is a set of $N$ quadrature weights and points for the unit disc \cite{engel}, and:
\begin{equation}
\psi(r,\theta)=\langle\phi\rangle_{\partial\Omega}-\frac{1}{2\pi}\int_0^{2\pi}\ln(1+r^2-2r\cos(\theta-\theta'))\left.\frac{\partial\phi}{\partial r'}\right|_{r'=1}d\theta'.
\end{equation}
The actual number of quadrature weights and points used in the reconstruction was $N=172$.

\begin{figure}[H]
\centering
\subfigure[$\phi_1^{\rm reg}({\bm x})$]{\includegraphics[height=5cm,angle=0]{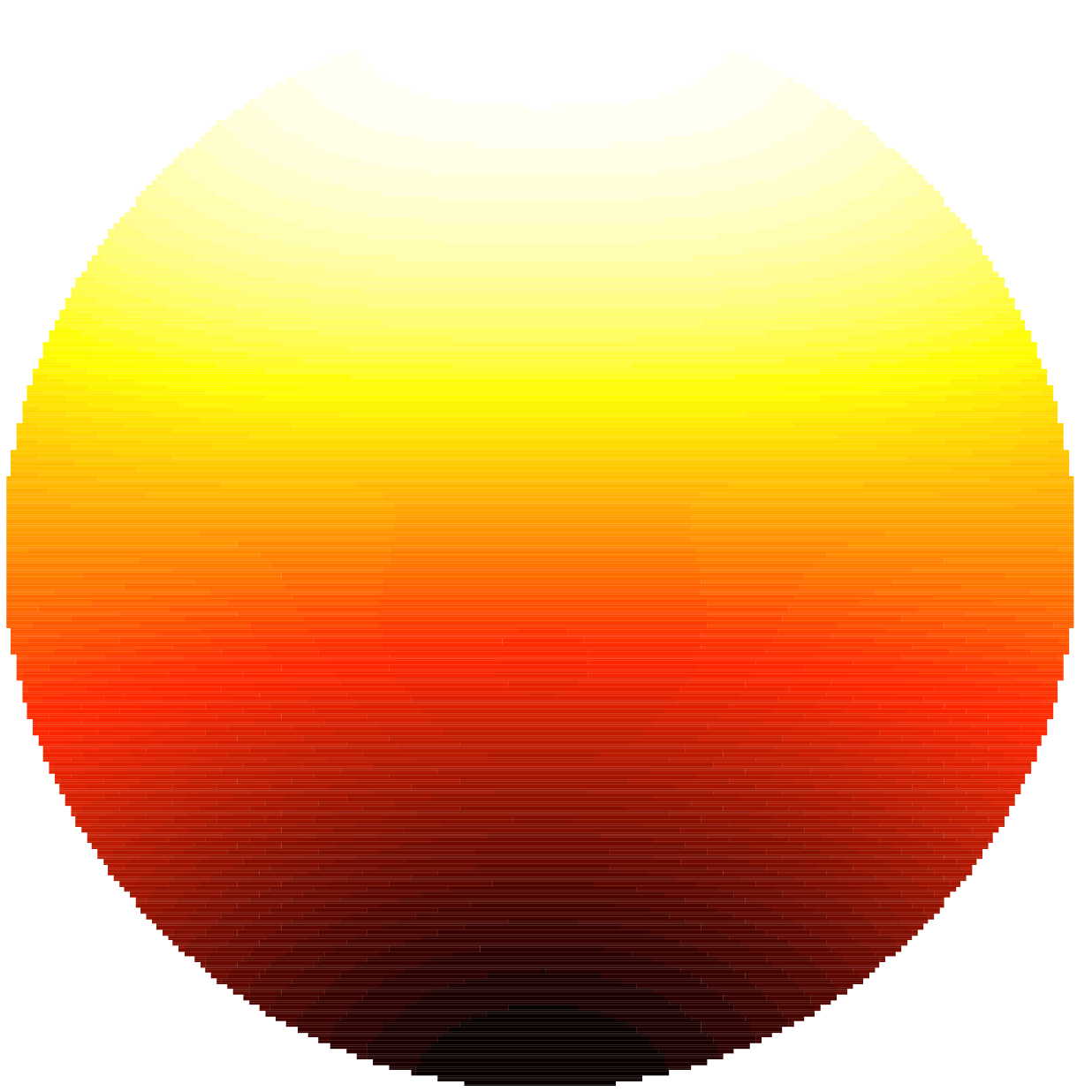}}
\subfigure[$\phi_2^{\rm reg}({\bm x})$]{\includegraphics[height=5cm,angle=0]{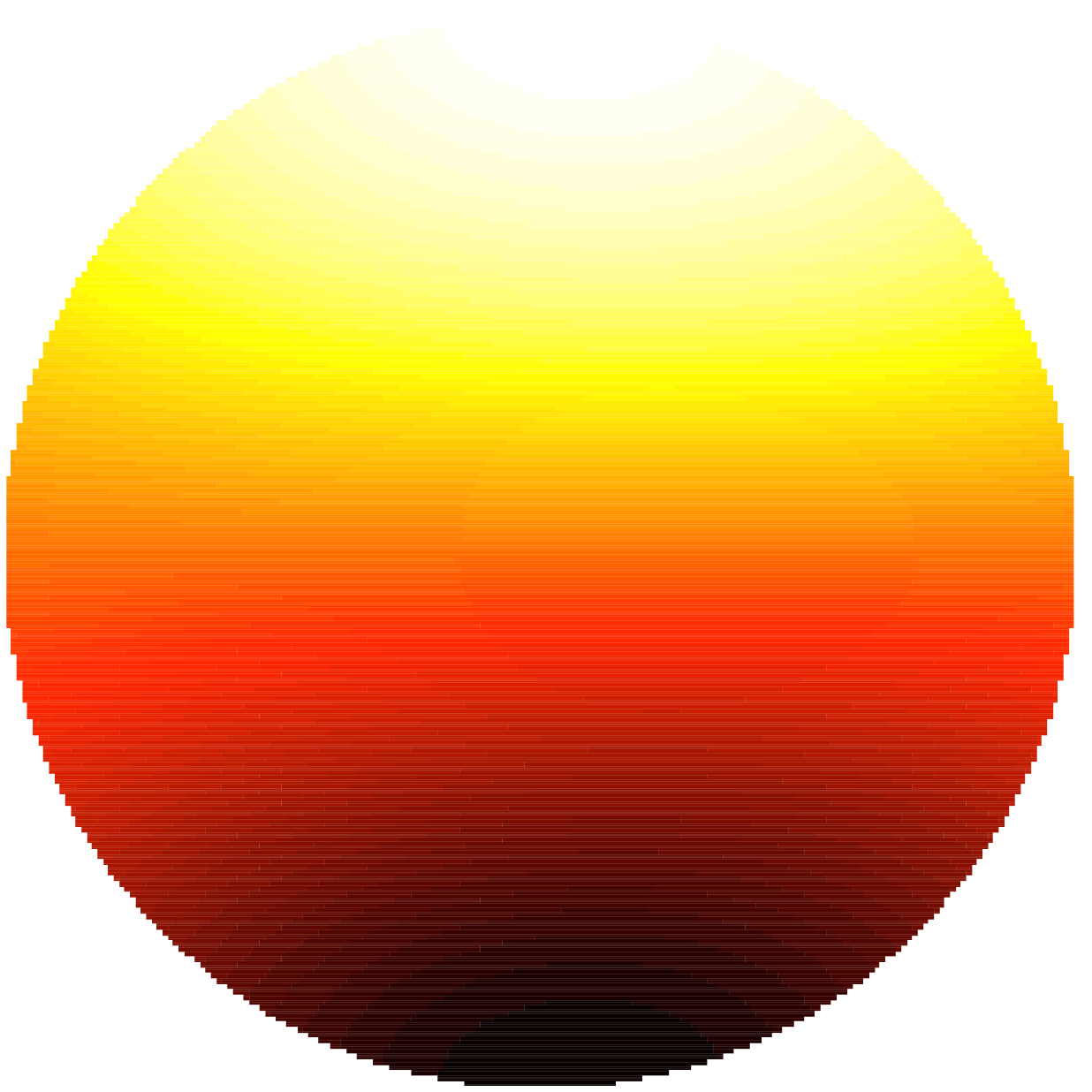}}
\caption[Reconstruction from a single set of measurements: Reconstructed $\phi_{\rm reg}({\bm x})$ for the two considered examples.]{Reconstructed $\phi_{\rm reg}({\bm x})$ for input current $\sin(\theta)$ [Eq.(\ref{inputcurrents})]. Reconstructions for the two model conductivities from Fig.\ref{exactsig} are shown.}
\label{recpotplot}
\end{figure}

The reconstructed potential is shown in Fig.\ref{recpotplot}. One can conclude that there is no improvement, even worse, in the reconstruction. We believe that by going further and performing also the last step we will not gain more.

Apart from the unsatisfactory reconstruction presented above, the treatment of the inverse conductivity problem with a single set of measurements on two dimensional domains is from practical reasons not advisable. In this case, the information one can use to reconstruct the conductivity is a single set of potentials and currents on the boundary of the domain. For an ideal case this information would be of infinite order, i.e., the potentials and currents wold be known at any point of the boundary, but in practice one knows only a finite number of values, corresponding to the number of electrodes. In our numerical example we have used 32 electrodes (to be in accord with practice), so that a good quality of the reconstruction was, in the first place, not to be expected. Besides these practical reasons, the reconstruction algorithm has also a mathematical drawback. In Eq.(\ref{inteq3}) the radial part is integrated out, i.e., we loose any radial information which might be present in $K$. Eq.(\ref{inteq3}) maps the domain to the boundary, that is it describes a mapping from $n$ to $(n-1)$ dimensions. The reconstruction algorithm needs then to make the inverse transition, from $(n-1)$ to $n$ dimensions, which is in principle impossible without additional information, like a model. 

Even though the radial information is lost, the angular information present in $Y_{\rm reg}$ can be further used as a qualitative model for similar reconstructions, where this {\em a priori} information is taken into account. Such an algorithm can be found for example in \cite{sebu1}.

\section{Reconstruction from more measurements}
\label{recalgorithm}

The linearisation of the problem allows us to use the information from more than one single set of measurements for the reconstruction of the conductivity distribution. Using as input more and different current configurations one gets a system of integral equations which can be easily solved by means of the generalised inverse (see Section \ref{generalisedsolution}). Once the regularisation parameter is fixed, the conductivity distribution is reconstructed by simple matrix multiplication. One needs to perform the matrix inversion only once since the measured data are not contained in the matrix. 

\subsection{Reconstruction by linearisation}

\subsubsection*{Reconstruction of $\delta\sigma$}

The starting point for the reconstruction of $\delta\sigma$ is the basic equation of EIT (cf. Eq.(\ref{basiceq2}))
\begin{equation}
 \bm\nabla\cdot\left[\sigma({\bm x})\bm\nabla\phi({\bm x})\right]=0.
\label{basicEIT}
\end{equation}
 
In the linear approximation, i.e. when the conductivity distribution differs only slightly from a constant
\begin{equation}
 \sigma({\bm x})=\sigma_0+\delta\sigma({\bm x}),
\label{lincond}
\end{equation}
 one will have for the potential
\begin{equation}
 \phi({\bm x})=\phi_0({\bm x})+\delta\phi({\bm x}),
\label{linpot}
\end{equation}
with $\phi_0({\bm x})$ the solution of (\ref{basicEIT}) for the constant conductivity $\sigma_0$. Inserting (\ref{lincond}) and (\ref{linpot}) into (\ref{basicEIT}) one finds:
\begin{equation}
 \bm\nabla\left[\sigma_0\bm\nabla\delta\phi({\bm x})\right]=-\bm\nabla\left[\delta\sigma({\bm x})\bm\nabla\phi_0({\bm x})\right].
\label{lineq}
\end{equation}
The Neumann boundary condition of the problem was already considered when computing $\phi_0({\bm x})$ (see Eq.(\ref{phi0})), so that one needs to solve (\ref{lineq}) together with $\partial(\sigma_0\delta\phi)/\partial n=0$. Considering $\sigma_0=1$, the solution will be:
\begin{equation}
 \delta\phi({\bm x})=\int_\Omega G_N({\bm x},{\bm x}')\bm\nabla\left[\delta\sigma({\bm x'})\bm\nabla\phi_0({\bm x'})\right]d^nx'.
\label{deltasigsol}
\end{equation}
The integrand of Eq.(\ref{deltasigsol}) can be rewritten as
\begin{equation}
\begin{array}{ll}
\displaystyle G_N({\bm x},{\bm x}')\bm\nabla\left[\delta\sigma({\bm x'})\bm\nabla\phi_0({\bm x'})\right]=&\bm\nabla\left[G_N({\bm x},{\bm x}')\delta\sigma({\bm x'})\bm\nabla\phi_0({\bm x'})\right]\\
\\
\displaystyle&-\bm\nabla G_N({\bm x},{\bm x}')\delta\sigma({\bm x'})\bm\nabla\phi_0({\bm x'}).
\end{array}
\label{integrand}
\end{equation}
Taking $\delta\sigma$ to vanish on the boundary, the first term in (\ref{integrand}) can be omitted, since it appears as a surface term in (\ref{deltasigsol}). The solution will then become:
\begin{equation}
 \delta\phi({\bm x})=-\int_\Omega \bm\nabla G_N({\bm x},{\bm x}')\bm\nabla\phi_0({\bm x'})\delta\sigma({\bm x'})d^nx'.
\end{equation}
Multiplying both sides with the boundary eigenfunctions, $u_i$, of $G_N({\bm x},{\bm x}')$ and integrating on the boundary, one obtains an integral equation for $\delta\sigma$:
\begin{equation}
 \langle\delta\phi,u_i\rangle_{\partial\Omega}=-\int_\Omega\bm\nabla\phi_0^i({\bm x})\bm\nabla\phi_0({\bm x})\delta\sigma({\bm x})d^nx.
\label{inteqdeltasig}
\end{equation}

\subsubsection*{Reconstruction of $\ln\sigma$}

Alternatively, the first iteration of integral equation (\ref{inteq1}) can be used, i.e.,
\begin{equation}
 \phi({\bm x})=\phi_0({\bm x})+\int_\Omega G_N({\bm x},{\bm x}')\bm\nabla\phi_0({\bm x}')\bm\nabla\ln\sigma({\bm x}')d^nx'.
\label{firstiteration}
\end{equation}
Here, the ground was chosen such that $\langle\phi\rangle_{\partial\Omega}=0$. As before, one can transform the integrand and, assuming $\ln\sigma$ vanishes on the boundary, Eq.(\ref{firstiteration}) becomes:
\begin{equation}
 \delta\phi({\bm x})=\phi({\bm x})-\phi_0({\bm x})=-\int_\Omega \bm\nabla G_N({\bm x},{\bm x}')\bm\nabla\phi_0({\bm x}')\ln\sigma({\bm x}')d^nx'.
\end{equation}
Note that here no linearisation of the conductivity was performed.

Taking the scalar product of $\delta\phi$ with the boundary eigenfunctions of Neumann Green's function, $u_i$, one finds an integral equation for $\ln\sigma$ similar to (\ref{inteqdeltasig})
\begin{equation}
 \langle\delta\phi,u_i\rangle_{\partial\Omega}=-\int_\Omega\bm\nabla\phi_0^i({\bm x})\bm\nabla\phi_0({\bm x})\ln\sigma({\bm x})d^nx.
\label{inteqlnsig}
\end{equation}

\subsubsection*{Comparison: $\delta\sigma$ versus $\ln\sigma$}

The kernels of the two integral equations (\ref{inteqdeltasig}) and (\ref{inteqlnsig}) are the same, so the quantities to be reconstructed will contain the same information. Indeed, in the linear approximation one has:
\begin{equation}
 \ln\sigma({\bm x})=\ln\left(1+\delta\sigma({\bm x})\right)\approx\delta\sigma({\bm x}).
\end{equation}
However, the reconstruction of $\ln\sigma$ is preferred since in this way the positivity of the reconstructed conductivity distribution is assured.

\subsubsection*{Discretisation}

One can perform a discretisation of (\ref{inteqlnsig}) and of the domain $\Omega$:
\begin{equation}
 \langle\delta\phi,u_i\rangle_{\partial\Omega}=-\sum_{k=1}^{N_p}\bm\nabla\phi_0^i({\bm x}_k)\bm\nabla\phi_0({\bm x}_k)\ln\sigma({\bm x}_k)\delta V({\bm x}_k),
\end{equation}
where $N_p$ is number of discretisation points ${\bm x}_k$. For $N_m$ sets of measured data, $\delta\phi_j$, and for $N_t$ ``test functions'' $u_i$, one finds the system of equations:
\begin{equation}
 \delta\phi_{ij}=\sum_k^{N_p}K_{ijk}\ln\sigma_k
\label{systemofeq}
\end{equation}
with the short-hand notations
\begin{equation}
 \begin{array}{l}
\displaystyle  \delta\phi_{ij}=\langle\delta\phi_j,u_i\rangle_{\partial\Omega},\\
\\
\displaystyle K_{ijk}=-\bm\nabla\phi_0^i({\bm x}_k)\bm\nabla\phi_0({\bm x}_k)\delta V({\bm x}_k),\\
\\
\displaystyle \ln\sigma_k=\ln\sigma({\bm x}_k).
 \end{array}
\label{short:system}
\end{equation}

The matrix inversion will be performed by means of the generalised inverse (see Section \ref{generalisedsolution}) and the truncated singular value decomposition used as a regularisation scheme. In this way, very small singular values will be cut out and will not enter in the reconstruction. This is a rather obvious choice since the measurement errors are reinforced by small singular values. The regularisation parameter $\lambda$, also called cut-off parameter, will be chosen in such a way that the errors are not reinforced but still the reconstruction is close to the true one.

For numerically solving Eq.(\ref{inteqlnsig}), a MATLAB code was written. The important steps of this code are summarised below. The entire code can be found in \cite{almasy1}.
\vspace{0.5cm}

\begin{algorithm}[H]
 \caption[Algorithm for reconstructing the conductivity distribution from more than one set of measurements.]{\sf Algorithm for reconstructing the conductivity distribution from more than one set of measurements.}
\dontprintsemicolon
\KwData{Geometry and data files.}
\KwResult{Reconstructed conductivity distribution.}
\Begin{
compute $\delta\phi$ [Eq.(\ref{short:system})]\;
compute $K$ [Eq.(\ref{short:system})]\;
\Repeat{$|\sigma-\sigma_{\rm exact}|<\varepsilon$,}{
choose $\lambda$\;
compute the generalised inverse of $K$ $\Rightarrow$ $K^+$\;
calculate $\ln\sigma=K^+\delta\phi$\;
calculate $\sigma={\rm exp}(\ln\sigma)$}($\varepsilon$ small)}
\end{algorithm}

\subsection{Numerical results}

As in Section \ref{ex:unitdisc}, we have performed tests for our algorithm on a two-dimensional domain, the unit disc. For this specific case, the kernel of the integral equations (\ref{inteqdeltasig}) and (\ref{inteqlnsig}) becomes:
\begin{equation}
 K_{ij}(r,\theta)=\frac{r^{|i|+|j|-2}}{\pi}\left\{
\begin{array}{l}
\cos\left((|i|-|j|)\theta\right),\ i,j>0\ {\rm or}\ i,j<0,\\
\sin\left((|i|-|j|)\theta\right),\ i>0,j<0,|j|\ge|i|,\\
\sin\left((|j|-|i|)\theta\right),\ i<0,j>0,|j|\ge|i|.\\
\end{array}\right.
\end{equation}
The ``test functions'' are (see Appendix \ref{app:green}):
\begin{equation}
 u_i(\theta)=\frac{1}{\sqrt\pi}\left\{
\begin{array}{l}
 \sin(|i|\theta),\ i>0,\\
\cos(|i|\theta),\ i<0,
\end{array}\right.
\end{equation}
and the potential due to the constant conductivity $\sigma_0$ and boundary condition $u_i$:
\begin{equation}
 \phi_0^i(r,\theta)=\frac{r^{|i|}}{|i|}u_i(\theta).
\end{equation}
For the tests we have used several conductivity distributions similar to (\ref{modelsigma}):
\begin{equation}
 \begin{array}{l}
\displaystyle  \sigma_1({\bm x})=1+\delta\sigma_1({\bm x}),\\
\\
\displaystyle  \sigma_2({\bm x})=1+\delta\sigma_1({\bm x})+\delta\sigma_2({\bm x}),\\
\\
\displaystyle  \sigma_3({\bm x})=1+\delta\sigma_1({\bm x})+\delta\sigma_2({\bm x})+\delta\sigma_3({\bm x}),
 \end{array}
\label{testconductivity}
\end{equation}
with
\begin{equation}
 \begin{array}{l}
\displaystyle  \delta\sigma_1({\bm x})=\exp\left[-1500\left(x^2+(y-0.4)^2\right)^2\right],\\
\\
\displaystyle  \delta\sigma_2({\bm x})=-0.5\exp\left[-2500\left((x-0.5)^2+(y+0.2)^2\right)^2\right],\\
\\
\displaystyle  \delta\sigma_3({\bm x})=2\exp\left[-1000\left((x+0.3)^2+(y+0.2)^2\right)^2\right].
 \end{array}
\end{equation}

The forward problem has been solved by means of the finite element method described in Section \ref{direct:FEM}. The simulated data, i.e.\ the potential on the boundary, contain no further errors, being exact up to computer accuracy. A first test was performed using the ``exact'' potential and later, additional errors were added to check the stability of the reconstruction. For the potential $\phi_0$ due to a constant conductivity, Eq.(\ref{phi0}), the theoretically calculated one was used. For the reconstruction, the first 10 input currents were used, that is $m=1,2,...,10$ in Eq.(\ref{inputcurrents}), and both sinus and cosine were considered. The number of electrodes was set to 32.

\begin{figure}[H]
\centering
\includegraphics[height=4.5cm,angle=0]{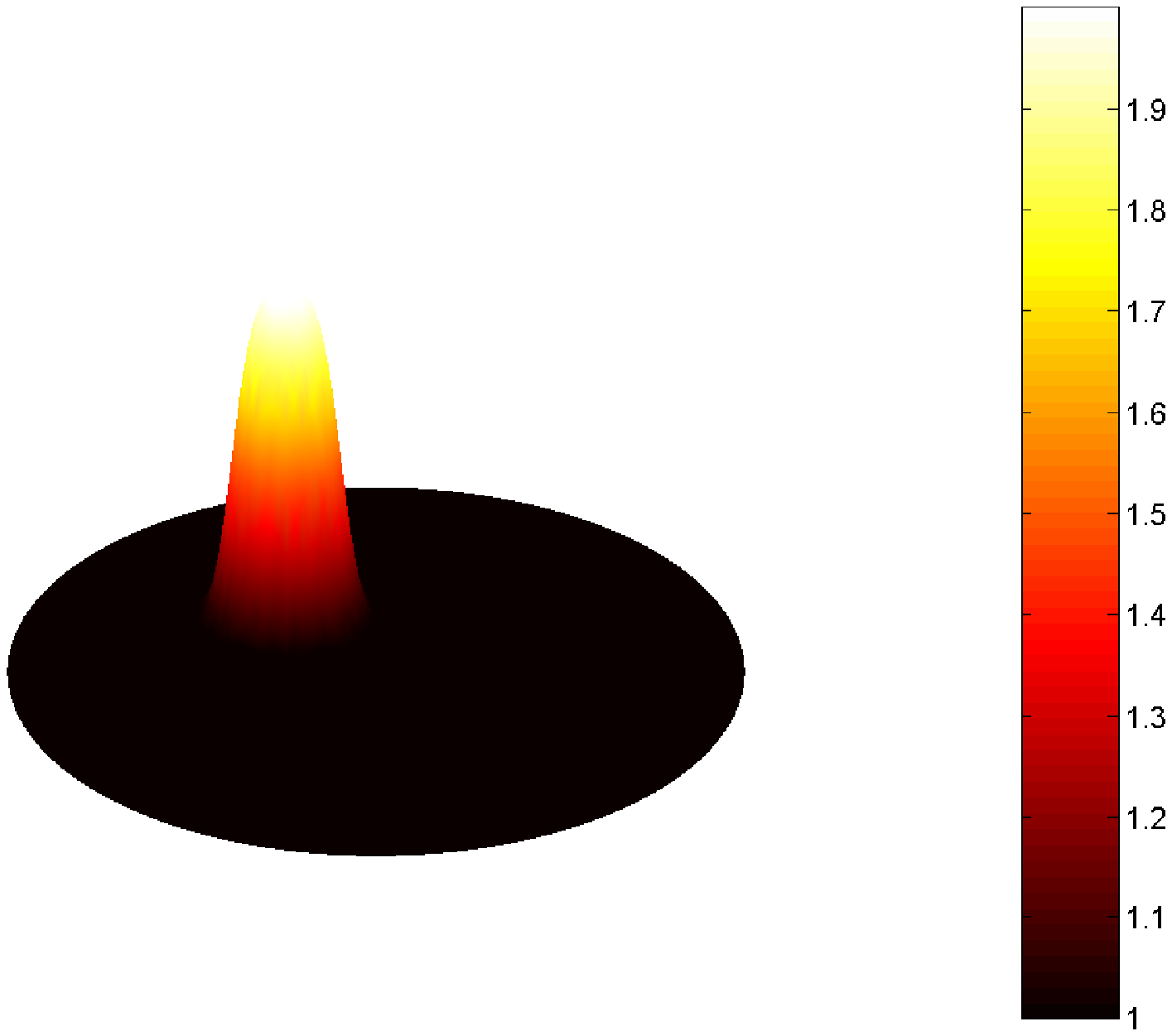}\includegraphics[height=4.5cm,angle=0]{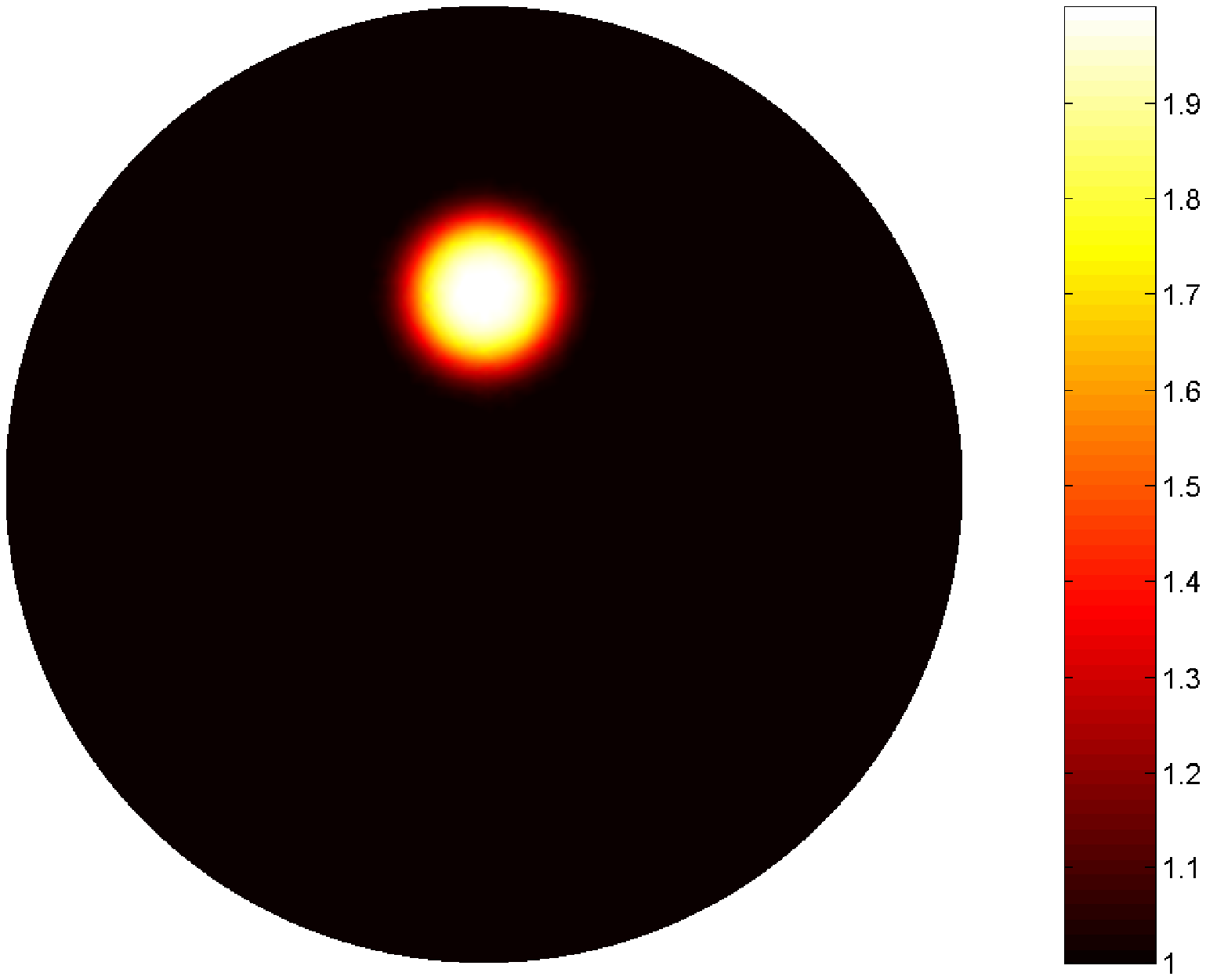}
\includegraphics[height=4.5cm,angle=0]{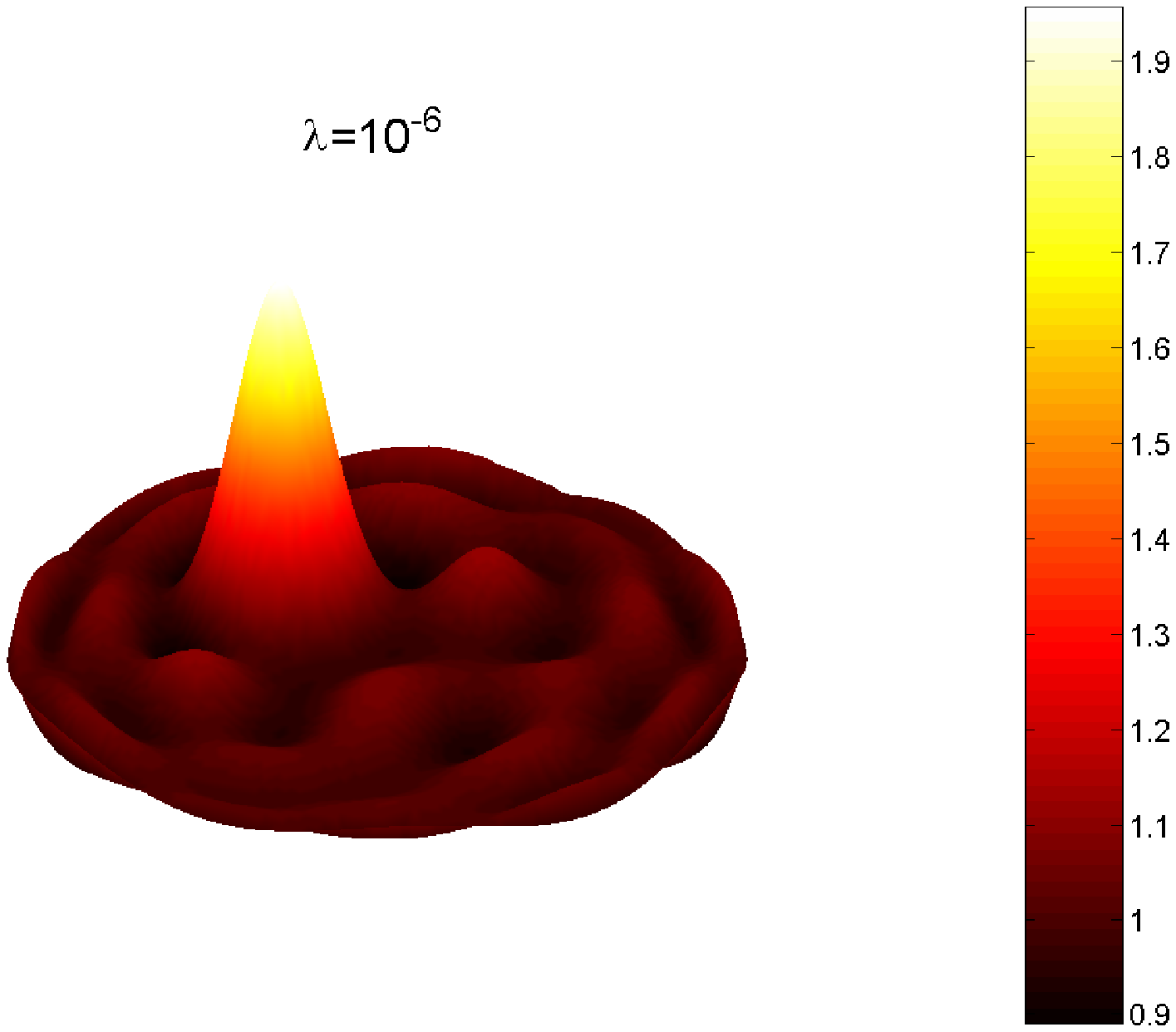}\includegraphics[height=4.5cm,angle=0]{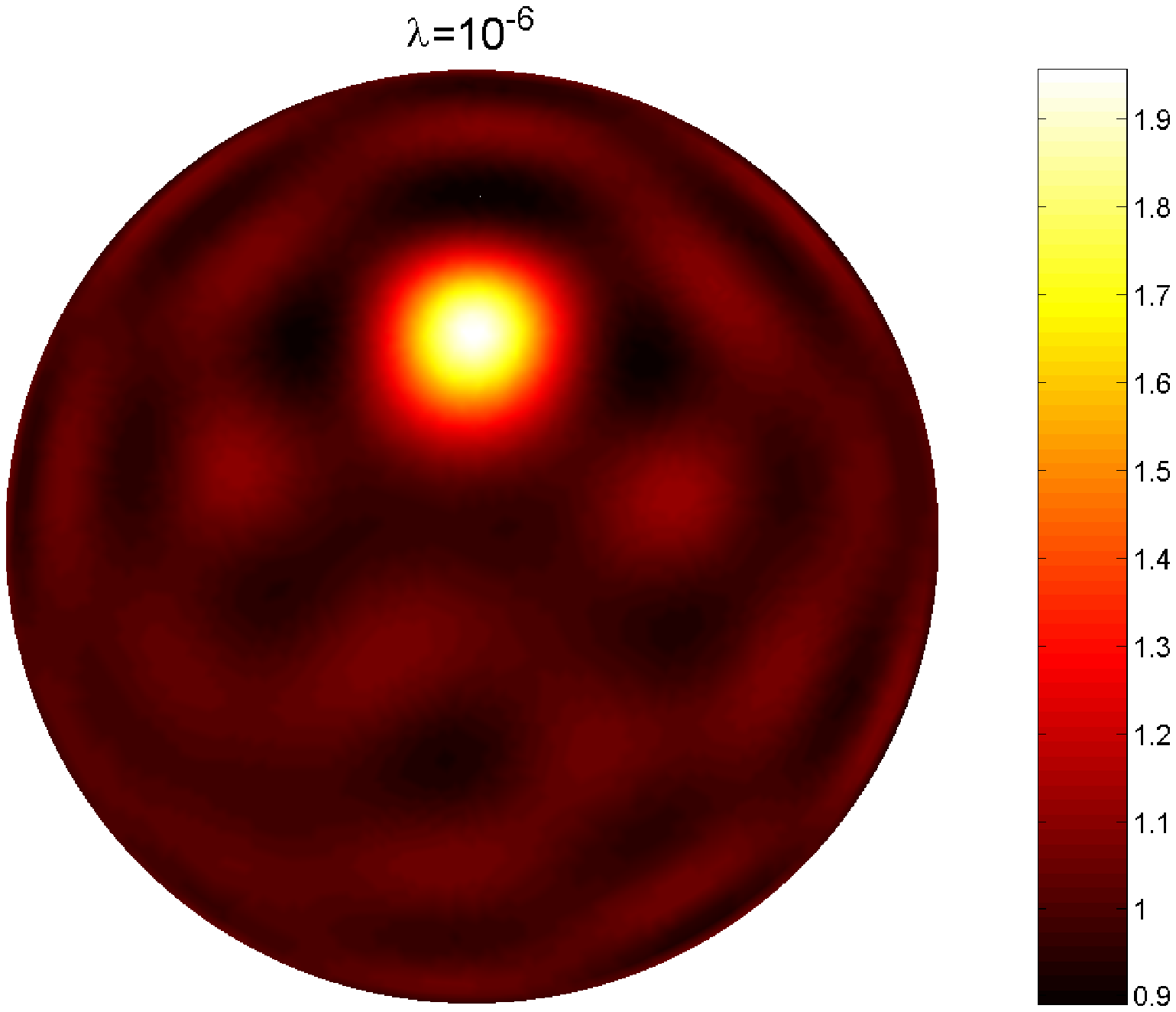}
\caption[Reconstruction from more measurements: Model conductivity $\sigma_1$ and its reconstruction.]{In the upper part, two- and three-dimensional plots of the model conductivity $\sigma_1$,  Eq.(\ref{testconductivity}), are displayed. The corresponding reconstructed conductivity distribution is shown in the lower part. The regularisation parameter was chosen to be $\lambda=10^{-6}$.}
\label{sig1rec}
\end{figure}
\clearpage

\begin{figure}[p!]
\centering
\includegraphics[height=4.5cm,angle=0]{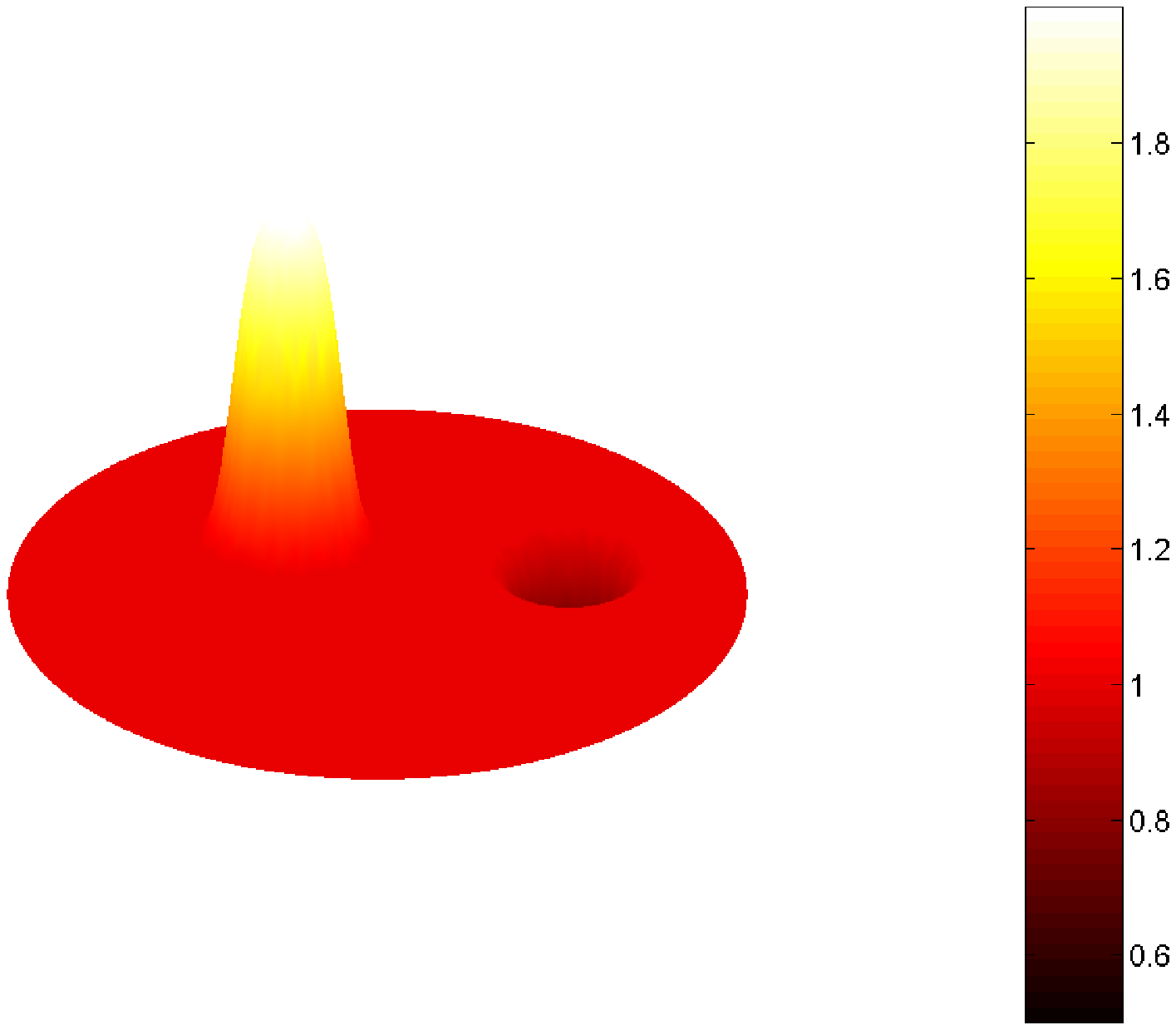}\includegraphics[height=4.5cm,angle=0]{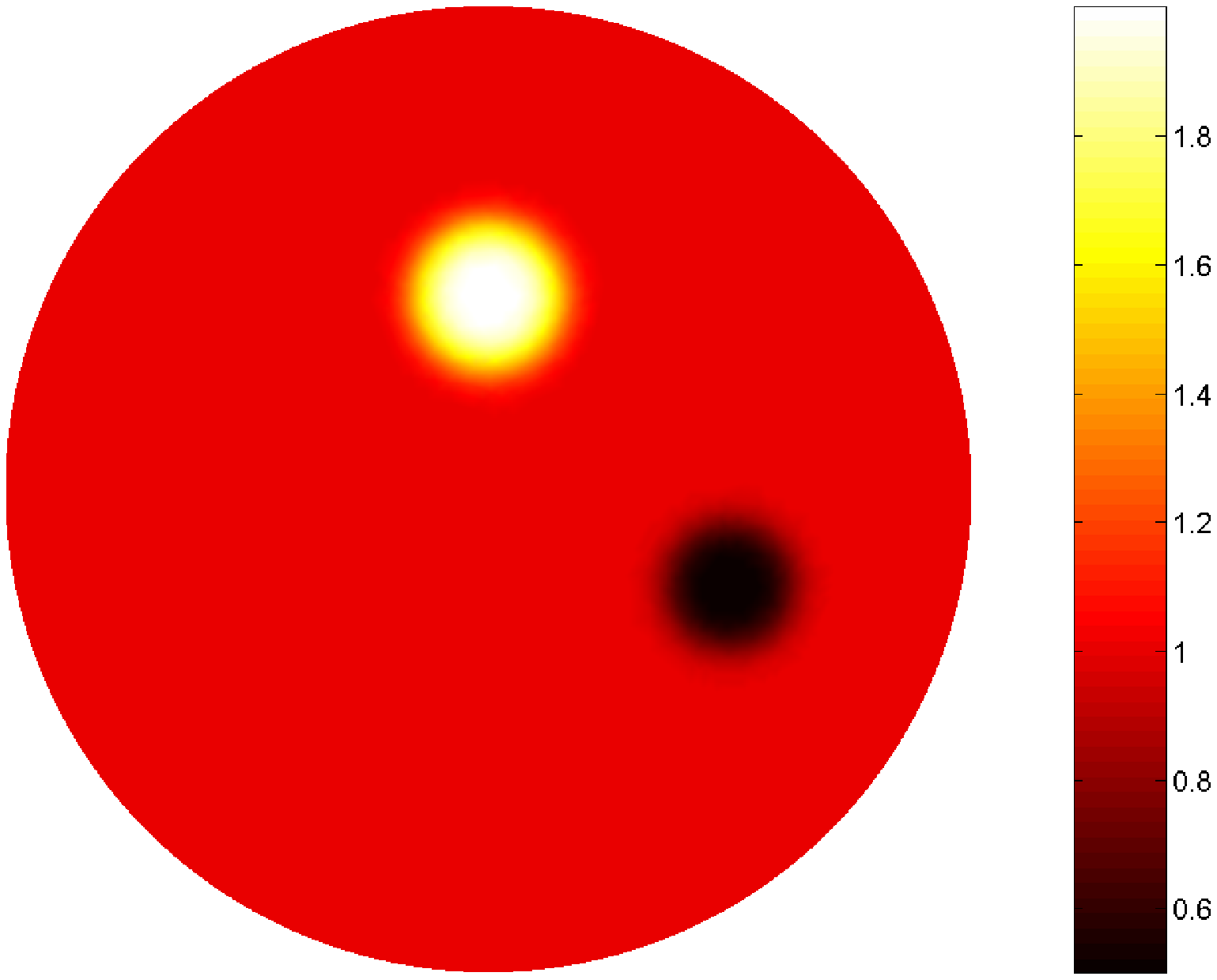}
\includegraphics[height=4.5cm,angle=0]{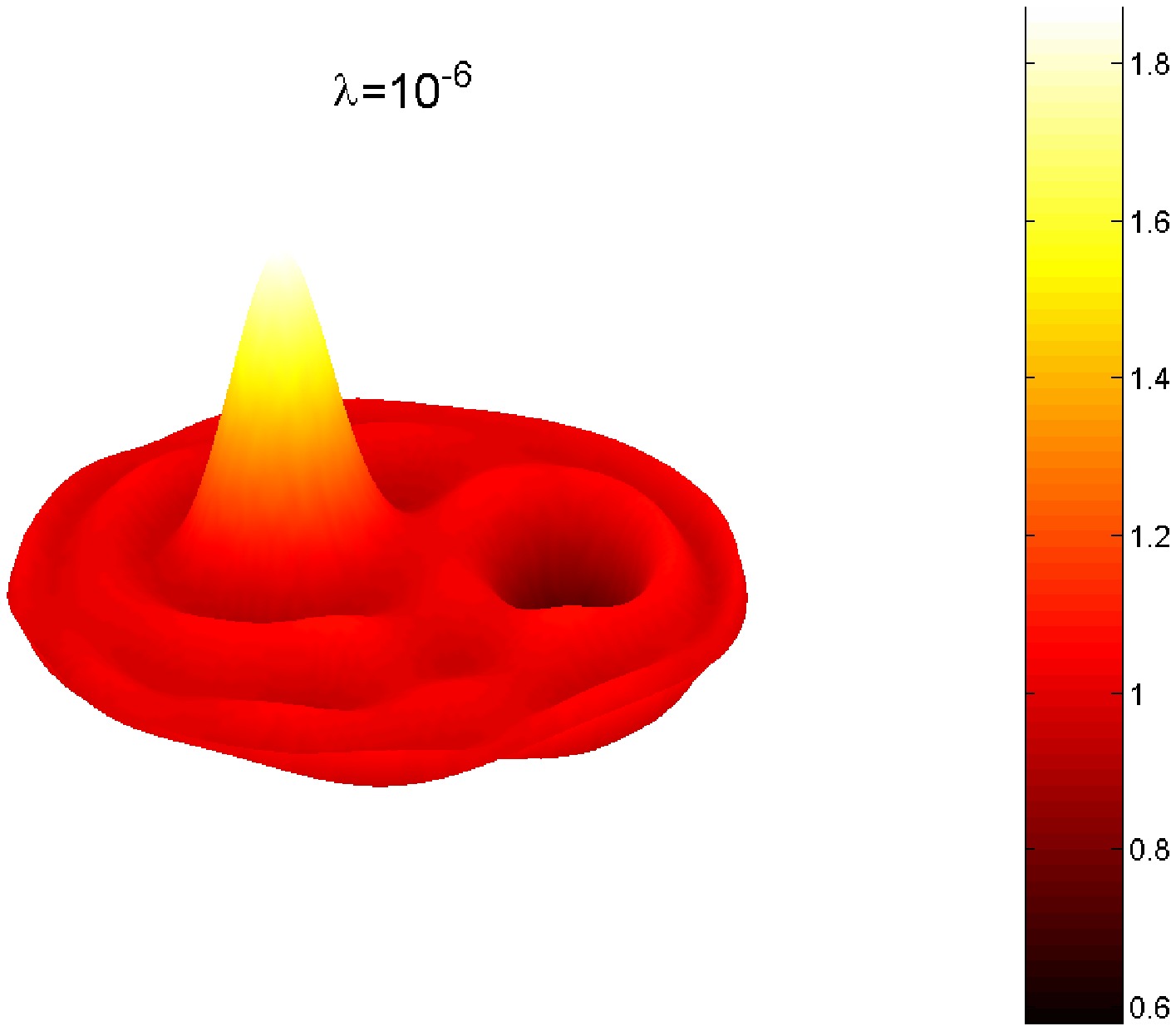}\includegraphics[height=4.5cm,angle=0]{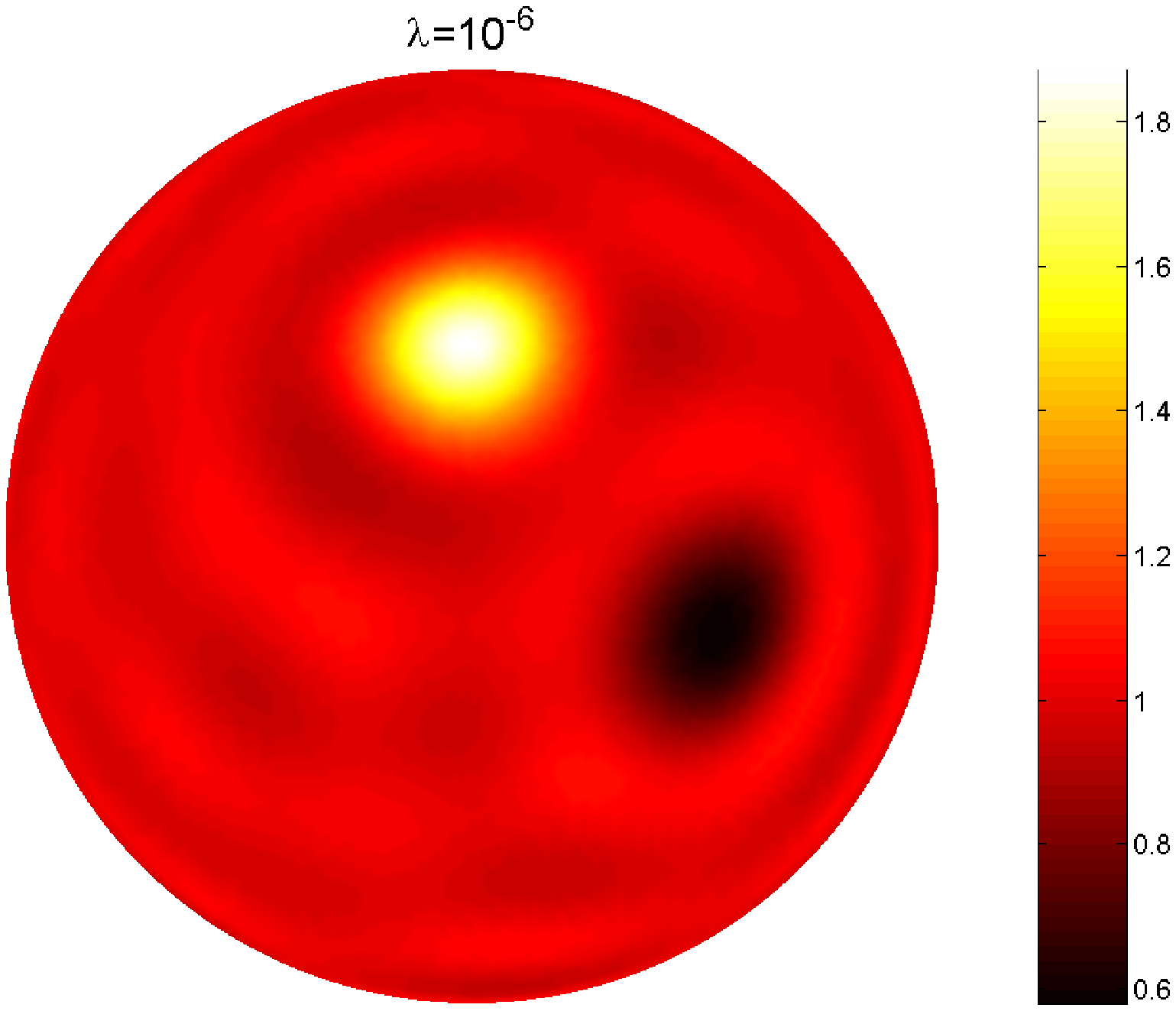}
\caption[Reconstruction from more measurements: Model conductivity $\sigma_2$ and its reconstruction.]{In the upper part, two- and three-dimensional plots of the model conductivity $\sigma_2$,  Eq.(\ref{testconductivity}), are displayed. The corresponding reconstructed conductivity distribution is shown in the lower part. The regularisation parameter was chosen to be $\lambda=10^{-6}$.}
\label{sig2rec}
\centering
\includegraphics[height=4.5cm,angle=0]{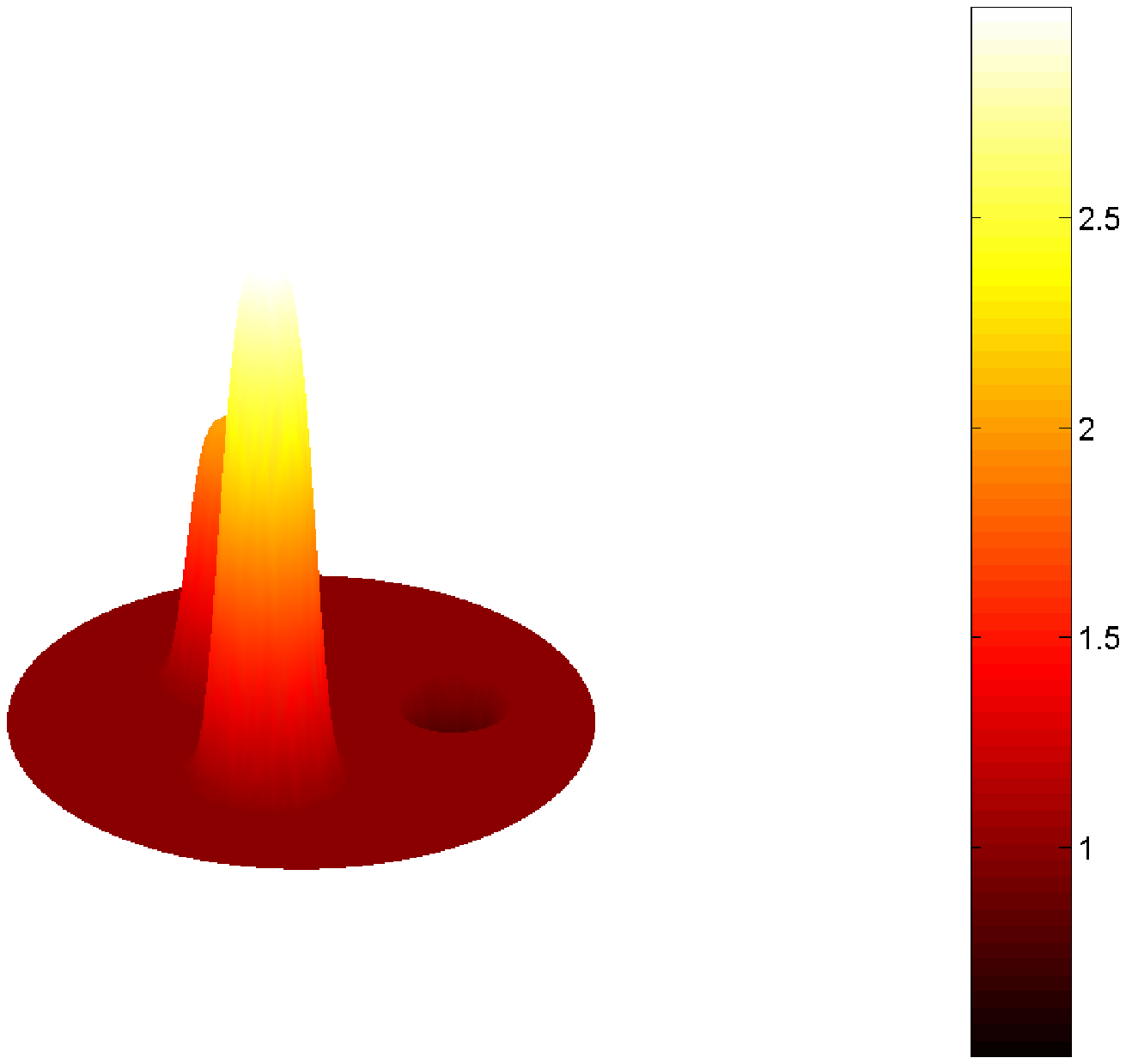}\includegraphics[height=4.5cm,angle=0]{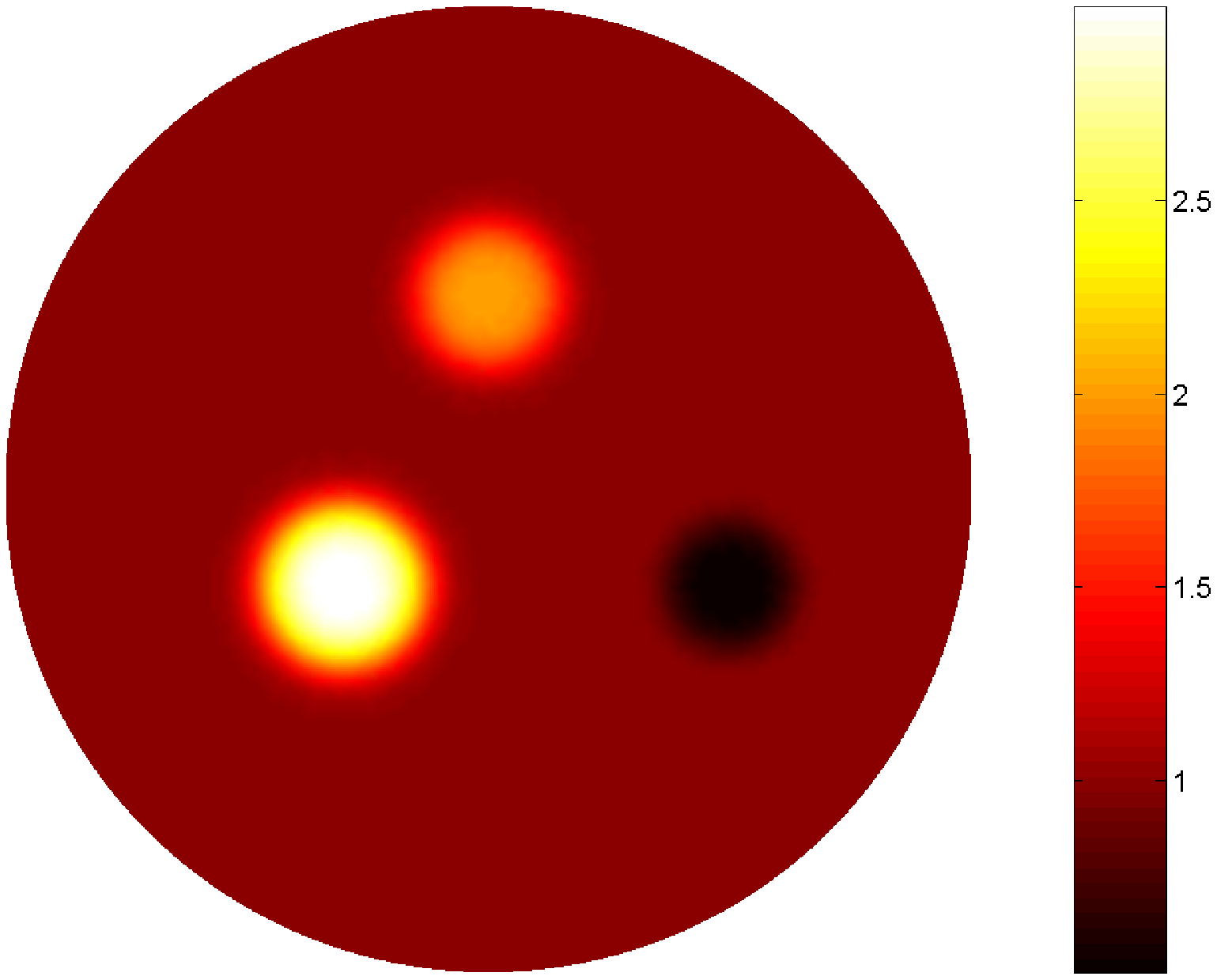}
\includegraphics[height=4.5cm,angle=0]{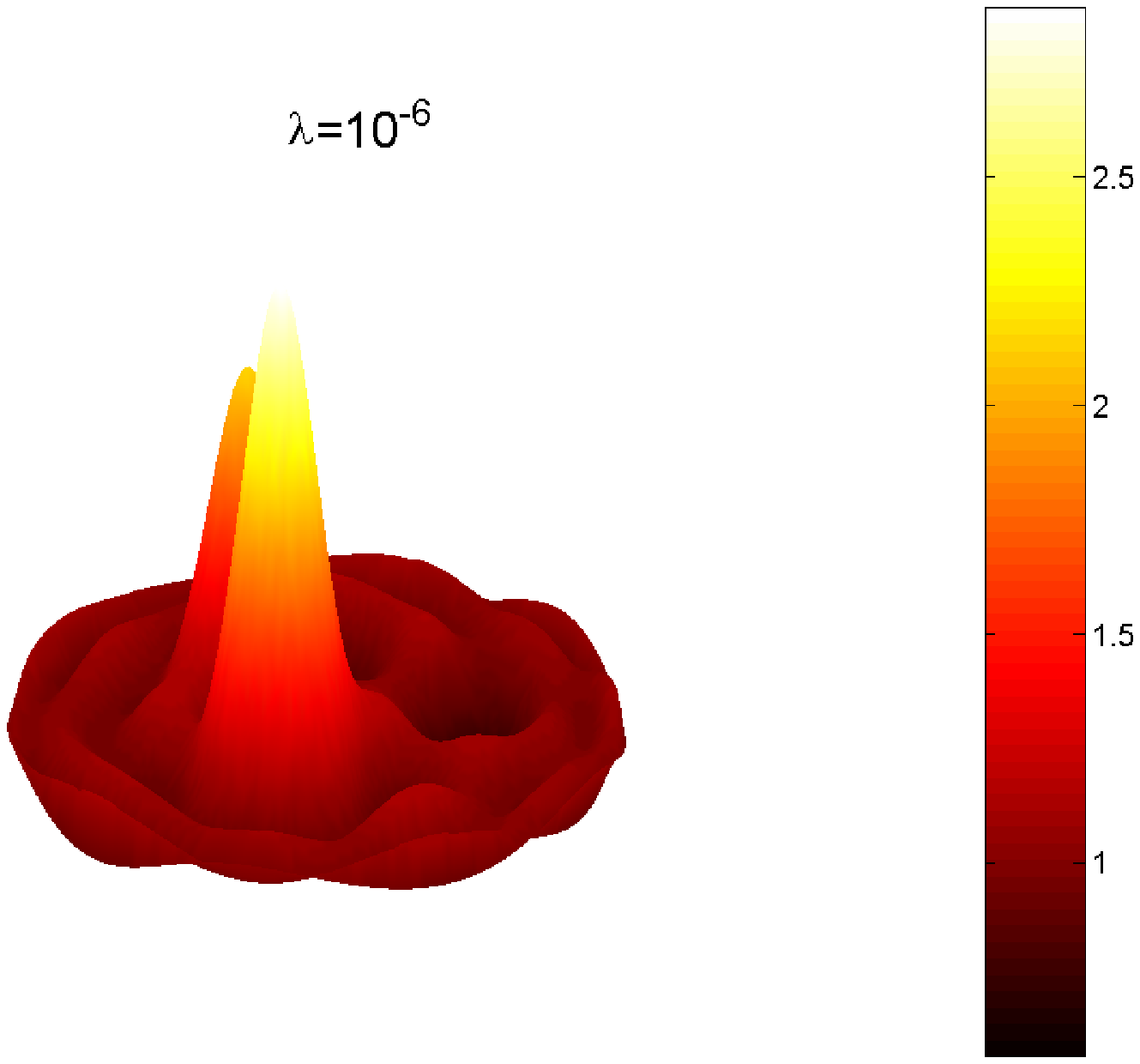}\includegraphics[height=4.5cm,angle=0]{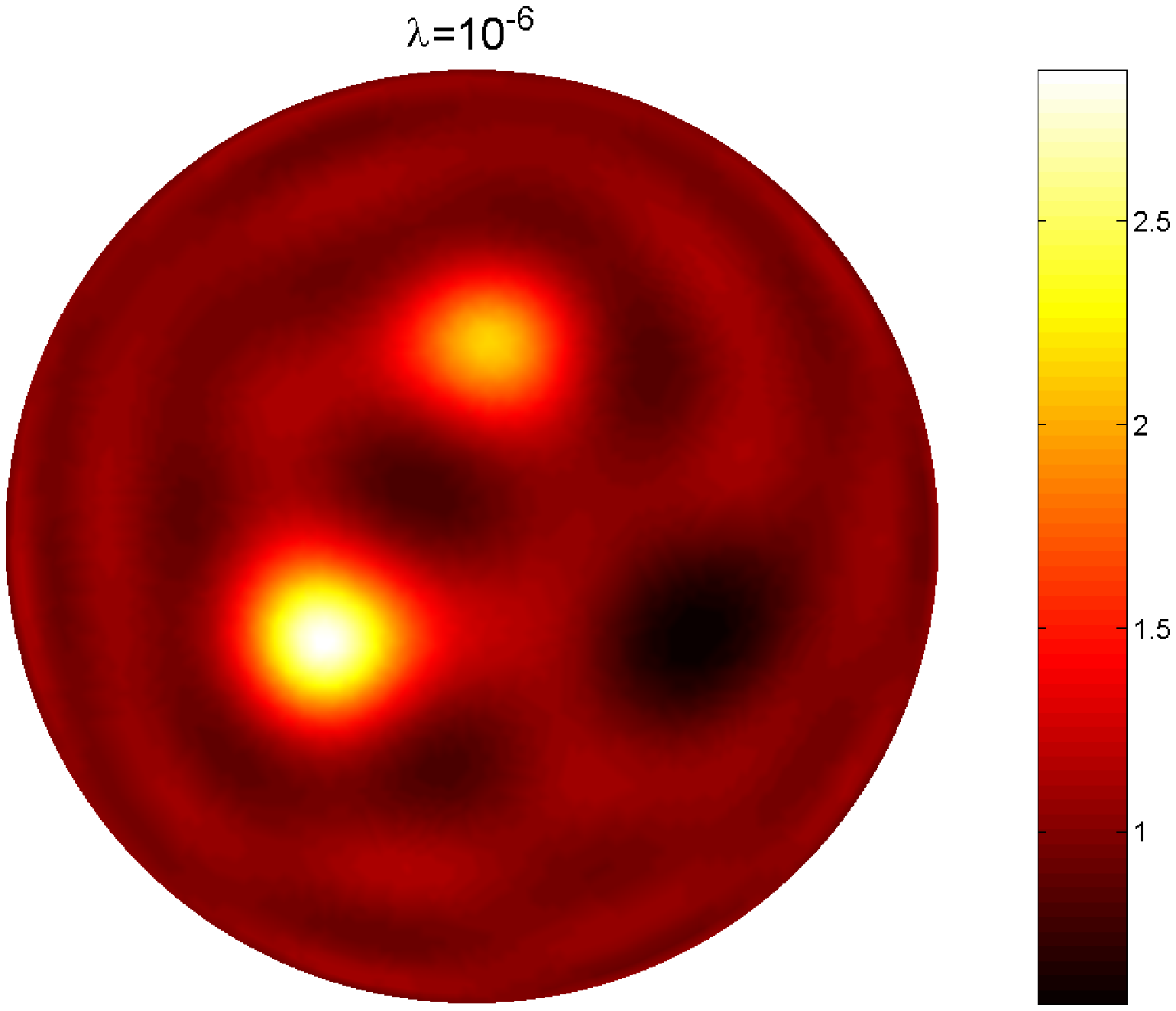}
\caption[Reconstruction from more measurements: Model conductivity $\sigma_3$ and its reconstruction.]{In the upper part, two- and three-dimensional plots of the model conductivity $\sigma_3$,  Eq.(\ref{testconductivity}), are displayed. The corresponding reconstructed conductivity distribution is shown in the lower part. The regularisation parameter was chosen to be $\lambda=10^{-6}$.}
\label{sig3rec}
\end{figure}
\clearpage

\begin{figure}[p!]
\vspace{0.5cm}
\centering
\includegraphics[height=4.5cm,angle=0]{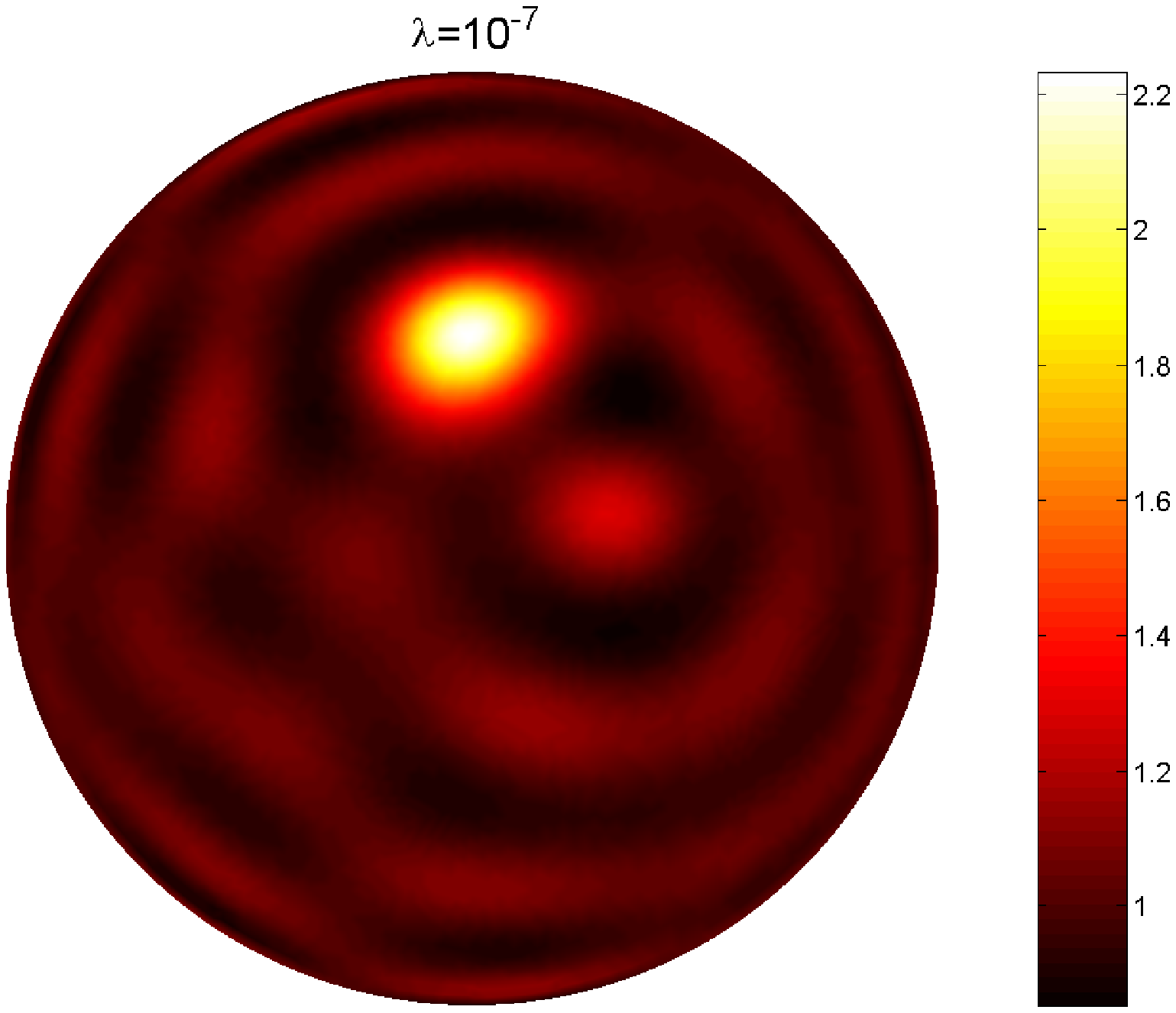}\includegraphics[height=4.5cm,angle=0]{figures/eit-moremes/2Dsig1-6.eps}
\includegraphics[height=4.5cm,angle=0]{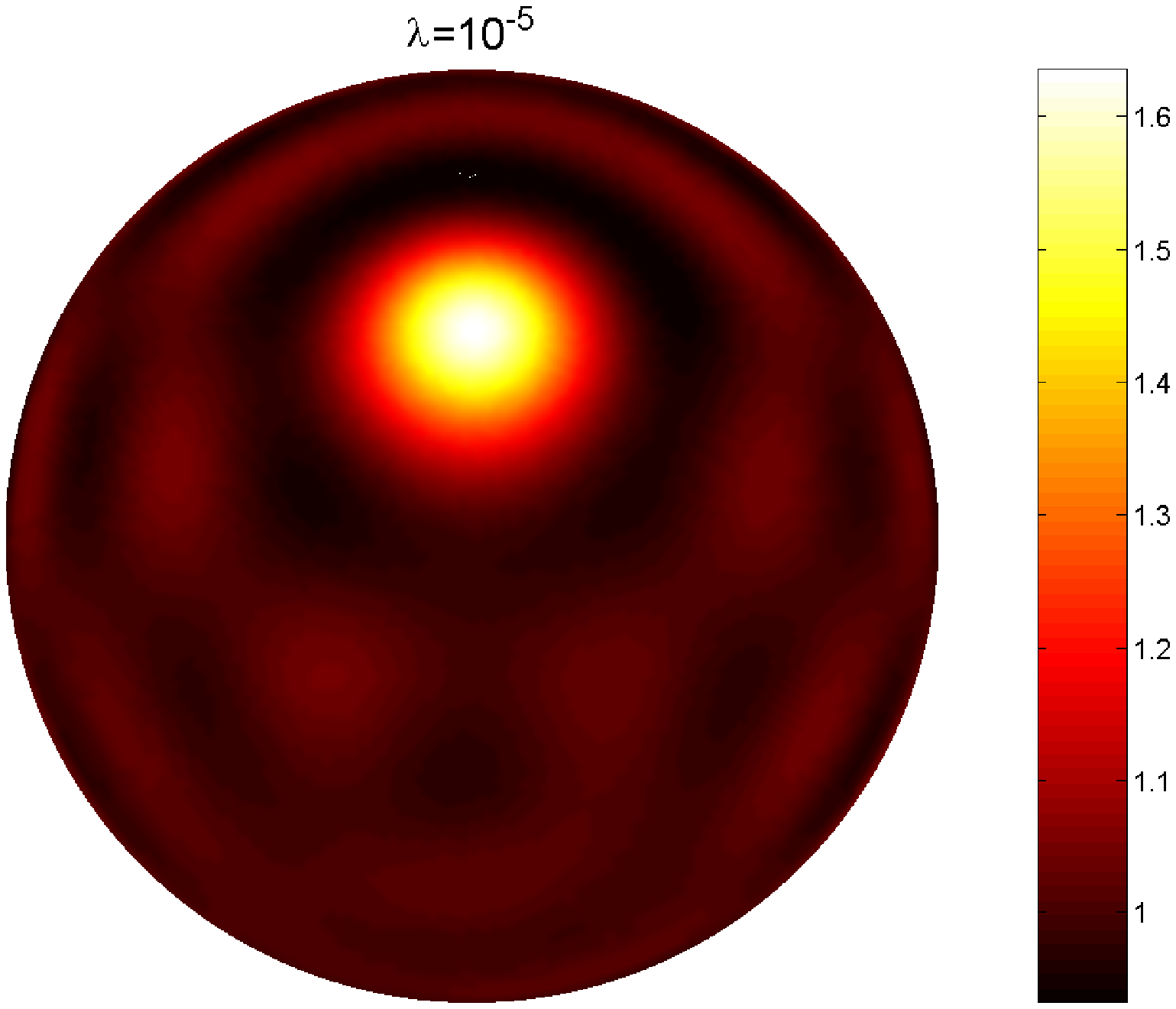}\includegraphics[height=4.5cm,angle=0]{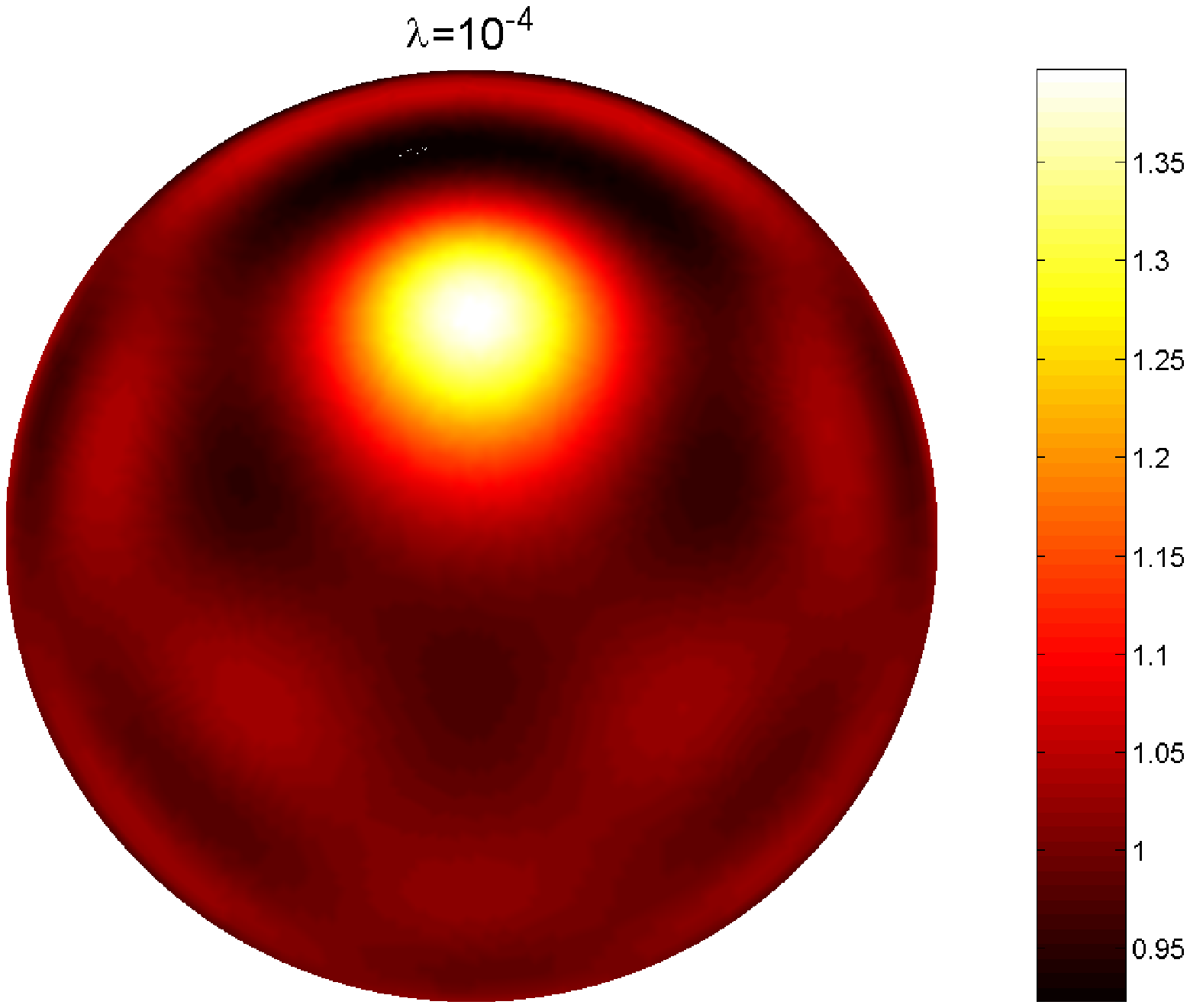}
\caption[Reconstruction from more measurements: Reconstructed conductivity distribution $\sigma_1$ for several values of the regularisation parameter $\lambda$.]{Reconstructed conductivity distribution $\sigma_1$ for several values of the regularisation parameter $\lambda$.\label{lamsig1rec}}
\centering
\includegraphics[height=5cm,angle=0]{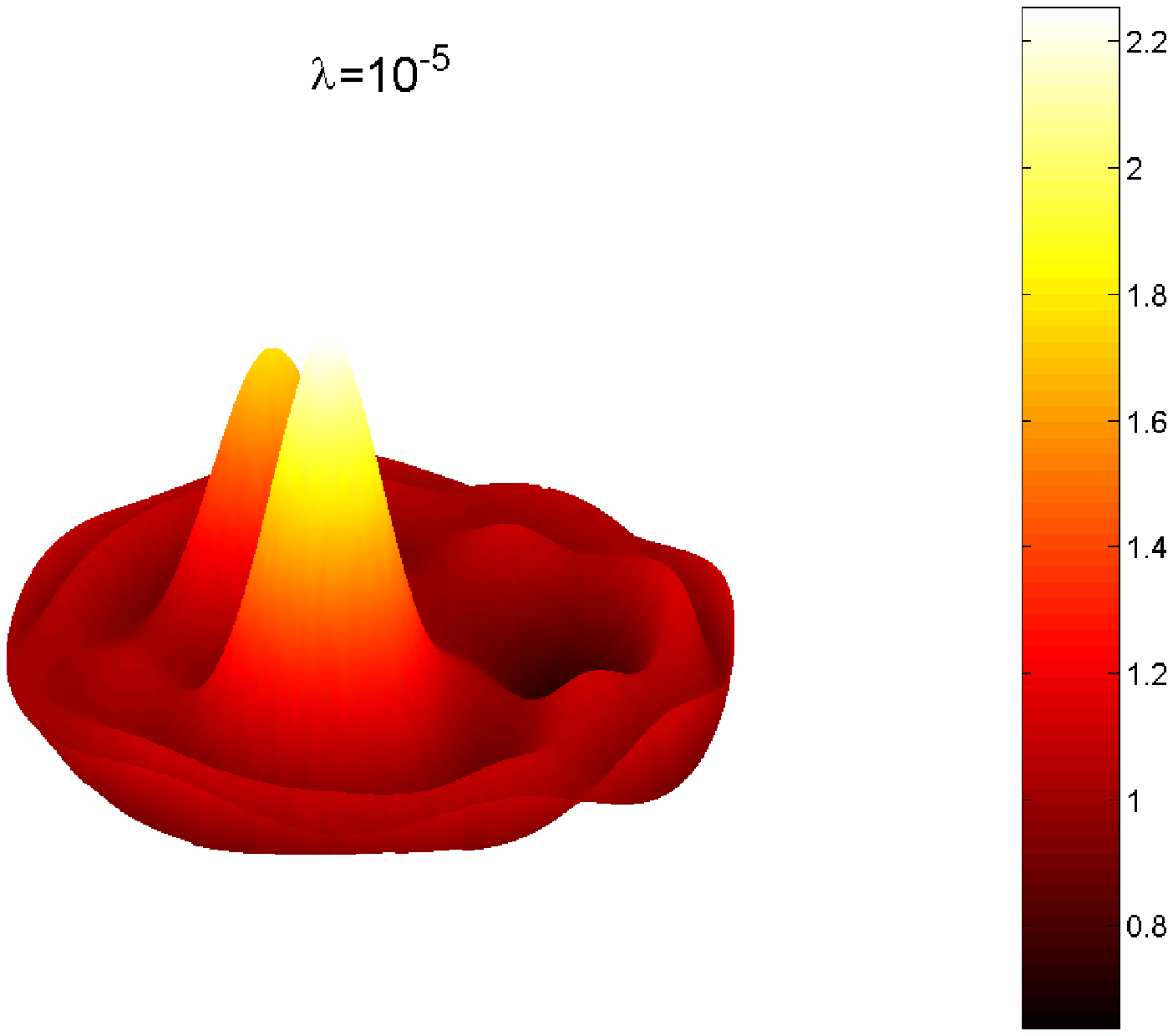}\includegraphics[height=5cm,angle=0]{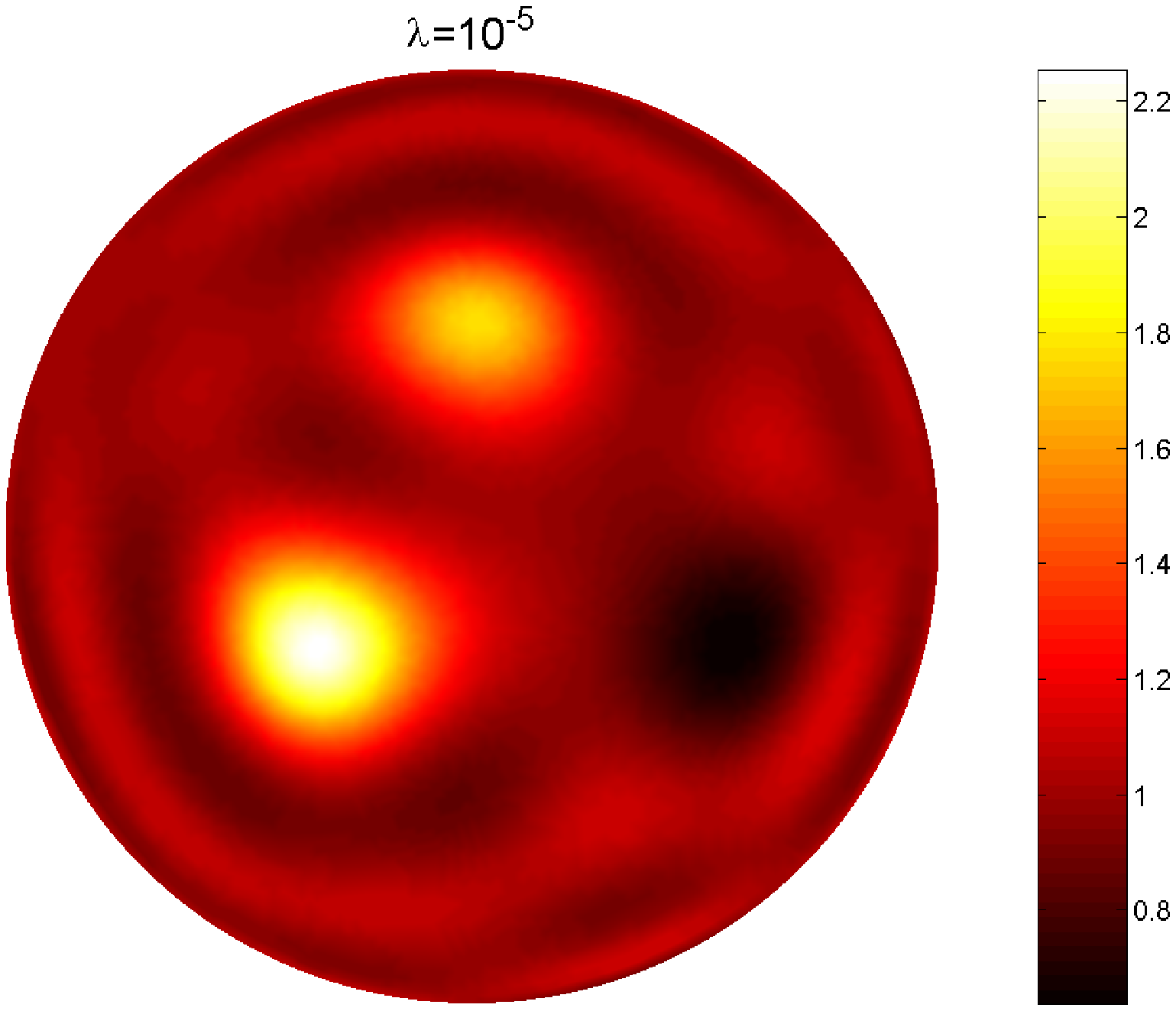}
\includegraphics[height=5cm,angle=0]{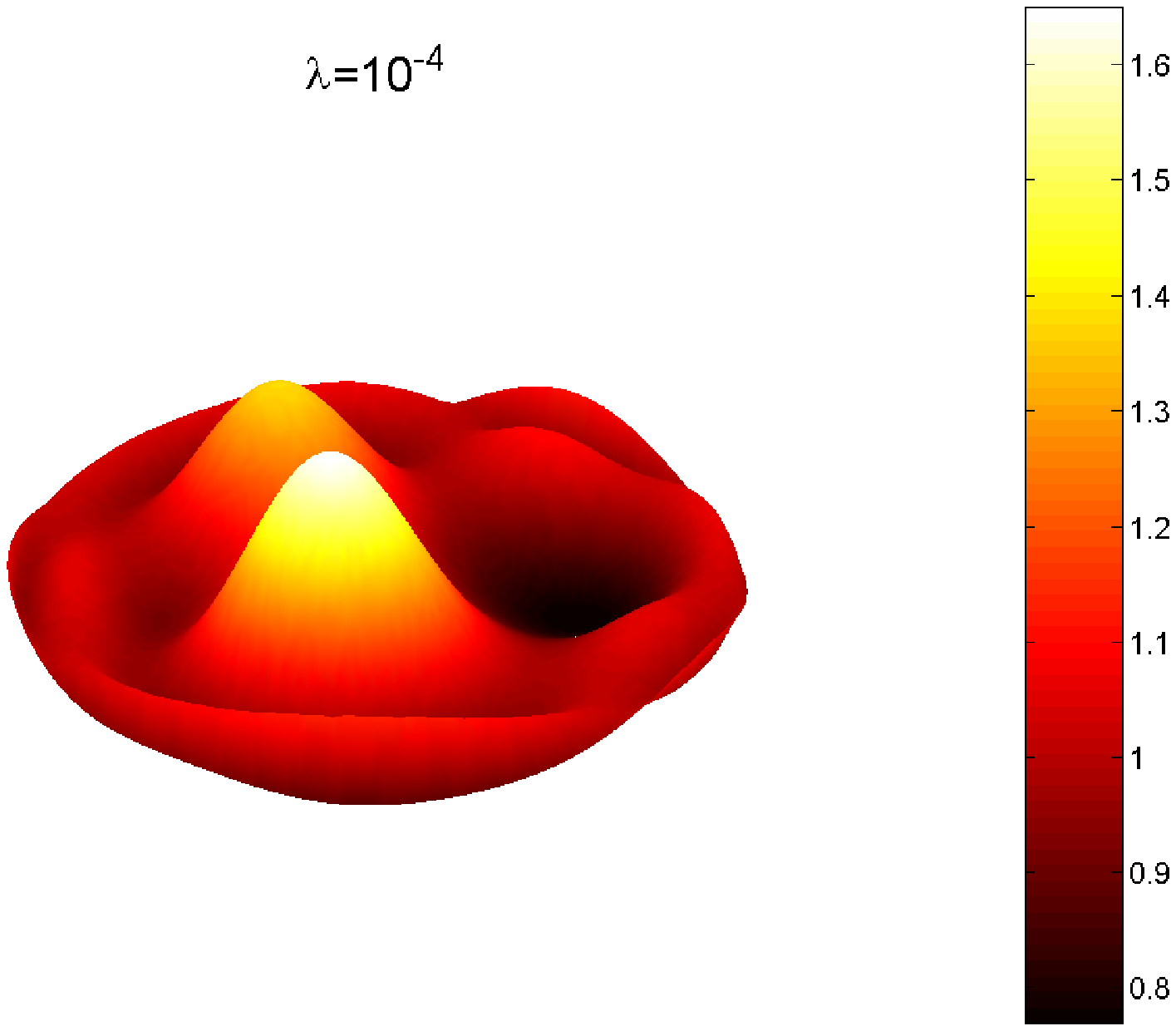}\includegraphics[height=5cm,angle=0]{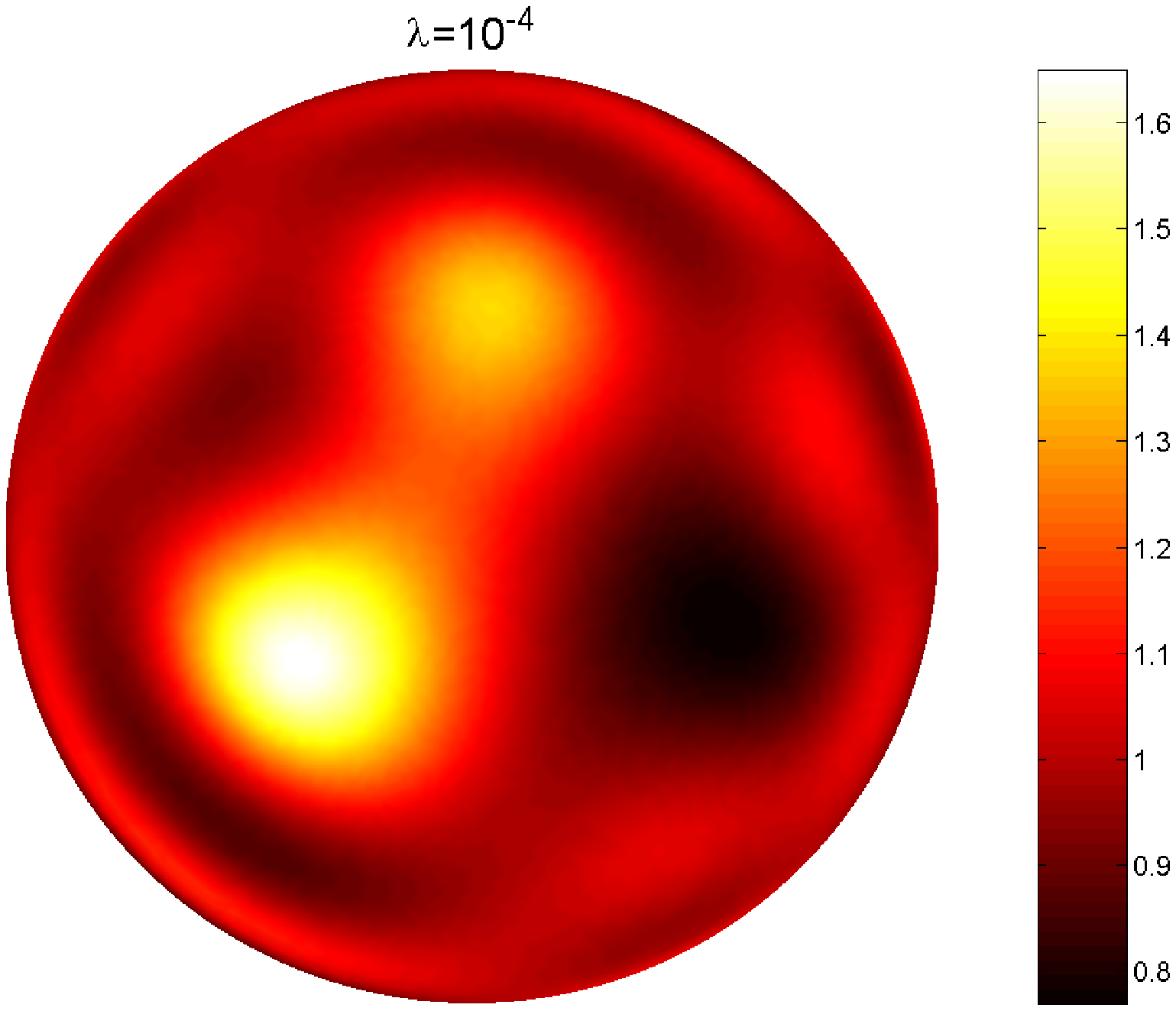}
\caption[Reconstruction from more measurements: Reconstructed conductivity distribution $\sigma_3$ from error affected data.]{Reconstructed conductivity distribution $\sigma_3$ from error affected data: $5\%$ errors (upper part) and $10\%$ errors (lower part).}
\label{ersig3rec}
\end{figure}
\clearpage

Figs.\ \ref{sig1rec}, \ref{sig2rec} , and \ref{sig3rec} show the three test conductivity distributions from Eq.(\ref{testconductivity}) and their reconstruction. One can see that the positions of the perturbations, as well as their height, are relatively well reconstructed. However, since small singular values are cut-off, i.e., there is some information missing, one cannot reconstruct neither of them exactly. The reconstruction is stable for a well-defined interval of the regularisation parameter, $\lambda\in[10^{-7},10^{-4}]$, as depicted in Fig.\ref{lamsig1rec}. There exists no rigorous mathematical reasoning for defining this interval, but one can find it by looking at the changes in the reconstructed conductivity distribution when scanning $\lambda$. It is obvious that for too small or too large values, the reconstruction will become very bad. This is the case since for too small values of $\lambda$ the errors on the data will be reinforced, while for too large values the actual signal gets suppressed.

To check further the stability of our method, we have also added additional randomly distributed errors on the data. This is realistic since in practice the measurements are always accompanied by errors. For big errors the algorithm looses its stability, i.e.\ the conductivity distribution will be not so well reconstructed. Indeed, one can observe in Fig.\ref{ersig3rec}, lower part, that the reconstructed perturbations become flatter as compared to the error free data (Fig.\ref{sig3rec}) or to the one where the errors on the data are smaller (Fig\ref{ersig3rec}, upper part). Also, the position of these perturbations is imprecisely reconstructed. These changes are due to the fact that the regularisation parameter needs to be increased, i.e., one has to cut more singular values as in the case of error free data, and hence loose more information.

\clearpage{\thispagestyle{empty}\cleardoublepage}

\chapter{Reconstructions based on real data}

One medical problem for which knowledge of internal electrical properties would be useful is the detection of pulmonary emboli, or blood clots in the lungs. The development of pulmonary emboli is a common, and often serious, complication of surgery. Unfortunately, at present the diagnosis is rather involved, one possibility being the inhalation of radioactive gas in order to determine the ventilated lung region. This is followed by injection of a radio-opaque dye or a dissolved radioactive substance into a vein to make an image of the blood circulation. The image of the circulation in the lungs is compared with the image of the ventilated region; areas that are ventilated but not perfused by blood indicate the presence of emboli.

Another way to determine the location of gas and blood within the body would be to map the internal electric conductivity and susceptivity. These electrical properties are very different for air, tissue, and blood (see Table \ref{tissue2}, Chapter \ref{EITintorduction}); moreover, they vary on different time scales. Thus a time-varying map of the electrical properties should show lung regions that are ventilated but not perfused by blood.

Furthermore, determining the presence of pulmonary emboli from a map of the electrical properties would have a number of advantages over present techniques. It would require no exposure to X-rays or radioactive material. It would be done at the bedside, with a relatively small and inexpensive electrical system.

Information about the internal electrical properties of a body could have many other medical uses. Such information could potentially be used for the following: monitoring for lung problems such as accumulating fluid or a collapsed lung, noninvasive monitoring of heart function and blood flow, monitoring for internal bleeding, screening for breast cancer, studying emptying of the stomach, studying pelvic fluid accumulation as a possible cause of pelvic pain, quantifying severity of premenstrual syndrome by determining the amount of intracellular vs.\ extracellular fluid, determining the boundary between dead and living tissue, measuring local internal temperature increases associated with hyperthermia treatments, and improving electrocardiograms and electroencephalograms.

To map the electric conductivity inside a body, a tomograph was constructed in collaboration with Dr. K.H. Georgi and N. Schuster. The electrical measurements were used to reconstruct and display approximate pictures of the electric conductivity inside the body, based on the reconstruction algorithm described in Section \ref{recalgorithm}. 

\section{The tomograph}

\begin{figure}[H]
\centering
\includegraphics[height=6cm,angle=0]{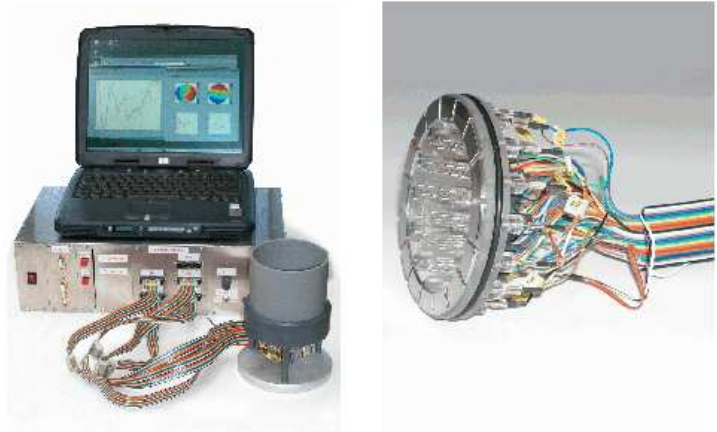}
\caption{The tomograph.}
\label{tomograph}
\end{figure}

The tomograph consists of three parts: a sensing head, an electronic device to apply and measure electric potential and current patterns, and a computer for the image reconstruction (Fig.\ref{tomograph}). It was designed for mammography applications. For a description of this tomograph we will follow Ref.\cite{azzouz}. More details can be found in \cite{azzouz2}.

The sensing head has a diameter of 10 cm and consists of 16 large electrodes, arranged on the outer ring of a disk. Through these electrodes the current is injected and both current and voltages can be measured. There is another set of 64 small high-impedance electrodes placed in the interior which can be used to measure additional voltages, however, these measurements have not yet been used to solve the inverse problem.

\begin{figure}[H]
\centering
\includegraphics[height=7cm,angle=0]{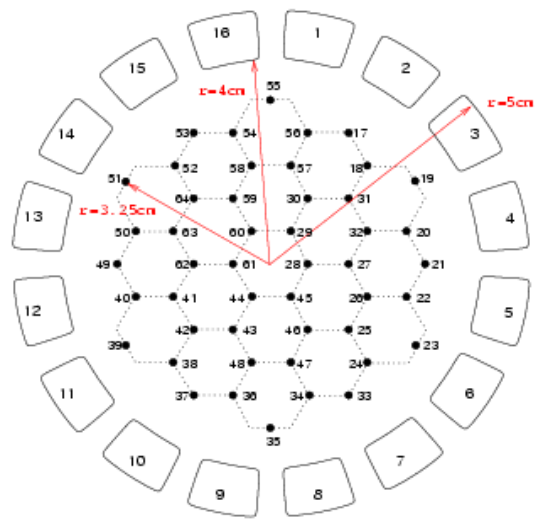}
\caption{The sensing had.}
\label{grid}
\end{figure} 

The data acquisition device consists of five modules. The first module is a micro-controller to facilitate the communication between the measuring device and external components via a RS232 serial interface. The second module generates sinusoidal voltages of frequencies in the range of 5-50KHz which can be used to drive 16 (or 32) current injecting electrodes. The amplitudes can be set positive or negative by 32 DACs to 16-bit accuracy and can be modulated to any desired amplitude pattern. The resulting current at each injection electrode passes through a precision resistor and a special operational amplifier to facilitate the simultaneous measurement of the current. The voltages on the interior electrodes are measured with 16-bit accuracy with the help of the third module. The fourth module serves to read the data and to measure the signal by a peak detector via 8 multiplexers of 16 channels each. The measured value of the peak detector is subsequently digitalised to 16-bit accuracy. In the fifth module, finally, the sign of the modulation is defined. A more detailed description of these five modules is found in \cite{azzouz2}.

There are $N-1$ independent measurements for $N$ current injecting electrodes. In most reconstruction algorithms it is assumed that currents are prescribed and voltages are measured. Because of the much simpler electronic implementation, voltages are applied and currents are measured on the injectors (voltages are also measured in the interior) in our device. It is, however, no problem to convert to current-driven data by linear combination of the various voltage-driven data. The current-driven data is preferred since, as stated in Section \ref{inverseformulation}, the Neumann-to-Dirichlet map (Eq.(\ref{NtDmap})) is better behaved then the Dirichlet-to-Neumann one (Eq.(\ref{DtNmap})) for noisy measurements. For optimal resolution it is of advantage to apply trigonometric voltage or current patterns of different frequencies \cite{hewell}.

\section{Measurements in a test tank}

So far we have no clinical data at our disposal. Instead, we have placed the sensing head into the bottom of an appropriate container of large lateral dimensions and filled with a conducting liquid. The level of the liquid in the tank has been kept very low so as to approximate a two-dimensional situation. Various objects have been immersed into the liquid. Measurements have been taken with and without the immersed bodies. The latter serve as reference measurements where necessary.

\begin{figure}[H]
\centering
\includegraphics[height=5cm,angle=0]{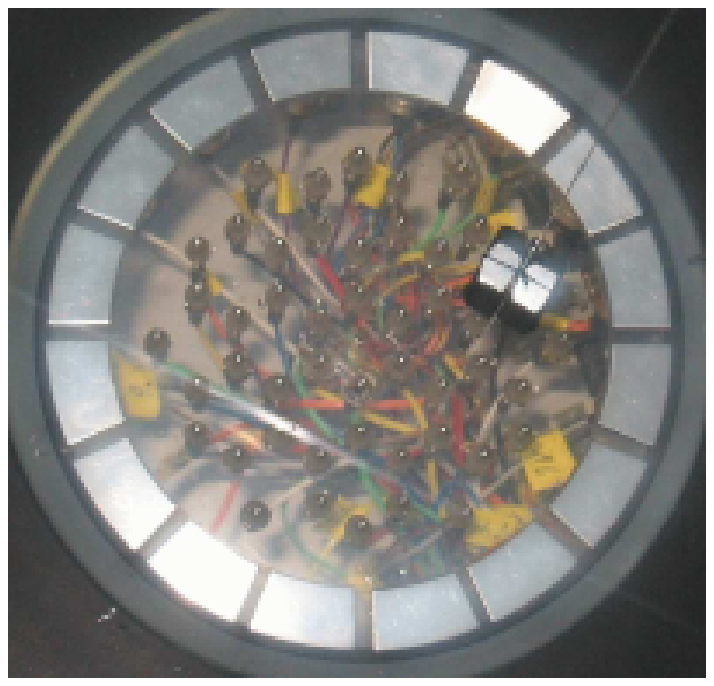}
\includegraphics[height=5cm,angle=0]{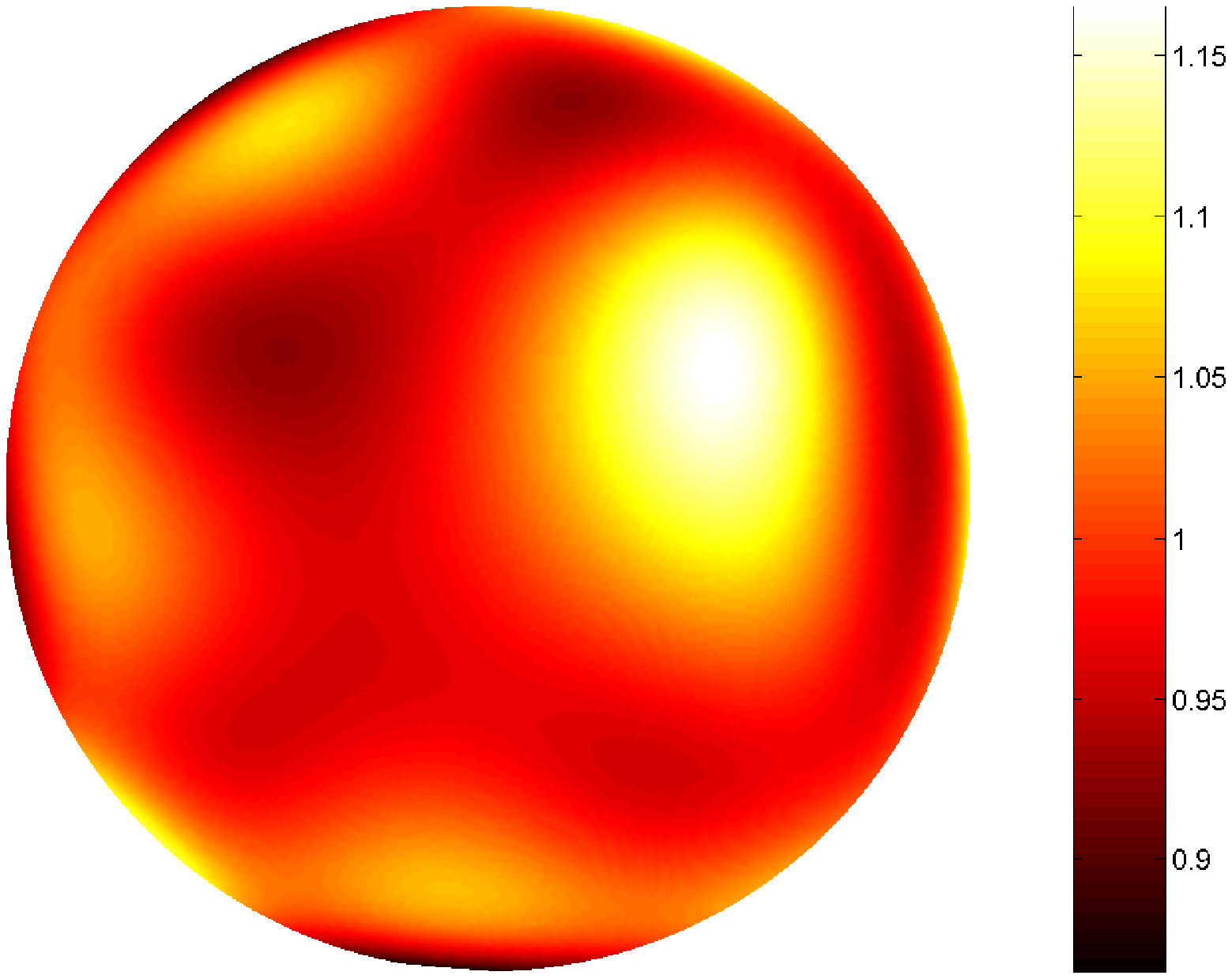}\includegraphics[height=5cm,angle=0]{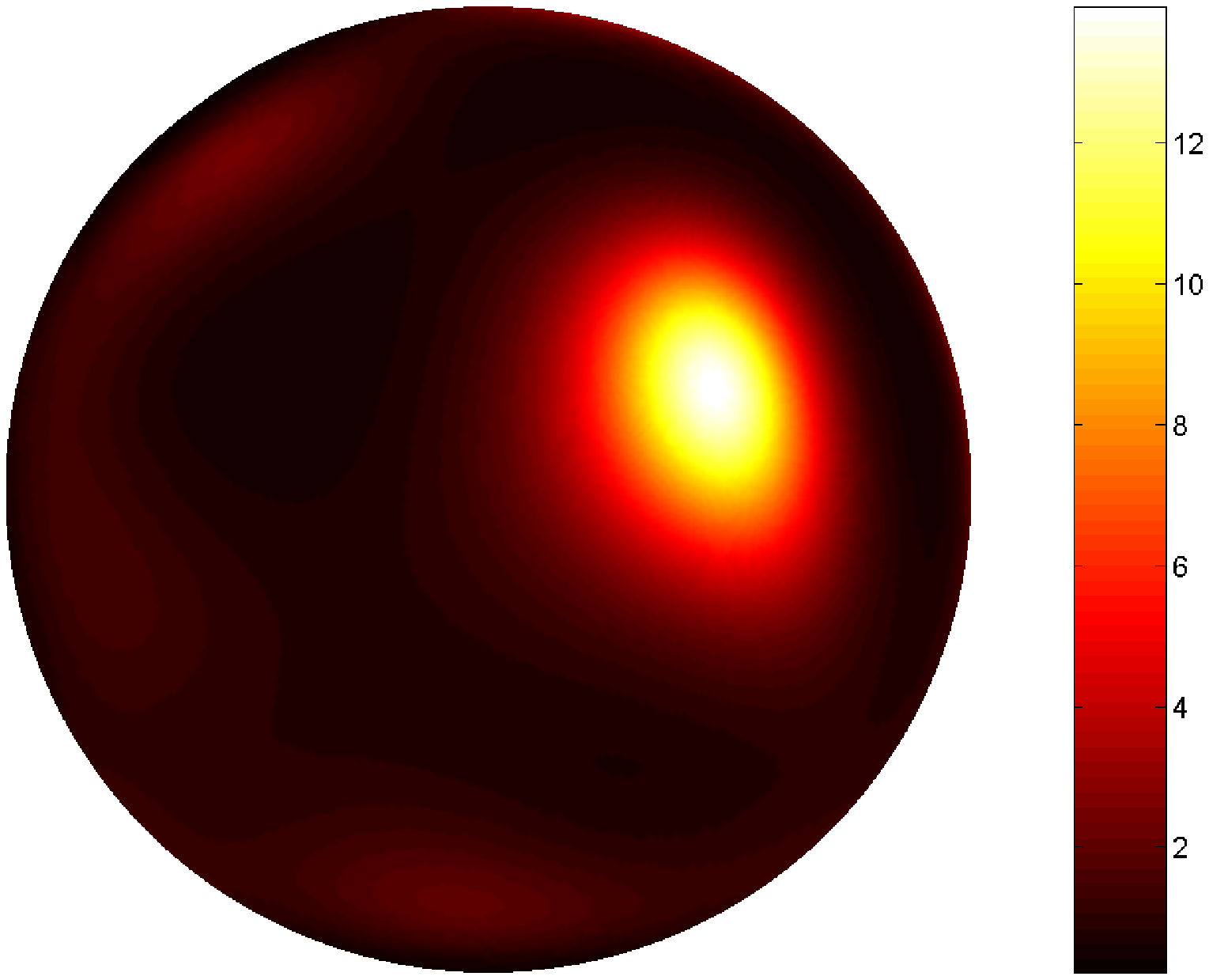}
\caption[Measurements in a test tank: A phantom and its reconstruction.]{The phantom and its reconstruction. The reconstruction on the left is from ``absolute'' data, i.e., no measured reference potential available, while the one on the right is the reconstruction from so called ``difference'' data, i.e., the measured reference potential was used.}
\label{phantomrec}
\end{figure}

Fig.\ref{phantomrec} shows such a phantom and the resulting reconstruction \cite{azzouz}. A metal object of roughly $12\times13{\rm  mm}$ had been immersed into the tank. The right-hand reconstruction has been computed from a measured reference potential $\phi_0$ corresponding to a tank with no object immersed. The one on the left has been computed to simulate a situation where only ``absolute'' data, i.e., the potentials $\phi$ are available. Here $\delta\phi$ has been approximated by eliminating the frequency of the injected current from the Fourier spectrum of $\phi$. Both reconstructions are fairly good, although the one with ``absolute'' data is only qualitatively correct. Note that potentials and currents are only known on the 16 planar electrodes that are clearly visible in the photo of the phantom. 

\begin{figure}[H]
\centering
\includegraphics[height=7cm,angle=0]{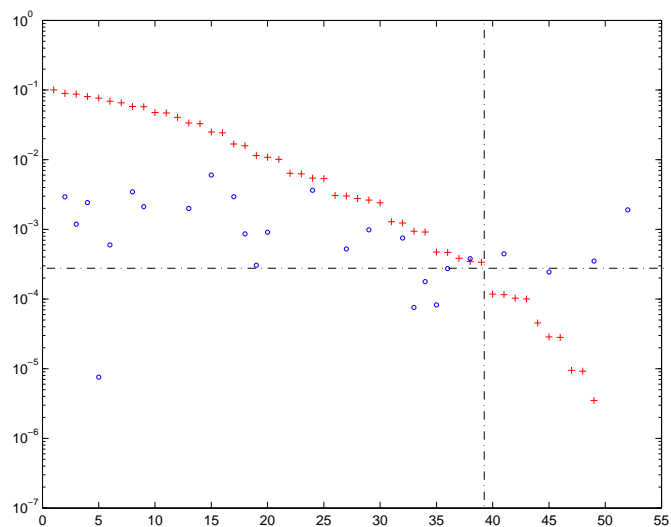}
\caption[Measurements in a test tank: Singular values and singular components.]{Singular values of the matrix (\ref{short:system}), displayed by the '{\tiny +}' symbols, and the corresponding singular components of the l.h.s.\ of (\ref{systemofeq}), displayed by the '{\tiny o}' symbols, for the set of data used in the above reconstruction.}
\label{singcut}
\end{figure} 
\clearpage

For the regularisation, the truncated SVD was implemented. To choose a value for the cut-off parameter $\lambda$ the interactive method can be used since we exactly know what we want to reconstruct. However, in medical applications one does not have, in general, a picture of the expected conductivity distribution. For this reason we have tried to develop a more mathematically rigorous way to fix $\lambda$. The form of the general solution to inverse problems (cf. Eq.(\ref{seriessolution})),
\begin{equation}
 f^+=\sum_{j=1}^\infty\frac{1}{\sigma_j}(g,u_j)_{\cal Y}v_j,
\end{equation}
suggests that one should compare the singular values $\sigma_j$ with the singular components $(g,u_j)_{\cal Y}$. As long as the singular values are larger than the singular components, the errors contained in $g$ (data errors) will be suppressed and when the singular values become smaller, the errors will be reinforced. This behaviour is a hint to choose the cut-off parameter in such a way that the singular values and singular components balance each other. In practice, this is done by plotting $\sigma_j$ and $(g,u_j)_{\cal Y}$ against $j$ and choose for $\lambda$ the point where the two of them cross each other. For the case of the measurements in a test tank considered in this section, this plot is shown in Fig.\ref{singcut}. Here one can recognise the usual behaviour of the singular values, i.e., they decrease, and the behaviour of the singular components which are expected to oscillate. The value of $\lambda$ used in the truncation is indicated by the dashed lines. However, as seen in the previous Chapter, there exists a well-defined interval of allowed values for the cut-off parameter, centred at the value chosen above.

\section{Measurements on a human chest}

To make measurements on patients, a belt of electrodes is placed around their chest. It is important that the electrodes are equally spaced, since the reconstruction algorithm is dependent on their exact position. 

For such measurements, there exist difficulties with the contact impedance. If currents are applied and voltages are measured on the same electrodes, the voltage measurements will be sensitive to the contact impedance which is formed at the electrode. To overcome this inconvenience, we have separated the electrodes measuring currents and potentials. If this technique is employed, no current flows through the voltage measurement electrodes, which means that no voltage is dropped across these electrodes, leading to more accurate voltage measurements across the sample.

\begin{figure}[ht]
\centering
\includegraphics[height=6cm,angle=0]{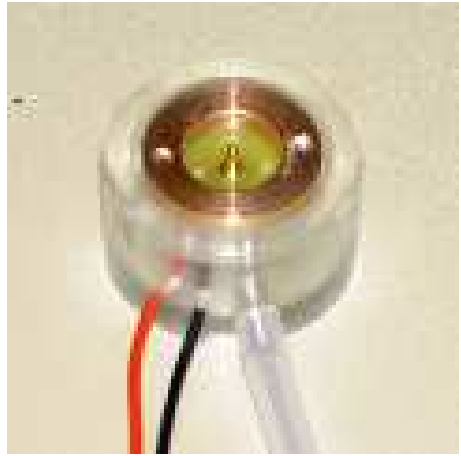}
\caption{The double-electrode.}
\label{electrode}
\end{figure}

The same tomograph was used, only now the sensing head was changed into a 16 double-electrode array. The double-electrodes consist of an outer gold-plated ring for current measurements and a middle, also gold-plated, pin for potential measurements (see Fig.\ref{electrode}). 

Once again, the reconstruction algorithm from Section \ref{recalgorithm} was used to produce pictures of the conductivity distribution inside the human body. Here, we have used ``absolute'' data, i.e., $\delta\phi$ (see Eq.(\ref{short:system})) was approximated by eliminating the frequency of the injected current from the Fourier spectrum of $\phi$. As a regularisation method the truncated singular value decomposition was implemented, with $\lambda$ chosen like in the previous section. 

\begin{figure}[p!]
\centering
\subfigure{\includegraphics[height=7.0cm,angle=0]{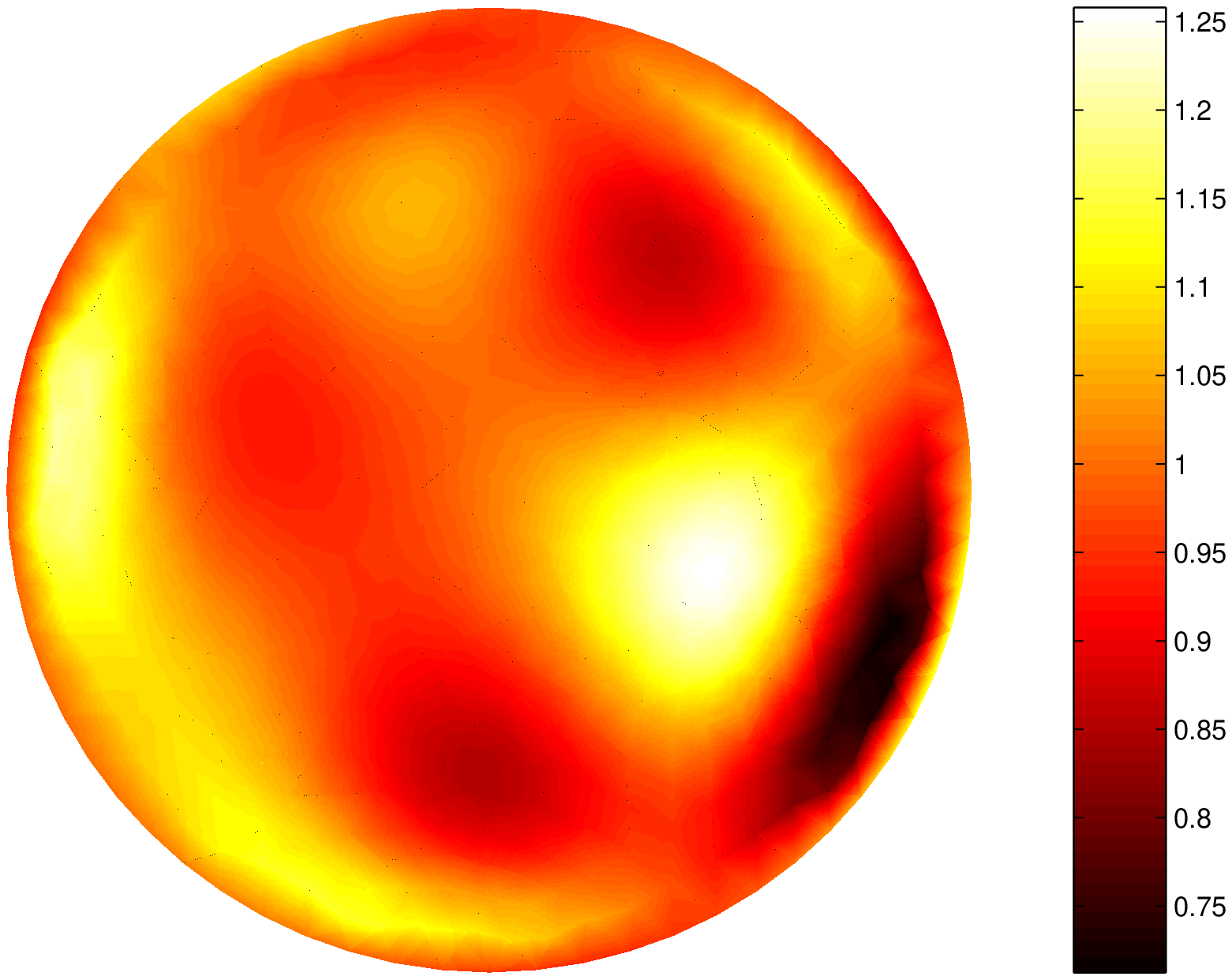}}
\caption{Chest reconstruction on a human volunteer.\label{breathed}}
\vspace{2cm}
\subfigure{\includegraphics[height=7cm,angle=0]{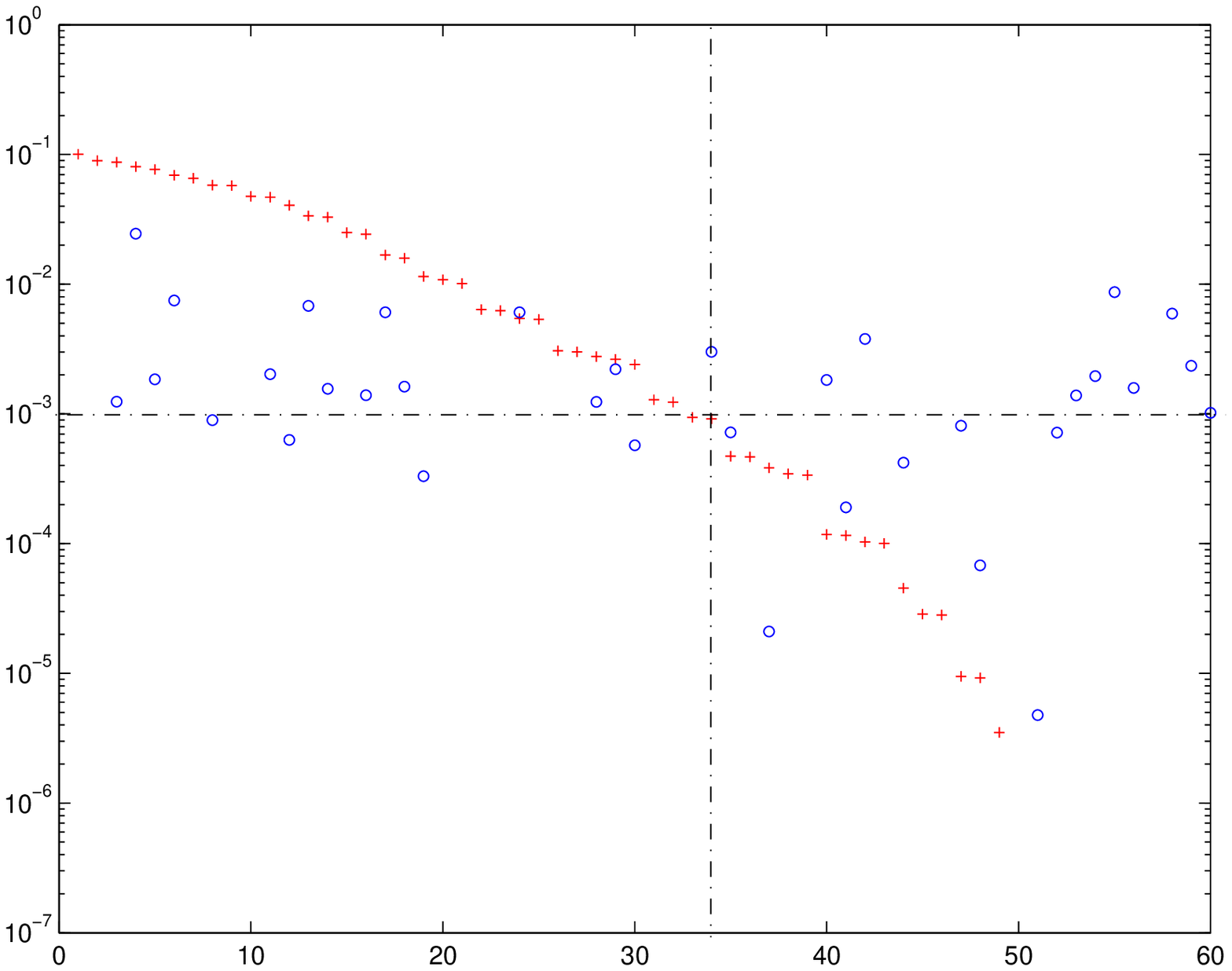}}
\caption[Singular values and singular components for the measurements on the chest of a human volunteer.]{Singular values of the matrix (\ref{short:system}), displayed by the '{\tiny +}' symbols, and the corresponding singular components of the l.h.s.\ of (\ref{systemofeq}), displayed by the '{\tiny o}' symbols, for the above set of data.\label{breathed:sing}}
\end{figure}

One such reconstruction is shown in Fig.\ref{breathed}. The corresponding plot of the singular values and singular components is shown in Fig.\ref{breathed:sing}, where the dashed lines correspond to the values chosen for $\lambda$. One can recognise the heart (high conductivity) and the lungs (low conductivity). Note that this is just a qualitative result. Remember that no reference measurement is at hand and that the background constant conductivity was approximated to 1, which is not the case in medical applications. Other potential sources of error are the inaccuracies in the location of electrodes. These inaccuracies may be in the non-uniformity in the spacing between the electrodes and/or in the shape of the boundary. However, these reconstructions may be useful even though the human volunteer is neither two-dimensional nor circular.

There is still a long way to go until actual medical applications, such as monitoring for lung problems, are possible. The resolution of the reconstruction should be improved, e.g.\ by using more electrodes and/or a more complex modelling of them (see Section \ref{electrodesmodeling}). Also the time needed for a full cycle of measurements needs to be improved. For applications like the so-called ``perfusion'', i.e., monitoring the blood flow to the lung tissue, a synchronisation of the measurements with the heart beat is needed. 

\clearpage

\clearpage{\thispagestyle{empty}\cleardoublepage}

\chapter*{Conclusions and further work}
\addcontentsline{toc}{chapter}{Conclusions and further work}

\section*{QCD condensates from $\tau$-decay data}

Although QCD has been with us for three decades, the knowledge of the values of the various fundamental or effective parameters of the theory (with the possible exception of the coupling constant), such as quark masses and condensates, is still astonishingly limited. The precise data on $\tau$-decay obtained by the ALEPH \cite{aleph05} and OPAL \cite{opal} collaborations have offered an opportunity for new studies, which range from an extraction of the strange quark mass \cite{gupta} to the determination of various condensate parameters. The dimension $d=6$ condensate was the most studied one \cite{aleph,almasy,bordes,cirigliano1,ciulli,dominguez,ioffe,narison1,opal,rojo,zyablyuk}.

We have used a functional method which allows the extraction of condensates from experiment within rather general assumptions. It is obvious that it is an inverse problem: we want to determine constants of the theory which describe the hadronic structure, from indirect measurements. The data used for this purpose was the $\tau$-decay data from the ALEPH collaboration \cite{aleph05} at LEP. One has to relate error affected data in the time-like region to asymptotic QCD in the space-like region. This task of analytic continuation constitutes, mathematically, an ill-posed problem. In fact, extracting condensates from data is highly sensitive to data errors. Not surprisingly, results from different collaborations have not been always consistent.

The method consists in the comparison of the time-like experimental data with the asymptotic space-like QCD prediction in an $L^2$-norm manner. Such a method must be well defined from a mathematical point of view and physically meaningful. One must give explicitly a quantitative estimate of the errors including both experimental and theoretical ones (truncation of the perturbative and operator product expansions). Then, a set of parameters is said to be {\em admissible} (or {\em compatible}, or {\em consistent}) if there exists at least one function with the required analytical properties which goes through the $L^2$-error corridors. Answering to this question can be reduced to a minimisation process over a finite number of variables and to a sequence of Fredholm equations of the second kind, which leads to feasible computations.

We have performed 1- and 2-parameter fits using the data available for the $V-A$ channel. The first one was used for the extraction of the 4-quark condensate ($d=6$) both at leading and next-to-leading order in $\alpha_s$. The results are in agreement, within errors, with those from conventional analyses based on finite energy sum rules. An important result is obtained from the 2-parameter fit, where the free parameters are the 4-quark ($d=6$) and quark-gluon ($d=8$) condensates. The strong correlation found allows one to determine a linear combination of the two condensates with a rather small error, in the order of $3\%$. 

We have also performed 1- and 2-parameter fits for the $V$, $A$ and $V+A$ channels separately. Here, the 1-parameter fit estimates the gluon condensate ($d=4$) while from the 2-parameter fit a correlation between the gluon and 4-quark condensates is found. Again, on the basis of this correlation, one can determine combinations of the two condensates with rather small errors.

We would like to continue the research on this line and:
\begin{itemize}
 \item Perform 3-parameter fits in all channels.
 \item Include contributions omitted in the present analysis like heavy quarks, broken chiral symmetry and duality violation. Such contributions may become important for a better stabilisation of the algorithm, i.e., for a more stable and continuous dependence of the algorithm (and results) on parameters like the end points of the space-like interval $\Gamma_L$ and the number $N$ of data points actually used in the analysis.
 \item To improve the precision and test the consistency, we would like to merge the results, find overall correlations and allowed ranges;
 \item It would be also interesting to compare, by means of the algorithm presented in this thesis, the $\tau$-decay data with the data from $e^+e^-$-annihilation and, possibly, resolve the discrepancy between them.
\end{itemize}

\section*{The inverse conductivity problem}

We have developed an approach for solving the inverse conductivity problem based on reformulating it in terms of integral equations and applied it to a two-dimensional domain, the unit disc, using no {\em a priori} information. Unfortunately, the information gained on the reconstructed conductivity distribution is restricted to its angular dependence (no radial information is present). One can only hope that using some {\em a priori} information could improve the reconstruction.

Also an algorithm based on linearisation was presented and tested both on ``exact'' (data produced by solving the forward problem) and real data. A promising result is that measurements on a human chest, taken with the tomograph constructed in collaboration with Dr.\ K.H.\ Georgi and N.\ Schuster, yield fairly good reconstructions. These results, even though they are just qualitative, encouraged us in believing that such an algorithm is useful in medical applications like monitoring for lung problems and thus intend to continue our research in this direction:
\begin{itemize}
 \item We would like to improve the resolution of the reconstruction by, e.g., using more electrodes and/or a more complex modelling for them (e.g.\ the complete model);
 \item Improve the time needed for a full measurement cycle and for the reconstructions;
 \item Synchronise the measurements with the heart beat, so that one can monitor the blood flow to the lung tissue.
\end{itemize}
 
Both approaches presented are geometrically constrained by requiring the availability of Neumann Green's function for the Laplace equation, which essentially restricts us to rectangular, circular, spherical, cylindrical or half-plane geometries. It is a challenging idea to remove this constraint and develop similar approaches using the free Green's function.

\clearpage{\thispagestyle{empty}\cleardoublepage}

\thispagestyle{empty}

\begin{center}
\vspace*{9cm}
{\Huge\bf Appendices}

\addcontentsline{toc}{part}{Appendices}

\end{center}

\clearpage{\thispagestyle{empty}\cleardoublepage}

\appendix
\chapter{Linear integral equations}
\label{integrlequation}

Integral equations are equations in which an unknown function appears under one or more integral signs. They occur naturally in many fields of mathematical physics. They also arise as representation formulas for the solutions of differential equations. Indeed, a differential equation can be replaced by an integral equation which incorporates its boundary conditions. Then, each solution of the integral equation automatically satisfies these boundary conditions. Integral equations also form one of the most useful tools in many branches of pure analysis, such as the theories of functional analysis and stochastic processes.

We shall first give a classification of integral equations and afterwards concentrate on Fredholm integral equations which are among the most popular ones. We will consider some specific cases of Fredholm equations of the second kind and aim to find their solution. For the Fredholm equations of the first kind, only the case of symmetric kernels is considered. Based on the solution for this specific case, we argue that integral equations of the first kind can have properties characteristic to ill-posedness. For a more detailed theory one can consult for example \cite{cour,wyld}.

The short theory of integral equations presented here was very useful throughout the entire work. In Chapter \ref{condensates} we introduced an integral equation of the second kind which needed to be solved for some specific cases (Chapters \ref{V-Aanalysis} and \ref{V+Aanalysis}). When we studied the kernel of this equation in more detail, it turned out that we deal with Fredholm integral equations of the second kind. Later on, in Chapter \ref{directproblem}, Fredholm equations of second kind arose again when solving the direct problem of EIT. As concerning Fredholm equations of the first kind and their ill-posedness, they were needed for the reconstruction of conductivities inside a specified domain from a single set of measurements (see Section \ref{singelrec}). 

\section{Types of integral equations}

Fredholm equations represent one of the most important classes of linear integral equations. We shall distinguish between Fredholm equations of the first and second kind. Equations of the {\em second kind} are the most interesting and important in applications. The simplest formulation for such equations is
\begin{equation}
f(x)=u(x)-\lambda\int_a^b K(x,t)u(t)dt.
\label{basicinteq}
\end{equation}
Here, the unknown function $u(x)$ is a function of the real variable $x$ which, along with $t$, is defined on the interval $[a,b]$. The interval $[a,b]$ may be finite or infinite. The functions $K(x,t)$ and $f(x)$ are known and defined almost everywhere on $a\le x,t\le b$ and $a\le x\le b$, respectively. $K(x,t)$ is called the kernel of (\ref{basicinteq}) and $\lambda$ is a parameter. For Fredholm equations, $K(x,t)$ satisfies
\begin{equation}
\int_a^b\int_a^b\left|K(x,t)\right|^2dx\, dt<\infty,
\label{fredkernel}
\end{equation}
and $f(x)$ satisfies
\begin{equation}
 \int_a^b\left|f(x)\right|^2dx<\infty.
\label{fredfreeterm}
\end{equation}
If $f(x)\equiv0$ (or, more precisely, if $f(x)$ vanishes almost everywhere on $[a,b]$), then Eq.(\ref{basicinteq}) is said to be {\em homogeneous}, otherwise {\em non-homogeneous}.

Fredholm equations of the {\em first kind} differ from the second kind in that the isolated term, $u(x)$, in (\ref{basicinteq}) is absent. The simplest form for it is:
\begin{equation}
f(x)=\lambda\int_a^b K(x,t)u(t)dt,
\label{firsttype}
\end{equation}
where $K(x,t)$ and $f(x)$ satisfy Eqs.\ (\ref{fredkernel}) and (\ref{fredfreeterm}), respectively.

If the kernel of the integral equation has the form
\begin{equation}
K(x,t)=\frac{H(x,t)}{|x-t|^\alpha},
\end{equation}
where $H(x,t)$ is bounded and $0<\alpha<1$, Eq.(\ref{basicinteq}) will be called an {\em integral equation with a weak singularity}.

When the kernel under consideration is of the form
\begin{equation}
 K(x,t)=\frac{A(x,t)}{x-t},
\end{equation}
where the numerator $A(x,t)$ is a differentiable function of $x$ and $t$ and if the divergent integral is interpreted in the sense of its principal value, Eq.(\ref{basicinteq}) is called a {\em singular integral equation}.

Volterra equations of the first, homogeneous and second kinds are defined precisely as above, except that $b=x$ is the variable upper limit of integration. There exist also other classes of integral equations, e.g.\ equations with convolution kernels, Wiener-Hopf equations, dual equations, integral transforms and non-linear integral equations. However, we shall consider here only the case of Fredholm equations. 

\section{Equations of the second kind}

We will start with integral equations of the second kind since they appear to be the easiest in the discussion. We will study a few specific cases, as concerning the properties of the kernel, for which a compact solution can be found and also give a solution by means of Neumann series.

\subsection{Degenerate kernels}

A kernel, which can be written as a finite sum of products of linearly independent functions of $x$ and linearly independent functions of $t$, 
\begin{equation}
K(x,t)=\sum_{i=1}^pa_i(x)b_i(t),
\end{equation}
is called a degenerate kernel. With the help of this representation and exchanging summation with integration, Eq.(\ref{basicinteq}) becomes:
\begin{equation}
 f(x)=u(x)-\lambda\sum_{i=1}^pa_i(x)\int_a^b b_i(t)u(t)dt.
\label{degenerate1}
\end{equation}
One can show that the solution of this Fredholm integral equation reduces to solving a system of linear equations. Indeed, if we define $c_i$ as the integral in (\ref{degenerate1})
\begin{equation}
 c_i=\int_a^b b_i(t)u(t)dt,
\end{equation}
multiply both sides of (\ref{degenerate1}) by $b_m(x)$ and integrate over $x$ from $a$ to $b$, we find:
\begin{equation}
 f_m=c_m-\lambda\sum_{i=1}^p a_{mi}c_i,\ \ \ m=1,2,...,p,
\end{equation}
where
\begin{equation}
 f_m=\int_a^b b_m(x)f(x)dx,
\end{equation}
and
\begin{equation}
 a_{mi}=\int_a^b b_m(x)a_i(x)dx.
\label{systemofeq2}
\end{equation}
Eq.(\ref{systemofeq2}) is a set of $p$ linear equations in $c_1,c_2,...,c_p$ which can be solved by usual methods.

\subsection{Symmetric kernels}

The theory of integral equations can be developed in greater detail if the 
kernel $K(x,t)$ is symmetric, i.e., if it satisfies the relation 
\begin{equation}
K(x,t)=K(t,x).
\end{equation}

\begin{theorem}
Every continuous symmetric kernel that does not vanish identically possesses 
eigenvalues and eigenfunctions; their number is denumerably infinite if and 
only if the kernel is non-degenerate. All the eigenvalues of a real symmetric 
kernel are real.
\end{theorem}
{\bf{\it Proof}:} The proof is made in several steps:
\begin{itemize}
\item[1.] Prove the existence of an eigenvalue of a symmetric kernel;
\item[2.] Consider the totality of eigenfunctions and eigenvalues;
\item[3.] Finally prove that all the eigenvalues are real.
\end{itemize}
For further details see \cite{cour}.\hfill$\Box$

\begin{theorem}
Every continuous function $g(x)$ which is an integral transform with symmetric
 kernel $K(x,t)$ of a piecewise continuous function $h(t)$ 
\begin{equation}
g(x)=\int_a^b K(x,t)h(t)dt,
\end{equation}
can be expanded in a series in the eigenfunctions of $K(x,t)$
\begin{equation}
g(x)=\sum_{i=1}^\infty g_i\varphi_i(x),\ \ \ g_i=(g,\varphi_i)=\frac{(h,
\varphi_i)}{\lambda_i};
\end{equation}
this series converges uniformly and absolutely.\hfill$\Box$
\end{theorem}
  
Setting now $h(y)=K(y,t)$, for the 'iterated kernel'
\begin{equation}
K^{(2)}(x,t)=\int_a^bK(x,y)K(y,t)dy
\end{equation}
we then have the expansion
\begin{equation}
K^{(2)}(x,t)=\sum_{i=1}^\infty\frac{\varphi_i(x)}{\lambda_i}\int_a^bK(y,
t)\varphi_i(y)dy
\end{equation}
or
\begin{equation}
K^{(2)}(x,t)=\sum_{i=1}^\infty\frac{\varphi_i(x)\varphi_i(t)}{\lambda_i^2}
\end{equation}
In the same way, the subsequent iterated kernels
\begin{equation}
K^{(n)}(x,t)=\int_a^bK^{(n-1)}(x,y)K(y,t)dy
\end{equation}
admit the expansions
\begin{equation}
K^{(n)}(x,t)=\sum_{i=1}^\infty\frac{\varphi_i(x)\varphi_i(t)}{\lambda_i^n} \ 
\ (n=2,3,...),
\label{expansion}
\end{equation}
all of which converge absolutely and uniformly both in $x$ and in $t$, and 
uniformly in $x$ and $t$ together.

However, such an expansion for the full kernel,
\begin{equation}
K(x,t)=\sum_{i=1}^\infty\frac{\varphi_i(x)\varphi_i(t)}{\lambda_i},
\label{eigenvalueexpansion}
\end{equation}
does not always exist.

We can now derive the formula for the solution of the inhomogeneous integral equation (\ref{basicinteq}) for a symmetric kernel $K(x,t)$ which admits the eigenfunction expansion (\ref{eigenvalueexpansion}). We assume initially that the parameter $\lambda$ is not equal to any of the eigenvalues $\lambda_i$. If the continuous function $u$ with the expansion coefficients $(u,\varphi_i)$ were a solution of the integral equation, then, by the expansion 
theorem applied to $\lambda u(t)$, the function $u(x)-f(x)=g(x)$ 
would be given by the uniformly and absolutely convergent series
\begin{equation}
g(x)=u(x)-f(x)=\sum_{i=1}^\infty c_i\varphi_i(x)=\lambda\int_a^bK(x,t)u(t)dt,
\end{equation}
where 
\begin{equation}
c_i=(g,\varphi_i)=\lambda\int_a^b\int_a^bK(x,t)\varphi_i(x)u(t)dxdt=
\frac{\lambda}{\lambda_i}(\varphi_i,u)=\frac{\lambda}{\lambda_i}(\varphi_i,f)+\frac{\lambda}{\lambda_i}(\varphi_i,g)
\end{equation}
from which it follows that
\begin{equation}
c_i=f_i\frac{\lambda}{\lambda_i-\lambda},\ \ \ (f_i=(\varphi_i,f)).
\end{equation}
We thus obtain the series expansion for $u$,
\begin{equation}
u(x)=f(x)+\lambda\sum_{i=1}^\infty\frac{f_i}{\lambda_i-\lambda}\varphi_i(x),
\end{equation}
which must represent the solution of (\ref{basicinteq}). This solution fails only if $\lambda=\lambda_i$ is an eigenvalue; it remains valid even in this case if $f(x)$ is orthogonal to all eigenfunctions $\varphi_i$ belonging to 
the eigenvalue $\lambda_i$.

\subsection{Neumann series and the reciprocal kernel}

The theorems of integral equations given above imply a prescription for actually computing solutions with arbitrary accuracy. It does not give the solutions, however, in an elegant closed form as in the previous section. 

We write the integral equation (\ref{basicinteq}), inserting in place of $u(t)$ in the integral the expression obtained from (\ref{basicinteq}), and iterate this procedure. With the aid of the iterated kernels we thus write (\ref{basicinteq}) in the form
$$\begin{array}{lll}
u(x) & \displaystyle=f(x)+\lambda\int_a^bK(x,t)f(t)dt & \displaystyle+\lambda^2\int_a^bK^{(2)}(x,t)u(t)dt\\
& \displaystyle=f(x)+\lambda\int_a^bK(x,t)f(t)dt & \displaystyle+\lambda^2\int_a^bK^{(2)}(x,t)f(t)dt\\
& & +\displaystyle\lambda^3\int_a^bK^{(3)}(x,t)u(t)dt\\
&=...& 
\end{array}$$
and see that the solution is given by the infinite series
\begin{equation}
u(x)=f(x)+\lambda\int_a^bK(x,t)f(t)dt+\lambda^2\int_a^bK^{(2)}(x,t)f(t)dt+...,
\label{neumann1}
\end{equation}
provided that this series converges uniformly. If, moreover, the series
\begin{equation}
{\cal K}(x,t)=K(x,t)+\lambda K^{(2)}(x,t)+\lambda^2K^{(3)}(x,t)+...
\label{neumann2}
\end{equation}
also converges uniformly then the solution of the integral equation (\ref{basicinteq}), 
\begin{equation}
f(x)=u(s)-\lambda\int_a^bK(x,t)u(t)dt,
\end{equation}
is represented by the {\em 'reciprocal integral equation'}
\begin{equation}
u(x)=f(x)+\lambda\int_a^b{\cal K}(x,t)f(t)dt.
\end{equation}
The function ${\cal K}(x,t)={\cal K}(x,t;\lambda)$ is therefore called the {\em reciprocal} or {\em resolvent} kernel.

Series (\ref{neumann1}) and (\ref{neumann2}) are known as {\em Neumann's series}. They converge uniformly if $|\lambda|$ is sufficiently small. Thus, for sufficiently small $|\lambda|$, the reciprocal kernel is an analytic 
function of $\lambda$. 

If the kernel $K(x,t)$ is symmetric we can find a remarkably simple form for the resolvent kernel. Assuming again $|\lambda|$ to be sufficiently small, we make use of the series expansion  (\ref{expansion}) for the symmetric kernels $K^{(2)}(x,t),K^{(3)}(x,t),...$ and take the sum of the geometric series occurring
 in (\ref{neumann2}); we immediately obtain
\begin{equation}
{\cal K}(x,t;\lambda)=K(x,t)+\lambda\sum_{i=1}^\infty\frac{\varphi_i(x)\varphi_i(t)}{\lambda_i(\lambda_i-\lambda)}.
\label{reciprocalkernel}
\end{equation}
We see that the series on the right converges for every value of $\lambda$ which is not an eigenvalue, and the convergence is uniform in $x$ and $t$.

Relation (\ref{reciprocalkernel}) which has been proved only for sufficiently small $|\lambda|$ provides the analytic continuation of the resolvent ${\cal K}(x,t;\lambda)$ into the entire complex $\lambda$-plane, with the 
eigenvalues $\lambda_i$ appearing as simple poles. Thus {\em the resolvent of a symmetric kernel is a meromorphic function of $\lambda$ which possesses simple poles at the eigenvalues of the integral equation}. Its residues at the
 poles $\lambda_i$ provide the eigenfunctions belonging to these eigenvalues.

\section{Equations of the first kind}
\label{firstkindinteq}

Before trying to solve Fredholm integral equations of the first kind (\ref{firsttype}) we may first ask whether a solution does exist at all. We emphasise this point, since the theory is rather restrictive for the existence of solutions if compared to that of Fredholm equations of the second kind.

The difficulty of finding a solution for the Fredholm integral equation (\ref{firsttype}) stems from the fact that the integration operation over the sought solution $u(t)$, is a smoothing process, especially, when combined with a nicely behaved kernel $K(x,t)$. This means, for example, that if the solution $u(x)$ is piecewise continuous, then the above integration operation on the r.h.s.\ of (\ref{firsttype}), with a continuous kernel $K(x,t)$, would result in a smoother, i.e., continuous output $f(x)$ of (\ref{firsttype}), as the given function at hand. So, if we are given a continuous function $f(x)$, we cannot, in general guarantee an answer in the search for a solution $u(x)$ among the class of continuous functions! In other words, if we look at the r.h.s.\ of (\ref{firsttype}) as an integral transform of piecewise continuous functions, then this transform maps such a class of functions to a more restrictive one, in this case continuous functions. Indeed, for more smooth kernels, i.e., if $K(x,t)$ is differentiable, the class of piecewise or even integrable functions $u(t)$ is mapped into differentiable functions $f(x)$. Hence, for a continuous kernel $K(x,t)$ (in both $x$ and $t$) and a continuous output $f(x)$, the integral equation of the first kind (\ref{firsttype}) cannot, in general, be solved by a continuous function $u(t)$. Of course, if $K(x,t)$ is not very regular, then it is possible that this irregularity is combined with the smoothness of the integration operation and a continuous solution $u(t)$, to produce a continuous output $f(x)$.

\begin{theorem}
 For a continuous real and symmetric kernel, and a continuous $f(x)$, a solution to (\ref{firsttype}) exists only if the given function $f(x)$ can be expressed in a series of the eigenfunctions of the kernel $K(x,t)$, i.e., only if
\begin{equation}
 f(x)=\sum_{k=1}^\infty a_k\varphi_k(x),\ \ a_k=\int_a^bf(x)\varphi_k(x)dx,
\end{equation}
 where we have used the orthonormal set of eigenfunctions $\left\{\varphi_k(x)\right\}$ on $[a,b]$.\hfill$\Box$
\end{theorem}

With the condition stated in this theorem, the solution of (\ref{firsttype}) takes the form
\begin{equation}
u(x)=\sum_{k=1}^\infty\lambda_k a_k \varphi_k(x).
\label{seriessol1}
\end{equation}

We may note here that, although the existence of the (continuous) solution of (\ref{firsttype}) in the form of the series (\ref{seriessol1}) is guaranteed, it is by no means a unique solution. This is the case, since if we add to the series in (\ref{seriessol1}) a function $\psi(x)$ that is orthogonal to the kernel $K(x,t)$, i.e.,
\begin{equation}
 \int_a^b K(x,t)\psi(t)dt=0,
\end{equation}
and substitute in (\ref{firsttype}), we obtain the same output $f(x)$. So, for a unique solution $u(x)$ in (\ref{seriessol1}), we must insist that there are no functions $\psi(x)$ that are orthogonal to the symmetric kernel $K(x,t)$.   

The general case is covered by a theorem of Picard \cite{picard} which gives necessary and sufficient conditions for the existence of a square integrable solution $u(t)$ of an equation of the first kind with an arbitrary (possibly unsymmetric) kernel.

It is also useful to remark that Fredholm integral equations of the first kind can easily show the characteristics of ill-posedness defined in Section \ref{illposedprob}. We have just established a unique solution for Fredholm equations of the first kind for the specific case of symmetric kernels. What remains, for the present discussion, is the question whether the third quality, the stability of the solution of the problem, can also be established. Unfortunately, from the solution $u(x)$ in (\ref{seriessol1}) we can show that the problem is not stable. When perturbing the given data $f(x)$ by a small $\delta f(x)$ we expect the solution $u(x)$, in its series representation (\ref{seriessol1}), to be perturbed by a series representation of $\delta f(x)$ (or a constant multiple of it). However, one finds a magnification, i.e., a much larger corresponding change $\delta u$ in $u(x)$. This is due to the eigenvalues $\lambda_k$ factor in (\ref{seriessol1}), where they are increasing. If we write (\ref{seriessol1}), using the explicit form for $a_k$, we have
\begin{equation}
 u(x)=\sum_{k=1}^\infty\lambda_k\varphi_k(x)\int_a^bf(y)\varphi_k(y)dy.
\end{equation}
Now, if we perturb $f(x)$ by $\delta f(x)$ we get
\begin{equation}
 \begin{array}{lcl}
\displaystyle  u(x)+\delta u(x)&=&\displaystyle\sum_{k=1}^\infty\lambda_k\varphi_k(x)\int_a^b\left[f(y)+\delta f(y)\right]\varphi_k(y)dy\\
\\
&=&\displaystyle\sum_{k=1}^\infty\lambda_k\varphi_k(x)\int_a^bf(y)\varphi_k(y)dy+\sum_{k=1}^\infty\lambda_k\varphi_k(x)\int_a^b\delta f(y)\varphi_k(y)dy,
 \end{array}
\end{equation}
\begin{equation}
 \delta u(x)=\sum_{k=1}^\infty\lambda_k\epsilon_k\varphi_k(x),\ \ \epsilon_k=\int_a^b\delta f(y)\varphi_k(y)dy,
\label{perturbationmag}
\end{equation}
where $\epsilon_k$ is the above expansion coefficient of the small perturbation $\delta f(x)$. In (\ref{perturbationmag}) we see that the expansion coefficients corresponding to the perturbation $\delta u$ are magnified by a multiplicative factor $\lambda_k$, increasing with $k$, which clearly will cause $\delta u(x)$ to be a larger change compared to the given small change $\delta f(x)$. Hence, it seems that the solution $u(x)$ is not stable in the Fredholm equation of the first kind (\ref{firsttype}). This possible ill-posedness adds to other difficulties for the existence of solutions for problems with more general kernels. It is no surprise that Fredholm integral equations of the first kind are in the forefront with regard to the research priority in the field of integral equations.

\clearpage{\thispagestyle{empty}\cleardoublepage}

\chapter{Green's function}
\label{app:green}

Green's functions provide a useful tool for dealing with source terms in differential equations. They are named after the British mathematician George Green, who first developed the concept in the 1830s. In this short overview we will present the general theory concerning the ordinary differential equations. The theory for partial differential equations is similar and it will not be treated here. Instead, some examples of the Green's function for the Laplace equation are presented. Finally, the Neumann Green's function for the unit disc is discussed in more detail. An elaborate theory of Green's functions can be found for example in \cite{cour,wyld}.

Throughout the present work, the method of Green's function was used to transform the partial differential equation governing the EIT into an integral equation (Chapters \ref{directproblem} and \ref{reconstructions}). Furthermore, the Neumann Green's function is needed in Chapter \ref{reconstructions} to solve these equations for the particular case of the unit disc.

\section{General theory}

Let us consider a linear homogeneous self-adjoint differential expression of second order 
\begin{equation}
Lu=\frac{d}{dx}\left[p\frac{du}{dx}\right]-qu,
\label{lindiffeq}
\end{equation}
for the function $u(x)$ in the fundamental domain $\Omega$: $a\leq x\leq b$, where $p$, $dp/dx$ and $q$ are continuous functions of $x$ and $p>0$. The associated non-homogeneous differential equation is of the form
\begin{equation}
Lu=\varphi(x),
\label{lop}
\end{equation}
where $\varphi(x)$ denotes a piecewise continuous function defined in $\Omega$. We are concerned with the boundary value problem: {\em Find a solution of Eq.(\ref{lop}) which satisfies given homogeneous boundary conditions at the boundary of $\Omega$}.

For $L$ we have the Green's theorem identity:
\begin{equation}
 \int_a^b\, dx\left[vLu-uLv\right]=\left[p\left(v\frac{du}{dx}-u\frac{dv}{dx}\right)\right]_a^b.
\label{greenth}
\end{equation}
In order to solve Eq.(\ref{lop}) for a general source $\varphi(x)$ we introduce the so called Green's function $G(x,x')$, which is the solution for a point source at $x'$:
\begin{equation}
 LG(x,x')=\delta(x-x'),
\label{greendef}
\end{equation}
$\delta(x-x')$ being the Dirac delta function. We then use the linear character of (\ref{lindiffeq}) to find the solution for a source $\varphi(x)$ by superposition:
\begin{equation}
 u(x)=\int_a^b\, dx'G(x,x')\varphi(x').
\end{equation}
 
In order to elaborate on the previous remarks, with suitable attention to the boundary conditions, let us suppose that the solution of (\ref{greendef}) can be found. First, let us apply Green's theorem (\ref{greenth}) to $u$ and $v=G$:
\begin{equation}
\begin{array}{l}
 \displaystyle \int_a^b\, dx\left[G(x,x')Lu(x)-u(x)LG(x,x')\right]\\
\\
\displaystyle\hspace{2cm} =\left[p(x)\left(G(x,x')\frac{du(x)}{dx}-u(x)\frac{d}{dx}G(x,x')\right)\right]_a^b\\
\\
\displaystyle\hspace{2cm} =\int_a^b\, dxG(x,x')\varphi(x)-u(x'),
\end{array}
\end{equation}
where we have used (\ref{lop}), (\ref{greendef}) and the integral properties of the delta function. Interchanging $x$ and $x'$ we find
\begin{equation}
 u(x)=\int_a^b\, dx'G(x',x)\varphi(x')-\left[p(x')\left(G(x',x)\frac{du(x')}{dx'}-u(x')\frac{d}{dx'}G(x',x)\right)\right]_a^b.
\label{genarlsolution}
\end{equation}
We can now consider several possibilities for the boundary conditions on $u(x)$ at $x=a,b$.

\subsubsection*{1. $u(a)$ and $u(b)$ given.}

For this case choose as boundary conditions on $G(x',x)$
\begin{equation}
 G(a,x)=G(b,x)=0.
\end{equation}
This has the desirable advantage of eliminating from Eq.(\ref{genarlsolution}) the unknown quantities $du/dx'$ at $x'=a,b$. In fact (\ref{genarlsolution}) reduces to
\begin{equation}
 u(x)=\int_a^b\, dx'G(x',x)\varphi(x')+\left[p(x')u(x')\frac{d}{dx'}G(x',x)\right]_a^b.
\end{equation}
This is the solution of Eq.(\ref{lop}) in terms of known quantities.

\subsubsection*{2. $u(a)$ and $u'(b)$ given.}

For this case choose as boundary conditions on $G(x',x)$
\begin{equation}
 G(a,x)=0,\ \left.\frac{d}{dx'}G(x',x)\right|_{x'=b}=0.
\end{equation}
Again the unknown quantities, in this case $u(b)$ and $du/dx$ at $x=a$ are eliminated form Eq.(\ref{genarlsolution}) and we obtain the solution of (\ref{lop}) in terms of known quantities:
\begin{equation}
 u(x)=\int_a^b\, dx'G(x',x)\varphi(x')-p(a)u(a)\left.\frac{d}{dx'}G(x',x)\right|_{x'=a}-p(b)G(b,x)\left.\frac{du}{dx'}\right|_{x'=b}.
\end{equation}

\subsubsection*{3. $Au(a)+Bu'(a)=X$ and $Cu(b)+Du'(b)=Y$ given.}

Choose
\begin{equation}
 \begin{array}{l}
\displaystyle  AG(a,x)+B\left.\frac{d}{dx}G(x',x)\right|_{x'=a}=0,\\
\\
\displaystyle  CG(b,x)+D\left.\frac{d}{dx}G(x',x)\right|_{x'=b}=0.
 \end{array}
\label{genralboundcond}
\end{equation}
The quantity needed in (\ref{genarlsolution}) at $x'=a$ is then
\begin{equation}
\begin{array}{l}
\displaystyle G(a,x)\left.\frac{du(x')}{dx'}\right|_{x'=a}-u(a)\left.\frac{d}{dx'}G(x',x)\right|_{x'=a}\\
\\
\displaystyle\hspace{2cm} =G(a,x)\left.\frac{du(x')}{dx'}\right|_{x'=a}-u(a)\left[-\frac{A}{B}G(a,x)\right]\\
\\
\displaystyle\hspace{2cm} =\frac{G(a,x)}{B}\left[Au(a)+B\left.\frac{du(x')}{dx'}\right|_{x'=a}\right]\\
\\
\displaystyle\hspace{2cm} =\frac{G(a,x)}{B}X\\
\\
\displaystyle\hspace{2cm} =-\frac{1}{A}\left.\frac{d}{dx'}G(x',x)\right|_{x'=a}X,
\end{array}
\end{equation}
which is a known quantity. A similar manipulation shows that the quantity needed at $x'=b$ in (\ref{genarlsolution}) can be expressed in terms of the given quantity $Y$.

In each case we can choose the boundary conditions on $G$ so as to eliminate unknown values of $u$ and $u'$ at the boundaries. In this way we obtain a solution of the inhomogeneous differential equation (\ref{lop}) once we find the Green's function $G(x',x)$.

We can prove that the Green's function is symmetric in its two arguments. To see this, apply the Green's theorem (\ref{greenth}) to $u(x)=G(x,x')$ and $v(x)=G(x,x'')$. Using Eq.(\ref{greendef}) we find
\begin{equation}
 G(x',x'')-G(x'',x')=\left[p(x)\left(G(x,x'')\frac{d}{dx}G(x,x')-G(x,x')\frac{d}{dx}G(x,x'')\right)\right]_a^b.
\label{symproof}
\end{equation}
For the general case with boundary conditions (\ref{genralboundcond}) the r.h.s.\ of Eq.(\ref{symproof}) vanishes and we find
\begin{equation}
 G(x',x'')=G(x'',x').
\end{equation}
 
Let us now return to the differential equation (\ref{greendef}) for $G(x,x')$:
\begin{equation}
\frac{d}{dx}\left[p(x)\frac{d}{dx}G(x,x')\right]-q(x)G(x,x')=\delta(x-x').
\label{diffeqG}
\end{equation}
For $x\neq x'$ this reduces to the homogeneous equation
\begin{equation}
\frac{d}{dx}\left[p(x)\frac{d}{dx}G(x,x')\right]-q(x)G(x,x')=0,\ \ x\neq x',
\label{homogeneouseq}
\end{equation}
so we can use whatever techniques are available for homogeneous equations in solving for $G(x,x')$ in the two regions $x>x'$ and $x<x'$. At $x=x'$ we must match the two solutions so obtained. The delta function $\delta(x-x')$ on the r.h.s.\ of (\ref{diffeqG}) is to be interpreted as due to a discontinuity in $dG(x,x')/dx$ at $x=x'$. In fact, integrating (\ref{diffeqG}) over an interval $x=x'-\varepsilon$ to $x=x'+\varepsilon$ of infinitesimal length about $x=x'$ and assuming $q(x)G(x,x')$ is finite in this region so that
\begin{equation}
 \int_{x'-\varepsilon}^{x'+\varepsilon}dx\, q(x)G(x,x')\underset{\varepsilon\rightarrow0} {\longrightarrow} 0,
\end{equation}
we find
\begin{equation}
 \left.\frac{d}{dx}G(x,x')\right|_{x'-\varepsilon}^{x'+\varepsilon}=\frac{1}{p(x')},
\label{discontinuity}
\end{equation}
provided $p(x)$ is continuous at $x=x'$. On the other hand, we must take $G(x,x')$ to be continuous at $x=x'$:
\begin{equation}
 G(x'+\varepsilon,x')=G(x'-\varepsilon,x').
\label{continuity}
\end{equation}
If there were a discontinuity in $G(x,x')$ itself, the first derivative $dG(x,x')/dx$ would contain a term proportional to $\delta(x-x')$ and the second derivative $d^2G(x,x')/dx^2$ a term proportional to $d\delta(x-x')/dx$, in contradiction to the defining Eq.(\ref{diffeqG}).

The homogeneous Eq.(\ref{homogeneouseq}) in the two regions $x<x'$ and $x>x'$ together with the two conditions (\ref{discontinuity}) and (\ref{continuity}) at $x=x'$ and the boundary conditions at $x=a$ and $x=b$ discussed previously, completely determine the Green's function, the four conditions fixing the two pairs of integration constants obtained on solving the differential equation in the two regions.

In the case of partial differential equations with homogeneous boundary conditions, Green's function can again be introduced as the kernel of an equivalent integral equation with the same properties as in the case of the ordinary differential equations \cite{cour}. We will not repeat the theory, but rather give some examples. 

\section{Examples of Green's function for partial differential equations}

Let us consider the Laplace equation for a two-dimensional domain $\Omega$
\begin{equation}
 \Delta\phi(\bm x)=0,\ \mbox{ with }\ \Delta=\frac{\partial^2}{\partial x^2}+\frac{\partial^2}{\partial y^2},
\end{equation}
the Laplace operator. In the following we will consider a few special cases:
\begin{itemize}
\item Choose $G_0$ to be the fundamental (free) Green's function, satisfying:
\begin{equation}
\Delta_{\bm x}G_0({\bm x},{\bm x}')=-\delta({\bm x}-{\bm x}')\ \ \mbox{for}\ {\bm x},{\bm x}'\in\Omega.
\end{equation}
This function is independent of the shape of $\Omega$ and for a two-dimensional space it is:
\begin{equation}
G_0({\bm x},{\bm x}')=-\frac{1}{2\pi}\log\left|{\bm x}-{\bm x}'\right|.
\end{equation}
\item Define $G_D$ to be the Dirichlet Green's function given by
\begin{equation}
\left\{
\begin{array}{ll}
\displaystyle \Delta_{\bm x}G_D({\bm x},{\bm x}')=-\delta({\bm x}-{\bm x}')&\mbox{for}\ {\bm x},{\bm x}'\in\Omega \\
\displaystyle G_D({\bm x},{\bm x}')=0 & \mbox{for}\ {\bm x}\in\partial\Omega\ \mbox{and}\ {\bm x}'\in\Omega
\end{array}
\right.
\end{equation}
In contrast to $G_0$, $G_D$ depends on the shape of $\Omega$ and the difference between them is a harmonic term. For a disc of radius
$R$, $G_D$ is given by the formula
\begin{equation}
G_D(r,\theta;r',\theta')=-\frac{1}{4\pi}\log\left[\frac{r^2+{r'}^2-2rr'
\cos(\theta-\theta')}{R^2+\frac{r^2}{R^2}{r'}^2-2rr'\cos(\theta-\theta')}\right],
\end{equation}
where $(r,\theta)$ and $(r',\theta')$ are the polar coordinates associated to ${\bm x}$ and ${\bm x}'$ respectively. 
\item  Now consider $G_N$ to be the Neumann Green's function which obeys the conditions:
\begin{equation}
\left\{
\begin{array}{ll}
\displaystyle \Delta_{\bm x}G_N({\bm x},{\bm x}')=-\delta({\bm x}-{\bm x}')+\frac{1}{|\Omega|}&\mbox{for}\ {\bm x},{\bm x}'\in\Omega \\
\displaystyle \frac{\partial G_N}{\partial n}({\bm x},{\bm x}')=0 & \mbox{for}\ {\bm x}\in\partial\Omega\ \mbox{and}\ {\bm x}'\in\Omega
\end{array}
\right.
\label{neumanncond}
\end{equation}
where $|\Omega|$ is the volume of the domain $\Omega$. Like $G_D$, $G_N$ also depends on the shape of the domain $\Omega$ and differs from $G_0$ by a harmonic term. For a disc of radius $R$, $G_N$ reads:
\begin{equation}
\begin{array}{ll}
G_N(r,\theta;r',\theta')=&-\displaystyle\frac{1}{4\pi}\log\left[r^2+{r'}^2-2rr'\cos(\theta-\theta')\right]\\
\\
&-\displaystyle\frac{1}{4\pi}\log\left[R^2+\frac{r^2}{R^2}{r'}^2-2rr'\cos(\theta-\theta')\right],
\end{array}
\label{neumannprob}
\end{equation}
where by $(r,\theta)$ and $(r',\theta')$ we denoted again the polar coordinates associated to ${\bm x}$ and ${\bm x}'$ respectively.  
\end{itemize}

\section{Neumann Green's function for the unit disc}
\label{app:neumann}

The eigenvalue problem
\begin{equation}
\left\{\begin{array}{l}
\Delta v+\lambda v=0,\\
\displaystyle \left.\frac{\partial v}{\partial r}\right|_{r=1}=0,
\end{array}\right.\ 
\label{eigendisc}
\end{equation}
defines the eigenfunctions $v_n$ of the Neumann Green's function, Eq.(\ref{neumannprob}), assumed to have the eigenvalue expansion:
\begin{equation}
G_N({\bm x},{\bm x}')=\sum_n\frac{v_n({\bm x})v_n({\bm x}')}{\lambda_n}.
\end{equation}

The Laplacian in polar coordinates reads:
\begin{equation}
\Delta=\frac{1}{r}\left[\frac{\partial}{\partial r}\left(r\frac{\partial}{\partial r}\right)+\frac{\partial}{\partial\theta}\left(\frac{1}{r}\frac{\partial}{\partial\theta}\right)\right].
\end{equation}
Let us search for solutions of (\ref{eigendisc}) of the form $v(r,\theta)=f(r)u(\theta)$:
\begin{equation}
\left\{\begin{array}{l}
\displaystyle\frac{d^2u}{d\theta^2}+l^2u=0\\
\\
\displaystyle\frac{1}{r}\frac{df}{dr}+\frac{d^2f}{dr^2}+\left(\lambda-\frac{l^2}{r^2}\right)f=0,\ \ \left.\frac{df}{dr}\right|_{r=1}=0
\end{array}\right.; \ \ l\in\N_+.
\label{eigenprobdisc}
\end{equation} 

The solutions of the first equation are just the well-known eigenfunctions for the unit circle
\begin{equation}
u_l^j(\theta)=\frac{1}{\sqrt{\pi(1+\delta_{l0})}}\left\{\begin{array}{lcl}
\sin l\theta &\mbox{for}& j=1\\
\cos l\theta &\mbox{for}& j=2
\end{array}\right.; \ \ l\in\N_+.
\end{equation} 

To solve the second equation in (\ref{eigenprobdisc}), we make the change of variable:
\begin{equation}
kr=\rho\ \ \mbox{with} \ \ \lambda=k^2,
\end{equation}
and find Bessel's equation
\begin{equation}
\frac{1}{\rho}f'(\rho)+f''(\rho)+\left(1-\frac{l^2}{\rho^2}\right)f(\rho)=0,
\end{equation}
which has the regular solution
\begin{equation}
f(\rho)=J_l(\rho)=J_l(kr),
\end{equation}
where $J_l$ are the Bessel functions.
Imposing the boundary condition (see Eq.(\ref{neumanncond})), one finds
\begin{equation}
kJ'_l(k)=0
\end{equation}
and thus $k$ are the zeros of the first derivative of $J_l$, usually denoted $j^*_{lm}$. The eigenfunctions corresponding to the eigenvalues $\lambda=(j_{lm}^*)^2$ will then be 
\begin{equation}
f_l(r)=J_l(j_{lm}^*r).
\end{equation}
One still has to normalise them. Recalling that
\begin{equation}
\int_0^1J_l^2(j_{lm}^*r)dr=\frac{1}{2}\left(1-\frac{l^2}{j_{lm}^{*2}}\right)J_l^2(j_{lm}^*),
\end{equation}
the normalisation constants become
\begin{equation}
C_{lm}=\sqrt{\frac{2}{1-l^2/j_{lm}^{*2}}}\ \frac{1}{|J_l(j_{lm}^*)|}.
\end{equation}

Finally, the eigenfunctions corresponding to the eigenvalues $\lambda=(j_{lm}^*)^2$, defined by the boundary eigenvalue problem (\ref{eigendisc}), are:
\begin{equation}
v_{lm}^j(r,\theta)=\frac{C_{lm}}{\sqrt{\pi(1+\delta_{l0})}}J_l(j_{lm}^*r)
\left\{\begin{array}{lcl}
\sin l\theta &\mbox{for}& j=1\\
\cos l\theta &\mbox{for}& j=2
\end{array}\right.; \ \ l\in\N_+,
\end{equation} 
or shortly
\begin{equation}
v_{lm}^j(r,\theta)=C_{lm}J_l(j_{lm}^*r)u_l^j(\theta),\ \ l\in\N_+,
\end{equation} 
and the Neumann Green's function is
\begin{equation}
G_N(r,\theta;r',\theta')=\sum_{l,m=1}^\infty\sum_{j=1}^2 \frac{v_{lm}^j(r,\theta)v_{lm}^j(r',\theta')}{\lambda_{lm}}.
\label{seriesdisc}
\end{equation}
On the boundary, i.e., on the unit circle, one has
\begin{equation}
G_N(1,\theta;1,\theta')=\sum_{l=1}^\infty\sum_{j=1}^2u_l^j(\theta)u_l^j(\theta')\sum_{m=1}^\infty\frac{2}{j_{lm}^{*2}-l^2}\, .
\end{equation}
Recalling that:
\begin{equation}
\sum_{m=1}^\infty \frac{2}{j_{lm}^{*2}-l^2}=\frac{1}{l},
\end{equation}
one finds:
\begin{equation}
G_N(1,\theta;1,\theta')=\sum_{l=1}^\infty\sum_{j=1}^2\frac{1}{l}u_l^j(\theta)u_l^j(\theta').
\end{equation}

\clearpage{\thispagestyle{empty}\cleardoublepage}
\chapter{SVD of an integral operator}
\label{appSVD}

One of the most fruitful tools in the theory of linear inverse problems is the singular value decomposition of a matrix and its extension to certain classes of linear operators. As we have seen in Chapter \ref{regularisationmethods}, SVD is basic both for understanding the ill-posedness of inverse problems and for describing the effect of the regularisation methods. Here, we will restrict the discussion to the SVD of integral operators with square-integrable kernel, needed in the EIT application of Section \ref{singelrec}.

Let us define an integral operator of the following form:
\begin{equation}
(Af)({\bm x})=\int_{\Omega}K({\bm x},{\bm x}')f({\bm x}')d{\bm x}',\ \ {\bm x}\in{\Omega}'
\label{intimagprob}
\end{equation}
where ${\Omega}$ and ${\Omega}'$ are the object and the image domain respectively. 

We assume, for simplicity, that both the object and the image are square-integrable functions of the space variables, so that we have ${\cal X}=L^2({\Omega})$ and ${\cal Y}=L^2({\Omega}')$. Then Eq.(\ref{intimagprob}) defines an operator from $L^2({\Omega})$ into $L^2({\Omega}')$. This operator is continuous if the integral kernel is square-integrable, i.e.
\begin{equation}
||K||^2=\int_{{\Omega}'}d{\bm x}\int_{\Omega}d{\bm x}'|K({\bm x},{\bm x}')|^2<\infty.
\label{knorm}
\end{equation}
An integral operator satisfying this condition is usually called an integral operator of the {\em Hilbert-Schmidt class}.

In order to show that this operator is continuous we apply the Schwartz inequality to the r.h.s.\ of Eq.(\ref{intimagprob}) which, for fixed ${\bm x}$, results in the scalar product of two square integrable functions. It follows
\begin{equation}
|(Af)({\bm x})|^2\leq\left(\int_{\Omega}|K({\bm x},{\bm x}')|^2d{\bm x}'\right)
\left(\int_{\Omega}|f({\bm x}')|^2d{\bm x}'\right).
\end{equation}
If we integrate both sides of this inequality with respect to ${\bm x}$, and take the square root of the result, we get
\begin{equation}
||Af||_{\cal Y}\leq||K||\cdot||f||_{\cal X},
\label{afnorm}
\end{equation}
i.e., the operator is bounded. This property implies the continuity of the operator.

The adjoint $A^*$ of the operator $A$ is given by
\begin{equation}
(A^*g)({\bm x}')=\int_{{\Omega}'}K^*({\bm x},{\bm x}')g({\bm x})d{\bm x},\ \ {\bm x}'\in{\Omega}.
\end{equation}
Then we can introduce the operators ${\bar A}=A^*A$ and ${\tilde A}=AA^*$. Both are integral operators with integral kernels given by
\begin{equation}
{\bar A}:\ \ {\bar K}({\bm x},{\bm x}')=\int_{{\Omega}'}K^*({\bm x}'',{\bm x})K({\bm x}'',{\bm x}')d{\bm x}'';\ \ {\bm x},{\bm x}'\in{\Omega},
\label{kbar}
\end{equation}
\begin{equation}
{\tilde A}:\ \ {\tilde K}({\bm x},{\bm x}')=\int_{{\Omega}}K({\bm x},{\bm x}'')K^*({\bm x}',{\bm x}'')d{\bm x}'';\ \ {\bm x},{\bm x}'\in{\Omega}'.
\label{ktilde}
\end{equation}

The integral operators ${\bar A}$, ${\tilde A}$ have the following properties:
\begin{itemize}
\item Both operators are self-adjoint, i.e., for any pair of functions $f$, $h$ in ${\cal X}$ and any pair of functions $g$, $w$ in ${\cal Y}$
\begin{equation}
({\bar A}f,h)_{\cal X}=(f,{\bar A}h)_{\cal X}, \ \ ({\tilde A}g,w)_{\cal Y}=(g,{\tilde A}w)_{\cal Y}.
\end{equation}
This property is a consequence of the following relations
\begin{equation}
{\bar K}^*({\bm x},{\bm x}')={\bar K}({\bm x}',{\bm x}), \ \ {\tilde K}^*({\bm x},{\bm x}')={\tilde K}({\bm x}',{\bm x})
\end{equation}
which can be easily checked by means of Eqs.(\ref{kbar}) and (\ref{ktilde});
\item Both operators are of the Hilbert-Schmidt class because their integral kernels are square-integrable (the proof of this result is similar to the proof of the inequality (\ref{afnorm}); the starting point is the application of the Schwartz inequality to the r.h.s.\ of Eqs.(\ref{kbar}) and (\ref{ktilde}));
\item Both operators are positive semi-definite
\begin{equation}
({\bar A}f,f)_{\cal X}\geq0, \ \ ({\tilde A}g,g)_{\cal Y}\geq0.
\end{equation}
\end{itemize}
\begin{remark}
We sketch the proof of this property in the case of ${\bar A}$. Indeed, from Eq.(\ref{kbar}), by means of an exchange of the integration order we have
\begin{eqnarray}
({\bar A}f,f)_{\cal X}&=&\int_{\Omega}\left(\int_{\Omega}{\bar K}({\bm x},{\bm x}')f({\bm x}')d{\bm x}'\right)f^*({\bm x})d{\bm x}\nonumber\\
&=&\int_{{\Omega}'}\left|\int_{\Omega}K({\bm x}'',{\bm x})f({\bm x})d{\bm x}\right|^2d{\bm x}''\geq0.
\end{eqnarray}
A similar proof applies to the case of ${\tilde A}$.\hfill$\Box$
\end{remark}

According to the Hilbert-Schmidt theory \cite{mikh}, an integral operator with a symmetric and square-integrable kernel has real eigenvalues with finite multiplicity. Moreover the eigenfunctions associated with different eigenvalues are orthogonal. The eigenvalues form, in general, a countable set with zero as accumulation point. 

If ${\cal K}({\bm x},{\bm x}')$ is a symmetric and square-integrable kernel, let $\lambda_1,\lambda_2,\lambda_3,...$ be the sequence of the eigenvalues of the corresponding integral operator, ordered in such a way that $|\lambda_1|\geq|\lambda_2|\geq|\lambda_3|\geq...$, each eigenvalue being counted as many times as its multiplicity. Moreover, let $v_1({\bm x}), v_2({\bm x}), v_3({\bm x}),...$ be the sequence of the eigenfunctions associated with these eigenvalues. They constitute an orthonormal set of square-integrable functions. Then the basic result of the Hilbert-Schmidt theory is the following {\em spectral representation} of the kernel ${\cal K}({\bm x},{\bm x}')$
\begin{equation}
{\cal K}({\bm x},{\bm x}')=\sum_{k=1}^\infty \lambda_kv_k({\bm x})v^*_k({\bm x}'),
\label{kexpansion}
\end{equation}
the series being convergent in the sense of the $L^2$-norm. Accordingly, the eigenvalues $\lambda_k$ satisfy the following condition
\begin{equation}
\sum_{k=1}^\infty\lambda_k^2<\infty.
\end{equation}

The result stated above apply to the operators ${\bar A}$ and ${\tilde A}$, whose non-zero eigenvalues are positive because they are positive semi-definite operators. One can also show that the operators ${\bar A}$ and ${\tilde A}$ have the same eigenvalues with the same multiplicity.

Let us denote by $\sigma_k^2$ the positive eigenvalues of ${\bar A}$, with the ordering $\sigma_1^2\geq\sigma_2^2\geq...\geq\sigma_k^2\geq...$ and $\sigma_k^2\rightarrow0$ for $k\rightarrow\infty$. Again each eigenvalue has been counted as many times as its multiplicity, which is certainly finite as follows from the Hilbert-Schmidt theory.

A normalised eigenfunction $v_k$ is associated to each eigenvalue $\sigma_k^2$ and these eigenfunctions constitute an orthonormal system, i.e. $(v_k,v_j)_{\cal X}=\delta_{kj}$. Then, to each eigenfunction $v_k$ of ${\bar A}$ we can associate a function $u_k$ in ${\cal Y}$ defined by
\begin{equation}
u_k(\bm x)=\frac{1}{\sigma_k}(Av_k)(\bm x).
\label{uk}
\end{equation}
Using the relation ${\tilde A}A=A{\bar A}$, which can be easily proved by an exchange of integration order in the definition of the integral kernels, we obtain
\begin{equation}
({\tilde A}u_k)(\bm x)=\frac{1}{\sigma_k}(A{\bar A}v_k)(\bm x)=\sigma_k^2\left(\frac{1}{\sigma_k}(Av_k)(\bm x)\right)=\sigma_k^2u_k(\bm x)
\end{equation}
and also
\begin{eqnarray}
(u_k,u_j)_{\cal Y}&=&\frac{1}{\sigma_k\sigma_j}(Av_k,Av_j)_{\cal Y}\nonumber\\
&=&\frac{1}{\sigma_k\sigma_j}(v_k,{\bar A}v_j)_{\cal X}=\frac{\sigma_j}{\sigma_k}(v_k,v_j)_{\cal X}=\delta_{kj}.
\end{eqnarray}
Therefore all eigenvalues $\sigma_k^2$ of ${\bar A}$ are also eigenvalues of ${\tilde A}$ and the $u_k$ are the corresponding eigenfunctions, which constitute an orthonormal system in ${\cal Y}$. In order to show that in this way we have obtained all eigenvalues  and eigenfunctions of ${\tilde A}$ it is sufficient to repeat the same argument starting from ${\tilde A}$. If its eigenvalues and eigenfunctions are denoted by $\sigma_k$ and $u_k$ respectively, then we can show, by means of the relation ${\bar A}A^*=A^*{\tilde A}$, that the functions of ${\cal X}$ defined by
\begin{equation}
v_k(\bm x)=\frac{1}{\sigma_k}(A^*u_k)(\bm x)
\label{vk}
\end{equation}
are orthonormal eigenfunctions of ${\bar A}$ associated with the eigenvalues $\sigma_k^2$.

Eqs.(\ref{uk}) and (\ref{vk}) define the usual shifted eigenvalue problem, which we write now explicitly in terms of the integral operators
\begin{eqnarray}
\int_{\Omega}K({\bm x},{\bm x}')v_k({\bm x}')d{\bm x}'=\sigma_ku_k({\bm x}),\\
\int_{{\Omega}'}K^*({\bm x},{\bm x}')u_k({\bm x})d{\bm x}=\sigma_kv_k({\bm x}').
\end{eqnarray}
 
The proof of the SVD of the operators $A$ and $A^*$ requires the completeness of the Hilbert spaces ${\cal X}$ and ${\cal Y}$. We do not report this proof here (see, for instance, \cite{groet}). We only give the result, which takes the form:
\begin{equation}
(Af)(\bm x)=\sum_{k=1}^\infty\sigma_k(f,v_k)_{\cal X}u_k(\bm x)
\label{asvd}
\end{equation}
and also
\begin{equation}
(A^*f)(\bm x)=\sum_{k=1}^\infty\sigma_k(g,u_k)_{\cal Y}v_k(\bm x),
\end{equation}
the series being convergent with respect to the norms of ${\cal Y}$ and ${\cal X}$, respectively.

We also remark that the SVD of $A$ is equivalent to the following series expansion of the kernel $K({\bm x},{\bm x}')$
\begin{equation}
K({\bm x},{\bm x}')=\sum_{k=1}^\infty\sigma_ku_k({\bm x})v_k^*({\bm x}'),
\end{equation}
the series being convergent with respect to the $L^2$-norm. From this expansion (which is a generalisation of equation (\ref{kexpansion})), from the orthogonality of the singular functions and from the definition (\ref{knorm}) of $||K||^2$ it follows that
\begin{equation}
||K||^2=\sum_{k=1}^\infty \sigma_k^2
\label{knorm2}
\end{equation}
and therefore {\em the sum of squares of the singular values of a Hilbert-Schmidt integral operator is convergent}.

Using the SVD (\ref{asvd}) it is also possible to show that the norm of the integral operator $A$ is given by
\begin{equation}
||A||=\sigma_1.
\end{equation}
From this equation and (\ref{knorm2}), it follows that {\em the norm of the integral operator is never greater than the $L^2$-norm of the integral kernel; the two norms coincide if and only if the integral operator has rank one}.

\clearpage{\thispagestyle{empty}\cleardoublepage}

\chapter{Statistics}
\label{statistics}

In statistics we are interested in using a given sample of data to make inferences about a probabilistic model, e.g., to assess the model's validity or to determine the values of its parameters. There are two main approaches to statistical inference, which we may call frequentist and Bayesian. In frequentist statistics, probability is interpreted as the frequency of the outcome of a repeatable experiment. Note that in frequentist statistics one does not define a probability for a hypothesis or for a parameter.

Frequentist statistics provides the usual tools for reporting objectively the outcome of an experiment without needing to incorporate prior beliefs concerning the parameter being measured or the theory being tested.

In Bayesian statistics however, the interpretation of probability is more general and includes the notion of a ``degree of belief''. One can then speak of a probability distribution function (p.d.f.) for a parameter, which expresses one's state of knowledge about where its true value lies. Bayesian methods allow for a natural way to include additional information such as physical boundaries and subjective information; in fact they {\em require} as input the {\em prior} p.d.f.\ for the parameters, i.e., the degree of belief about the parameters' values before carrying out the measurement.

Bayesian techniques are often used to treat systematic uncertainties, where the subjective beliefs about, say, the accuracy of the measuring device may enter. Bayesian statistics also provides a useful framework for discussing the validity of different theoretical interpretations of the data.

For many inference problems, the frequentist and Bayesian approaches give the same numerical answers, even though they are based on fundamentally different interpretations of probability. For small data samples, however, and for measurements of a parameter near a physical boundary, the different approaches may yield different results, so we are forced to make a choice. For a discussion of Bayesian vs. non-Bayesian methods, see references written by a statistician \cite{efron}, by a physicist \cite{cousins}, or the more detailed comparison in Ref.\cite{stuart}.

In what follows we will concentrate on frequentist statistics and briefly discuss least-squares parameter estimation, confidence regions and the $\chi^2$ test of goodness-of-fit. In Chapters \ref{condensates}, \ref{V-Aanalysis} and \ref{V+Aanalysis} we aimed to determine values of some fundamental QCD parameters, the so called condensates, in a least-squares sense. The errors on these parameters were then defined through constant $\chi^2$ boundaries as CLs around the values found. It was also important to define the level of agreement between data and theory, for which we have made use of the $\chi^2$ test.

\section{Least-squares parameter estimation}
\label{leastsquaresmin}

Consider a situation in which $N$ data pairs, $x_i$, $y_i$, with $i=1,2,...,N$, are to be modeled by a functional relationship
\begin{equation}
 F({\bm x})=F({\bm x},{\bm\theta}),\ y_i=F(x_i,{\bm\theta}),
\end{equation}
where $\bm\theta$ is the vector of $M$ adjustable parameters $\theta_j$. The goal is to determine the ``best'' values of $\bm\theta$ that describe the true values ${\bm\theta}_0$.

Given a particular data set of $x_i$'s and $y_i$'s, we have the intuitive feeling that some parameters $\bm\theta$ are very unlikely (those for which the model function $F({\bm x})$ looks {\em nothing like} the data) while others may be very likely (those that closely resemble the data). How can we quantify this intuitive feeling? How can we select fitted parameters that are {\em most likely} to be correct? It is not meaningful to ask the question, ``What is the probability that a particular set of fitted parameters $\bm\theta$ is correct?'' The reason is that there is no statistical universe of models from which the parameters are drawn. There is just one model, the correct one, and a statistical universe of data sets that are drawn from it.

That being the case, we can, however, turn the question around, and ask, ``Given a particular set of parameters, what is the probability that this data set could have occurred?'' If the $y_i$'s take on continuous values, the probability will always be zero unless we add the phrase, ``...plus or minus some fixed $\Delta y$ on each data point''. So let's always take this phrase as understood. If the probability of obtaining the data set is infinitesimally small, then we can conclude that the parameters under consideration are {\em unlikely} to be right. Conversely, our intuition tells us that the data set should not be too improbable for the correct choice of parameters.

In other words, we identify the probability of the data given the parameters as the {\em likelihood} of the parameters given the data. This identification is entirely based on intuition. It has no formal mathematical basis in and of itself. Once we made this intuitive identification, however, it is only a small further step to decide to fit for the parameters $\bm\theta$ precisely by finding those values that {\em maximise} the likelihood defined in the above way.

Suppose that each data point $y_i$ has a measurement error $\sigma_i$ and that $y_i$ are independently random and distributed as a normal (Gaussian) distribution around the {\em true} model $F({\bm x})$. Then the probability $P$ that the data set is described by $F({\bm x})$ is given by 
\begin{equation}
-\ln P\propto \chi^2(\bm\theta)=\sum_{i=1}^N\left(\frac{y_i-F(x_i;\bm\theta)}{\sigma_i}\right)^2.
\label{defchi}
\end{equation}

Maximising the joint probability function, $P$, is equivalent to minimising $\chi^2$. This is achieved by differentiating Eq.(\ref{defchi}) for each $\theta_j$
\begin{equation}
 \frac{\partial\chi^2}{\partial\theta_j}=-2\sum_{i=1}^N\left(\frac{y_i-F(x_i;\bm\theta)}{\sigma_i^2}\right)\left(\frac{\partial F(x_i;{\bm\theta})}{\partial\theta_j}\right),
\label{parameq}
\end{equation}
 and equating to zero. Obtaining Eq.(\ref{parameq}) for each $j$ yields a set of $M$ equations, one for each unknown $\theta_j$. Minimising $\chi^2$ thus reduces to solving a system of $M$ equations.

The curvature matrix
\begin{equation}
 \alpha_{jk}=\frac{1}{2}\frac{\partial^2\chi^2}{\partial\theta_j\theta_k}=\sum_{i=1}^N\frac{1}{\sigma_i^2}\left[\frac{\partial F(x_i;{\bm\theta})}{\partial\theta_j}\frac{\partial F(x_i;{\bm\theta})}{\partial\theta_k}\right],
\end{equation}
defines the covariance matrix for the parameters:
\begin{equation}
 V=\alpha^{-1}.
\end{equation}
Once $\chi^2$ is minimised, the variances in the parameters $\bm\theta$ are approximated by the diagonal matrix elements of $V$:
\begin{equation}
 \sigma_{\theta_j}^2=V_{jj}.
\end{equation}

\section{Confidence regions}
\label{confreg}

The full probability distribution is a function defined on the $M$-dimensional space of parameters $\bm\theta$. A {\em confidence region} (or {\em confidence interval}) is just a region of that $M$-dimensional space that contains a certain percentage of the total probability distribution. When pointing to a confidence region one can say, e.g., ``this is a $99\%$ chance that the measured values $y_i$ are obtained from the model with the parameter values ${\bm \theta}$ lying within this region''.

It is worth emphasising that one has to determine both the {\em confidence level} ($99\%$ in the above example), and the shape of the confidence region. The only requirement is that the region does include the stated percentage of probability. Certain percentages are, however, customary in scientific usage: $68.27\%$ (the lowest confidence worthy of quoting) also called $1\sigma$CL, $90\%$, $95.45\%$ ($2\sigma$CL), $99\%$ and $99.73\%$ ($3\sigma$CL). Higher CL's are conventionally ``ninety-nine point nine ... nine''. As for shape, obviously one wants a region that is compact and reasonably centred on ${\bm\theta}_0$, since the whole purpose of a confidence limit is to inspire confidence in that value. In one dimension, the convention is to use a line segment centred on the measured value; in higher dimensions, ellipses or ellipsoids are most frequently used. 

\begin{figure}[h]
\centering
 \includegraphics[height=.6\textwidth,angle=0]{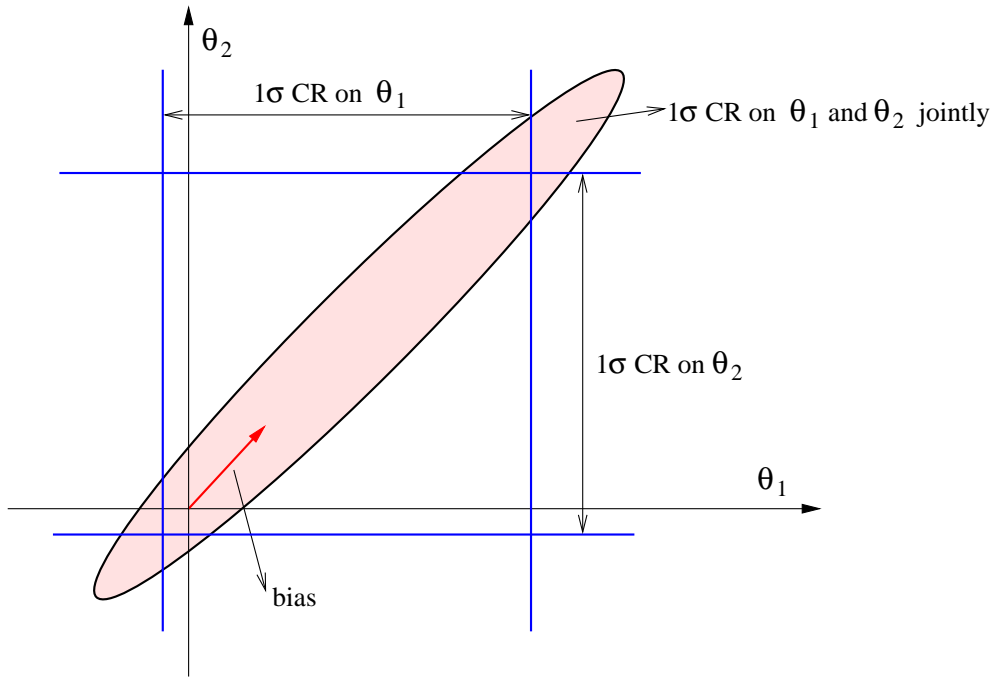}
\caption[Confidence intervals in 1 and 2 dimensions.]{Confidence intervals in 1 and 2 dimesions. The same fraction of measured points (here $68.27\%$) lies (i) between the two vertical lines, (ii) between the two horizontal lines, (iii) within the ellipse.}
\label{confidence0}
\end{figure}

The numbers $68.27\%$, $95.45\%$ and $99.73\%$ and the use of ellipsoids are connected with a Gaussian distribution, however not allways relevant. In general, the probability distribution of the parameters will not be Gaussian, and the above numbers, used as levels of confidence, are purely matters of convention. In these cases, the simple quotation of error ranges like $\bm\theta_0\pm\Delta\bm\theta$ do not provide information about the shape of the p.d.f.

Fig.\ref{confidence0} sketches a possible probability distribution for the case $M=2$. Shown are three different confidence regions, all at the same confidence level. The two vertical lines enclose a band (horizontal interval) which represents the $1\sigma$CR for the variable $\theta_1$ without regard to the value of $\theta_2$. Similarly the horizontal lines enclose a $1\sigma$CR for $\theta_2$. The ellipse shows a joint $1\sigma$CR for pairs of values $\theta_1$ and $\theta_2$.

\section{Constant $\chi^2$ boundaries as CL}
 
When the method used to estimate the parameters is $\chi^2$-minimisation, as in Section \ref{leastsquaresmin}, then there is a natural choice for the shape of confidence intervals, whose use is almost universal. For the observed data, the value of $\chi^2$ at ${\bm\theta}_0$ is a minimum. Call this minimum $\chi_{\rm min}^2$. If the vector $\bm\theta$ of parameter values is perturbed away from ${\bm\theta}_0$, then $\chi^2$ increases. The region within which $\chi^2$ increases by no more than a chosen amount $\Delta\chi^2$ defines some $M$-dimensional confidence region around ${\bm\theta}_0$. If $\Delta\chi^2$ is set to be a large number, this will be a big region; if it is small, it will be small. Somewhere in between there will be choices of $\Delta\chi^2$ that cause the region to contain, variously, $1\sigma$, $2\sigma$, etc.\ of probability distributions for $\bm\theta$, as defined above. These regions are taken as the confidence regions for the parameters ${\bm\theta}_0$. Values of $\Delta\chi^2$ for different confidence regions and different numbers of parameters are listed in Table \ref{deltachival}.

Very frequently one is interested not in the full $M$-dimensional confidence region, but in individual confidence regions for some smaller number $\nu$ of parameters. For example, one might be intersted in the confidence interval of each parameter taken separately (the bands in Fig.\ref{confidence0}), in which case $\nu=1$. In that case, the natural confidence regions in the $\nu$-dimensional subspace of the $M$-dimensional parameter space are the {\em projections} of the $M$-dimensional regions defined by fixed $\Delta\chi^2$ into the $\nu$-dimensional spaces of interest.

\begin{table}[h]
\centering
\begin{tabular}[h!]{||c|c|c|c||}\hline\hline
 CR(\%)&$M=1$&$M=2$&$M=3$\\ \hline\hline
68.27&1.00&2.30&3.53\\ \hline
90.00&2.71&4.61&6.25\\ \hline
95.45&4.00&6.18&8.03\\ \hline
99.00&6.63&9.21&11.34\\ \hline
99.73&9.00&11.83&14.16\\ \hline\hline
\end{tabular}
 \caption{$\Delta\chi^2$ corresponding to a confidence region CR, for joint estimation of $M$ parameters based on Gaussian (normal) p.d.f.}
\label{deltachival}
\end{table}

\section{The $\chi^2$ test}

{\em \hfill 'Don't ask what it means, but ruther how it is used.'}

{\sf \hfill L. Wittgenstein}

\vspace{1cm}

Often one wants to quantify the level of agreement between the data and a hypothesis without explicit reference to alternative hypotheses. This can, in many cases, be done by the $\chi^2$ test invented in 1900 by Karl Pearson \cite{pearson}.

The goodness-of-fit is quantified by giving the {\em observed significance level} also called $p$-value. The $p$-value is defined as the probability to find $\chi^2$ in the region of equal or smaller compatibility with the considered hypothesis than the level of compatibility observed with the actual data. Since $\chi^2$ is defined such that large values correspond to poor agreement with the hypothesis, the $p$-value will be
\begin{equation}
 p=\int_{\chi^2}^\infty f(z;,n)dz
\end{equation}
where $\chi^2$ is the value obtained in the minimisation process (e.g., in the present work the value of $\chi_{L,{\rm min}}^2$, Eq.(\ref{chiL2}), was used to estimate the agreement between experimental data and the theoretical model, i.e., QCD). $f(z;n)$ is the $\chi^2$ p.d.f.\ with $n$ degrees of freedom\footnote{\sf The term {\sl degrees of freedom} is a measure of the number of independent pieces of information on which the precision of a parameter estimate is based. The degrees of freedom for an estimate equals the number of observations (values) minus the number of additional parameters estimated for that calculation. As we have to estimate more parameters, the degrees of freedom available decreases. It can also be thought of as the number of observations (values) which are freely available to vary given the additional parameters estimated.},
\begin{equation}
 f(z;n)=\frac{1}{\Gamma(n/2)2^{n/2}}z^{n/2-1}e^{-z/2},
\end{equation}
where $z\ge0$ and $f(z;n)=0$ for $z\le0$. Here $\Gamma$ denotes the Gamma function.

\begin{figure}[H]
\centering
 \includegraphics[height=.7\textwidth,angle=0]{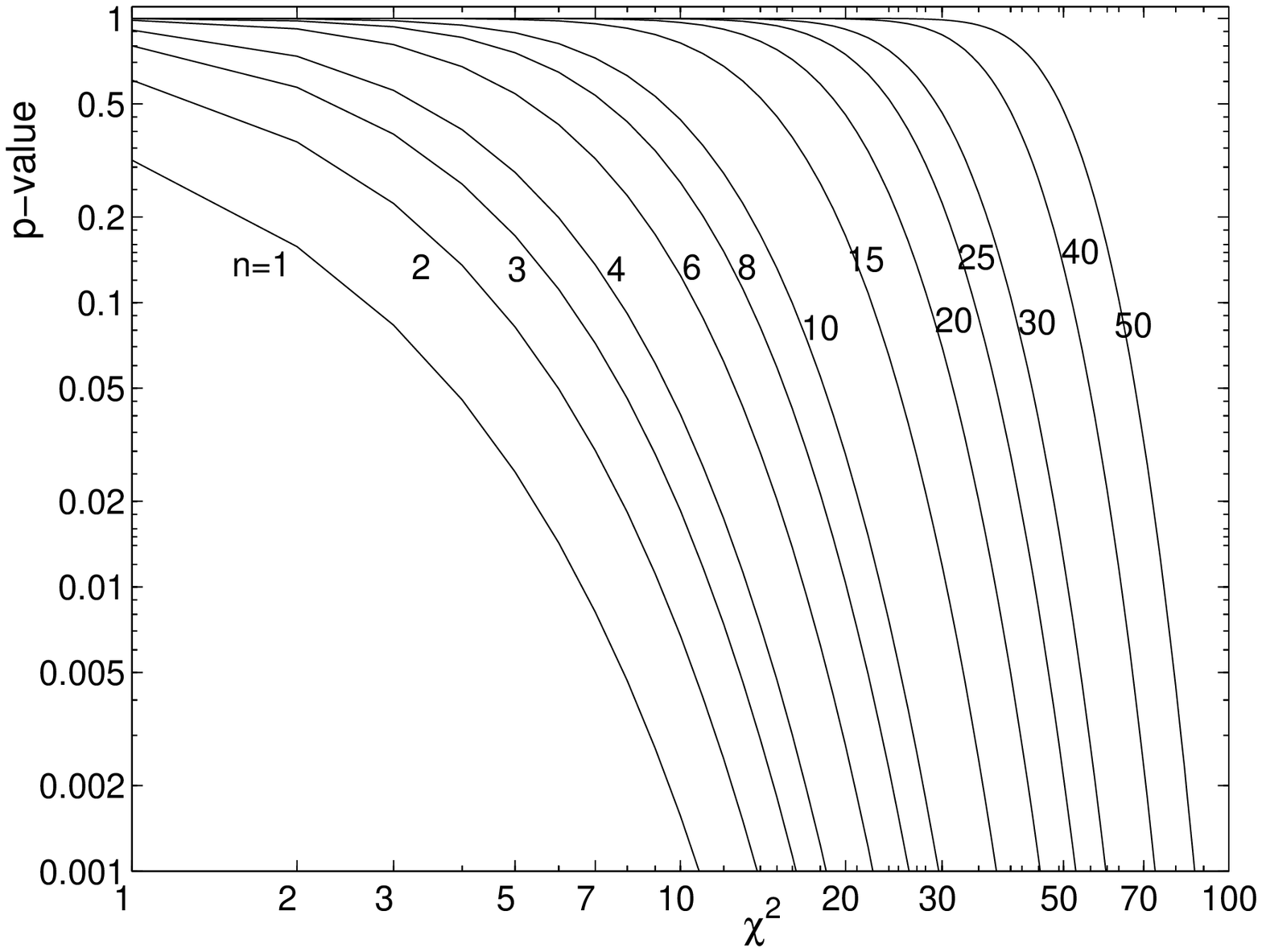}
\caption[The $p$-value for the $\chi^2$ goodness-of-fit test]{The $p$-value as a function of $\chi^2$ for $n$ degrees of freedom.}
\label{pvaluechi}
\end{figure}

 Values of $p$ can be obtained from Fig.\ref{pvaluechi}. If the conditions for using the $\chi^2$ p.d.f.\ do not hold, the $p$-value can still be defined as before, but the p.d.f., $f$, must be determined by other means, e.g., using a Monte Carlo calculation.

If one finds a $\chi^2$ value much greater than $n$ and a correspondingly small $p$-value, one may be tempted to expect a high degree of uncertainty for any fitted parameter. Although this may be true for systematic errors in the parameters, it is not in general the case for statistical uncertainties. If, for example, the error bars (or covariance matrix) used in constructing the $\chi^2$ are underestimated, then this will lead to underestimated statistical errors for the fitted parameters. But in such a case an estimate ${\bm\theta}_0$ can differ from the true value $\bm\theta$ by an amount much greater than its estimated  statistical error. The standard deviations of estimators that one finds reflect how widely the estimates would be distributed as if one were to repeat the measurement many times, assuming that the measurement errors used in $\chi^2$ are also correct. They do not include the systematic error which may result from an incorrect hypothesis or incorrectly estimated measurements errors in the $\chi^2$.

\begin{figure}[H]
\centering
\includegraphics[height=.7\textwidth,angle=0]{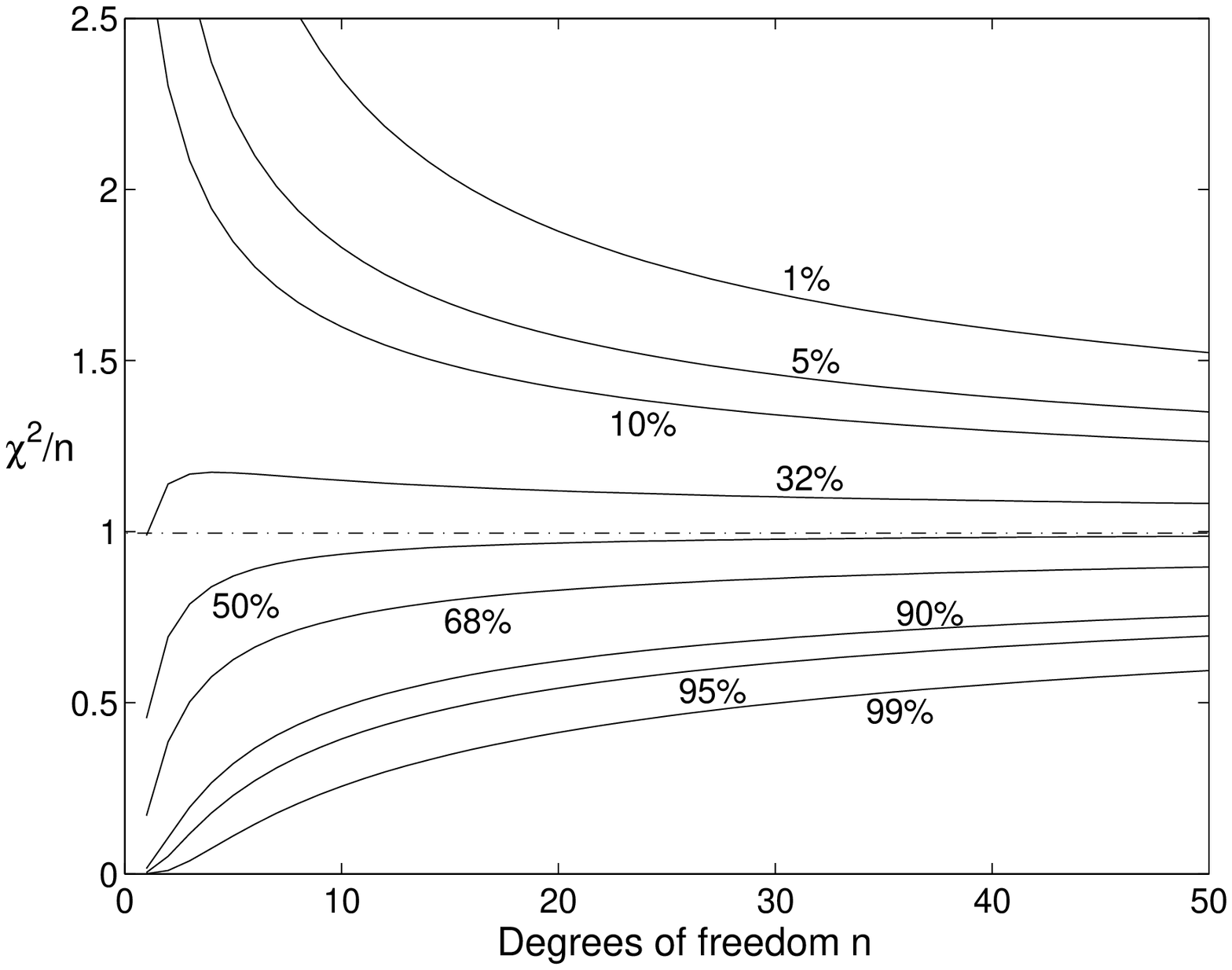}
\caption[The ``reduced'' $\chi^2$.]{The ``reduced'' $\chi^2$, equal to $\chi^2/n$, for $n$ degrees of freedom. The curves show $\chi^2/n$ that corresponds to a given $p$-value as a function of $n$.}
\label{dofpvaluechi}
\end{figure}

Since the mean of the $\chi^2$ distribution is equal to $n$, one expects in a ``reasonable'' experiment to obtain $\chi^2\approx n$. Hence the quantity $\chi^2/n$ is sometimes reported. Since the probability distribution function of $\chi^2/n$ depends on $n$, however, one must report $n$ as well in order to make a meaningful statement. The $p$-values obtained for different values of $\chi^2/n$ are shown in Fig.\ref{dofpvaluechi}.

\clearpage

 \clearpage{\thispagestyle{empty}\cleardoublepage}

\pagestyle{plain}

\clearpage{\thispagestyle{empty}\cleardoublepage}

\end{document}